\documentclass[aps,a4paper,showkeys,nofootinbib,longbibliography,twocolumn,notitlepage]{revtex4-2} 
\usepackage[utf8]{inputenc}
\usepackage{filecontents}
\usepackage{natbib} 
\usepackage{amsmath,amssymb,bm,amsthm}
\usepackage{subfigure}
\usepackage{graphicx} 
\usepackage{dcolumn} 
\usepackage{bm} 
\usepackage[mathlines]{lineno} 
\usepackage{booktabs}
\usepackage{color}
\newcounter{one}
\usepackage{url}

\usepackage{textcase}

\usepackage{mathrsfs}
\usepackage{colortbl}
\usepackage{tabularx}
\usepackage{verbatim}
\usepackage{multirow}

\usepackage[T1]{fontenc} 
\usepackage{lmodern}
\usepackage{bbm}
\usepackage[utf8]{inputenc}
\usepackage{amsfonts}
\usepackage{array}

\usepackage{bbm}
\usepackage{enumitem}
\usepackage{umoline} 
\usepackage[usenames,svgnames]{xcolor}
\usepackage{natbib}
\usepackage[hyperindex,breaklinks]{hyperref} 
\hypersetup{
     colorlinks=true,       		 
     linkcolor=Navy,          	 
     citecolor=Navy,             
     filecolor=Navy,      		 
     urlcolor=Navy,           	 
    runcolor=cyan,
 } 
\usepackage{graphicx}
\usepackage{amsfonts}
\usepackage{amssymb}
\usepackage{amsmath}
\usepackage{bbm}
\usepackage{enumitem}

\newcommand{\ave}[1]{\left \langle #1 \right \rangle}
\newcommand{\bra}[1]{\langle #1 |}
\newcommand{\ket}[1]{| #1 \rangle}

\newcommand{\tr}[0]{ {\rm tr}}
\newcommand{\floor}[1]{\left \lfloor #1  \right \rfloor}
\newcommand{\ceil}[1]{\left \lceil #1  \right \rceil}
\newcommand{\half}[1]{{ \rm h}}
\newcommand{\Oorderof}{\mathcal{O}}
\newcommand{\orderof}[1]{\Oorderof(#1)} 

\newcommand{\for}[0]{\quad \textrm{for} \quad}

\newcommand{\dist}{d}

\newcommand{\co}{{\rm c}}

\newcommand{\diam}{{\rm diam}}

\newcommand{\tc}{{\rm t}}
\newcommand{\Gs}{\Omega}

\newcommand{\poly}{{\rm poly}}

\newcommand{\ad}{{\rm ad}}

\newcommand{\AND}{\quad {\rm and} \quad}

\def\beq{\begin{equation}}
\def\eeq{\end{equation}}
\def\nbeq{\begin{equation*}}
\def\neeq{\end{equation*}} 
\def\<{\langle}
\def\>{\rangle}

\def\tr{{\rm tr}}

\newcommand{\mF}{{\mathcal{F}}}
\newcommand{\mE}{{\mathcal{E}}}
\newcommand{\mA}{{\mathcal{A}}}
\newcommand{\mB}{{\mathcal{B}}}

\newcommand{\mG}{{\mathcal{G}}}
\newcommand{\msC}{\mathscr{C}}

\newcommand{\mD}{{\mathcal{D}}}

\newcommand{\mQ}{{\mathcal{Q}}}
\newcommand{\mJ}{{\mathcal{J}}}
\newcommand{\var}{{\rm Var}}

\newcommand{\bo}[0]{{\rm B}}

\newcommand{\ma}[0]{\mathfrak{a}}
\newcommand{\mb}[0]{\mathfrak{b}}
\newcommand{\mc}[0]{\mathfrak{c}}

\newtheorem{theorem}{Theorem}
\newtheorem{subtheorem}{Subtheorem}
\newtheorem{lemma}{Lemma}
\newtheorem{corol}[lemma]{Corollary}
\newtheorem{assump}[lemma]{Assumption} 
\newtheorem{definition}{Definition}   
\newtheorem{prop}[lemma]{Proposition}

\newcommand{\bal}[2]{#1[#2]}

\newcommand{\nb}{\hat{n}}

\newcommand{\br}[1]{\left( #1 \right)}
\newcommand{\abs}[1]{\left | #1 \right|}
\newcommand{\brr}[1]{\left[ #1 \right]}
\newcommand{\brrr}[1]{\left\{ #1 \right\}}
 \newcommand{\norm}[1]{\left \|  #1 \right \|}

\makeatletter

\newcommand{\supponecolumnfootnotes}{
	\onecolumn@grid@setup
	\let\set@footnotewidth\set@footnotewidth@one
	\let\compose@footnotes\compose@footnotes@one
}

\newcommand{\startsupplementarycontents}{
	\let\mainaddcontentsline\addcontentsline
	\renewcommand{\addcontentsline}[3]{
		\def\tempa{##1}
		\def\tempb{toc}
		\ifx\tempa\tempb
		\mainaddcontentsline{stoc}{##2}{##3}
		\else
		\mainaddcontentsline{##1}{##2}{##3}
		\fi
	}
}

\newcommand{\supplementarytableofcontents}{
	\begingroup
	\section*{\tocname}
	\@starttoc{stoc}
	\endgroup
}

\makeatother

\usepackage{mathtools}
\def\multiset#1#2{\ensuremath{\left(\kern-.3em\left(\genfrac{}{}{0pt}{}{#1}{#2}\right)\kern-.3em\right)}}

\renewcommand\thefootnote{*\arabic{footnote}} 
\begin{document}

\title{Entanglement area law in interacting bosons from the Bose-Hubbard model to $\phi$4 theory and beyond}
 
\author{Donghoon Kim$^{1}$}
\email{donghoon.kim@riken.jp}

\author{Tomotaka Kuwahara$^{1,2,3}$}
\email{tomotaka.kuwahara@riken.jp} 
\affiliation{$^{1}$ 
Analytical quantum complexity RIKEN Hakubi Research Team, RIKEN Center for Quantum Computing (RQC), Wako, Saitama 351-0198, Japan
}

\affiliation{$^{2}$ 
RIKEN Cluster for Pioneering Research (CPR), Wako, Saitama 351-0198, Japan
}
\affiliation{$^{3}$
PRESTO, Japan Science and Technology (JST), Kawaguchi, Saitama 332-0012, Japan}

\begin{abstract}
	The entanglement area law is a universal principle that characterizes quantum many-body phases and underpins tensor network algorithms. Traditionally, its validity has been limited to systems with short-range interactions and bounded local energy. Achieving a complete generalization that removes both of these constraints has been a longstanding goal in quantum many-body theory, especially for interacting boson systems where unbounded energy presents intrinsic difficulties. In this work, we rigorously prove the area law for one-dimensional interacting boson systems with long-range interactions, covering broad models including the Bose-Hubbard and $\phi4$ classes. Furthermore, we establish an efficiency guarantee for Matrix-Product-State approximations of the ground states, offering a practical route to numerical simulation. One of our main technical contributions is a general method for Hilbert space dimension reduction, whose applicability extends to arbitrary spatial dimensions. These results address two major challenges simultaneously and provide important foundations for simulating long-range cold atomic systems.

\end{abstract} 

\maketitle

\begin{figure*}[tt]
	\centering
	{\includegraphics[clip, scale=0.324]{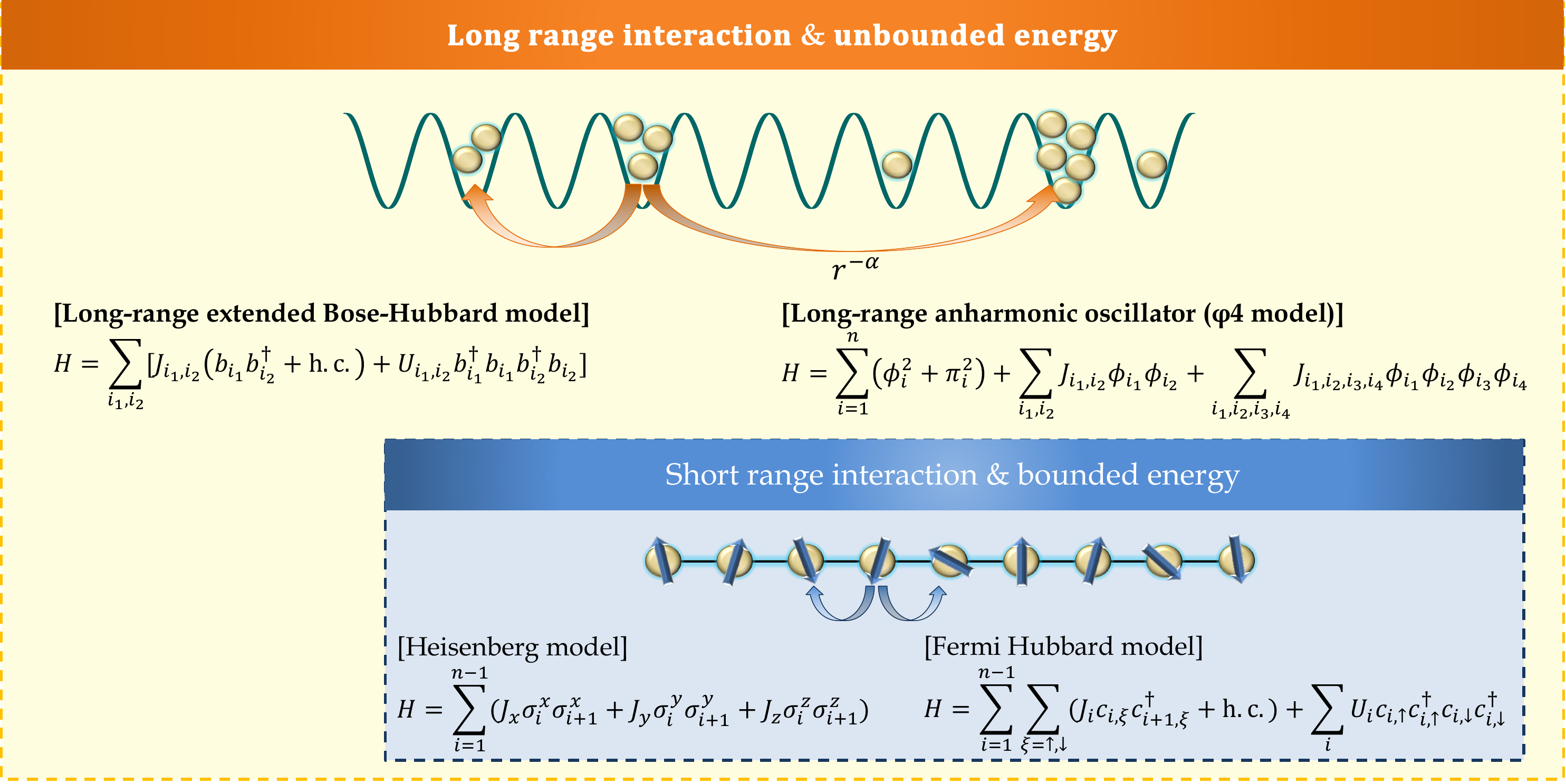}}
	\caption{Entanglement area law for one-dimensional quantum systems: limitations and generalization.
		(Inner box): Short-range interactions and bounded local energies.
		Previous rigorous results on the entanglement area law have been limited to systems with short-range interactions and bounded local Hilbert space dimensions or local energies. Paradigmatic examples include spin chains and fermionic models. In these settings, the area law ensures that the entanglement entropy of ground states scales with the boundary, not the volume, of a subsystem. 
		(Outer box): Long-range interactions and unbounded local energies.
		Lifting these foundational constraints is crucial for understanding a broad class of physically relevant systems, such as bosonic quantum fields in cold atom experiments and lattice quantum field theories, where both long-range interactions and unbounded local Hilbert spaces naturally arise. In this work, we rigorously establish the entanglement area law for one-dimensional bosonic systems with power-law decaying interactions and unbounded local energies (Result~2 below), providing a rigorous step toward addressing this longstanding challenge. Our results also provide a theoretical foundation for the efficient and accurate representation of ground states using Matrix Product States (MPS), even in these general settings (Result~3 below).}
	\label{fig:Fig1}
\end{figure*}

\section*{Introduction}

In modern physics, one of the biggest challenges is figuring out how to accurately simulate systems with many interacting particles. This problem shows up in various fields, like condensed matter physics, high-energy physics, and statistical mechanics. To tackle this, researchers study a field called Hamiltonian complexity~\cite{doi:10.1137/S0097539704445226,0034-4885-75-2-022001,gharibian2015quantum}, which tries to reveal general rules that explain how particles in these systems possess complex structures from the information-theoretic viewpoint. A main focus in this field is the ground state, which is the system's lowest-energy state at absolute zero temperature, where quantum effects are most pronounced. One of the key discoveries about the ground state is the entanglement area law~\cite{PhysRevLett.90.227902,RevModPhys.82.277}. It states that when a system is divided into two parts,  the entanglement entropy scales with the surface area of the partition rather than the volume. The area law is deeply related to the structural complexity of ground states~\cite{PhysRevB.76.035114} and is a fundamental ansatz in tensor network algorithms~\cite{RevModPhys.77.259,ORUS2014117}, which are some of the most widely used tools for running simulations. Proving the area law and understanding the complexity of ground states have become landmark achievements in quantum information theory.

Over the past two decades, there has been significant progress in our understanding of one-dimensional (1D) ground states. The first rigorous proof of the entanglement area law in 1D systems was provided by Hastings in 2007, marking a major milestone~\cite{Hastings_2007}. Since then, qualitative improvements have been made, particularly by Arad and colleagues, who refined the approach~\cite{10.1145/1536414.1536472,PhysRevB.85.195145,arad2013area}. Additionally, Brand\~ao and collaborators made notable attempts to prove the area law based solely on the exponential decay of correlations in these systems~\cite{brandao2013area,Brandao2015}. Currently, the proof of the area law for gapped ground states is predominantly achieved through the formalism known as Approximate Ground State Projection (AGSP)~\cite{PhysRevB.85.195145,arad2013area}. This formalism became a crucial stepping stone, leading to the development of a polynomial-time algorithm for computing ground states~\cite{landau2015polynomial,Arad2017}. As of now, 1D systems represent the most successful application of these advancements.

Previous proofs of the area law have primarily relied on two basic conditions~\cite{Hastings_2007,PhysRevB.85.195145,arad2013area,10.1145/3357713.3384292,10.1145/3519935.3519962}: i) the short-range interactions, and ii) bounded local energy, as summarized in Fig.~\ref{fig:Fig1}. 
While these conditions are universally valid in common spin and fermion systems, it is well-known that physical systems violating these conditions are also ubiquitous. As the violation of condition i), long-range interactions are characterized by power-law decay of the interaction strength, which, for instance, have been experimentally observed in cold atomic systems~\cite{RevModPhys.82.2313,richerme2014non,Landig2016,Baier2016,bernien2017probing,zhang2017observation,Lagoin2022,Su2023}. Understanding how these interactions increase complexity compared to short-range interactions has been a fundamental challenge in Hamiltonian complexity~\cite{chen2019finite,PhysRevX.10.031010,Kuwahara2020arealaw,PhysRevX.11.031016,PhysRevLett.129.150604,kim2025thermal,achutha2025provably}. 
On the other hand, as the violation of condition ii), systems with unbounded local energy are well-represented by interacting boson systems. These systems remain largely unexplored, though recent advancements have been made in understanding them from the perspective of information propagation~\cite{PhysRevLett.127.070403,PhysRevX.12.021039,Tong2022provablyaccurate,kuwahara2022optimal}. The most representative example of interacting boson systems is the Bose-Hubbard model, which serves as the minimal model for describing cold atomic systems~\cite{PhysRevLett.81.3108}. Another notable example is the $\phi4$ model, a fundamental model in lattice quantum field theory~\cite{doi:10.1126/science.1217069,preskill2018,PRXQuantum.4.027001}. This model is equivalent to anharmonic oscillators, including nonlinear terms~\cite{PhysRev.184.1231,doi:10.1098/rspa.1978.0086,PhysRevLett.77.4114}, and is one of the most well-studied models in high-energy theory and statistical mechanics.

Proving the area law in systems that break the aforementioned limitations has long been a major goal, and various studies have been conducted in this direction. For example, in one-dimensional spin systems, generalization to long-range interacting systems has been achieved~\cite{Kuwahara2020arealaw,PhysRevLett.119.050501}. However, many unresolved aspects remain when it comes to boson systems. This is partly due to the fact that there exist counter-examples that break the area law conjecture, such as the Bose-Hubbard model with attractive interactions. 
For example, by considering a Bose-Hubbard model such that 
$H= J( b_1^\dagger b_{2} + {\rm h.c.}) -  V(\nb_1+ \nb_{2} )^2 + \sum_{i\neq 1,2} U \nb_i (\nb_i-1)$ ($V,U>0$), all the bosons concentrate on the sites $1$ and $2$. Hence, as long as the total boson number is proportional to the system size $n$, the entanglement entropy between the site $1$ and the other sites can be as large as $\log(n)/2$.
Previous research on the boson area law has been largely limited to non-interacting boson systems~\cite{PhysRevA.73.012309,PhysRevA.74.052326}. Recent breakthroughs have demonstrated that interactions between bosons and other bounded fields (but excluding boson-boson interactions) can be efficiently handled~\cite{Tong2022provablyaccurate}, from which the area law can be proven under finite-range interactions~\cite{PhysRevA.108.042422}. 
Still, a unified understanding of generic interacting bosonic systems, encompassing key models such as the Bose-Hubbard and $\phi^4$ classes, remains a completely open problem.

In this study, we resolve the one-dimensional area law conjecture for general interacting boson systems, including the Bose-Hubbard and $\phi4$ classes. Our results hold even in the presence of long-range interactions, where the interaction decay rate must be faster than $r^{-2}$ for the area law to hold. This allows us to establish the area law in the most general cases without the two primary conditions---short-range interactions and bounded local energy. Our results quantitatively show how the boson parameters and long-range interaction parameters, such as the power-law decay rate, influence the entanglement structures [see Ineq.~\eqref{area_law_boson} below]. In addition, we provide a general upper bound on the bond dimension required to describe these ground states using Matrix Product States (MPS). 
Our result serves as a theoretical foundation for simulating long-range interacting cold atomic systems, 
which constitute a major platform for modern quantum technologies~\cite{doi:10.1126/science.1229957,Bluvstein2024}.

A central challenge in analyzing interacting bosonic systems is the infinite dimensionality of the local Hilbert space. Indeed, as highlighted in several prior works~\cite{doi:10.1126/science.1217069,PhysRevLett.127.070403,Tong2022provablyaccurate,kuwahara2022optimal}, providing a rigorous and general framework for dimension reduction---with explicit control over the induced error---has long been recognized as one of the most fundamental and difficult tasks in the field. To rigorously establish the area law, it is essential to quantitatively control the ground-state boson number distribution and to implement dimension reduction in a controlled way. Although truncating the local Hilbert space is a standard prescription in both numerical and theoretical studies, it has traditionally been applied without rigorous guarantees on its validity or accuracy.
One of our main technical contributions is to develop a general framework for dimension reduction that can accommodate different types of boson statistics and interaction structures. In particular, we establish analytic tools tailored to both the Bose-Hubbard and the $\phi^4$ models (Results~1.1 and 1.2 below), each of which requires distinct theoretical approaches due to their differing physical and mathematical structures.

Crucially, when combining long-range interactions with bosonic degrees of freedom, an additional layer of difficulty arises. In such systems, a straightforward dimension reduction fails because preserving the ground state demands an excessively large cutoff. As a result, energy and entanglement entropy grow logarithmically with the cutoff, making the area law inapplicable.
To address this, our framework implements a site-dependent and carefully controlled dimension reduction scheme. We further reconstruct and generalize the effective Hamiltonian theory developed in Ref.~\cite{Arad_2016}, adapting it to account for the non-uniformity and correlations induced by long-range interactions. This allows us to rigorously maintain the validity of the area law even for bosonic systems with power-law decaying interactions, overcoming obstacles that are absent in either short-range or spin systems.

Our work opens the unified theoretical framework capable of addressing the combined challenges of long-range interactions and unbounded local energy in quantum many-body systems. By integrating these two sources of complexity, our approach enables rigorous analysis and simulation techniques that were previously out of reach.

\section*{Results}

\subsection*{System setup} 

We here describe the system setups for the main results that will be shown in the subsequent sections. 
We consider a quantum system comprising $n$ sites in an arbitrary-dimensional lattice and denote the set of total sites by $\Lambda$, where $|\Lambda| = n$.
We denote the boson creation and annihilation operators at a site $i\in \Lambda$ by $b^\dagger_i$ and $b_i$, respectively.  
Then, the most general form of the interacting boson Hamiltonians up to $k$-th degree is given by:
\begin{equation}
	\label{Ham_def_boson}
	H= \mF_H (\mathbf{b},\mathbf{b}^\dagger)= \sum_{Z: |Z|\le k} h_Z(\mathbf{b}_Z,\mathbf{b}_Z^\dagger),
\end{equation} 
where $\mF_H (\mathbf{b},\mathbf{b}^\dagger) $ is an arbitrary $k$-th degree polynomial of $\mathbf{b}=\{b_i\}_{i\in \Lambda}$ and $\mathbf{b}^\dagger=\{b^\dagger_i\}_{i\in \Lambda}$, and $h_Z(\mathbf{b}_Z,\mathbf{b}_Z^\dagger)$ with $\mathbf{b}_Z=\{b_i\}_{i\in Z}$ acts on the subset $Z \subset \Lambda$.

We define $\Pi_{i, < N}$ as the projection operator onto the space such that the boson number at the site $i$ is smaller than $N$. 
As a generalization, we denote $\Pi_{\Lambda, < N}$ by $\Pi_{\Lambda, < N}= \bigotimes_{i\in \Lambda} \Pi_{i, < N}$, which truncates the boson number at each site to below $N$. 
We adopt the parameter $g$ such that 
\begin{equation}
	\sum_{Z: Z\ni i} \norm{ h_Z(\mathbf{b}_Z,\mathbf{b}_Z^\dagger)\Pi_{\Lambda, < N}} \le g N^{k/2}  
\end{equation} 
with $g=\orderof{1}$; that is, as long as the boson number is truncated by $\orderof{1}$, their interaction is also upper-bounded by an $\orderof{1}$ constant.
We denote the ground state and the spectral gap by $\ket{\Omega}$ and $\Delta$, where we assume the non-degeneracy of the ground-state energy.

Our purpose is to derive a general upper bound for the entanglement entropy $S_L(\Omega)$ for any bipartition of the total system as $\Lambda=L\sqcup R$.  
The entanglement entropy is described by $-\tr \br{\rho_L \log(\rho_L)}$ with $\rho_L$ the reduced density matrix of the ground state on the subset $L$.

\subsection*{Bose-Hubbard classes}

The Bose-Hubbard class is represented as follows in an integrated manner: 
\begin{equation}
	\label{Boson_most_general_intro}
	H= H_p(\mathbf{b},\mathbf{b}^\dagger)+\sum_{i\in \Lambda} U_i \nb_i^{k/2}\quad (U_i>0,\  \forall i\in \Lambda),
\end{equation}
where $\nb_i=b_i^\dagger b_i$, and $H_p(\mathbf{b},\mathbf{b}^\dagger)$ is an arbitrary $p$th degree polynomial of $\mathbf{b},\mathbf{b}^\dagger$ with $p\le k-1$. 
Now, $k$ denotes the highest degree of the boson number (potential) terms, while $p$ indicates the highest degree among off-diagonal terms. It is important to emphasize that our class of Hamiltonians generally allows for particle-number non-conserving terms, including those responsible for pair creation and annihilation processes (e.g., $\sum_{i_1,i_2,i_3,i_4}J_{i_1,i_2,i_3,i_4} ( b_{i_1}b_{i_2}b_{i_3}b_{i_4} + {\rm h.c.})$ with $p=4$). This goes beyond the standard Bose-Hubbard model and encompasses a much wider variety of physically relevant interactions.
For the Bose--Hubbard class, representative gapped phases include the Mott insulating phase at integer filling and the density-wave insulating phase stabilized by repulsive density--density interactions between neighboring sites.

We note that the standard Bose-Hubbard model, i.e.,
\begin{equation}
	H= \sum_{i,i'} J_{i,i'} (b_ib_{i'}^\dagger + {\rm h.c.}) + \sum_{i\in \Lambda} U\nb_i(\nb_i -1)  
\end{equation}
with $U>0$, 
corresponds to the case of $k=4$ with $p=2$, and the above condition is satisfied. 
If we consider the case of  $p=k$ in Eq.~\eqref{Boson_most_general_intro}, there exists a competition between the repulsive and attractive interactions of bosons, and hence, the condition becomes more nontrivial (see Assumption~1 in Supplementary Note 2).
For example, in the Hamiltonian $J(b_i^2 b_j^2 + \mathrm{h.c.}) + U(n_i^2 + n_j^2)$, whether the ground state is well-defined depends on the relative strength of $J$ and $U$: if $J$ becomes large, the attractive effect can dominate and potentially lead to a divergent boson number in the ground state.

We emphasize that our analysis is restricted to the repulsive regime, where on-site interactions suppress clustering tendencies and keep the boson number suitably controlled. In contrast, attractive interactions can favor the accumulation of particles on a single site, leading to collapse- or clustering-type behavior. Such a gathering qualitatively changes the boson number distribution, preventing the effective Hilbert space dimension from remaining small, and may invalidate the concentration bounds underlying our proof. Consequently, the attractive regime admits ground states with a qualitatively different entanglement structure from the repulsive case, and there exist ground states that violate the entanglement area law.

Notably, our model includes interaction-induced tunneling terms of the form $b_i(\nb_i+\nb_{i'})b_{i'}^\dagger$, where the tunneling amplitude depends explicitly on the occupation numbers. Such interactions are known to violate the Lieb-Robinson bound (Theorem~3 in Ref.~\cite{Vu2024optimallightcone}), meaning that information propagation can, in principle, occur at infinite speed. 
Surprisingly, however, our results show that even in the presence of these interactions---where dynamical locality is lost---the ground state can still satisfy the entanglement area law. This highlights a fundamental distinction between the locality of information propagation (dynamical locality) and the static locality properties of ground-state entanglement.
In fact, the Lieb-Robinson bound evaluated with respect to a specific state, such as a gapped ground state, can exhibit much slower effective propagation than the state-independent bound, and its violation is determined by a different criterion. This state-dependent perspective on dynamical locality helps explain why a distinction emerges between the loss of general dynamical locality and the possibility that the ground-state entanglement can still remain at an area-law scale.

\subsection*{$\phi4$ classes}
For the $\phi4$ classes, we introduce the $\phi$ operator and $\pi$ operator as
\begin{equation}
	\phi =  \frac{b + b^\dagger}{\sqrt{2}} \AND \pi=-i \frac{b - b^\dagger}{\sqrt{2}} .
\end{equation}
They correspond to the position and momentum operators, respectively. 
Then, we treat the following general Hamiltonian as the $\phi4$ class: 
\begin{equation}
	H = \sum_{i\in \Lambda} \mu_i \pi_i^2 + \mF(\boldsymbol{\phi}),
	\label{phi_4_type_model_intro}
\end{equation}
where $\{\mu_i\}_{i\in\Lambda}$ are real parameters. Throughout this work, we restrict attention to parameter regimes in which the Hamiltonian is bounded from below and admits a unique gapped ground state.
Also, $\mF(\boldsymbol{\phi})$ is an arbitrary $k$-degree even function of $\boldsymbol{\phi}=\{\phi_i\}_{i\in \Lambda}$, e.g., $\sum_{i_1,i_2,i_3,i_4}f_{i_1,i_2,i_3,i_4}\phi_{i_1}\phi_{i_2}\phi_{i_3}\phi_{i_4}$.  
Note that the Hamiltonian satisfies the parity symmetry, i.e., invariant under $\boldsymbol{\phi} \to -\boldsymbol{\phi}$. 
Clearly, the standard $\phi4$ model, i.e., 
\begin{equation}
	\label{phi4_mode_intro}
	H =\sum_{i\in \Lambda}\br{\pi_i^2 + \phi_i^2 + \lambda \phi_i^4} +\gamma \sum_{\ave{i,i'} }   \phi_i \phi_{i'} ,
\end{equation}
is a specific case of the above Hamiltonian~\eqref{phi_4_type_model_intro}. 
Here, the highest-order term of the boson operator is $\sum_{i\in \Lambda}\lambda \phi_i^4$. 
We note that the parity symmetry is a physically essential assumption underlying our analysis.
By forbidding odd-power terms, it prevents bosonic degrees of freedom from concentrating around large-field configurations.
This allows uniform control of boson-number expectation values, which is essential for establishing the concentration bounds used in our proof. 
Indeed, if this assumption is violated, explicit counterexamples that break the entanglement area law can be constructed (see the discussion following Assumption~2 in the Supplementary Information).
This assumption is compatible with standard realizations of $\phi4$-type systems, including superconducting Kerr oscillators, symmetric double-well optical-lattice potentials, and the conventional lattice $\phi4$ field theory, where the effective potential remains even unless an external bias or magnetic field induces a linear term that breaks the symmetry.
In the $\phi 4$ classes with the parity symmetry, the gapped phase exhibits the physical character of a massive disordered (paramagnetic) state: the field fluctuates around the symmetric configuration and exhibits no long-range order, so correlations remain short-ranged.

As an analytical challenge, we can see that the term $\phi_i^4$ allows an infinite number of bosons to sit on a single site at the lowest energy state. 
Indeed, even with an infinitesimally small perturbation, the ground state of the $\phi4$ model becomes unstable in the sense that an infinite number of bosons sit on a single site. For example, if we consider a single-site model ($n=1$) with the Hamiltonian $\phi^4 + \phi^2+\pi^2+\epsilon (b^\dagger \nb b^\dagger+b \nb b)$ for an arbitrary $\epsilon>0$, the ground state is not well-defined due to the instability. 
This prohibits any perturbative approaches to analyze the boson number distribution in the $\phi4$ class, and hence we employ completely different analyses from the case of the Bose-Hubbard class. 

As a nontrivial feature of the $\phi^4$ model, only the existence of a spectral gap guarantees low boson density at each site, but only in the presence of parity symmetry. In contrast, without parity symmetry, a spectral gap alone cannot prevent the accumulation of an extensive number of bosons on a single site. This underlines the crucial role of symmetry in ensuring the stability and localization properties of the ground state (see also the discussion below Theorem~2 in Supplementary Note 5).

\bigskip
\subsection*{Bosonic concentration bound} 
As our first main result, we show the concentration bound on the boson number distribution at an arbitrary site, where the tail of the distribution is characterized by (sub)exponential decay. 
In practical numerical simulations of bosonic systems, it is unavoidable to truncate the local boson number at each site, based on the empirical observation that the occupation number distribution decays rapidly. However, such truncations have traditionally lacked rigorous error bounds or theoretical justification. The concentration bounds established in this section provide a systematic and quantitative foundation for dimension reduction in bosonic models: they ensure that the numerical truncation is both justified and accurate.

We stress that all results presented in this section apply to bosonic systems of arbitrary spatial dimension and remain valid even when the interactions are long-range. Thus, our concentration bounds provide rigorous guarantees for a wide class of quantum lattice models far beyond conventional settings.

{\bf Result~1.1.} 
For the Bose-Hubbard class (Eq.~\eqref{Boson_most_general_intro}), we prove the following inequality: 
\begin{equation}
	\label{intro_BH_conc}
	\bra{\Omega}  \Pi_{i,\ge x} \ket{\Omega} \le e^{-2(x-M_{i,0})/k} ,
\end{equation} 
where $M_{i,0}$ is an $\orderof{1}$ constant as shown in Eq.~(S.151) of Corollary~12 in Supplementary Note 4. 
Here, the concentration bound is independent of the spectral gap, and the decay rate is characterized by exponential decay. 
Theorem~1 in Supplementary Note 4 generally treats the case of $p=k$ in Eq.~\eqref{Boson_most_general_intro}, where the repulsive and attractive interactions have the same order.  

{\bf Result~1.2.} 
On the other hand, for the $\phi4$ class (Eq.~\eqref{phi_4_type_model_intro}), we have a (sub)exponential concentration bound as follows (Theorem~2 in Supplementary Note 5): 
\begin{equation}
	\label{intro_phi4_conc}
	\bra{\Omega} \Pi_{i, >x} \ket{\Omega} \le 4e^{k} e^{-kx^{1/k}/(8e \tilde{C})},
\end{equation} 
where $\tilde{C}$ is a constant proportional to $1/\sqrt{\Delta}$ [see Eq.~(S.224) in Supplementary Note 5].
In the $\phi4$ class, the concentration bound depends on the spectral gap $\Delta$.
Here, the variance of the boson number in the ground state scales as $\Delta^{-k/2}$, which diverges in the limit of $\Delta\to 0$.

It is worth noting that the qualitative difference in the decay behavior between the Bose-Hubbard and $\phi^4$ models does not arise from any simple physical intuition. Instead, it reflects the fundamentally different mathematical and physical structures of the two models, which require distinct analytical techniques (see Supplementary Note 5).  In particular, the sub-exponential decay observed in the $\phi^4$ model~(Eq.~\eqref{phi4_mode_intro}) appears to be an intrinsic feature of the model, as revealed by our analysis and further supported by numerical calculations for single-site systems.

\subsection*{1D entanglement area law} 
Our second main result establishes an entanglement area law for 1D systems under the assumption of a general boson number concentration bound as follows:
\begin{equation}
	\norm{\Pi_{i,> N}  \ket{\Omega}} \le \mc e^{-\mb N^{1/\ma}}  \for \forall i\in \Lambda ,
	\label{concentration_bound_assmp}
\end{equation}
where $\mb$ depends on the spectral gap as $\mb \propto \Delta^{2\upsilon/(\ma k)}$ with $\upsilon$ a constant that depends on the system detail (see Supplementary Note 6).
This bound enables us to effectively reduce the local Hilbert space dimension and control the entanglement entropy, even in the presence of unbounded bosonic degrees of freedom and long-range interactions.
It is worth noting that we here consider the most general Hamiltonian in Eq.~\eqref{Ham_def_boson} and do not restrict ourselves to any specific model, such as the Hamiltonian classes in Eqs.~\eqref{Boson_most_general_intro} and \eqref{phi_4_type_model_intro}.

For the Hamiltonian in Eq.~\eqref{Ham_def_boson}, we allow general power-law decaying interactions.
We characterize the decay rate by
\begin{equation}
	\norm{ h_Z(\mathbf{b}_Z,\mathbf{b}_Z^\dagger)\Pi_{\Lambda, < N}} \le J_Z N^{k/2},
\end{equation}
and 
\begin{equation}
	\label{long_range_interaction}
	\sum_{Z:Z\ni \{i,i'\}} |J_Z| \le g \bar{J}(\dist_{i,i'}),
\end{equation}
where $\bar{J}(r) $ polynomially decays with the distance $r$ as $r^{-\alpha}$ ($\alpha>2$).

{\bf Result~2.} 
Under the above conditions, we prove the following upper bound on the entanglement entropy for an arbitrary bipartition $\Lambda=L\cup R$ (Theorem~3 in Supplementary Note 6):   
\begin{equation}
	\label{area_law_boson}
	S_L(\Omega)  \le  C_0 \mG_{\bar{\alpha},\upsilon,\chi}(\Delta)
\end{equation}
with $\chi=k\ma/2$ and $\bar{\alpha}=\alpha-2$, where 
\begin{equation}
	\mG_{\bar{\alpha},\upsilon,\chi}(\Delta)= \Delta^{-(1+2/\bar{\alpha})(\upsilon+1)} \brr{\log\br{1 /\Delta}}^{4+3/\bar{\alpha} +\chi(1+2/\bar{\alpha}) }  ,
\end{equation}
and $C_0$ is an $\orderof{1}$ constant. 
Here, $\bar{\alpha}$ characterizes the decay of the long-range interaction, while $\chi$ and $\upsilon$ capture the influence of the bosonic statistics and the spectral gap, respectively.

The derived area law in Eq.~\eqref{area_law_boson} depends on three parameters: $\bar{\alpha}$, $\chi$, and $\upsilon$. The parameter $\bar{\alpha}$ characterizes the strength of the long-range interaction, while $\chi$ and $\upsilon$ capture the bosonic properties. In the limits $\chi \to 0$ and $\upsilon \to 0$, we recover the result for the long-range area law of spin systems from Ref.~\cite{Kuwahara2020arealaw} [see the inequality~(S.351) in Supplementary Note 6, where $\upsilon=0$ is specifically considered]. Additionally, in the limit $\bar{\alpha} \to \infty$, our area law reduces to the original area law for short-range spin systems from Ref.~\cite{arad2013area}. This shows that our area law is a natural extension of the previous results, as it reduces to them in the appropriate limits.

Moreover, our results provide a rigorous efficiency guarantee for approximating the ground state with a Matrix Product State (MPS), a class of variational ansatzes that underlies many powerful tensor network algorithms for simulating quantum many-body systems. 

{\bf Result~3.} We prove that there exists an MPS $\ket{{\rm M}_{\mD}}$ that approximates the ground state in the sense that 
\begin{equation}
	\norm{\tr_{X^\co} \br{\ket{\Omega}\bra{\Omega}- \ket{{\rm M}_{\mD}}\bra{{\rm M}_{\mD}}}}_1 \le  \delta |X|
\end{equation}
for $\forall X\subseteq \Lambda$ by choosing the bond dimension $\mD$ as 
\begin{equation}
	\label{Intro_MPS_approx_Area_law}
	\mD=e^{C_1  \mG_{\bar{\alpha},\upsilon,\chi}(\Delta)+C_2 \Delta^{-(\upsilon+1)/2} \log^{\chi/2+5/2}\br{\frac{1}{\delta \Delta}}},
\end{equation}
where $C_1,C_2$ are $\orderof{1}$ constants and the error $\delta$ is arbitrarily chosen.  
In particular, for polynomially small error $\delta = 1/\mathrm{poly}(n)$ and constant gap $\Delta = \orderof{1}$, the required bond dimension $\mD$ scales only quasi-polynomially with system size $n$:
\begin{equation}
	\mD=\exp\brr{ C'_2 \log^{\chi/2+5/2}(n)}, \quad C'_2=\orderof{1}.
\end{equation}
If a many-body quantum state defined on $n$ sites admits a matrix product state representation with bond dimension $\mathcal{D}$, tensor-network algorithms evaluate local observables with cost $\mathcal{O}(n \mathcal{D}^{3})$; thus, a quasi-polynomial bound on $\mathcal{D}$ implies classical simulability with complexity close to polynomial time.
This result provides a rigorous guarantee for the efficiency of tensor network algorithms in simulating interacting bosonic systems, even when both long-range interactions and unbounded local Hilbert spaces are present.

By matching the parameters in the general concentration bound (Eq.~\eqref{concentration_bound_assmp}) to the explicit forms obtained for the Bose-Hubbard (Eq.~\eqref{intro_BH_conc}) and $\phi^4$ (Eq.~\eqref{intro_phi4_conc}) classes, we set
$\{\ma,\mb,\mc,\upsilon\} =\{1,1/k,e^{M_{i,0}/k},0\}$ for the Bose-Hubbard class, and
$\{\ma,\mb,\mc,\upsilon\} =\{k,k/(16e \tilde{C}),2e^{k/2} ,k^2/4\}$ for the $\phi^4$ class, where we use $\norm{\Pi_{i,> N}  \ket{\Omega}} = \bra{\Omega} \Pi_{i,> N}  \ket{\Omega}^{1/2}$.
This yields explicit area law statements for both models [see the inequalities~(S.354) and (S.356) in Supplementary Note 6 for the explicit forms].

\section*{Discussions}

In this work, we have established the entanglement area law and MPS efficiency guarantees for general interacting bosonic systems, including those with long-range interactions and unbounded local energies. Our results provide theoretical foundations for the validity of existing numerical methods in studying the Bose-Hubbard models~\cite{PhysRevB.61.12474,PhysRevB.58.R14741} and related quantum field theories~\cite{Sugihara_2004,PhysRevD.88.085030}. 

There are still several open questions. In the Bose-Hubbard class, we have focused on repulsive interactions, leaving the entanglement structure in the attractive case largely unknown. For models with conserved total boson number (e.g., the standard Bose-Hubbard model), there is a trivial logarithmic violation of the entanglement area law. However, it is unclear whether (sub)volume entanglement scaling can occur under a spectral gap assumption.
In the $\phi^4$ class, it remains an open problem to remove the parity symmetry assumption. As shown in Supplementary Information (Theorem~2 therein), a finite bound on the position operator $\{\phi_i\}_{i\in \Lambda}$ enables the derivation of a concentration bound (Eq.~\eqref{intro_phi4_conc}). A key challenge lies in proving this boundedness under only the spectral gap assumption.

Furthermore, the concentration bounds on the boson number probability apply to systems in arbitrary dimensions. 
Although the area law conjecture is still intractable in high dimensions, our results provide an essential foundation for future resolutions of the area law in high-dimensional boson systems. 
At the moment, by imposing additional assumptions---such as the existence of an adiabatic path to non-interacting models---we expect to prove a bosonic area law in higher dimensions by employing the Small-Incremental-Entangling (SIE) theorem~\cite{PhysRevLett.111.170501}. 

Finally, developing a quasi-polynomial-time algorithm for simulating interacting boson systems with long-range interactions is also an important goal. Currently, efficient algorithms for one-dimensional ground states exist only for Hamiltonians with short-range interactions and bounded local energy~\cite{landau2015polynomial,Arad2017}. Extending these algorithms to the general setting and resolving the time complexity for generic interacting boson systems remains a major open problem in quantum many-body physics.

Among the open directions, the most tractable with current techniques appears to be establishing an area law in higher dimensions under the assumption of an adiabatic path to a trivial gapped system. 
Since our work already proves suppression of boson number distributions in arbitrary dimensions, combining this with the SIE theorem provides a concrete route toward such a generalization.
Furthermore, beyond the entropy itself, analyzing the dynamics of the full entanglement spectrum may reduce the quasi-polynomial complexity of long-range interacting boson systems to polynomial complexity~\cite{kim2025spectral}, providing a potential pathway toward the development of polynomial-time algorithms for interacting bosonic systems.

\section*{Methods}

\subsection*{Proof sketch of Result~1.1 (Bose-Hubbard class)}
\begin{figure}[t]
	\begin{center}
		\includegraphics[width=0.485\textwidth]{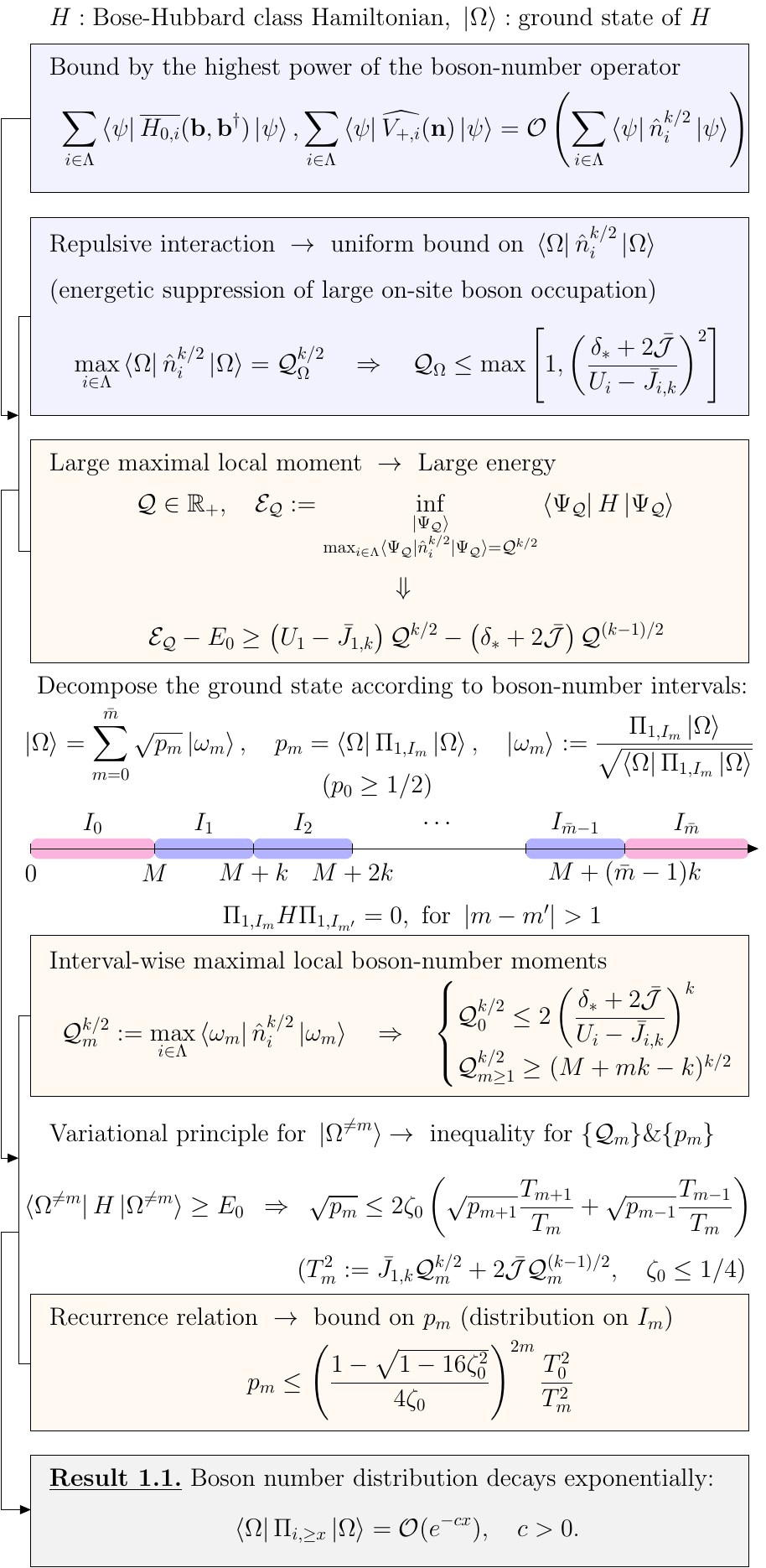}
	\end{center}
	\caption{Proof strategy for the exponential boson-number concentration bound in the Bose-Hubbard class.
			The diagram shows how the proof of Result~1.1 converts local repulsion into an exponential tail bound for the boson-number distribution. 
			The argument first bounds the non-diagonal and density-dependent parts of the Hamiltonian by the highest power of the local number operator. In the repulsive regime, this yields a uniform $\mathcal{O}(1)$ bound on the ground-state maximal local moment $\mathcal{Q}_{\Omega}$. 
			It then establishes an energetic penalty for any state with a large maximal local moment $\mathcal{Q}$, showing that such states lie high above the ground-state energy $E_0$. 
			Next, the ground state is decomposed into boson-number intervals $I_{m}$ at a fixed site, with weights $p_{m}$ and normalized components $\ket{\omega_{m}}$.
			Because the Hamiltonian changes the local occupation by at most $k$, only neighboring intervals are coupled. 
			Lower bounds on the interval-resolved moments $\mathcal{Q}_{m}$, together with the variational principle applied to the state with the $m$th component removed, yield a recurrence relation for $\{p_{m}\}$. 
			Solving this recurrence gives an exponential bound on $p_{m}$ and hence an exponential decay of the local boson-number tail, establishing Result~1.1.}
	\label{fig_main_BH_boson_number} 
\end{figure}

We outline the proof of the exponential concentration bound (Eq.~\eqref{intro_BH_conc}) for the Bose-Hubbard class of Hamiltonians (See Fig.~\ref{fig_main_BH_boson_number}).  

We first provide a more precise setup.
The Bose-Hubbard Hamiltonian (Eq.~\eqref{Boson_most_general_intro}) can be decomposed into three distinct components:
\begin{equation}
	H = H_{0}(\mathbf{b},\mathbf{b}^{\dagger}) + V_{+}(\mathbf{n}) + \sum_{i \in \Lambda} U_{i} \hat{n}_{i}^{k/2},
\end{equation}
where the first term, \( H_{0}(\mathbf{b},\mathbf{b}^{\dagger}) \), is a polynomial of degree at most \( k \) in the bosonic annihilation and creation operators \( \mathbf{b} = \{b_{i}\}_{i \in \Lambda} \) and \( \mathbf{b}^{\dagger} = \{b_{i}^{\dagger}\}_{i \in \Lambda} \), typically representing particle hopping processes or other interaction terms with possibly negative contributions. 
Its explicit form is given by
\begin{equation}
	\begin{aligned}
		H_{0}(\mathbf{b}, \mathbf{b}^{\dagger}) = &\sum_{k_{1} = 0}^{k - 1} \sum_{i_{1}, \ldots, i_{k_{1}} \in \Lambda} \sum_{s = 1}^{k_{1} + 1} J_{i_{1}, \ldots, i_{k_{1}}}^{(s)} \\
		&\times b_{i_{k_{1}}}^{\dagger} \cdots b_{i_{s + 1}}^{\dagger} 
		b_{i_{s}}^{\dagger} b_{i_{s - 1}} \cdots b_{i_{2}} b_{i_{1}} .
	\end{aligned}
\end{equation}
The second term, \( V_{+}(\mathbf{n}) \), is a positive semi-definite polynomial of degree at most $k/2$ in the local number operators \( \mathbf{n} = \{\hat{n}_{i}\}_{i \in \Lambda} \).
The third term represents the leading-order local repulsive interaction, \( \sum_{i \in \Lambda} U_{i} \hat{n}_{i}^{k / 2} \), which contains the highest-degree (order \( k \)) contributions in the bosonic operators.

For notational convenience, we define for any region \( X \subset \Lambda \) the restricted and supported Hamiltonians by
\begin{equation}
	H_{X} := \sum_{Z \subset X} h_{Z}, \quad
	\widehat{H_{X}} := \sum_{Z : Z \cap X \neq \emptyset} h_{Z} ,
\end{equation}
and adopt the same definitions analogously for the components \( H_{0} \) and \( V_{+} \).
To analyze local contributions from \( H_0 \), using the definition \( \abs{O} := \sqrt{O^{\dagger} O} \) of the modulus of an operator \( O \), we define $\overline{H_{0,i}}(\mathbf{b},\mathbf{b}^{\dagger})$ as the modulus of the site-supported component \( \widehat{H_{0,i}}(\mathbf{b},\mathbf{b}^{\dagger}) \): 
\begin{widetext}
	\begin{equation}
		\overline{H_{0,i}}(\mathbf{b},\mathbf{b}^{\dagger}) = \sum_{\substack{i_{1}, \ldots, i_{k_{1}} \in \Lambda \\ \{i_{1}, \ldots, i_{k_{1}}\} \ni i}} \sum_{s = 1}^{k_{1} + 1} \abs{J_{i_{1}, \ldots, i_{k_{1}}}^{(s)}} \cdot \abs{b_{i_{k_{1}}}^{\dagger} \cdots b_{i_{s+1}}^{\dagger} b_{i_{s}}^{\dagger} b_{i_{s-1}} \cdots b_{i_{2}} b_{i_{1}} } .
	\end{equation}
\end{widetext}
To quantify the strength of this interaction, we introduce the parameter $\bar{J}_{i,k_{1}} $ as follows:
\begin{equation}
	\label{Def_bar_j_i_k_1}
	\bar{J}_{i,k_{1}} := \sum_{\substack{i_{1}, \ldots, i_{k_{1}} \in \Lambda \\ \{i_{1}, \ldots, i_{k_{1}}\} \ni i}} \sum_{s = 1}^{k_{1}} \abs{J_{i_{1}, \ldots, i_{k_{1}}}^{(s)}} .
\end{equation}
This picks up all the interaction coefficients that include the site $i$.  
We assume \( U_{i} > 5 \bar{J}_{i,k} \) for all \( i \in \Lambda \), where \( \bar{J}_{i,k} \) is defined by setting \( k_1 = k \) in Eq.~\eqref{Def_bar_j_i_k_1}.  
This ensures that the leading-order repulsion dominates (see Assumption~1 in the Supplementary Information).
If \( H_{0} \) contains terms only up to degree \( k - 1 \) as has been shown in Eq.~\eqref{Boson_most_general_intro}, then \( \bar{J}_{i,k} = 0 \) and we only have to assume $U_i>0$.

The first core idea of the proof is to initiate the analysis from the maximum local moment with respect to the ground state \( \ket{\Omega} \) of the Hamiltonian \( H \) across the lattice:
\begin{equation}
	\max_{i \in \Lambda} \bra{\Omega} \hat{n}_{i}^{k/2} \ket{\Omega} =: \mathcal{Q}_{\Omega}^{k / 2},
\end{equation}
which serves as a central quantity for bounding the concentration behavior. 
We aim to clarify the following two points.
\begin{enumerate}
	\item {} In the ground state, the quantity $\mathcal{Q}_{\Omega}$ is upper-bounded by a constant value as in the inequality~\eqref{Q_bound} below. 
	\item {} Any quantum state with a large maximum local moment must have a large energy, as will be shown in the inequality~\eqref{Energy_bound3}. 
\end{enumerate}

We begin with the first point. 
Without loss of generality, we label the site \( i = 1 \) such that \( \langle \hat{n}_{1}^{k/2} \rangle = \mathcal{Q}_{\Omega}^{k/2} \), and denote its complement by \( \Lambda_{1} := \Lambda \setminus \{1\} \). The Hamiltonian can then be decomposed as $H = \widehat{H}_{1} + H_{\Lambda_{1}}$ where the supported part \( \widehat{H_{1}} \) admits the decomposition $\widehat{H_{1}} = \widehat{H_{0,1}}(\mathbf{b}, \mathbf{b}^{\dagger}) + \widehat{V_{+,1}}(\mathbf{n}) + U_{1} \hat{n}_{1}^{k/2}$.
Since \( \ket{\Omega} \) is the ground state of \( H \), the variational principle implies
\begin{equation}
	\bra{\Omega} \widehat{H}_{1} \ket{\Omega} \leq \delta_{*} \mathcal{Q}_{\Omega}^{(k-1)/2},
\end{equation}
where \( \delta_{*} = 0 \) when the total boson number is not conserved (see Eq.~(S.90) in the Supplementary Information), and $\delta_{*} \propto (N / |\Lambda|)^{k/2}$
when the total boson number is fixed to \( N \) (see Eq.~(S.111) and Eq.~(S.132) in the Supplementary Information).
Here, we note that the analysis required to obtain the above inequality becomes more complicated in the total-boson-number-conserving case than in the nonconserving case.
In the latter situation, the local occupation numbers can fluctuate independently, which allows us to construct simple variational reference states when applying the variational principle for the ground-state energy.
In contrast, when the total boson number is conserved, the allowed variational states must satisfy a global constraint on the total particle number.
This constraint correlates local occupation numbers across different sites, making the construction of suitable variational states more involved.
Using the bounds \( \widehat{H_{0,1}}(\mathbf{b}, \mathbf{b}^{\dagger}) \succeq -\overline{H_{0,1}}(\mathbf{b}, \mathbf{b}^{\dagger}) \) and \( \widehat{V_{+,1}}(\mathbf{n}) \succeq 0 \), we obtain
\begin{equation}
	\begin{aligned}
		\delta_{*} \mathcal{Q}_{\Omega}^{(k-1)/2} 
		&\geq \bra{\Omega} \widehat{H}_{1} \ket{\Omega} \\
		&\geq U_{1} \bra{\Omega} \hat{n}_{1}^{k/2} \ket{\Omega} 
		- \bra{\Omega} \overline{H_{0,1}}(\mathbf{b}, \mathbf{b}^{\dagger}) \ket{\Omega} . \label{Energy_bound1}
	\end{aligned}
\end{equation}
Exploiting the extremality of \( \mathcal{Q}_{\Omega}^{k/2} \), we derive the following bound on \( \overline{H_{0,1}} \) (see Proposition 4 in the Supplementary Information):
\begin{equation}
	\bra{\Omega} \overline{H_{0,1}}(\mathbf{b}, \mathbf{b}^{\dagger}) \ket{\Omega} 
	\leq \bar{J}_{1,k} \, \mathcal{Q}_{\Omega}^{k/2} 
	+ 2 \bar{\mathcal{J}} \, \mathcal{Q}_{\Omega}^{(k-1)/2}, \label{Energy_bound2}
\end{equation}
where the quantity \( \bar{\mathcal{J}} \) is defined as $\bar{\mathcal{J}} := \max_{i \in \Lambda} \sum_{k_{1} = 1}^{k} \bar{J}_{i,k_{1}} \left( 1 + (2k_{1})^{k_{1}} \right)$.
Substituting Eq.~\eqref{Energy_bound2} into Eq.~\eqref{Energy_bound1}, we obtain the following upper bound on \( \mathcal{Q}_{\Omega} \) (see Propositions 6 and 10 in the Supplementary Information):
\begin{equation}
	\mathcal{Q}_{\Omega} \leq \max \left\{ 1, \left( \frac{\delta_{*} + 2 \bar{\mathcal{J}}}{U_{1} - \bar{J}_{1,k}} \right)^{2} \right\} . \label{Q_bound}
\end{equation}
This implies that the \( k/2 \)-th moment of the local boson number operator is bounded by an \( \mathcal{O}(1) \) constant in the ground state, provided that the total boson number satisfies \( N = \mathcal{O}(|\Lambda|) \).

Conversely, we establish an energetic penalty for large local moments. 
Specifically, let us consider the lowest possible energy $\mathcal{E}_{\mathcal{Q}}$ among all states whose maximal local $k/2$-th moment satisfies $\max_{i \in \Lambda} \bra{\Psi} \hat{n}_{i}^{k/2} \ket{\Psi} = \mathcal{Q}^{k/2}$. 
We show that if $\mathcal{Q}$ is large, then the energy difference $\mathcal{E}_{\mathcal{Q}} - E_0$ above the ground state grows rapidly with $\mathcal{Q}$. 
In other words, quantum states exhibiting large local boson-number moments are necessarily separated from the ground state by a large energy gap, and therefore, such states are only possible at high energy above the ground state.
More precisely (see Corollaries 7 and 11 in the Supplementary Information),
\begin{equation}
	\mathcal{E}_{\mathcal{Q}} - E_{0} \geq \left( U_{1} - \bar{J}_{1,k} \right) \mathcal{Q}^{k/2} - \left( \delta_{*} + 2 \bar{\mathcal{J}} \right) \mathcal{Q}^{(k-1)/2} . \label{Energy_bound3}
\end{equation}

The second crucial idea is to partition the boson number into several intervals and derive a recurrence relation for the concentration probability within each interval, based on the previously established energy bounds.
We divide the nonnegative real line into intervals as follows: let \( I_{0} := [0, M) \),  $I_{m} := [M + (m - 1)k,\, M + mk)$ for $1 \leq m < \bar{m}$, $I_{\bar{m}} := [M + (\bar{m} - 1)k,\, \infty)$ for some control parameter \( M \in \mathbb{N} \). 

Using these intervals, we decompose the ground state \( \ket{\Omega} \) via projections \( \Pi_{1, I_{m}} \), which project onto the subspace where the boson number at site \( i = 1 \) lies in \( I_{m} \). This yields:
\begin{equation}
	\ket{\Omega} = \sum_{m = 0}^{\bar{m}} \Pi_{1, I_{m}} \ket{\Omega} = \sum_{m = 0}^{\bar{m}} p_{m}^{1/2} \ket{\omega_{m}},
\end{equation}
where \( p_{m} := \bra{\Omega} \Pi_{1, I_{m}} \ket{\Omega} \) represents the probability of the boson number at site 1 falling within the interval \( I_{m} \), and the normalized vector is defined as \( \ket{\omega_{m}} := p_{m}^{-1/2} \Pi_{1, I_{m}} \ket{\Omega} \).
Since the Hamiltonian can only change the local boson number by at most \( k \), the constructed intervals ensure that
\begin{equation}
	\Pi_{1, I_{m}} H \Pi_{1, I_{m'}} = 0 \qquad \text{for all} \quad |m - m'| > 1 ,
\end{equation}
i.e., the Hamiltonian does not couple nonadjacent intervals.

As in the previous argument, we define
\begin{equation}
	\mathcal{Q}_{m}^{k/2} := \max_{i \in \Lambda} \bra{\omega_{m}} \hat{n}_{i}^{k/2} \ket{\omega_{m}},
\end{equation}
which quantifies the maximum local \( k/2 \)-moment within each component state \( \ket{\omega_{m}} \). By construction, for all \( m \geq 1 \),
\begin{equation}
	\mathcal{Q}_{m}^{k/2} \geq (M + mk - k)^{k/2} . \label{Q_inequality1}
\end{equation}
We choose the parameter \( M \in \mathbb{N} \) such that the initial weight satisfies \( p_{0} \geq 1/2 \), and the corresponding moment bound obeys (see Lemma 13 in the Supplementary Information)
\begin{equation}
	\mathcal{Q}_{0}^{k/2} \leq 2 \left( \frac{\delta_{*} + 2 \bar{\mathcal{J}}}{U_{i} - \bar{J}_{i,k}} \right)^{k} .\label{Q_inequality2}
\end{equation}
The final goal is to derive a recurrence relation among the weights \( \{p_{m}\}_{m=0}^{\bar{m}} \), using the moment bound on \( \mathcal{Q}_{m} \) to control the tail distribution of the local boson number.

To investigate the decay of the concentration tail, we consider the modified state
\begin{equation}
	\ket{\Omega^{\neq m}} := \frac{1}{\sqrt{1 - p_{m}}} \sum_{s \neq m} p_{s}^{1/2} \ket{\omega_{s}},
\end{equation}
which is obtained from the ground state \( \ket{\Omega} \) by removing the \( \ket{\omega_{m}} \) component. We define the associated energy difference as
\begin{equation}
	p_{m} \Delta E_{m} := E_{0} - (1 - p_{m}) \bra{\Omega^{\neq m}} H \ket{\Omega^{\neq m}},
\end{equation}
where \( \Delta E_{m} \leq E_{0} \) follows from the variational principle.
Due to the fact that the Hamiltonian only couples adjacent intervals, the quantity \( p_{m} \Delta E_{m} \) depends solely on the states \( \ket{\omega_{m}} \) and \( \ket{\omega_{m \pm 1}} \). Substituting the decompositions of \( \ket{\Omega} \) and \( \ket{\Omega^{\neq m}} \), and applying the Cauchy–Schwarz inequality, we obtain the following lower bound:
\begin{widetext}
	\begin{equation}
		p_{m} \Delta E_{m} \geq p_{m} \bra{\omega_{m}} H \ket{\omega_{m}} 
		- 2 \sum_{x = \pm 1} \sqrt{p_{m} p_{m+x}} \left( \bra{\omega_{m+x}} \overline{H_{0,1}} \ket{\omega_{m+x}} 
		\bra{\omega_{m}} \overline{H_{0,1}} \ket{\omega_{m}} \right)^{1/2} .\label{recurrence_lemma1}
	\end{equation}
\end{widetext}
Analogously to Eq.~\eqref{Energy_bound2}, we use the bound
\begin{equation}
	\begin{aligned}
		\bra{\omega_{m}} \overline{H_{0,1}} \ket{\omega_{m}} 
		&\leq \bar{J}_{1,k} \mathcal{Q}_{m}^{k / 2} + 2 \bar{\mathcal{J}} \mathcal{Q}_{m}^{(k - 1) / 2} \\
		&=: T_{m}^{2}, \quad \forall m \label{recurrence_lemma2}.
	\end{aligned}
\end{equation}
Furthermore, from the ground-state energy bound (cf. Eq.~\eqref{Energy_bound1}), we have
\begin{equation}
	\begin{aligned}
		&\bra{\omega_{m}} H \ket{\omega_{m}}   \\
		&\geq (U_{1} - \bar{J}_{1,k}) \mathcal{Q}_{m}^{k / 2} - (\delta_{*} + 2 \bar{\mathcal{J}}) \mathcal{Q}_{m}^{(k - 1) / 2} + E_{0} \\
		&\geq \frac{1}{\zeta_{0}} T_{m}^{2} + E_{0}  \label{recurrence_lemma3}
	\end{aligned}
\end{equation}
for some \( \mathcal{O}(1) \) constant \( \zeta_{0} \leq 1/4 \) (see Eq. (S.179) in the Supplementary Information).
Combining Eqs.~\eqref{recurrence_lemma1}–\eqref{recurrence_lemma3}, we obtain the recurrence inequality:
\begin{equation}
	\sqrt{p_{m}} \leq 2 \zeta_{0} \left( \sqrt{p_{m+1}} \frac{T_{m+1}}{T_{m}} + \sqrt{p_{m-1}} \frac{T_{m-1}}{T_{m}} \right).
\end{equation}
By solving the recurrence relation (as presented in Lemma~15 of the Supplementary Information) and applying Eqs.~\eqref{Q_inequality1} and \eqref{Q_inequality2}, we obtain
\begin{equation}
	\label{recurrence_lemma4}
	p_{m} \leq \left( \frac{1 - \sqrt{1 - 16 \zeta_{0}^{2}} }{4 \zeta_{0}} \right)^{2m},
\end{equation}
which establishes the exponential bound of the boson-number concentration in the Bose–Hubbard class.

\subsection*{Proof sketch of Result~1.2 ($\phi4$ class)}
\begin{figure}[t]
	\begin{center}
		\includegraphics[width=0.485\textwidth]{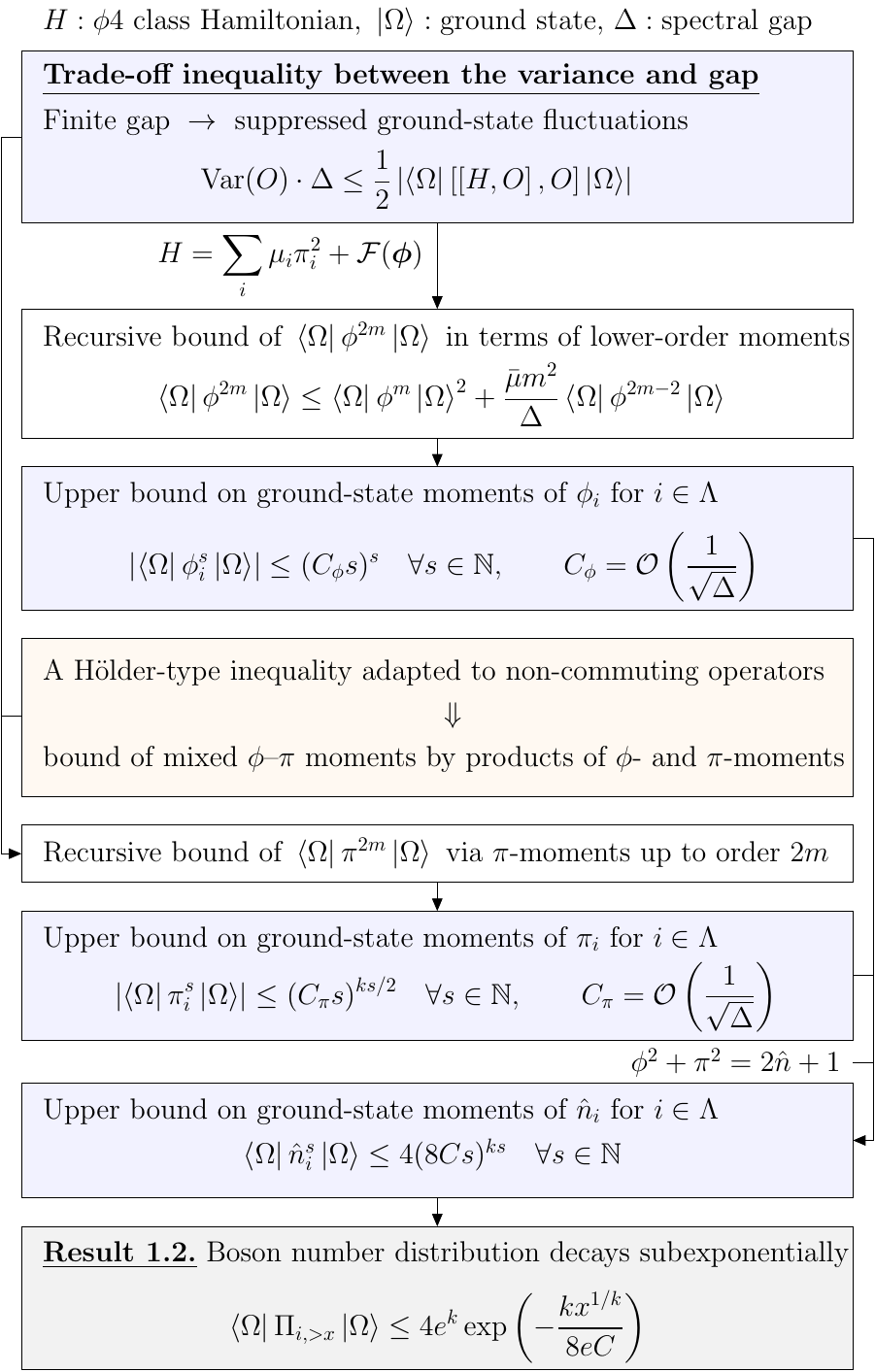}
	\end{center}
	\caption{Proof strategy for the subexponential boson-number concentration bound in the $\phi 4$ class.
			The diagram illustrates how a finite spectral gap leads to subexponential suppression of large boson-number fluctuations in Result~1.2. 
			The proof starts from the variance-gap trade-off inequality $\mathrm{Var}(O) \cdot \Delta \le |\bra{\Omega}[[H,O],O]\ket{\Omega}|/2$, which constrains ground-state fluctuations through the spectral gap $\Delta$. 
			Applying this identity to powers of the field operator $\phi_i$ recursively bounds higher $\phi$-moments by lower ones and yields $|\bra{\Omega}\phi_i^s\ket{\Omega}| \le (C_{\phi}s)^s$ with $C_{\phi} = \mathcal{O}(\Delta^{-1/2})$. 
			A central technical step is then to control mixed non-commuting moments involving both $\phi$ and $\pi$. 
			For this purpose, we prove a H\"{o}lder-type inequality adapted to non-commuting operators, which reduces mixed moments to controllable combinations of $\phi$- and $\pi$-moments. 
			This closes the recursion for $\pi$-moments and gives $|\bra{\Omega}\pi_i^s\ket{\Omega}| \le (C_{\pi}s)^{ks/2}$. 
			Combining the resulting $\phi$- and $\pi$-estimates with $\phi^2+\pi^2=2\hat n+1$ yields upper bounds on the moments of the local boson-number operator.
			Markov's inequality then converts these moment bounds into a subexponential tail estimate for the local boson-number distribution, establishing Result~1.2.}
	\label{fig_main_phi4_boson_number} 
\end{figure}

We present a sketch of the proof for the subexponential concentration bound (Eq.~\eqref{intro_phi4_conc}) applicable to the $\phi^4$ class of Hamiltonians (Eq.~\eqref{phi_4_type_model_intro}) (See Fig.~\ref{fig_main_phi4_boson_number}). In contrast to the Bose-Hubbard case, our argument here relies on the assumption of a finite spectral gap. 
Indeed, if the gap condition is not imposed, there exist models where the ground state allows arbitrarily many bosons to occupy a single site.
To connect the gap condition to the concentration of the boson number distribution, the key idea is to utilize the following universal trade-off inequality between the spectral gap and the variance of an arbitrary operator (cf. Ref.~\cite{Kuwahara_2017}):
\begin{equation}
	\mathrm{Var}(O) \cdot \Delta \leq \frac{1}{2} \left| \bra{\Omega} [[H,O],O] \ket{\Omega} \right|. \label{variance_gap}
\end{equation}
Here, $\mathrm{Var}(O)$ denotes the variance of an operator $O$ with respect to the ground state $\ket{\Omega}$, defined as $\mathrm{Var}(O) = \bra{\Omega} O^{2} \ket{\Omega} - \bra{\Omega} O \ket{\Omega}^{2}$, and $\Delta$ has been defined as the spectral gap.
This inequality expresses that a finite spectral gap constrains the extent to which an observable can fluctuate in the ground state. In particular, if an operator has a small double commutator with the Hamiltonian, so that it is approximately conserved or only weakly affected by the dynamics, then its ground-state fluctuations must be correspondingly suppressed. Conversely, large fluctuations of an observable necessarily imply that it couples strongly to excitations across the gap, as quantified by its noncommutativity with the Hamiltonian.

We apply the inequality~\eqref{variance_gap} to the operator $\phi^m$, where we omit the lattice index $i$ for simplicity.
In the $\phi^4$ class of Hamiltonians, fluctuations of the field operator are constrained solely through the kinetic term $\pi^2$, which is the only term that mediates the noncommutativity underlying the inequality~\eqref{variance_gap}.
Evaluating the resulting double commutator using the canonical commutation relation $[\phi,\pi]=i$, we obtain
\begin{equation}
	\begin{aligned}
		\label{phi_recursive_ineq}
		\mathrm{Var}(\phi^{m}) &= \left( \bra{\Omega} \phi^{2m} \ket{\Omega} - \bra{\Omega} \phi^{m} \ket{\Omega}^{2} \right) \cdot \Delta \\
		&\leq \frac{\bar{\mu}}{2} \abs{\bra{\Omega} [[\pi^{2}, \phi^{m}], \phi^{m}] \ket{\Omega}} \\
		&\leq \bar{\mu} m^{2} \abs{\bra{\Omega} \phi^{2m - 2} \ket{\Omega}},
	\end{aligned}
\end{equation}
where $\bar{\mu} := \max_{i \in \Lambda} |\mu_{i}|$.  
As a consequence, ground-state fluctuations of $\phi$ are hierarchically constrained, with higher moments recursively bounded in terms of lower ones.
Solving this recurrence (see Proposition~16 in the Supplementary Information) yields the bound
\begin{equation}
	\left| \bra{\Omega} \phi^{s} \ket{\Omega} \right| \leq (C_{\phi} s)^{s},
\end{equation}
where $C_{\phi} = \mathcal{O} \left( \sqrt{\bar{\mu}/\Delta} \right)$.

A similar overall approach can be applied to the $\pi$ operator, but the situation becomes significantly more complicated. 
An important distinction arises between the analysis of $\phi$ and $\pi$ operators. 
For the moments of $\phi$, all relevant expectations involve only powers of $\phi$, so the moment bound can be closed solely in terms of $\phi$ itself as in~\eqref{phi_recursive_ineq}. 
In contrast, when estimating moments of $\pi$, the high-order terms with respect to $\phi$ inevitably lead to expectation values involving mixed products of both operators, such as $\abs{\bra{\Omega} \pi^{2m-s} \phi^{s'} \ket{\Omega}}$. 
If one could somehow separate the $\phi$ and $\pi$ contributions, it would be possible to apply the established moment bounds for $\phi$ and derive a closed recurrence inequality for $\pi$.
However, since H\"older's inequality is not available for non-commuting operators, this separation is highly nontrivial and poses a major technical challenge.
To address this issue, we establish a non-commutative version of H\"older's inequality (see Proposition~20 in the Supplementary Information), which enables us to control mixed moments of $\pi$ and $\phi$ in terms of the moments of $\pi$ alone.

While the detailed calculations are deferred to the Supplementary Information (see Subtheorem~1 therein), the key outcome is that we ultimately obtain a simple closed-form inequality for the moments of the $\pi$ operator, analogous to that for $\phi$:
\begin{equation}
	\left| \bra{\Omega} \pi^{s} \ket{\Omega} \right| \leq (C_{\pi} s)^{ks/2},
\end{equation}
where $C_{\pi} = \mathcal{O} \left( \sqrt{\bar{\mu}/\Delta} \right)$.
Finally, using the identity $\phi^{2} + \pi^{2} = 2 \hat{n} + 1$ that relates the fields to the boson number operator, we derive (see Proposition~18 in the Supplementary Information) the bound
\begin{equation}
	\bra{\Omega} \hat{n}^{s} \ket{\Omega} \leq (C s)^{ks}, \label{boson_number_bound}
\end{equation}
where $C = \mathcal{O} \left( \sqrt{\bar{\mu}/\Delta} \right)$.

Based on the moment bound of the boson number operator (Eq.~\eqref{boson_number_bound}), we can estimate the tail of its probability distribution via Markov's inequality:
\begin{equation}
	\bra{\Omega} \Pi_{i, > x} \ket{\Omega} \leq \min_{s \in \mathbb{N}} \frac{(C s)^{k s}}{x^{s}}.
\end{equation}
Choosing the optimizing parameter as $s = \left\lfloor x^{1/k} / (eC) \right\rfloor$, we obtain the subexponential concentration bound for the boson number:
\begin{equation}
	\bra{\Omega} \Pi_{i, > x} \ket{\Omega} \lesssim \exp\left( - \frac{k}{eC} x^{1/k} \right),
\end{equation}
which completes the proof.

\subsection*{Proof sketch of Result~2 (entanglement area law)}
\begin{figure}[t]
	\begin{center}
		\includegraphics[width=0.485\textwidth]{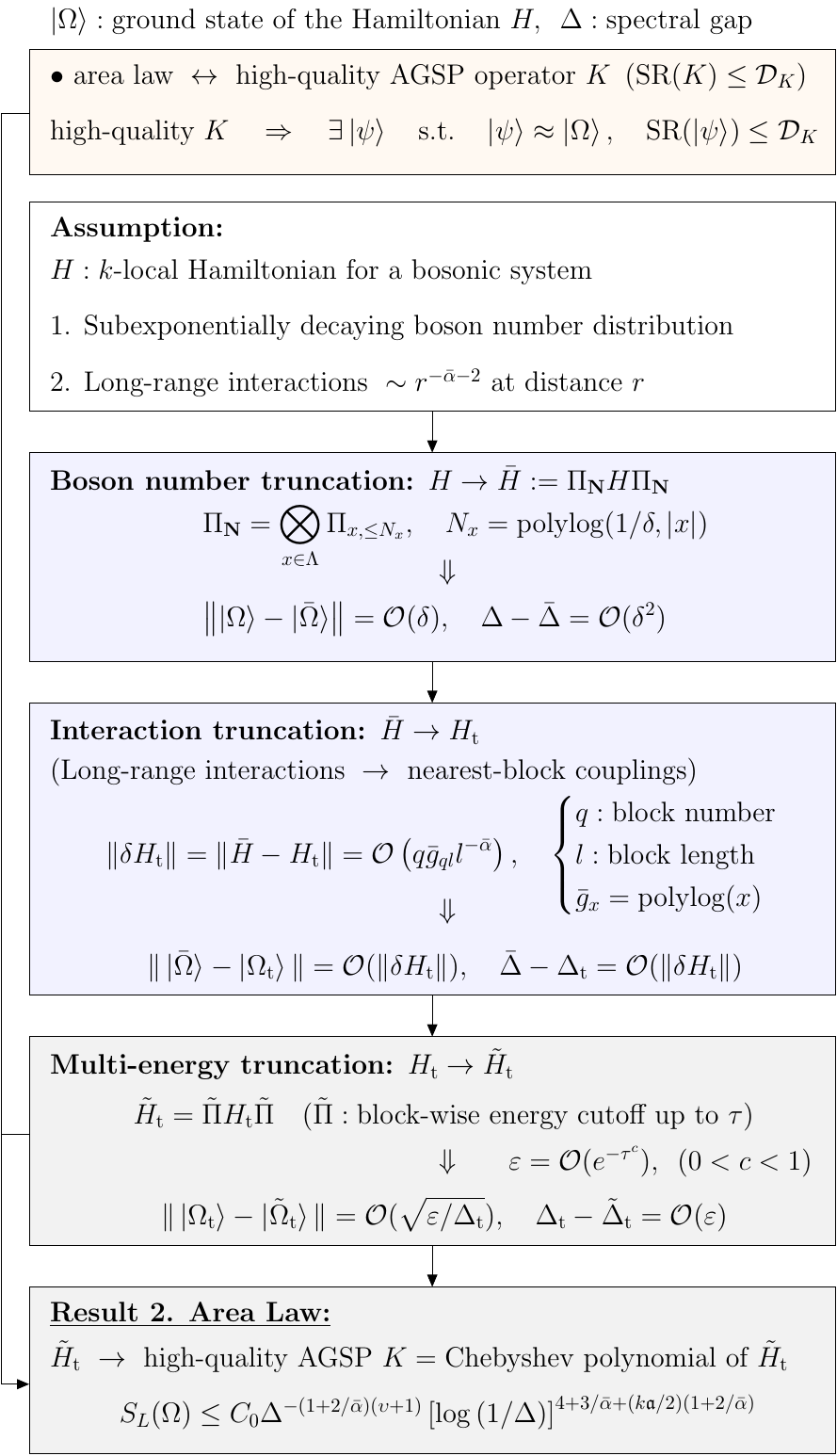}
	\end{center}
	\caption{Proof strategy for the one-dimensional bosonic area law with long-range interactions.
			The diagram outlines how the proof of Result~2 reduces a long-range bosonic Hamiltonian with unbounded local Hilbert spaces to a bounded effective model amenable to the approximate ground-state projector (AGSP) method. 
			The top box recalls the AGSP principle, namely that a sufficiently accurate AGSP with controlled Schmidt rank implies an entanglement area law. 
			The proof begins from a general $k$-local bosonic Hamiltonian with a subexponentially decaying local boson-number distribution and power-law interactions.
			First, a site-dependent boson-number truncation $H \to \bar{H}$ is introduced, with cutoffs $N_{x}$ growing polylogarithmically with the distance from the bipartition boundary; this controls the local boson number while preserving the ground state and the spectral gap up to controlled error.
			Second, an interaction truncation $\bar{H} \to H_{\mathrm{t}}$ reduces the Hamiltonian to block-local terms and nearest-neighbor inter-block couplings, with the induced error controlled by the block number $q$, the block length $l$, and the interaction decay.
			Third, a block-wise multi-energy truncation $H_{\mathrm{t}} \to \tilde{H}_{\mathrm{t}}$ makes the effective Hamiltonian bounded while preserving the low-energy sector up to a subexponentially small error in the cutoff $\tau$. 
			A Chebyshev-polynomial AGSP construction can then be applied to $\tilde{H}_{\mathrm{t}}$, yielding the quantitative area law in Result~2.}
	\label{fig_main_area_law} 
\end{figure}

In this section, we provide a brief summary of the proof of the area law for one-dimensional bosonic systems with long-range interactions (See Fig.~\ref{fig_main_area_law}). We consider a one-dimensional lattice defined on $\mathbb{Z}$, where each site $i$ is labeled by its position $x \in \mathbb{Z}$. The system is bipartitioned into left and right semi-infinite subsystems, $L = (-\infty, 0]$ and $R = [1, \infty)$, and we study the entanglement entropy of the ground state $\ket{\Omega}$ across this bipartition.

Our approach is based on the framework of approximate ground state projectors (AGSPs), originally introduced for systems with short-range interactions~\cite{PhysRevB.85.195145, arad2013area}, and subsequently extended to long-range interacting systems~\cite{Kuwahara2020arealaw}. An AGSP operator $K$ approximates the ground state projector $\ket{\Omega}\bra{\Omega}$, exhibiting a trade-off between its Schmidt rank and approximation accuracy (see Supplementary Note 7.A). The central objective is to construct an AGSP with low Schmidt rank and high fidelity, which enables the construction of a low-entangled approximation to the ground state (see Lemma~24 in the Supplementary Information).
In this work, we further extend the AGSP framework to encompass unbounded bosonic Hamiltonians, under the assumption that the boson number distribution exhibits subexponential decay (Eq.~\eqref{concentration_bound_assmp}).

For this purpose, we perform a sequence of controlled truncations that systematically reduce the original long-range, unbounded bosonic Hamiltonian to a bounded, quasi-local, and gapped form suitable for the application of standard area law techniques. Concretely, we transform the original Hamiltonian $H$ through the following stages:
\begin{equation}
	\begin{aligned}
		\label{three_step_truncation}
		H \xrightarrow{\text{boson-number truncation}}
		&\bar H \\
		\xrightarrow{\text{interaction truncation}} 
		&H_{\mathrm t} \\
		\xrightarrow{\text{multi-energy truncation}}
		& \tilde{H}_{\mathrm t}.
	\end{aligned}
\end{equation}
First, a boson-number truncation is performed to bound the local particle number on each site, effectively rendering the local Hilbert space finite-dimensional. 
Next, we apply an interaction truncation to remove long-range couplings beyond adjacent blocks near the bipartition boundary, thereby localizing the relevant interactions. Finally, a multi-energy truncation is used to constrain the intra-block energies at a prescribed cutoff, ensuring that the total Hamiltonian becomes bounded.
At each step, the truncation parameters are carefully chosen to control the approximation error and to preserve a positive spectral gap, as rigorously quantified in Propositions~25--26 and Theorem 4 in the Supplementary Information. 

Once this reduction is achieved, the resulting Hamiltonian $\tilde{H}_{\mathrm t}$ is both bounded and quasi-local, allowing us to apply the standard Chebyshev-polynomial-based AGSP construction developed for short-range systems. This ultimately establishes the one-dimensional entanglement area law for the original class of long-range, unbounded bosonic Hamiltonians.

We begin with the first truncation in~Eq.~\eqref{three_step_truncation} to address the unbounded local Hilbert space.
A central idea is to introduce a site-dependent boson-number truncation, where the cutoff $N_x$ grows poly-logarithmically with the distance from the boundary:
\begin{equation}
	\begin{aligned}
		&\bar{\Pi} = \Pi_{\mathbf{N}} := \bigotimes_{x \in \Lambda} \Pi_{x, \leq N_{x}},  \\
		& N_{x} = \mathcal{O}\left( \log^{\mathfrak{a}}(\epsilon^{-1}) + \log^{\mathfrak{a}} \left( |x|^3 + 1 \right) \right),
	\end{aligned}
\end{equation}
where the parameter $\epsilon$ is appropriately chosen below so that the ground state is preserved up to a sufficiently small error. 
We then define the truncated Hamiltonian as $\bar{H} := \bar{\Pi} H \bar{\Pi}$. This choice of $N_x$ ensures that both the spectral gap $\bar{\Delta}$  and the ground state $\ket{\bar{\Omega}}$ of $\bar{H}$ approximate those of the original Hamiltonian (see Proposition~25 in the Supplementary Information):
\begin{equation}
	\bar{\Delta} \geq \Delta \left( 1 - \frac{3}{2} \delta^2 \right), \quad \| \ket{\Omega} - \ket{\bar{\Omega}} \| \leq \delta ,
\end{equation}
provided that the parameter $\epsilon$ is chosen as $\epsilon = \mathcal{O} \left( \delta^2 \log^{-\mathfrak{a}k/2}(\delta^{-1}) \right)$. This truncation partially resolves the issue of unboundedness: while the local energy near the boundary becomes bounded due to the stricter cutoff, it remains unbounded in regions far from the boundary. The remaining unboundedness will be addressed by subsequent truncation procedures.

In the next truncation of Eq.~\eqref{three_step_truncation}, we perform an interaction truncation localized near the boundary. If all long-range interactions across the entire system were truncated uniformly, the resulting Hamiltonian $H_{\mathrm{t}}$ would differ significantly from the previously truncated Hamiltonian $\bar{H}$, with an operator norm distance $\| \bar{H} - H_{\mathrm{t}} \| = \mathcal{O}(|\Lambda|)$. Such a global modification would substantially alter the ground state and degrade its fidelity. To avoid this, we confine the truncation of long-range interactions to a finite region around the interface between the subsystems $L$ and $R$.
Specifically, we partition the neighborhood near the boundary into $q$ blocks $\{ B_s \}_{s = 1}^{q}$, each of length $l$, and define the remaining outer regions as $B_0$ and $B_{q+1}$ on the left and right, respectively. The interaction truncation is implemented such that only adjacent blocks are allowed to interact, and all longer-range interactions across non-neighboring blocks are eliminated. The resulting truncated Hamiltonian is given by
\begin{equation}
	H_{\mathrm{t}} = \sum_{s = 0}^{q+1} h_{s} + \sum_{s = 0}^{q} h_{s,s+1},
\end{equation}
where $h_{s}$ denotes the internal Hamiltonian acting within block $B_s$, and $h_{s,s+1}$ represents the interaction between adjacent blocks $B_s$ and $B_{s+1}$.
The truncation error, measured by $\| \bar{H} - H_{\mathrm{t}} \|$, can be effectively controlled by choosing the number of blocks $q$ and their length $l$ appropriately (see Proposition~26 in the Supplementary Information):
\begin{equation}
	\| \bar{H} - H_{\mathrm{t}} \| \leq \mathcal{O} \left( q \log^{\mathfrak{a}k/2}(q l) \cdot l^{2} \bar{J}(l) \right),
\end{equation}
where $\bar{J}(l)$ characterizes the decay of long-range interactions, defined in Eq.~\eqref{long_range_interaction}.
Moreover, the spectral gap $\Delta_{\mathrm{t}}$ and the ground state $\ket{\Omega_{\mathrm{t}}}$ of $H_{\mathrm{t}}$ remain close to those of $\bar{H}$, with the following bounds:
\begin{equation}
	\begin{aligned}
		&	\Delta_{\mathrm{t}} \geq \bar{\Delta} - 2 \| \bar{H} - H_{\mathrm{t}} \|, \\
		&	\| \ket{\bar{\Omega}} - \ket{\Omega_{\mathrm{t}}} \| \leq \frac{\| \bar{H} - H_{\mathrm{t}} \|}{\bar{\Delta} - 4 \| \bar{H} - H_{\mathrm{t}} \|}.
	\end{aligned}
\end{equation}

The third step in Eq.~\eqref{three_step_truncation} involves applying multi-energy truncations within all blocks---not just at the boundaries---in order to control the operator norm of the truncated Hamiltonian. For each intra-block Hamiltonian $\{h_s\}_{s = 0}^{q+1}$, we truncate its spectrum by projecting onto the subspace of eigenvalues less than or equal to $\tau_s := E_{s,0} + \tau$, where $E_{s,0}$ denotes the ground-state energy of $h_s$ and $\tau$ is a tunable cutoff parameter.
Let $\Pi_{\leq \tau_s}^{(s)}$ be the corresponding spectral projection for block $B_s$, and define the total projection as
\begin{equation}
	\tilde{\Pi} := \bigotimes_{s = 0}^{q + 1} \Pi_{\leq \tau_s}^{(s)}.
\end{equation}
The fully truncated Hamiltonian is then given by $\tilde{H}_{\mathrm{t}} := \tilde{\Pi} H_{\mathrm{t}} \tilde{\Pi}$.
At this stage, we must also truncate the boundary blocks $B_0$ and $B_{q+1}$, which still remained unbounded. Ensuring the preservation of the ground state and spectral gap under this truncation is technically the most delicate part of the analysis.

A major difficulty arises from the fact that the unbounded Hamiltonian $H_{\mathrm{t}}$ renders the imaginary-time evolution $e^{\beta H_{\mathrm{t}}} O e^{-\beta H_{\mathrm{t}}}$ ill-defined due to divergence. 
For bounded local Hamiltonians, robustness of the ground state under truncation is typically proven by controlling imaginary-time evolution, bounding expressions of the form $\norm{e^{\beta H_{\mathrm{t}}} O e^{-\beta H_{\mathrm{t}}}}$ via convergent expansions in multi-commutators~\cite{Arad_2016}. 
This approach crucially relies on the fact that  $\norm{\mathrm{ad}_{H_{\mathrm{t}}}^{m}(O)}$ grows at most factorially (or more generally, in a way that still yields a finite bound after summing $\sum_{m} \beta^{m} \norm{\mathrm{ad}_{H_{\mathrm{t}}}^{m}(O)} / m!$).
In bosonic systems, however $H_{\mathrm{t}}$ is unbounded, and multi-commutators exhibit super-factorial growth of the form $\norm{\mathrm{ad}_{H_{\mathrm{t}}}^{m}(O)} \lesssim (T_{m} m)^{m}$ with $T_{m}$ increasing in $m$, so the imaginary-time dressing $e^{\beta \mathrm{ad}_{H_{\mathrm{t}}}}(O)$ fails to be well controlled and may diverge.
To overcome this difficulty, we follow the alternative strategy of Ref.~\cite{Anshu_2021}, which avoids imaginary-time evolution altogether.
Instead of using $e^{\beta H_{\mathrm{t}}}$, the analysis employs a polynomial energy filter $(E + \theta - H_{\mathrm{t}})^{m}$, where $\theta > 0$ denotes the energy separation between the low-energy sector $\Pi_{\mathrm{t}, \leq E}$ and the excited sector under consideration. 
Expanding this filter in multi-commutators and optimizing over the degree $m$ yields explicit bounds on matrix elements of the form $\norm{\Pi_{\mathrm{t},[E + \theta, E + \theta + T_{0}]} O \Pi_{\mathrm{t},\leq E}}$.
These bounds decay subexponentially in $\theta$.
They are sufficient to show that the multi-energy truncated projector $\tilde{\Pi}$ captures almost all of the ground-state weight.
Consequently, the effective Hamiltonian $\tilde{H}_{\mathrm{t}} = \tilde{\Pi} H_{\mathrm{t}} \tilde{\Pi}$ preserves the ground state and the spectral gap up to a subexponentially small error in the cutoff parameter $\tau$.

While the preservation of ground-state properties under bounded Hamiltonians is well understood~\cite{Arad_2016}, the unbounded setting introduces significant analytical challenges.
To overcome this issue, we adopt an alternative approach developed in Ref.~\cite{Anshu_2021}.

This leads to a modified robustness theorem for the ground state and spectral gap of $\tilde{H}_{\mathrm{t}}$ (see Theorem~4 in the Supplementary Information):
\begin{equation}
	\tilde{\Delta}_{\mathrm{t}} \geq \Delta_{\mathrm{t}} - 3 \epsilon', \quad 
	\| \ket{\Omega_{\mathrm{t}}} - \ket{\tilde{\Omega}_{\mathrm{t}}} \| \leq \sqrt{ \frac{2 \epsilon'}{ \Delta_{\mathrm{t}} } },
\end{equation}
where
\begin{equation}
	\epsilon' = \mathcal{O} \left( 
	\exp\left[ 
	- q \left( \tau + \log^{\mathfrak{a}k/2}(q l) \right)^{1 / (1 + \mathfrak{a}k/2)} 
	\right] 
	\right). 
\end{equation}

We point out that Ref.~\cite{PhysRevA.108.042422} directly applied the robustness theorem from Ref.~\cite{arad2013area}---originally proved for bounded Hamiltonians---to the case of unbounded bosonic systems. However, such an application does not directly extend to the unbounded setting: the key technical assumptions and estimates used in the bounded case do not necessarily hold for unbounded Hamiltonians, which require different analytical tools (see Supplementary Note 10.B for a detailed discussion).
This is because the proof techniques for bounded Hamiltonians rely crucially on norm estimates and operator inequalities that can break down in the unbounded setting, necessitating a different analytical approach.
In our work, we resolve this issue by employing a framework tailored to unbounded Hamiltonians~\cite{Anshu_2021}, which allows us to rigorously control the ground state and spectral gap under energy truncation in our setup. As a consequence, the truncation error $\epsilon'$ in our setting exhibits subexponential decay in the energy cutoff $\tau$, in contrast to the exponential decay observed in bounded cases. This difference reflects the essential analytical distinction between bounded spin models and unbounded bosonic models and underscores the necessity of a refined treatment in the present setting.

Finally, with the effective Hamiltonian $\tilde{H}_{\mathrm{t}}$ at hand, we follow the same strategy as in Refs.~\cite{arad2013area,Kuwahara2020arealaw} to establish the area law by constructing an AGSP based on Chebyshev polynomials. The approximation error of the AGSP scales as $e^{-m\sqrt{\Delta / \| \tilde{H}_{\mathrm{t}} \|}}$ (see Lemma~27 in the Supplementary Information), where $m$ is the degree of the polynomial and $\Delta$ denotes the spectral gap. The corresponding Schmidt rank is also shown to remain within acceptable bounds (see Lemma~28). By optimizing all relevant parameters---including the boson-number cutoff profile, the number and size of the blocks, and the polynomial degree---we construct an AGSP that satisfies the desired approximation properties (see Lemma~24 and Proposition~29). Iterating this AGSP yields an upper bound on the entanglement entropy (Lemma~33) and establishes a quantitative bound on the Schmidt rank required to approximate the ground state (Proposition~30), thereby completing the proof of the entanglement area law for one-dimensional bosonic systems with long-range interactions.

\section*{Data availability}
The data that support the findings of this study are available within the paper and its Supplementary Information.

\def\bibsection{\section*{References}}

\bibliography{boson_area_law}

\section*{acknowledgments}

Both authors acknowledge the Hakubi projects of RIKEN.  T.K. is supported by JST PRESTO (Grant No. JPMJPR2116), Exploratory Research for Advanced Technology (Grant No. JPMJER2302), and JSPS Grants-in-Aid for Scientific Research (No. JP23H01099, JP24H00071), Japan.

\section*{Author contributions}

D.K. and T.K. contributed to the conceptualization, methodology, data analysis, and drafting of the manuscript.

\section*{Competing interests}

The authors declare no competing interests.

\section*{Additional information} 
\textbf{Supplementary information} is available for this paper.

\clearpage

\onecolumngrid 
\supponecolumnfootnotes

\renewcommand\thefootnote{*\arabic{footnote}}

\setcounter{page}{1}
\begin{center}
{\large \bf Supplementary Information for  ``Entanglement area law in interacting bosons from the Bose-Hubbard model to $\phi$4 theory and beyond''}\\
\vspace*{0.5cm}
Donghoon Kim$^{1}$ and Tomotaka Kuwahara$^{1,2,3}$ \\
\vspace*{0.3cm}

$^{1}${\small \it 
Analytical quantum complexity RIKEN Hakubi Research Team, \\ RIKEN Center for Quantum Computing (RQC), Wako, Saitama 351-0198, Japan
}

$^{2}${\small \it 
RIKEN Cluster for Pioneering Research (CPR), Wako, Saitama 351-0198, Japan
}

$^{3}${\small \it 
PRESTO, Japan Science and Technology (JST), Kawaguchi, Saitama 332-0012, Japan}
\end{center}

\supplementarytableofcontents
\startsupplementarycontents

\renewcommand\thefootnote{*\arabic{footnote}}

\setcounter{equation}{0}
\setcounter{figure}{0}

\renewcommand{\figurename}{Supplementary Figure}
\renewcommand{\theequation}{S.\arabic{equation}}

\renewcommand{\tablename}{Supplementary Table}
\renewcommand{\thetable}{\arabic{table}}

\renewcommand{\thesection}{Supplementary Note~\arabic{section}}

\newpage

\section{Set up and general notations}

\subsection{Setup}

Consider a quantum system on a $D$-dimensional lattice with $n$ sites, with $\Lambda$ representing the set of all sites. 
For any arbitrary partial set $X\subseteq \Lambda$, we denote the cardinality (number of sites in $X$) as $|X|$. The complementary subset of $X$ is denoted by $X^\co := \Lambda \setminus X$.   
For subsets $X$ and $Y$ of $\Lambda$, the distance $\dist_{X,Y}$ is defined as the shortest path length on the graph connecting $X$ and $Y$, with $\dist_{X,Y}=0$ if $X\cap Y \neq \emptyset$. When $X$ comprises only one element (i.e., $X=\{i\}$), we use $\dist_{i,Y}$ to represent $\dist_{\{i\},Y}$ for simplicity. 
We also define $\diam(X)$ as follows: 
\begin{align}
	\diam(X) := 1 + \max_{i,i' \in X} (\dist_{i,i'}).
\end{align}
The surface subset of $X$ is denoted by 
\begin{align}
	\partial X := \{ i\in X \, | \, \dist_{i,X^\co}=1\} . \label{surface_definition_partial_X}
\end{align} 

For a subset $X\subseteq \Lambda$, the extended subset $\bal{X}{r}$ is defined as
\begin{align}
	\bal{X}{r} := \{i\in \Lambda \, | \, \dist_{X,i} \le r \}, \label{def:bal_X_r}
\end{align}
where $\bal{X}{0} = X$, and $r$ is an arbitrary positive number (i.e., $r\in \mathbb{R}^+$). We introduce a geometric parameter $\gamma$ determined solely by the lattice structure, satisfying $\gamma = \orderof{1}$, which fulfills the inequalities:
\begin{align}
	\max_{i \in \Lambda} \br{|\partial i[r] |} \le \gamma (r^{D-1} +1), \quad \max_{i \in \Lambda} |i[r]| \le  \gamma (r^D +1), \label{def:parameter_gamma}
\end{align}
where $r \ge 1$.

\subsection{Boson operators}

We define $b_i$ and $b_i^\dagger$ as the annihilation and the creation operators of a boson, respectively.
We also define $\nb_i$ as the number operator of bosons on a site $i \in \Lambda$, namely $\nb_i\coloneqq b_i^\dagger b_i$.
For an arbitrary set $\mathbf{N}=\{N_i\}_{i\in \Lambda}$, we define the Mott state $\ket{\mathbf{N}}$ as such 
\begin{align}
	\nb_i \ket{\mathbf{N}}=N_i  \ket{\mathbf{N}} \for \forall i\in \Lambda. 
\end{align}

We adopt the notation of $\Pi_{i,N}$ ($i\subseteq \Lambda$) as the projection onto the eigenspace of $\nb_i$ with the eigenvalue $N$:
\begin{align}
	\label{def_Pi_X_q}
	\nb_i \Pi_{i,N} = N \Pi_{i,N}.
\end{align}
We also define $\Pi_{i,\ge N}$ ($\Pi_{i, < N}$) as the projection operator onto the space such that the boson number at the site $i$ is larger than (smaller than) $N$: 
\begin{align}
	\label{Pi_X_I_definition}
	\Pi_{i, \ge N} =\sum_{N_1 \ge N} \Pi_{i,N_1},\quad  \Pi_{i, < N}=  \sum_{N_1 < N} \Pi_{i, N_1}.  
\end{align}
Here, for an arbitrary subset $X\subseteq \Lambda$, we denote $\Pi_{X, \ge N}$ ($\Pi_{X, < N}$) as 
\begin{align}
	\Pi_{X, \ge N}= \bigotimes_{i\in X} \Pi_{i, \ge N} ,\quad \Pi_{X, < N}= \bigotimes_{i\in X} \Pi_{i, < N} . 
\end{align}

\subsection{General bosonic Hamiltonians}

We often decompose the Hamiltonian in the form of 
\begin{align}
	\label{Boson_most_general_k-local}
	H= \sum_{Z:Z\subset \Lambda} h_Z   ,
\end{align}
where $h_Z$ composed of the boson operators $\{b_i\}_{i\in Z}$ and $\{b^\dagger_i\}_{i\in Z}$.
We adopt the notation of the subset Hamiltonian $H_X$ as follows:
\begin{align}
	\label{Boson_most_general_k-local_subset}
	H_X= \sum_{Z:Z\subseteq X} h_Z   .
\end{align}
We also denote $\widehat{H_X}$ by 
\begin{align}
	\label{Boson_most_general_k-local_subset_wide}
	\widehat{H_X}= \sum_{Z:Z\cap X\neq \emptyset} h_Z   ,
\end{align}
which gives 
\begin{align}
	H= \widehat{H_X} + H_{X^\co} . 
\end{align}

We introduce the parameters $g$ and $k$ as follows:
\begin{definition}[Parameters $g$ and $k$]
	Under the decomposition of Eq.~\eqref{Boson_most_general_k-local}, we define $g$ and $k$ by the constants that do not depend on the system size $|\Lambda|$ and satisfy 
	\begin{align}
		\label{Boson_most_general_k-local_g_ext}
		\max_{i: i\in \Lambda} \sum_{Z:Z\ni i} \norm{ h_Z \Pi_{\Lambda,\le N} } \le g N^{k/2} ,
	\end{align}
	where $g$ is an $\orderof{1}$ constant. 
\end{definition}

Also, we define the ground state $\ket{\Omega}$ and the spectral gap $\Delta$ as follows:
\begin{definition}[Ground state]
	We define the state $\ket{\Omega}$ as the minimum energy state:
	\begin{align}
		\bra{\Omega} H \ket{\Omega} \le \bra{\psi} H \ket{\psi} 
	\end{align}
	for an arbitrary quantum state $\ket{\psi}$. 
	We define the spectral gap $\Delta$ as the energy difference between the ground state and the first excited state:
	\begin{align}
		\Delta \le \bra{\psi_\bot} H \ket{\psi_\bot}  - \bra{\Omega} H \ket{\Omega} 
	\end{align}
	for an arbitrary quantum state $\ket{\psi_\bot}$ such that $\langle \Omega\ket{\psi_\bot}=0$. 
	In particular,  if the ground state is degenerate, we let $\Delta=0$.  
\end{definition}

{~} \\

{~} \\

\part{Boson number distribution}

\section{Bose-Hubbard classes and $\phi4$ class}

\subsection{Assumption for the Bose-Hubbard class} \label{sec:Bose-Hubbard class}

To treat the bosonic Hamiltonian, we have to restrict the class of the Hamiltonian $H$ in Eq.~\eqref{Boson_most_general_k-local}.  
We consider the most general form of the interacting boson systems up to $k$th degree:
\begin{align}
	H= \mF_H (\mathbf{b},\mathbf{b}^\dagger) ,
\end{align}
where $\mF_H (\mathbf{b},\mathbf{b}^\dagger) $ is an arbitrary $k$-degree polynomial of $\mathbf{b}=\{b_i\}_{i\in \Lambda}$ and $\mathbf{b}^\dagger=\{b^\dagger_i\}_{i\in \Lambda}$.
We decompose the term that commutes with the boson number operator $\{\nb_i\}_{i\in \Lambda}$ and rewrite the above form as 
\begin{align}
	\label{Boson_most_general}
	H= H_0(\mathbf{b},\mathbf{b}^\dagger)+ V_+(\mathbf{n}) + \sum_{i\in \Lambda} U_i \nb_i^{k/2}, 
\end{align}
where $U_i>0$ for $\forall i\in \Lambda$, $H_0(\mathbf{b},\mathbf{b}^\dagger)$ is a $k$-degree polynomial of $\mathbf{b}=\{b_i\}_{i\in \Lambda}$ and $\mathbf{b}^\dagger=\{b^\dagger_i\}_{i\in \Lambda}$, and $V_+(\mathbf{n})$ is a positive $(k/2)$-degree polynomial function with respect to $\mathbf{n}=\{\nb_i\}_{i\in \Lambda}$. 
Here, the Hamiltonian $H_0(\mathbf{b},\mathbf{b}^\dagger)$ consists of operators that have negative eigenvalues (e.g., $-n_in_j$, $b_ib_j^\dagger+ {\rm h.c.}$, $n_ib_j b_k+ {\rm h.c.}$ and so on). 
To consider the non-trivial boson-boson interactions, we treat the case of $k\ge 3$. 
The case of $k=2$ is the so-called bilinear Hamiltonian, and its properties, including the entanglement area law, have been extensively investigated~\cite{PhysRevA.73.012309}. 

Most of the interesting boson models are reduced to Eq.~\eqref{Boson_most_general} by appropriately choosing the degree $k$\footnote{Typically, it is enough to consider $k=4$.}.
For example, by letting $k=4$ and choosing in Eq.~\eqref{Boson_most_general} as 
\begin{align}
	\label{Bose_hubbard_example}
	&H_0(\mathbf{b},\mathbf{b}^\dagger)\to  \sum_{\ave{i,j}} J_{i,j}  \br{ b_i b_j^{\dagger} + {\rm h.c.} } - U \sum_{i\in \Lambda} \nb_i  , \quad V_+(\mathbf{n})\to 0, \quad U_i \to  U  , 
\end{align}
we obtain the Bose-Hubbard model.

{\bf Remark.} In the previous studies~\cite{Tong2022provablyaccurate,PhysRevA.108.042422}, boson models such that $H_0(\mathbf{b},\mathbf{b}^\dagger)$ is linear function with respect to $\mathbf{b},\mathbf{b}^\dagger$ are considered~[see Ineq. (6a), (6b) and (6c) in Ref.~\cite{Tong2022provablyaccurate}]\footnote{More precisely, the interaction between bosons and fermions, e.g., $b_i (c_i^{\dagger} c_i)+{\rm h.c.}$ with $c_i,c_i^\dagger$ the fermion operators at the site $i\in \Lambda$, can be included.}.
Hence, we cannot apply the method to a standard boson hopping operator like $b_ib_j^\dagger$ and the $\phi^4$ terms in Eq.~\eqref{phi4_Hamiltonian_example}, which has been thought to be a significant and challenging open question.

We need to exclude the possibility that the ground state is not well-defined in the thermodynamic limit, where an infinite number of bosons may accumulate on a single site. 
It indeed happens in the standard Bose-Hubbard model~\eqref{Bose_hubbard_example} when the on-site boson-boson interaction is attractive.
We, therefore, need an assumption to exclude the possibility. 

In the case where $H_0(\mathbf{b},\mathbf{b}^\dagger)$ is given by a polynomial up to $(k-1)$th order, the assumption of $U_i>0$ ($i\in \Lambda$) in Eq.~\eqref{Boson_most_general} is enough (see Assumption~\ref{assump:Repulsive condition} below).
However, when the repulsive and the attractive interactions have the same order, the condition becomes highly non-trivial.  
As a convenient notation, we first adopt the following definition:
\begin{definition} \label{Def:Bar_J_i_k}
	Let us pick up the $k_1$th order terms in $H_0(\mathbf{b},\mathbf{b}^\dagger)$ and denote them by
	\begin{align}
		\label{H_0_k_1_vec_b_vec_b_dagger}
		H_0^{(k_1)}(\mathbf{b},\mathbf{b}^\dagger)= \sum_{i_1,i_2,\ldots, i_{k_1} \in \Lambda} \sum_{s=1}^{k_1+1} J^{(s)}_{i_1,i_2,\ldots,i_{k_1}} b_{i_{k_1}}^\dagger \cdots b_{i_{s+1}}^\dagger  b_{i_s}^\dagger b_{i_{s-1}}   \cdots  b_{i_2}b_{i_1}  ,
	\end{align}
	where the site indices $i_1,i_2,\ldots, i_{k_1}$ can be identical to each other\footnote{We note that any interaction forms can be expressed in the form of Eq.~\eqref{H_0_k_1_vec_b_vec_b_dagger}.}. 
	We then define the parameters $\bar{J}_{i,k_1}$ and $\bar{J}_{k_1}$ as 
	\begin{align}
		\label{bar_J_i_k_1}
		&\bar{J}_{i,k_1}:=  \sum_{\substack{i_1,i_2,\ldots, i_{k_1} \in \Lambda \\ \{ i_1,i_2,\ldots, i_{k_1}\} \ni i}} \sum_{s=1}^{k_1} 
		\abs{J^{(s)}_{i_1,i_2,\ldots,i_{k_1}}} , \notag \\
		&\bar{J}_{k_1} :=  \max_{i\in \Lambda}(\bar{J}_{i,k_1}) .
	\end{align}
	We note if $H_0(\mathbf{b},\mathbf{b}^\dagger)$ includes up to $(k_1-1)$-degree, we have $\bar{J}_{i,k_1}=0$. 
\end{definition}

Using the parameter $\bar{J}_i$, we write the assumption as follows:
\begin{assump}[Repulsive condition] \label{assump:Repulsive condition}
	In the Hamiltonian~\eqref{Boson_most_general}, we call the interaction repulsive if 
	\begin{align}
		U_i >5 \bar{J}_{i,k}  \for \forall i\in \Lambda.
	\end{align}
	Note that the condition reduces to $U_i>0$ in the case where $H_0(\mathbf{b},\mathbf{b}^\dagger)$ is given by a $(k-1)$-degree polynomial.
\end{assump}
The specific numerical constant appearing here is chosen for technical convenience; the analysis depends only on the repulsive on-site term being dominant over the interaction terms, and different choices within this regime affect only the numerical bounds, not the qualitative physical conclusions.

Also, for the convenience of our analyses, 
we define a similar decomposition for $V_+(\mathbf{n})$ In a similar way to Def.~\ref{Def:Bar_J_i_k}:  
\begin{definition}
	The operator $V_+(\mathbf{n})$ is generally written as 
	\begin{align}
		V_+(\mathbf{n})=\sum_{k_1=1}^{k/2} \sum_{\substack{i_1,i_2,\ldots, i_{k_1} \in \Lambda} 
		}V_{+,i_1,i_2,\ldots,i_{k_1}} \nb_{i_1}\nb_{i_2} \cdots \nb_{i_{k_1}},  \quad V_{+,i_1,i_2,\ldots,i_{k_1}}>0.
	\end{align}
	We also define the parameter $\bar{v}$ as 
	\begin{align}
		\label{def_bar/v}
		&\bar{v}:=\sum_{k_1=1}^{k/2} \bar{v}_{k_1}, \notag \\
		&\bar{v}_{k_1}:=\max_{i\in \Lambda}  \sum_{\substack{i_1,i_2,\ldots, i_{k_1} \in \Lambda \\ \{ i_1,i_2,\ldots, i_{k_1}\} \ni i}} 
		V_{+,i_1,i_2,\ldots,i_{k_1}} .
	\end{align}
\end{definition}

\subsection{Assumption for the $\phi 4$ class}

In the $\phi4$-class of the interacting boson systems, we consider the following $\phi$ operator and $\pi$ operator:
\begin{align}
	\phi = \frac{1}{\sqrt{2}} (b + b^\dagger),\quad \pi= \frac{-i}{\sqrt{2}} (b - b^\dagger),
	\label{phi_pi_definition}
\end{align}
where we omit the site index. 
From the bosonic commutator relation as $[b, b^\dagger] = 1$, 
the field operator $\phi$ and its conjugate momentum $\pi$ satisfy the following canonical commutation relation:
\begin{align}
	[\phi, \pi] = i .
\end{align}

Using the notation $\phi$ and $\pi$, we consider the following form of the Hamiltonian:
\begin{align}
	&H = \sum_{i\in \Lambda} \mu_i \pi_i^2 + \mF(\boldsymbol{\phi}) ,  
	\label{phi_4_type_model}
\end{align}
where $\mF(\boldsymbol{\phi})$ is an arbitrary $k$-degree function of $\boldsymbol{\phi}=\{\phi_i\}_{i\in \Lambda}$. 
We denote the upper bound of $\abs{\mu_i}$ by $\bar{\mu}$: 
\begin{align}
	\label{def_bar_mu_phi}
	\bar{\mu} := \max_{i\in \Lambda} (\abs{\mu_i} ) .
\end{align}
The representative case is the $\phi4$ Hamiltonian as follows:
\begin{align}
	\label{phi4_Hamiltonian_example}
	&H = \sum_{i\in \Lambda}\br{\pi_i^2 + \phi_i^2 + \lambda \phi_i^4} +\gamma \sum_{\ave{i,i'} }   \phi_i \phi_{i'}
\end{align}
By replacing $\phi_i$ and $\pi_i$ using Eq.~\eqref{phi_pi_definition}, the Hamiltonian becomes:
\begin{align}
	H = \sum_i \left[2 b_i^\dagger b_i  + 1+ \frac{\lambda}{4}(b_i + b_i^\dagger)^4 \right] + 
	\frac{ \gamma}{2}\sum_{\langle i,i' \rangle} (b_i + b_i^\dagger)(b_{i'} + b_{i'}^\dagger) . 
\end{align}
As a significant problem, we can see that Assumption~\ref{assump:Repulsive condition} is not satisfied in the $\phi 4$ class.
To see the point, we look at the highest-order term $(b_i + b_i^\dagger)^4$, whose ground state has infinite boson density. 
Hence, we can prove that there exists a product state $\ket{\psi}=\bigotimes_{i\in \Lambda}\ket{\mu_i}$ such that 
\begin{align}
	\bra{\psi}(b_i + b_i^\dagger)^4\ket{\psi} \approx 0 , \quad {\rm but} \quad \bra{\psi}\nb_i \ket{\psi} \gg1 . 
\end{align}
This means that the operator $(b_i + b_i^\dagger)^4$ does not satisfy the repulsive-interaction condition as in Assumption~\ref{assump:Repulsive condition}. 

We need a qualitatively different approach to the $\phi4$ class. 
Significantly, we only need the following simple condition for the Hamiltonian class as in Eq.~\eqref{phi_4_type_model}. 
\begin{assump}[Parity symmetry] \label{assump:Parity symmetry}
	We assume the Hamiltonian is invariant under $\boldsymbol{\phi} \to -\boldsymbol{\phi}$. This is satisfied when $ \mF(\boldsymbol{\phi})$ is given by an even function as follows:
	\begin{align}
		\label{mF_phi_explicit}
		\mF(\boldsymbol{\phi})=\sum_{k_1=1}^{k/2} \sum_{i_1,i_2,\ldots, i_{2k_1} \in \Lambda} f_{i_1,i_2,\ldots, i_{2k_1} } \phi_{i_1}\phi_{i_2} \cdots \phi_{i_{2k_1}} ,
	\end{align}
	where $k$ is an even integer, and $i_1,i_2,\ldots, i_{2k_1}$ can be identical to each other.  
	Notably, we do not need any additional constraints on the coefficients $\{f_{i_1,i_2,\ldots, i_{2k_1}}\}$, such as the repulsive interactions. 
\end{assump}

{\bf Remark.} The assumption is automatically satisfied for the $\phi4$ Hamiltonian~\eqref{phi4_Hamiltonian_example}.
Hence, only the gap condition excludes the possibility of the boson concentration on a single site. 
By imposing a similar assumption to Assumption~\ref{assump:Repulsive condition} for $\boldsymbol{\phi}$ operators, we will be able to extend the theory to systems with parity violation. 
Throughout the paper, we only consider the case of $k\ge 2$ in Eq.~\eqref{mF_phi_explicit}, which gives non-trivial behaviours.

We stress that Assumption 2 (parity symmetry) is essential for our area-law proof; without it, the area law can fail even in gapped, local bosonic systems.
If odd powers of $\phi$ are allowed, the potential need not be centered at the origin and may instead develop a minimum at large field values.
In such a situation, bosonic degrees of freedom can macroscopically condense around this minimum, leading to a large expectation value of the local boson number.
This condensation strongly induces anisotropic fluctuations among collective modes and can generate arbitrarily large bipartite entanglement, thereby obstructing any uniform entanglement bound.

To illustrate the role of parity symmetry explicitly, we consider the following two-site bosonic model whose Hamiltonian contains odd powers of $\phi$:
\begin{align}
	H = \pi_1^2+\pi_2^2 + \left(\epsilon \phi_{1}^{4} - \phi_{1}^{3}\right) + \left(\epsilon \phi_{2}^{4} - \phi_{2}^{3}\right) + (\phi_1 - \phi_2)^2 (\phi_1^2 + \phi_2^2), \qquad \epsilon>0. \label{eq:odd_parity_H}
\end{align}
As before, the canonical variables are defined as
\begin{align}
	\phi_i=\frac{1}{\sqrt2} \br{b_i+b_i^\dagger}, \qquad \pi_i = \frac{-i}{\sqrt2} \br{b_i-b_i^\dagger}, \qquad [\phi_i,\pi_j]=i\delta_{ij}, \qquad (i,j\in\{1,2\}). \label{eq:odd_parity_canonical}
\end{align}
The onsite potential
\begin{align}
	V(\phi) = \epsilon \phi^{4} - \phi^{3} \label{eq:odd_parity_potential}
\end{align}
explicitly breaks the parity symmetry $\phi\to-\phi$ and admits a nontrivial classical minimum determined by
\begin{align}
	V'(\phi) = \phi^2(4\epsilon\phi-3) = 0,
\end{align}
namely
\begin{align}
	\phi_0 = \frac{3}{4\epsilon}. \label{eq:odd_parity_phi0}
\end{align}
As $\epsilon$ decreases, the classical minimum $\phi_{0}$ shifts to larger field values, and bosonic degrees of freedom concentrate around this configuration.

Writing $\phi_i=\phi_0+x_i$ and expanding the Hamiltonian to quadratic order in the fluctuations $x_i$, one finds
\begin{align}
	H = E_{\mathrm{cl}}(\epsilon) + \pi_{1}^{2} + \pi_{2}^{2} + \frac{1}{2}V''(\phi_{0})(x_{1}^{2} + x_{2}^{2}) + 2 \phi_{0}^{2} (x_{1} - x_{2})^{2} + \mathcal{O}(x^3), 	\label{eq:odd_parity_Hquad}
\end{align}
with
\begin{align}
	V''(\phi_0)=\frac{9}{4\epsilon}, \qquad E_{\mathrm{cl}}(\epsilon) = - \frac{27}{128 \epsilon^3}. \label{eq:odd_parity_Ecl}
\end{align}
By defining symmetric and antisymmetric normal modes, $x_{\pm} = \frac{x_{1} \pm x_{2}}{\sqrt2}$ and $p_{\pm} = \frac{\pi_{1} \pm \pi_{2}}{\sqrt{2}}$, we can rewrite the quadratic Hamiltonian as
\begin{align}
	H = E_{\mathrm{cl}}(\epsilon) + \br{p_{+}^{2} + \Omega_{+}^{2} x_{+}^{2}} + \br{p_{-}^{2} + \Omega_{-}^{2} x_{-}^{2}} + \mathcal{O}(x^3), \label{eq:odd_parity_Hmodes}
\end{align}
where the normal-mode frequencies are
\begin{align}
	\Omega_{+}^{2} = \frac{9}{8\epsilon}, \qquad \Omega_{-}^{2} = \frac{9(\epsilon+2)}{8\epsilon^2}. \label{eq:odd_parity_Omegas}
\end{align}
As $\epsilon \rightarrow 0$, these frequencies scale as
\begin{align}
	\Omega_{+} \sim \epsilon^{-1/2}, \qquad \Omega_{-} \sim \epsilon^{-1}, \label{eq:odd_parity_Omegas_asymp}
\end{align}
showing a parametrically large separation of energy scales.

This separation controls the low-energy properties of the system. 
At quadratic order, the ground-state energy is
\begin{align}
	E_{0}(\epsilon) \simeq E_{\mathrm{cl}}(\epsilon) + \Omega_{+}(\epsilon) + \Omega_{-}(\epsilon) = - \frac{27}{128 \epsilon^3} + \sqrt{\frac{9}{8 \epsilon}} + \sqrt{\frac{9(\epsilon + 2)}{8 \epsilon^{2}}}, \label{eq:odd_parity_E0}
\end{align}
while the lowest excitation gap is set by the softer symmetric mode,
\begin{align}
	\Delta(\epsilon) \simeq 2\Omega_+(\epsilon) = \frac{3}{\sqrt{2 \epsilon}}, \label{eq:odd_parity_gap}
\end{align}
so the system remains gapped even for small $\epsilon$.
Finally, the same mode anisotropy determines the bipartite entanglement structure.
Within the leading quadratic theory, tracing out one site yields a symplectic eigenvalue
\begin{align}
	\nu = \frac{1}{4} \sqrt{2 + \frac{\Omega_{-}}{\Omega_{+}} + \frac{\Omega_{+}}{\Omega_{-}}} \sim \frac{1}{4} \sqrt{\frac{\Omega_{-}}{\Omega_{+}}} \sim \epsilon^{-1/4}, \label{eq:odd_parity_nu}
\end{align}
which implies a logarithmically growing bipartite entanglement entropy between the two sites~\cite{PhysRevA.63.032312}:
\begin{align}
	S(\rho_1) = \left(\nu+\frac{1}{2}\right)\log\left(\nu+\frac{1}{2}\right) - \left(\nu - \frac{1}{2}\right) \log \left(\nu-\frac{1}{2}\right) \sim \frac{1}{4} \log  \left(\frac{1}{\epsilon}\right) + \mathcal{O}(1). \label{eq:odd_parity_entropy}
\end{align}
The analytical predictions above are supported by direct numerical simulations of the Hamiltonian~\eqref{eq:odd_parity_H}; see Supplementary Figure~\ref{fig_boson_phi3}.

This demonstrates explicitly that allowing odd-power terms enables boson concentration at large-field minima, leading to parametrically large entanglement even in a local and gapped system.
Parity symmetry excludes this mechanism by forbidding such terms and is therefore essential for establishing an entanglement area law.

\begin{figure}
	\begin{center}
		\includegraphics[width=\textwidth]{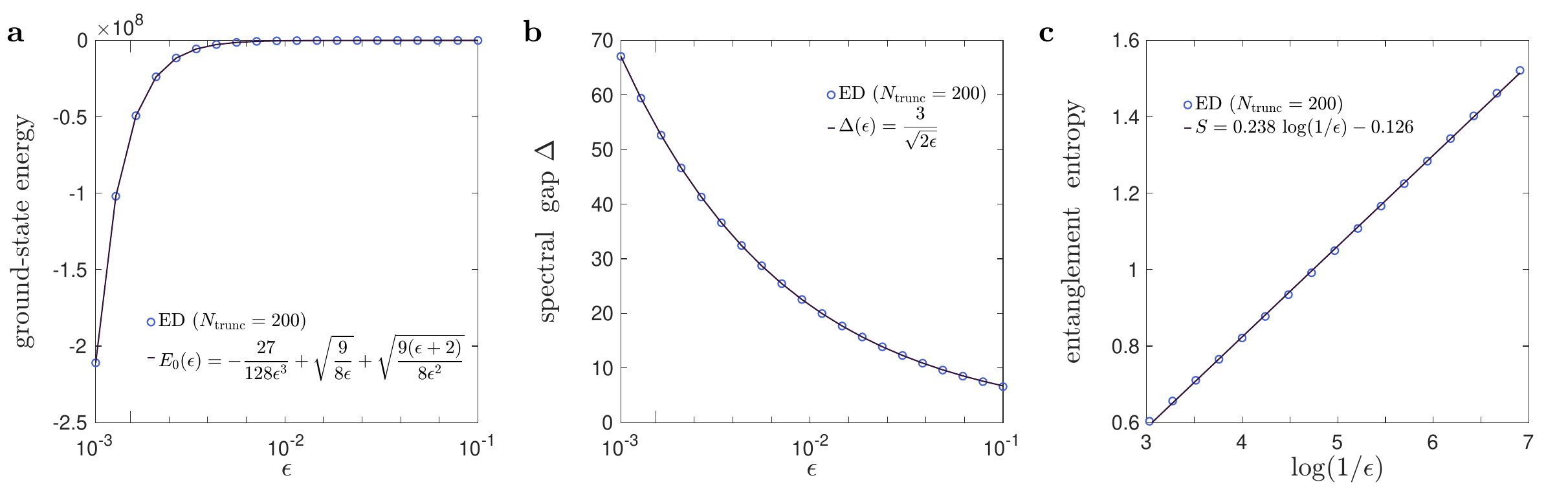}
	\end{center}
	\caption{Numerical results for the two-site bosonic Hamiltonian in Eq.~\eqref{eq:odd_parity_H}. The Hamiltonian is diagonalized in a displaced harmonic-oscillator basis defined around the classical minimum of the onsite potential, with the local Hilbert space at each site truncated to dimension $N_{\mathrm{trunc}} = 200$.
		Panel $\mathbf{a}$ shows the ground-state energy, and panel $\mathbf{b}$ shows the spectral gap between the first excited state and the ground state; in both cases, solid black lines indicate the leading-order predictions from the quadratic expansion around the classical minimum.
		Panel $\mathbf{c}$ shows the bipartite entanglement entropy between site 1 and site 2; the solid line represents a linear fit to the numerical data, exhibiting a logarithmic scaling with a slope close to $1/4$, as predicted by Eq.~\eqref{eq:odd_parity_entropy}.}
	\label{fig_boson_phi3}
\end{figure}

For the convenience of our analyses, we define the parameter $\bar{f}$ as 
\begin{align}
	\label{Definition/of_bar_f}
	\bar{f}= \max_{i\in \Lambda} \br{ 
		\sum_{k_1=1}^{k/2} \sum_{\substack{ i_1,i_2,\ldots, i_{2k_1} \in \Lambda \\ \{i_1,i_2,\ldots, i_{2k_1}\} \ni i}} \abs{ f_{i_1,i_2,\ldots, i_{2k_1} } } }.
\end{align}

\section{Maximum moment bounds for Bose-Hubbard class} \label{sec:Maximum moment bounds for Bose-Hubbard class}

In this section, we mainly treat the Bose-Hubbard class and aim to derive an upper bound for the $(k/2)$th order moment.
The result in this section plays a key role in proving the concentration bound on the boson number distribution in the subsequent section (\ref{sec:Boson number distribution for Bose-Hubbard class}). 
A schematic overview of the proof is presented in Supplementary Figure~\ref{fig_supp_BH_boson_number}.
In the following, we often separately consider the cases of i) the total boson number is not conserved and ii) the total boson number is conserved. 
In case (ii), we fix the total boson number to be $N$ and make the ground state dependent on the boson number.
The ground state and the ground energy are described by $\ket{\Omega_N}$ and $E_0(N)$, respectively.  

\begin{figure}
	\begin{center}
		\includegraphics[width=\textwidth]{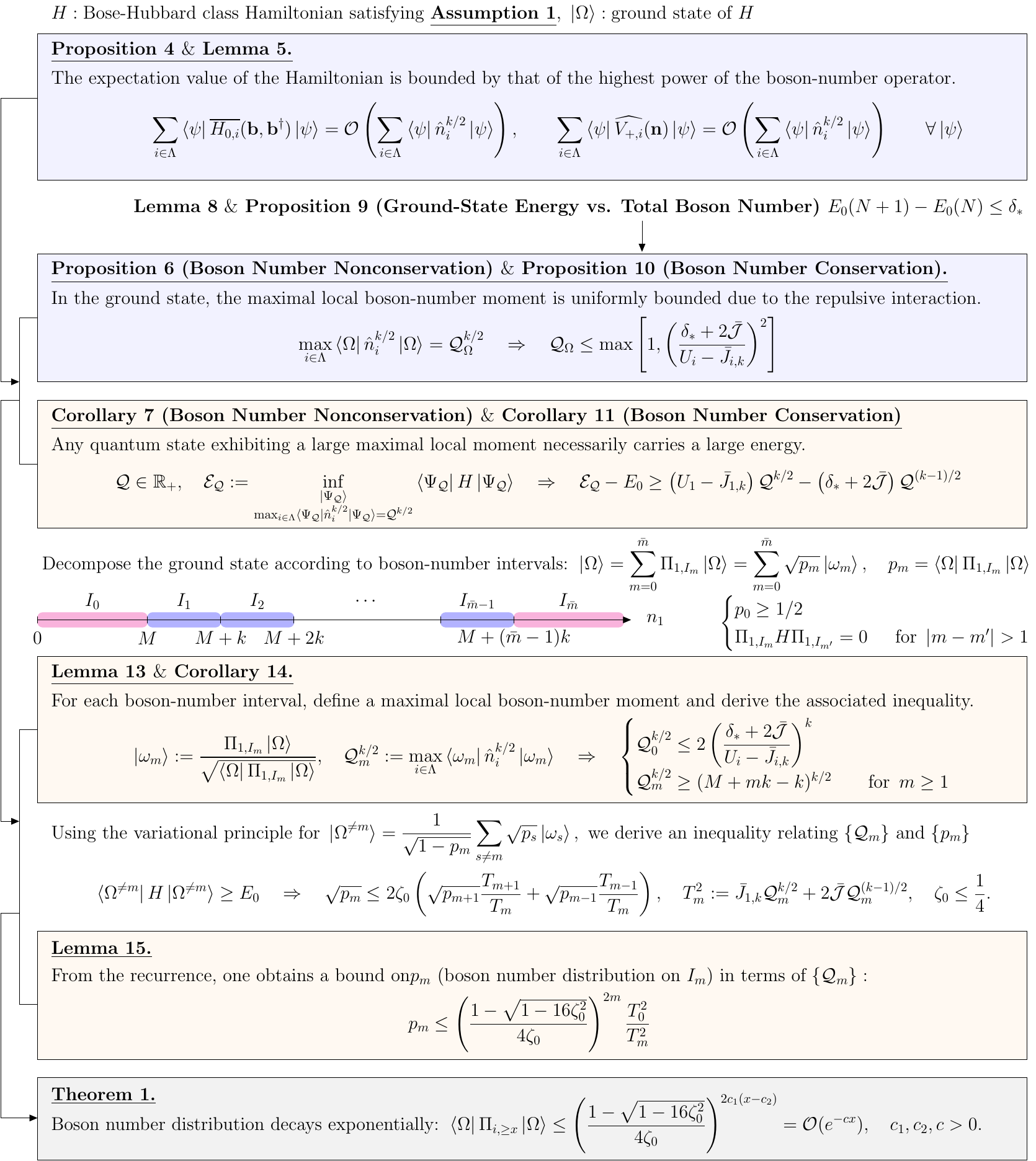}
	\end{center}
	\caption{This figure outlines the proof of Theorem 1, which establishes the exponential decay of the boson-number distribution for Bose--Hubbard class Hamiltonians.}
	\label{fig_supp_BH_boson_number}
\end{figure}

\subsection{Preliminary}
In this section, we show several essential ingredients for the proof.

We first prove that the ground energy of the subset Hamiltonian $H_{X}$ is no smaller than that of $H_{\Lambda}$.
In general, we prove the following lemma:
\begin{lemma} \label{lem:subset_ground_energy}
	For arbitrary two subsets $X$ and $\bar{X}$ such that $X \subseteq \bar{X}$, we have 
	\begin{align}
		E_{0,X}\ge E_{0,\bar{X}} ,
		\label{lem:subset_ground_energy_main_ineq_uncons}
	\end{align}
	where $E_{0,X}$ is the ground energy of $H_X$. 
	
	In the case where the Hamiltonian conserves the total boson number, we consider the $N$ dependence of the ground energy $E_{0,X}(N)$ for the subset Hamiltonian $H_X$ ($X\subseteq \Lambda$) with the total boson number equal to $N$. 
	We obtain 
	\begin{align}
		E_{0,X}(N) \ge E_{0,\bar{X}}(N) .
		\label{lem:subset_ground_energy_main_ineq_0}
	\end{align}
\end{lemma}

\textit{Proof of Lemma~\ref{lem:subset_ground_energy}.}
We consider the case where the boson number is conserved, but the same proof applies to the case where it is not.  
The proof is immediately followed by considering the quantum state as $\ket{\Phi}_{\bar{X}}= \ket{\Omega_N}_{X} \otimes \ket{0}_{\bar{X}\setminus X}$, where $\ket{\Omega_N}_{X}$ is the ground state of $H_X$. 
We then obtain 
\begin{align}
	\bra{\Phi} H_{\bar{X}} \ket{\Phi}_{\bar{X}}= \bra{\Omega_N} H_{X} \ket{\Omega_N}_{X}=E_{0,X}(N) .
\end{align}
On the other hand, we always obtain
\begin{align}
	\bra{\Phi} H_{\bar{X}} \ket{\Phi}_{\bar{X}} \ge \bra{\Omega_N} H_{\bar{X}} \ket{\Omega_N}_{\bar{X}} = E_{0,\bar{X}}(N). 
\end{align}
By combining the above two inequalities, we obtain the main inequality~\eqref{lem:subset_ground_energy_main_ineq_0}.
This completes the proof. $\square$

{~}

\hrulefill{\bf [ End of Proof of Lemma~\ref{lem:subset_ground_energy}]}

{~}

The second proposition plays an essential role in analyzing the Hamiltonian $H_0(\mathbf{b},\mathbf{b}^\dagger)$. 

\begin{prop} \label{prop_1_site_upper_bound}
	Let $\ket{\psi}$ be an arbitrary quantum state. 
	We also consider the operators of 
	\begin{align}
		\label{overline_H_0_i_vec}
		\overline{H_{0,i}}(\mathbf{b},\mathbf{b}^\dagger)= \sum_{\substack{i_1,i_2,\ldots, i_{k_1} \in \Lambda \\ \{ i_1,i_2,\ldots, i_{k_1}\} \ni i}}  \sum_{s=1}^{k_1+1} \abs{J^{(s)}_{i_1,i_2,\ldots,i_{k_1}}}\cdot \abs{b_{i_{k_1}}^\dagger \cdots b_{i_{s+1}}^\dagger  b_{i_s}^\dagger b_{i_{s-1}}   \cdots  b_{i_2}b_{i_1}}  ,
	\end{align}
	and 
	\begin{align}
		\widehat{V_{+,i}}(\mathbf{n})=\sum_{k_1=1}^{k/2}  \sum_{\substack{i_1,i_2,\ldots, i_{k_1} \in \Lambda \\ \{ i_1,i_2,\ldots, i_{k_1}\} \ni i}}V_{+,i_1,i_2,\ldots,i_{k_1}} \nb_{i_1}\nb_{i_2} \cdots \nb_{i_{k_1}}  ,
	\end{align}
	where $\abs{O}$ is defined by $\sqrt{O^\dagger O}$ for an arbitrary operator.  
	Then, we obtain 
	\begin{align}
		\label{prop_1_site_upper_bound/main1}
		\sum_{i \in \Lambda} \bra{\psi} \overline{H_{0,i}}(\mathbf{b},\mathbf{b}^\dagger) \ket{\psi}\le  
		\sum_{i \in \Lambda}  \brr{ \bar{J}_{i,k}\bra{\psi} \nb_{i}^{k/2} \ket{\psi} + \bar{\mJ} \br{\bra{\psi} \nb_{i}^{(k-1)/2}\ket{\psi}+1}}
		,
	\end{align}
	and 
	\begin{align}
		\label{prop_1_site_upper_bound/main2}
		\sum_{i \in \Lambda} \bra{\psi} \widehat{V_{+,i}}(\mathbf{n}) \ket{\psi}\le  \bar{v} \sum_{i \in \Lambda}\bra{\psi}\nb_i^{k/2} \ket{\psi}  ,
	\end{align}
	where we have defined $\bar{v}$ in Eq.~\eqref{def_bar/v}.
	Also, we define the parameter $\bar{\mJ}$ as follows:
	\begin{align}
		\label{def_bar_mJ0}
		\bar{\mJ}:= \max_{i\in \Lambda}\brrr{ \sum_{k_1=1}^k \bar{J}_{i,k_1}\brr{1+ (2k_1)^{k_1}}}.
	\end{align}
\end{prop}

{\bf Remark.} The operator $\overline{H_{0,i}}(\mathbf{b},\mathbf{b}^\dagger)$ satisfies the operator inequality as 
\begin{align}
	\widehat{H_{0,i}}(\mathbf{b},\mathbf{b}^\dagger)\preceq   \overline{H_{0,i}}(\mathbf{b},\mathbf{b}^\dagger) ,
\end{align}
where we remind that the notation $\widehat{H_{0,i}}$ has been defined in Eq.~\eqref{Boson_most_general_k-local_subset_wide}. 
Also, the inequality~\eqref{prop_1_site_upper_bound/main1} implies the operator inequality as 
\begin{align}
	\sum_{i \in \Lambda} \overline{H_{0,i}}(\mathbf{b},\mathbf{b}^\dagger) \preceq 
	\sum_{i \in \Lambda}  \brr{ \bar{J}_{i,k} \nb_{i}^{k/2} + \bar{\mJ} \br{ \nb_{i}^{(k-1)/2}+1}}.
\end{align}
This is helpful to connect the Hamiltonian $H_0(\mathbf{b},\mathbf{b}^\dagger)$ with the moment functions with respect to the boson number operators. 

\subsubsection{Proof of Proposition~\ref{prop_1_site_upper_bound}}

We first upper-bound the expectation $\bra{\psi}\widehat{V_{+,i}}(\mathbf{n})\ket{\psi}$.
Using the H\"older inequality, we obtain 
\begin{align}
	\bra{\psi}\nb_{i_1}\nb_{i_2} \cdots \nb_{i_{k_1}} \ket{\psi}
	&\le \prod_{j=1}^{k_1} \br{\bra{\psi}\nb_{i_j}^{k_1} \ket{\psi}}^{1/k_1}\le \frac{1}{k_1} \sum_{j=1}^{k_1} \bra{\psi}\nb_{i_j}^{k_1} \ket{\psi} .
\end{align}
Using the above inequality, we can derive 
\begin{align}
	\label{psi_V_+i_psi_up}
	\sum_{i \in \Lambda} \bra{\psi}\widehat{V_{+,i}}(\mathbf{n})\ket{\psi}
	&\le \sum_{k_1=1}^{k/2} \sum_{i \in \Lambda} \sum_{\substack{i_1,i_2,\ldots, i_{k_1} \in \Lambda \\ \{ i_1,i_2,\ldots, i_{k_1}\} \ni i}}
	\frac{V_{+,i_1,i_2,\ldots,i_{k_1}}}{k_1}\sum_{j=1}^{k_1} \bra{\psi}\nb_{i_j}^{k_1} \ket{\psi}  \notag \\
	&=\sum_{k_1=1}^{k/2} \sum_{i \in \Lambda} \sum_{\substack{i_1,i_2,\ldots, i_{k_1} \in \Lambda \\ \{ i_1,i_2,\ldots, i_{k_1}\} \ni i}}V_{+,i_1,i_2,\ldots,i_{k_1}} \bra{\psi}\nb_i^{k_1} \ket{\psi} \notag \\
	& \le \sum_{k_1=1}^{k/2}    \bar{v}_{k_1} \sum_{i \in \Lambda}\bra{\psi}\nb_i^{k/2} \ket{\psi} ,
\end{align}
where we use the definition~\eqref{def_bar/v} for $\bar{v}_{k_1}$.
Note that $\bra{\psi}\nb_i^p \ket{\psi} \le \bra{\psi}\nb_i^{p'} \ket{\psi}$ for $p\le p'$.
We thus prove one of the main inequalities, i.e., \eqref{prop_1_site_upper_bound/main2}. 

For the proof of the inequality~\eqref{prop_1_site_upper_bound/main1}, we utilize the following basic lemma (the proof is given in \ref{sec:Proof of Lemma_lem:Upper_bound_hopping}.):
\begin{lemma}\label{lem:Upper_bound_hopping}
	For an arbitrary multiset of $\{i_1,i_2,\ldots, i_k\}$, we obtain the following operator inequality:
	\begin{align}
		\abs{ b_{i_k}^\dagger \cdots b_{i_{s+1}}^\dagger  b_{i_s}^\dagger b_{i_{s-1}}   \cdots  b_{i_2}b_{i_1}  } \preceq 
		\prod_{j=1}^k (\nb_{i_j}+k)^{1/2}.
		\label{lem:Upper_bound_hopping/main}
	\end{align}
	Also, for an arbitrary quantum state $\ket{\psi}$, we derive 
	\begin{align}
		\label{corol_upper_hopping/main/ineq}
		\abs{\bra{\psi}  b_{i_k}^\dagger \cdots b_{i_{s+1}}^\dagger  b_{i_s}^\dagger b_{i_{s-1}}   \cdots  b_{i_2}b_{i_1} \ket{\psi} }
		\le \frac{1}{k} \sum_{j=1}^k \bra{\psi} \nb_{i_j}^{k/2} \ket{\psi} + \frac{ (2k)^{k} }{k} \sum_{j=1}^k \bra{\psi} \nb_{i_j}^{k/2-1}  \ket{\psi}  + k^{k/2} , 
	\end{align}
\end{lemma}

By applying the above lemma to the definition of Eq.~\eqref{overline_H_0_i_vec} for $\overline{H_{0,i}}(\mathbf{b},\mathbf{b}^\dagger)$, we obtain 
\begin{align}
	\label{corol_upper_hopping/main/ineq_2}
	&\bra{\psi} \overline{H_{0,i}}(\mathbf{b},\mathbf{b}^\dagger) \ket{\psi} \notag \\
	&\le \sum_{k_1=1}^k \sum_{\substack{i_1,i_2,\ldots, i_{k_1} \in \Lambda \\ \{ i_1,i_2,\ldots, i_{k_1}\} \ni i}}  \sum_{s=1}^{k_1+1} \abs{J^{(s)}_{i_1,i_2,\ldots,i_{k_1}}}  
	\brr{\frac{1}{k_1} \sum_{j=1}^{k_1} \bra{\psi} \nb_{i_j}^{k_1/2} \ket{\psi} + \frac{ (2k_1)^{k_1} }{k_1} \sum_{j=1}^{k_1} \bra{\psi} \nb_{i_j}^{k_1/2-1}  \ket{\psi}  + k_1^{k_1/2}} .
\end{align}
To further upper-bound the RHS of the above inequality, we first use a similar inequality to~\eqref{psi_V_+i_psi_up} and derive the following three inequalities:
\begin{align}
	\sum_{i \in \Lambda} \sum_{\substack{i_1,i_2,\ldots, i_{k_1} \in \Lambda \\ \{ i_1,i_2,\ldots, i_{k_1}\} \ni i}}  \sum_{s=1}^{k_1+1} \abs{J^{(s)}_{i_1,i_2,\ldots,i_{k_1}}}  
	\frac{1}{k_1} \sum_{j=1}^{k_1} \bra{\psi} \nb_{i_j}^{k_1/2} \ket{\psi} 
	&=
	\sum_{i \in \Lambda} \sum_{\substack{i_1,i_2,\ldots, i_{k_1} \in \Lambda \\ \{ i_1,i_2,\ldots, i_{k_1}\} \ni i}}  \sum_{s=1}^{k_1+1} \abs{J^{(s)}_{i_1,i_2,\ldots,i_{k_1}}}  
	\bra{\psi} \nb_{i}^{k_1/2} \ket{\psi}  \notag \\
	&\le  \sum_{i \in \Lambda}\bar{J}_{i,k_1}  \bra{\psi} \nb_{i}^{k_1/2} \ket{\psi}  ,
\end{align}
\begin{align}
	\sum_{i \in \Lambda} \sum_{\substack{i_1,i_2,\ldots, i_{k_1} \in \Lambda \\ \{ i_1,i_2,\ldots, i_{k_1}\} \ni i}}  \sum_{s=1}^{k_1+1} \abs{J^{(s)}_{i_1,i_2,\ldots,i_{k_1}}}  
	\frac{ (2k_1)^{k_1} }{k_1} \sum_{j=1}^{k_1} \bra{\psi} \nb_{i_j}^{k_1/2-1}  \ket{\psi}
	&\le  \sum_{i \in \Lambda}(2k_1)^{k_1} \bar{J}_{i,k_1}  \bra{\psi} \nb_{i}^{k_1/2-1} \ket{\psi}  ,
\end{align}
and 
\begin{align}
	\sum_{i \in \Lambda} \sum_{\substack{i_1,i_2,\ldots, i_{k_1} \in \Lambda \\ \{ i_1,i_2,\ldots, i_{k_1}\} \ni i}}  \sum_{s=1}^{k_1+1} \abs{J^{(s)}_{i_1,i_2,\ldots,i_{k_1}}}   k_1^{k_1/2}
	&\le 
	\sum_{i \in \Lambda} \bar{J}_{i,k_1} k_1^{k_1/2},
\end{align}
where the parameter $ \bar{J}_{i,k_1}$ was defined in Eq.~\eqref{bar_J_i_k_1}. 

By combining all the above inequalities, we reach the desired upper bound~\eqref{prop_1_site_upper_bound/main1} as follows: 
\begin{align}
	&\sum_{i \in \Lambda} \bra{\psi} \overline{H_{0,i}}(\mathbf{b},\mathbf{b}^\dagger) \ket{\psi} \notag \\
	&\le
	\sum_{i \in \Lambda}\sum_{k_1=1}^k \bar{J}_{i,k_1} \br{\bra{\psi} \nb_{i}^{k_1/2} \ket{\psi} +  (2k_1)^{k_1} \bra{\psi} \nb_{i}^{k_1/2-1} \ket{\psi} +k_1^{k_1/2}} \notag \\
	&\le \sum_{i \in \Lambda}  \brrr{ \bar{J}_{i,k}\bra{\psi} \nb_{i}^{k/2} \ket{\psi} +  \sum_{k_1=1}^k \bar{J}_{i,k_1} \brr{1+  (2k_1)^{k_1}}
		\br{\bra{\psi}  \nb_{i}^{(k-1)/2}\ket{\psi} +1}}  \notag \\
	&\le \sum_{i \in \Lambda}  \brr{ \bar{J}_{i,k}\bra{\psi} \nb_{i}^{k/2} \ket{\psi} + \bar{\mJ} \br{\bra{\psi} \nb_{i}^{(k-1)/2}\ket{\psi}+1}},
\end{align}
where we use the definition~\eqref{def_bar_mJ0} of $\bar{\mJ}$, and the second inequality is derived from $\bra{\psi} \nb_{i}^p \ket{\psi}\le \bra{\psi} \nb_{i}^{p'} \ket{\psi}$ for $p\le p'$.
This completes the proof of Proposition~\ref{prop_1_site_upper_bound}. $\square$

\subsubsection{Proof of Lemma~\ref{lem:Upper_bound_hopping}.} \label{sec:Proof of Lemma_lem:Upper_bound_hopping}
We denote $\hat{B}_k$ by 
\begin{align}
	\hat{B}_k=  b_{i_k}^\dagger \cdots b_{i_{s+1}}^\dagger  b_{i_s}^\dagger b_{i_{s-1}}   \cdots  b_{i_2}b_{i_1}  . 
\end{align}
Because of 
\begin{align}
	\brr{ \nb_i, \hat{B}^\dagger _k  \hat{B}_k }=0 \for \forall i\in \Lambda,
\end{align}
we can describe 
\begin{align}
	| \hat{B}_k|^2 = \hat{B}_k^\dagger \hat{B}_k  = f_{k}(\mathbf{n}), 
\end{align}
where $f_{k}(\mathbf{n})$ is an appropriate function with respect to $\mathbf{n}=\{\nb_i\}_{i\in \Lambda}$.
Our purpose is to prove 
\begin{align}
	\label{Upper_bound_hopping/induction}
	f_{k}(\mathbf{n}) \preceq \prod_{j=1}^k (\nb_{i_j}+k),
\end{align}
which also implies the inequality~\eqref{lem:Upper_bound_hopping/main} by taking the square root of the both sides. 

For $k=1$, the inequality~\eqref{Upper_bound_hopping/induction} trivially holds from $b_ib_i^\dagger=n_i+1$ and $b_i^\dagger b_i=\nb_i$.
We assume the target inequality up to a certain $k$ and prove the case of $k+1$.
We then generally consider the operators of
\begin{align}
	b_i^\dagger  \hat{B}_k^\dagger  \hat{B}_k  b_i ,\quad b_i \hat{B}_k^\dagger  \hat{B}_k  b_i^\dagger 
\end{align}
for $\forall i\in \Lambda$.
By using the assumption, we obtain 
\begin{align}
	\label{Upper_bound_hopping/induction_pr1}
	b_i^\dagger  \hat{B}_k^\dagger  \hat{B}_k  b_i \preceq b_i^\dagger \prod_{s=1}^k (\nb_{i_s}+k)  b_i ,\quad 
	b_i \hat{B}_k^\dagger  \hat{B}_k  b_i^\dagger  \preceq b_i \prod_{s=1}^k (\nb_{i_s}+k)  b_i^\dagger .
\end{align}
For an arbitrary integer $m$, we have 
\begin{align}
	\label{Upper_bound_hopping/induction_pr2}
	&b_i^\dagger  (\nb_i+k)^m  b_i =\nb_i(\nb_i+k-1)^m \preceq (\nb_i+k+1)^{m+1},\notag \\
	&b_i  (\nb_i+k)^m  b_i^\dagger = (\nb_i+1)(\nb_i+k+1)^m\preceq (\nb_i+k+1)^{m+1}, 
\end{align}
where we use the fact that each of the above two operators can be diagonalized by the Fock states on the site $i$.  
By combining the inequalities~\eqref{Upper_bound_hopping/induction_pr1} and \eqref{Upper_bound_hopping/induction_pr2}, we prove the operator inequalities of 
\begin{align}
	&b_i^\dagger  \hat{B}_k^\dagger  \hat{B}_k  b_i \preceq \prod_{j=1}^{k+1} (\nb_{i_j}+k+1) , \quad b_i \hat{B}_k^\dagger  \hat{B}_k  b_i^\dagger \preceq \prod_{j=1}^{k+1} (\nb_{i_j}+k+1),
\end{align}
which proves~\eqref{Upper_bound_hopping/induction}, where we let $i_{k+1}=i$. 

For the proof of the second inequality~\eqref{corol_upper_hopping/main/ineq}, we begin with the H\"older inequality since the operators $\{\nb_{i_j}+k\}_{j=1}^k$ commute with each other:  
\begin{align}
	\label{abs_psi_b_i_ket_psi}
	\abs{\bra{\psi}  b_{i_k}^\dagger \cdots b_{i_{s+1}}^\dagger  b_{i_s}^\dagger b_{i_{s-1}}   \cdots  b_{i_2}b_{i_1} \ket{\psi} }
	&\le  \bra{\psi} \prod_{j=1}^k (\nb_{i_j}+k)^{1/2} \ket{\psi}  \notag \\
	&\le  \prod_{j=1}^k \br{\bra{\psi}(\nb_{i_j}+k)^{k/2} \ket{\psi}}^{1/k}  \notag \\
	&\le \frac{1}{k} \sum_{j=1}^k \bra{\psi}(\nb_{i_j}+k)^{k/2} \ket{\psi} .
\end{align}

Finally, for an arbitrary site $i\in \Lambda$, we can prove the operator inequality of 
\begin{align}
	\label{bar_q_ineq_gen}
	(\nb_i+m)^{k/2}
	&\preceq   \nb_i^{k/2} + \nb_i^{k/2-1}\max\brr{ (1+m)^{k/2}-1, \frac{mk}{2}} + m^{k/2}\notag \\
	&\preceq  \nb_i^{k/2} + (2m)^{k}  \nb_i^{k/2-1} + m^{k/2}, 
\end{align}
where $k$ and $m$ are arbitrary non-negative integers.
The above inequality is proved from the following inequality (see below for the proof):
\begin{align}
	\label{x+y^z_upp}
	(x+y)^z \le x^z + x^{z-1} \max\brr{ (1+y)^z-1, yz} 
\end{align}
for $x\ge 1$, $y>0$ and $z>0$.
We applied the inequality~\eqref{x+y^z_upp} by letting $x\to \nb_i$, $y\to m$ and $z\to k/2$. 
By combining the inequalities~\eqref{abs_psi_b_i_ket_psi} and~\eqref{bar_q_ineq_gen} with $m=k$, we prove the second main inequality~\eqref{corol_upper_hopping/main/ineq} as follows:
\begin{align}
	\label{abs_psi_b_i_ket_psi_fin}
	\abs{\bra{\psi}  b_{i_k}^\dagger \cdots b_{i_{s+1}}^\dagger  b_{i_s}^\dagger b_{i_{s-1}}   \cdots  b_{i_2}b_{i_1} \ket{\psi} }
	&\le \frac{1}{k} \sum_{j=1}^k \bra{\psi} \nb_{i_j}^{k/2} \ket{\psi} + \frac{ (2k)^{k} }{k} \sum_{j=1}^k \bra{\psi} \nb_{i_j}^{k/2-1}  \ket{\psi}  + k^{k/2} .
\end{align}
This completes the proof. $\square$

{~} \\

\noindent
{\bf [Proof of the inequality~\eqref{x+y^z_upp}]}\\
For the proof, we consider an upper bound of 
\begin{align}
	f_{y,z}(x):= x(1+y/x)^z - x ,
\end{align}
which yields 
\begin{align}
	(x+y)^z - x^z  \le x^{z-1} \sup_{x:x\ge 1} \brr{f_{y,z}(x)}. 
\end{align}
We first consider the derivative of
\begin{align}
	f'_{y,z}(x):= -1 + \br{\frac{x+y}{x}}^z \br{1-\frac{yz}{x+y}} .
\end{align}
For $z\le 1$, we have $[1-y/(x+y)]^z \le 1- yz/(x+y)$, and hence 
\begin{align}
	f'_{y,z}(x)\ge -1 +\br{\frac{x+y}{x}}^z \br{1-\frac{y}{x+y}}^z=0,
\end{align}
which means that $f_{y,z}(x)$ is maximized at $x=\infty$. We obtain $\lim_{x\to \infty} f_{y,z}(x)= yz$. 
On the other hand, for $z> 1$, we have $[1-y/(x+y)]^z \ge 1- yz/(x+y)$
\begin{align}
	f'_{y,z}(x)\le -1 + \br{\frac{x+y}{x}}^z\br{1-\frac{y}{x+y}}^z =0 ,
\end{align}
which means that $f_{y,z}(x)$ is maximized at $x=1$ as $x\ge 1$. We then obtain $f_{y,z}(1)=(1+ y)^z-1$.

\subsection{Maximum moment bounds: the total boson number not conserved}

We here aim to derive the boson number distribution on a single site in the ground state $\ket{\Omega}$ for the Bose-Hubbard class under Assumption~\ref{assump:Repulsive condition}. 
We treat separately the cases in which the total boson number is not conserved and those in which it is. 
In the latter case, the ground state $\ket{\Omega_N}$ as well as the ground energy $E_0(N)$ depend on the total boson number, and the analyses become more complicated.
When there is no total-boson-number dependence, we simply denote the ground state and the ground energy by $\ket{\Omega}$ and $E_0$, respectively.

In this section, we begin by treating the case where the total boson number is not conserved.
Let us consider an upper bound of the maximum moment in the set $\Lambda$, i.e.,
\begin{align}
	\label{bar_q_def}
	\max_{i\in \Lambda}\bra{\Omega} \nb_i^{k/2} \ket{\Omega} = \mQ_\Omega^{k/2}  . 
\end{align}
Regarding the quantity $\mQ_\Omega$, we aim to prove the following proposition:
\begin{prop} \label{Prop:Moment_upper:bound_non-conserve}
	Under Assumption~\ref{assump:Repulsive condition}, the moment $\mQ_\Omega$ is bounded from above by
	\begin{align}
		\mQ_\Omega \leq \max\brr{1,   \br{\frac{2 \bar{\mJ} }{U_i- \bar{J}_{i,k}}}^2 },
	\end{align}
	where we define $\bar{\mJ}$ as in Eq.~\eqref{def_bar_mJ0}.
\end{prop}

\textit{Proof of Proposition~\ref{Prop:Moment_upper:bound_non-conserve}.}
It is enough to prove the case of $\mQ_\Omega \ge 1$. 
Without loss of generality, let us label the site $1$ such that $\ave{\nb_1^{k/2}}_\Omega=\mQ_\Omega^{k/2}$.
For arbitrary $k_1 \le k$ and $i\in \Lambda$, we obtain 
\begin{align}
	\ave{\nb_i^{k_1/2}}_\Omega \le \br{\ave{\nb_i^{k/2}}_\Omega }^{k_1/k} \le \mQ_\Omega^{k_1/2} .
	\label{Ineq_nb_i_k_1}
\end{align}

Then, we consider the Schmidt decomposition of 
\begin{align}
	\ket{\Omega} = \sum_{m=1}^\infty \lambda_m \ket{S_m}_1 \ket{\chi_m}_{\Lambda_1} , \quad  \Lambda_1= \Lambda\setminus \{1\}.
\end{align}
Using it, we adopt the reference quantum state $\tilde{\rho}$ as 
\begin{align}
	\tilde{\rho}
	&=\ket{0}\bra{0}_1 \otimes  \sum_{m=1}^\infty \lambda^2_m \ket{\chi_m}\bra{\chi_m} _{\Lambda_1}  \notag \\
	&=: \rho_1 \otimes \rho_{\Lambda_1} .
\end{align}
Note that $\rho_{\Lambda_1}$ are equivalent to the reduced density matrix of $\ket{\Omega}$ on the subset $\Lambda_1$. 

By decomposing the Hamiltonian as 
\begin{align}
	H= \widehat{H_1} + H_{\Lambda_1} ,
\end{align}
we have 
\begin{align}
	\label{ineq_H_tilde_rho_Omega}
	\tr\br{ H \tilde{\rho} } - \bra{\Omega} H \ket{\Omega}
	= \tr\br{  \widehat{H_1} \tilde{\rho} } - \bra{\Omega}  \widehat{H_1} \ket{\Omega}=- \bra{\Omega}  \widehat{H_1} \ket{\Omega} \ge 0,
\end{align}
where we use $\tr\br{H_{\Lambda_1} \tilde{\rho} } =\tr\br{H_{\Lambda_1} \rho_{\Lambda_1}}= \bra{\Omega} H_{\Lambda_1}  \ket{\Omega}$. 

In the following, we aim to upper-bound $- \bra{\Omega}  \widehat{H_1} \ket{\Omega}$ (or lower-bound $ \bra{\Omega}  \widehat{H_1} \ket{\Omega}$).
Because we have 
\begin{align}
	\widehat{H_1}=\widehat{H_{0,1}}(\mathbf{b},\mathbf{b}^\dagger)+ \widehat{V_{+,1}}(\mathbf{n}) + U_1 \nb_1^{k/2},
\end{align}
from the definition~\eqref{Boson_most_general} and the notation~\eqref{Boson_most_general_k-local_subset_wide}, 
we calculate
\begin{align}
	\label{wide_hat_H_1_Omega}
	\bra{\Omega}  \widehat{H_1} \ket{\Omega} 
	&\ge U_1  \bra{\Omega} \nb_1^{k/2} \ket{\Omega} -  \bra{\Omega} \overline{H_{0,1}}(\mathbf{b},\mathbf{b}^\dagger) \ket{\Omega},
\end{align}
where we have defined $\overline{H_{0,i}}(\mathbf{b},\mathbf{b}^\dagger)$ for $\forall i\in \Lambda$ in Eq.~\eqref{overline_H_0_i_vec} which satisfies
$\widehat{H_{0,1}}(\mathbf{b},\mathbf{b}^\dagger) \preceq \overline{H_{0,i}}(\mathbf{b},\mathbf{b}^\dagger)$. 
Note that $ \bra{\Omega}\widehat{V_{+,1}}(\mathbf{n}) \ket{\Omega} \ge 0$ from $\widehat{V_{+,1}}(\mathbf{n})\succeq 0$. 
By applying the inequality~\eqref{corol_upper_hopping/main/ineq_2} in Lemma~\ref{lem:Upper_bound_hopping}, we obtain 
\begin{align}
	\label{Omega_over_line_upp}
	&\bra{\Omega} \overline{H_{0,1}}(\mathbf{b},\mathbf{b}^\dagger) \ket{\Omega} \notag \\
	&\le 
	\sum_{k_1=1}^k \sum_{\substack{i_1,i_2,\ldots, i_{k_1} \in \Lambda \\ \{ i_1,i_2,\ldots, i_{k_1}\} \ni 1}}  \sum_{s=1}^{k_1+1} \abs{J^{(s)}_{i_1,i_2,\ldots,i_{k_1}}}  
	\br{\frac{1}{k_1} \sum_{j=1}^{k_1} \bra{\Omega} \nb_{i_j}^{k_1/2} \ket{\Omega} + \frac{ (2k_1)^{k_1} }{k_1} \sum_{j=1}^{k_1} \bra{\Omega} \nb_{i_j}^{k_1/2-1}  \ket{\Omega}  + k_1^{k_1/2}} \notag \\
	&\le 
	\sum_{k_1=1}^k \sum_{\substack{i_1,i_2,\ldots, i_{k_1} \in \Lambda \\ \{ i_1,i_2,\ldots, i_{k_1}\} \ni 1}}  \sum_{s=1}^{k_1+1} \abs{J^{(s)}_{i_1,i_2,\ldots,i_{k_1}}}   
	\br{\sum_{j=1}^{k_1} \mQ_\Omega^{k_1/2}   +  (2k_1)^{k_1} \sum_{j=1}^{k_1} \mQ_\Omega^{k_1/2-1}   + k_1^{k_1/2}}  \notag \\
	&\le 
	\sum_{k_1=1}^k   \bar{J}_{1,k_1}
	\br{\sum_{j=1}^{k_1} \mQ_\Omega^{k_1/2}   +  (2k_1)^{k_1} \sum_{j=1}^{k_1} \mQ_\Omega^{k_1/2-1}   + k_1^{k_1/2}}, 
\end{align}
where in the second inequality, we use the upper bound~\eqref{Ineq_nb_i_k_1}, and in the last inequality, we use the definition of $\bar{J}_{i,k_1}$ in Eq.~\eqref{bar_J_i_k_1}.

By using the parameter $\bar{\mJ}$ in Eq.~\eqref{def_bar_mJ0} and $\mQ_\Omega \ge 1$, we derive 
\begin{align}
	\sum_{k_1=1}^k\bar{J}_{1,k_1}\brr{ \mQ_\Omega^{k_1/2} + (2k_1)^{k_1} \mQ_\Omega^{k_1/2-1} +k_1^{k_1/2} } 
	&\le  \bar{J}_{1,k} \mQ_\Omega^{k/2} +  2 \sum_{k_1=1}^k\bar{J}_{1,k_1}\brr{1+ (2k_1)^{k_1}} \mQ_\Omega^{(k-1)/2} \notag \\
	&\le \bar{J}_{1,k} \mQ_\Omega^{k/2} +   2\bar{\mJ} \mQ_\Omega^{(k-1)/2} ,
\end{align}
where we use $\mQ_\Omega \ge 1$ to get $\mQ_\Omega^{k_1/2} \le \mQ_\Omega^{(k-1)/2}$ for $k_1\le k-1$. 
The above inequality reduces the inequality~\eqref{Omega_over_line_upp} to 
\begin{align}
	\label{Omega_over_line_upp_fin}
	&\bra{\Omega} \overline{H_{0,1}}(\mathbf{b},\mathbf{b}^\dagger) \ket{\Omega} 
	\le \bar{J}_{1,k} \mQ_\Omega^{k/2} + 2 \bar{\mJ} \mQ_\Omega^{(k-1)/2}  \notag \\
	&\longrightarrow  - \bra{\Omega}  \widehat{H_1} \ket{\Omega} \le - U_1  \mQ_\Omega^{k/2} + \br{ \bar{J}_{1,k} \mQ_\Omega^{k/2} +  2 \bar{\mJ} \mQ_\Omega^{(k-1)/2}} ,
\end{align}
where we use $\bra{\Omega} \nb_1^{k/2} \ket{\Omega} = \mQ_\Omega^{k/2}$ from the assumption, and in the second line, we use the inequality~\eqref{wide_hat_H_1_Omega}.  
Therefore, from the condition~\eqref{ineq_H_tilde_rho_Omega}, we obtain the main inequality as follows: 
\begin{align}
	&-U_1  \mQ_\Omega^{k/2}  + \br{ \bar{J}_{1,k} \mQ_\Omega^{k/2} + 2 \bar{\mJ} \mQ_\Omega^{(k-1)/2}} \ge 0 \notag \\
	&\longrightarrow  \mQ_\Omega \le  \br{\frac{2 \bar{\mJ} }{U_1- \bar{J}_{1,k}}}^2 . 
\end{align}
This completes the proof. $\square$

We have proved that the $(k/2)$th order moment of the local boson number is upper-bounded by $\orderof{1}$ constant in the ground state as long as $N=\orderof{|\Lambda|}$.
Conversely, we can prove that any quantum state with a large $(k/2)$th order moment has a large energy:  
\begin{corol} \label{corol:large_q_quantum_energy_unconserve}
	Let us define $\mE_\mQ$ as the minimum energy such that 
	\begin{align}
		\label{definition_mE_q} 
		\mE_\mQ:= \inf_{\ket{\Psi_\mQ}} \br{ \bra{\Psi_\mQ} H \ket{\Psi_\mQ}} \for \mQ \in \mathbb{R}_+ ,
	\end{align}
	where $\inf_{\ket{\Psi_\mQ}}$ is taken for the class of quantum state $\ket{\Psi_\mQ}$ satisfying 
	\begin{align}
		\max_{i \in \Lambda} \bra{\Psi_\mQ} \nb^{k/2}_i \ket{\Psi_\mQ}=\mQ^{k/2} .
	\end{align}
	We then obtain 
	\begin{align}
		\label{corol:large_q_quantum_energy_unconserve_main}
		\mE_\mQ - E_0 \ge  (U_1 -\bar{J}_{1,k}) \mQ^{k/2} -  2 \bar{\mJ} \mQ^{(k-1)/2}.
	\end{align}
\end{corol}

\textit{Proof of Corollary~\ref{corol:large_q_quantum_energy_unconserve}.}
Let us assume $\bra{\Psi_\mQ} \nb^{k/2}_1 \ket{\Psi_\mQ}=\mQ^{k/2}$ without loss of generality.
We then obtain 
\begin{align}
	\label{braPsi_qHketPsi_q=1}
	\bra{\Psi_\mQ} H \ket{\Psi_\mQ}&=\bra{\Psi_\mQ} \widehat{H_1} \ket{\Psi_\mQ} + \tr_{\Lambda_1} \brr{ H_{\Lambda_1} \tr_{1}\br{ \ket{\Psi_\mQ}\bra{\Psi_\mQ}}}\notag \\
	& \ge \bra{\Psi_\mQ} \widehat{H_1}  \ket{\Psi_\mQ}+E_{0,\Lambda_1} \ge 
	\bra{\Psi_\mQ} \widehat{H_1}  \ket{\Psi_\mQ}+E_0 ,
\end{align}
where $E_{0,\Lambda_1}$ has been defined as the ground energy of $H_{\Lambda_1}$, which was proven to be larger than $E_{0,\Lambda}$ in Lemma~\ref{lem:subset_ground_energy}. 
We can derive the same inequality as~\eqref{Omega_over_line_upp_fin} for $\bra{\Psi_\mQ} \widehat{H_1}  \ket{\Psi_\mQ}$, which yields the lower bound of 
\begin{align}
	\bra{\Psi_\mQ} \widehat{H_1}  \ket{\Psi_\mQ} \ge  U_1  \mQ^{k/2} - \br{ \bar{J}_{1,k} \mQ^{k/2} +  2 \bar{\mJ} \mQ^{(k-1)/2}} .
\end{align}
By applying the above inequality to~\eqref{braPsi_qHketPsi_q=1}, we prove the desired inequality~\eqref{corol:large_q_quantum_energy_unconserve_main}.
This completes the proof. $\square$

\subsection{Maximum moment bounds: the total boson number conserved}

\subsubsection{Preliminary lemmas}

We here consider the lower and upper bounds of the ground energy on a subsystem $X \subseteq \Lambda$, which is proven by the following lemma:
\begin{lemma} \label{lem:subset_ground_energy_low_up}
	For the ground energy $E_{0,X}(N)$, the following inequality holds in general:
	\begin{align}
		- \bar{\mJ}|X| \brr{1+\br{\frac{2 \bar{\mJ}}{\tilde{U}}}^{k-1} } + \frac{\tilde{U}}{2} \sum_{i\in X}  \bra{\Omega_{X,N}} \nb_{i}^{k/2} \ket{\Omega_{X,N}} 
		\le E_{0,X}(N) \le   g N_\ast^{k/2} |X|,
		\label{lem:subset_ground_energy_main_ineq_2} 
	\end{align}
	where $N_\ast:= \ceil {N/|X|}\le 1+ N/|X|$, $\ket{\Omega_{X,N}}$ is the ground state of the subset Hamiltonian $H_X$\footnote{When the ground state is degenerate, we can pick up an arbitrary state from the degenerate space.}, and
	\begin{align}
		\tilde{U}:=\min_{i\in\Lambda}(U_i- \bar{J}_{i,k}) . 
	\end{align}
	We remind that the parameters $g$ and $\bar{\mJ}$ have been defined in Eqs.~\eqref{Boson_most_general_k-local_g_ext} and~\eqref{def_bar_mJ0}, respectively.
\end{lemma}

\textit{Proof of Lemma~\ref{lem:subset_ground_energy_low_up}.}
It is enough to consider the case of $X=\Lambda$, and the generalization is straightforward.
For obtaining the upper bound, let us choose the Mott state $\ket{M}$, such that $\nb_i \ket{M} = n_i \ket{M}$ with $n_i=N_\ast$ or $N_\ast-1$, where
we adopt the definition of $N_\ast:=\ceil {N/|\Lambda|}\le 1+ N/|\Lambda|$. 
Then, using the inequality~\eqref{Boson_most_general_k-local_g_ext}, we obtain 
\begin{align}
	\bra{M} H \ket{M} \le \sum_{i\in \Lambda}
	\sum_{Z:Z\ni i} \norm{\Pi_{\Lambda,\le N_\ast} h_Z \Pi_{\Lambda,\le N_\ast} } \le g N_\ast^{k/2}|\Lambda|,
\end{align}
which yields the upper bound in~\eqref{lem:subset_ground_energy_main_ineq_2}:
\begin{align}
	E_{0,\Lambda}(N) \le\bra{M} H \ket{M} \le  g N_\ast^{k/2} |\Lambda|  .
\end{align}

We next consider the lower bound in~\eqref{lem:subset_ground_energy_main_ineq_2}. 
For this purpose, we use $V_+(\mathbf{n})\succeq 0$ to derive 
\begin{align}
	\label{ave/H_0_Omega_N__0}
	\bra{\Omega_N} H\ket{\Omega_N}\ge \bra{\Omega_N}H_0(\mathbf{b},\mathbf{b}^\dagger)\ket{\Omega_N} + \sum_{i\in \Lambda} U_i \bra{\Omega_N}\nb_i^{k/2} \ket{\Omega_N},
\end{align}
By using the inequality~\eqref{prop_1_site_upper_bound/main1} in Proposition~\ref{prop_1_site_upper_bound}, we lower-bound
$\bra{\Omega_N}H_0(\mathbf{b},\mathbf{b}^\dagger)\ket{\Omega_N}$ by
\begin{align}
	\label{ave/H_0_Omega_N_3}
	\bra{\Omega_N}H_0(\mathbf{b},\mathbf{b}^\dagger)\ket{\Omega_N} 
	&\ge - \sum_{i\in \Lambda}  \bra{\Omega_N} \overline{H_{0,i}}(\mathbf{b},\mathbf{b}^\dagger)  \ket{\Omega_N} \notag \\
	&\ge -  \sum_{i \in \Lambda}  \brr{ \bar{J}_{i,k}\bra{\Omega_N} \nb_{i}^{k/2} \ket{\Omega_N} + \bar{\mJ} \br{\bra{\Omega_N} \nb_{i}^{(k-1)/2}\ket{\Omega_N}+1}} .
\end{align}

By combining the inequalities~\eqref{ave/H_0_Omega_N__0} and \eqref{ave/H_0_Omega_N_3}, we can derive the desired upper bound as follows:
\begin{align}
	\label{ave/H_0_Omega_N__fin}
	&\bra{\Omega_N} H\ket{\Omega_N} \notag \\
	&\ge \sum_{i \in \Lambda}  \brr{(U_i- \bar{J}_{i,k}) \bra{\Omega_N} \nb_{i}^{k/2} \ket{\Omega_N} - \bar{\mJ} \br{\bra{\Omega_N} \nb_{i}^{(k-1)/2}\ket{\Omega_N}+1}}  \notag \\
	&\ge - \bar{\mJ} |\Lambda| + \sum_{i \in \Lambda}  \frac{U_i- \bar{J}_{i,k}}{2} \bra{\Omega_N} \nb_{i}^{k/2} \ket{\Omega_N}
	+ \sum_{i\in \Lambda} \br{ \frac{U_i- \bar{J}_{i,k}}{2} \bra{\Omega_N} \nb_{i}^{k/2} \ket{\Omega_N}- \bar{\mJ} \bra{\Omega_N} \nb_{i}^{(k-1)/2}\ket{\Omega_N}} \notag \\
	&\ge  - \bar{\mJ}|\Lambda| \brr{1+\br{\frac{2 \bar{\mJ}}{\tilde{U}}}^{k-1} } + \frac{\tilde{U}}{2} \sum_{i\in \Lambda}  \bra{\Omega_N} \nb_{i}^{k/2} \ket{\Omega_N} 
\end{align}
with the definition of $\tilde{U}:=\min_{i\in\Lambda}(U_i- \bar{J}_{i,k})$, 
where we use the following inequality by letting $\bra{\Omega_N} \nb_{i}^{k/2} \ket{\Omega_N}=x^{k/2}$ 
\begin{align}
	&\frac{U_i- \bar{J}_{i,k}}{2} \bra{\Omega_N} \nb_{i}^{k/2} \ket{\Omega_N}- \bar{\mJ} \bra{\Omega_N} \nb_{i}^{(k-1)/2}\ket{\Omega_N} \notag \\
	\ge & \frac{U_i- \bar{J}_{i,k}}{2} x^{k/2}- \bar{\mJ}x^{(k-1)/2} 
	= \frac{U_i- \bar{J}_{i,k}}{2} x^{(k-1)/2} \br{ x^{1/2}-\frac{2 \bar{\mJ}}{U_i- \bar{J}_{i,k}}} \notag \\
	\ge & - \frac{U_i- \bar{J}_{i,k}}{2} \br{\frac{2 \bar{\mJ}}{U_i- \bar{J}_{i,k}}}^k
	= -  \bar{\mJ}\br{\frac{2 \bar{\mJ}}{U_i- \bar{J}_{i,k}}}^{k-1}.
\end{align}
Note that $\bra{\Omega_N} \nb_{i}^{(k-1)/2}\ket{\Omega_N}\le \bra{\Omega_N} \nb_{i}^{k/2}\ket{\Omega_N}^{(k-1)/k}=x^{(k-1)/2}$.

We thus prove the main inequality~\eqref{lem:subset_ground_energy_main_ineq_2}. 
$\square$

{~}

\hrulefill{\bf [ End of Proof of Lemma~\ref{lem:subset_ground_energy_low_up}]}

{~}

Using Lemma~\ref{lem:subset_ground_energy_low_up}, we further consider the energy difference between $E_{0,X}(N+1)$ and $E_{0,X}(N)$.
From the previous lemma, the ground energy is proportional to the total number of bosons $N$, and hence, it is natural to expect that the energy cannot change drastically by adding one boson. 
The following proposition ensures the point:

\begin{prop} \label{lem:subset_ground_energy_dif}
	For the ground energy $E_{0,X}(N)$, the following inequality holds in general:
	\begin{align}
		E_{0,X}(N+1) -E_{0,X}(N) \le \delta_{N_\ast} , \quad N_\ast := \ceil {\frac{N}{|X|}}
		\label{lem:subset_ground_energy_main_ineq} ,
	\end{align}
	where the parameter $\delta_{z}$ depends on $z$ as  
	\begin{align}
		&\delta_{z}  = 2\bar{\mJ}_1 +\frac{4\bar{\mJ}_2}{\tilde{U}} \brr{ g z^{k/2}+\bar{\mJ} +\bar{\mJ}\br{\frac{2 \bar{\mJ}}{\tilde{U}}}^{k-1}  }
		\label{lem:subset_ground_energy_main_ineq_delta_q}  
	\end{align}
	with $\bar{\mJ}_1$ and $\bar{\mJ}_2$ defined as 
	\begin{align}
		\bar{\mJ}_1 := 2\bar{\mJ}+ \bar{v} + \max_{i\in \Lambda}\br{U_i+ \bar{J}_{i,k} } ,\quad 
		\bar{\mJ}_2:= 2^{k/2} \brr{\bar{\mJ}+ \bar{v} + \max_{i\in \Lambda}\br{U_i+ \bar{J}_{i,k} }},
	\end{align}
	respectively. 
\end{prop}

{\bf Remark.}
When we consider the subsystem ground energy from $E_{0,X}(0)$ to $E_{0,X}(N)$, we can derive the upper bound of 
\begin{align}
	E_{0,X}(N) -E_{0,X}(N-m) \le \sum_{s=1}^{m} E_{0,X}(N-s+1) -E_{0,X}(N-s) \le  \sum_{s=1}^{m} \delta_{q_{\ast,s}} , 
\end{align}
where $q_{\ast,s}=\ceil{(N-s)/|X|} \le \ceil{N/|X|}=: \bar{N}_\ast$. 
Therefore, as long as $\bar{N}_\ast$ is $\orderof{1}$, the energy difference between $E_{0,X}(m-1)$ to $E_{0,X}(m)$ $(m\le N)$ is always upper-bounded by an $\orderof{1}$ constant. 
This proves the following inequality:
\begin{align}
	E_{0,X}(N) -E_{0,X}(N-m) \le m \delta_{\bar{N}_\ast} . \label{lem:subset_ground_energy_main_ineq_m} 
\end{align}

\textit{Proof of Proposition~\ref{lem:subset_ground_energy_dif}.}
As in the proof of Lemma~\ref{lem:subset_ground_energy_low_up}, we consider the case of $X=\Lambda$ for simplicity. 
For the proof, we expand the ground state $\ket{\Omega_N}$ with respect to the boson number on the site $i$ as follows:
\begin{align}
	\ket{\Omega_N}= \sum_{m=0}^\infty a_m \ket{m}_i \ket{\chi_m}_{\Lambda_i} ,
\end{align}
where $\nb_i \ket{m}_i=m\ket{m}_i$, and we denote $\Lambda_i=\Lambda \setminus \{i\}$.
We then define the quantum state $\ket{\Phi_i}$ with the total boson number $N+1$ as 
\begin{align}
	\ket{\Phi_i}= \sum_{m=0}^\infty a_m \ket{m+1}_i \ket{\chi_m}_{\Lambda_i} ,
\end{align}
where we have $\sum_{i\in \Lambda}\bra{\Phi_i}\nb_i \ket{\Phi_i} =N+1$. 
Using $\{\ket{\Phi_i}\}_{i\in \Lambda}$, we set the reference state $\rho_{N+1}$ as 
\begin{align}
	\rho_{N+1} = \frac{1}{|\Lambda|} \sum_{i\in \Lambda} \ket{\Phi_i} \bra{\Phi_i}  ,
\end{align}
and analyze the upper bound of 
\begin{align}
	\label{average_by_rgo_N+1___0}
	E_{0,X}(N+1) -E_{0,X}(N)
	& \le \tr\br{H \rho_{N+1} } - \bra{\Omega_N} H \ket{\Omega_N}  \notag\\
	&= \frac{1}{|\Lambda|} \sum_{i\in \Lambda}\br{ \bra{\Phi_i} \widehat{H_i}\ket{\Phi_i}  - \bra{\Omega_N} \widehat{H_i} \ket{\Omega_N} },
\end{align}
where, in the last equation, we use the fact that the reduced density matrix of $\ket{\Phi_i}$ on $\Lambda_i$ is identical to that of $\ket{\Omega_N}$.

By applying Proposition~\ref{prop_1_site_upper_bound} to $\sum_{i\in \Lambda} \bra{\Phi_i} \widehat{H_i}\ket{\Phi_i}$, 
we obtain 
\begin{align}
	\label{Phi_H_i_upper_ineq}
	\sum_{i\in \Lambda} \bra{\Phi_i} \widehat{H_i}\ket{\Phi_i} 
	&\le  \sum_{i\in \Lambda}\brrr{ U_i \bra{\Phi_i} \nb_i^{k/2}\ket{\Phi_i} + 
		\bar{v} \bra{\Phi_i}\nb_i^{k/2} \ket{\Phi_i} 
		+ \brr{ \bar{J}_{i,k}\bra{\Phi_i} \nb_{i}^{k/2} \ket{\Phi_i} + \bar{\mJ} \br{\bra{\Phi_i} \nb_{i}^{(k-1)/2}\ket{\Phi_i}+1}}} \notag \\
	&\le \bar{\mJ}|\Lambda| + \sum_{i\in \Lambda} \br{U_i+ \bar{v}+ \bar{J}_{i,k} + \bar{\mJ} }\bra{\Omega_N} (\nb_{i}+1)^{k/2} \ket{\Omega_N} \notag \\
	&\le \bar{\mJ}|\Lambda| + \sum_{i\in \Lambda} \br{U_i+ \bar{v}+ \bar{J}_{i,k} + \bar{\mJ} } \br{2^{k/2}\bra{\Omega_N} \nb_{i}^{k/2} \ket{\Omega_N} +1} \notag \\
	&\le \bar{\mJ}_1 |\Lambda| +\bar{\mJ}_2  \sum_{i\in \Lambda}\bra{\Omega_N} \nb_{i}^{k/2} \ket{\Omega_N} ,
\end{align}
where we use $(\nb_{i}+1)^{k/2}\preceq (2\nb_{i})^{k/2}+1$ and the definitions for $\bar{\mJ}_1$ and $\bar{\mJ}_2$. 
On the other hand, for $-\bra{\Omega_N} \widehat{H_i} \ket{\Omega_N}$, we obtain the same upper bound as 
\begin{align}
	\label{Omega_H_i_upper_ineq}
	- \sum_{i\in \Lambda}  \bra{\Omega_N} \widehat{H_i} \ket{\Omega_N} 
	&\le  \sum_{i\in \Lambda}  \brr{ \bar{J}_{i,k}\bra{\Omega_N} \nb_{i}^{k/2} \ket{\Omega_N} + \bar{\mJ} \br{\bra{\Omega_N} \nb_{i}^{(k-1)/2}\ket{\Omega_N}+1}} \notag \\
	&\le \bar{\mJ}_1 |\Lambda| +\bar{\mJ}_2  \sum_{i\in \Lambda}\bra{\Omega_N} \nb_{i}^{k/2} \ket{\Omega_N}  ,
\end{align} 
where we use a similar inequality to~\eqref{Phi_H_i_upper_ineq}. 

Therefore, we derive 
\begin{align}
	\label{average_by_rgo_N+1}
	E_{0,X}(N+1) -E_{0,X}(N)
	& \le 2\bar{\mJ}_1 + \frac{2\bar{\mJ}_2 }{|\Lambda|} \sum_{i\in \Lambda}\bra{\Omega_N} \nb_{i}^{k/2} \ket{\Omega_N} .
\end{align}
From the relation~\eqref{lem:subset_ground_energy_main_ineq_2} in Lemma~\ref{lem:subset_ground_energy_low_up}, we obtain the upper bound of 
\begin{align}
	&- \bar{\mJ}|\Lambda| \brr{1+\br{\frac{2 \bar{\mJ}}{\tilde{U}}}^{k-1} } + \frac{\tilde{U}}{2} \sum_{i\in \Lambda}  \bra{\Omega_N} \nb_{i}^{k/2} \ket{\Omega_N} 
	\le  g N_\ast^{k/2} |\Lambda| \notag \\
	&\longrightarrow \sum_{i\in \Lambda}  \bra{\Omega_N} \nb_{i}^{k/2} \ket{\Omega_N} \le \frac{2|\Lambda|}{\tilde{U}} \brr{ g N_\ast^{k/2}+\bar{\mJ} +\bar{\mJ}\br{\frac{2 \bar{\mJ}}{\tilde{U}}}^{k-1}  }.
	\label{lem:subset_ground_energy_main_ineq_2_again} 
\end{align}
By combining the upper bounds~\eqref{average_by_rgo_N+1} and \eqref{lem:subset_ground_energy_main_ineq_2_again}, we prove the main inequality~\eqref{lem:subset_ground_energy_main_ineq}.
This completes the proof. $\square$

{~}

\hrulefill{\bf [ End of Proof of Proposition~\ref{lem:subset_ground_energy_dif}]}

{~}

\subsubsection{Main statement}

\begin{prop} \label{Prop:Moment_upper:bound_conserve}
	We adopt the same setup as Proposition~\ref{Prop:Moment_upper:bound_non-conserve}, and consider 
	\begin{align}
		\label{bar_q_de_def_re}
		\max_{i\in \Lambda}\bra{\Omega} \nb_i^{k/2} \ket{\Omega} = \mQ_\Omega^{k/2}  .
	\end{align}
	Then, under Assumption~\ref{assump:Repulsive condition}, the moment $\mQ_\Omega$ is bounded from above by
	\begin{align}
		\mQ_\Omega \le \max\brr{1, \br{ \frac{\delta_{2N/|\Lambda|}+2 \bar{\mJ}}{U_i-\bar{J}_{i,k}}}^2 },
	\end{align}
	where we define $\bar{\mJ}$ and $\delta_z$ for $z \ge 0$ as in Eqs.~\eqref{def_bar_mJ0} and \eqref{lem:subset_ground_energy_main_ineq_delta_q}, respectively. 
\end{prop}

\textit{Proof of Proposition~\ref{Prop:Moment_upper:bound_conserve}.}
The proof is similar to the one for Proposition~\ref{Prop:Moment_upper:bound_non-conserve}.
We consider the case of $\mQ_\Omega \ge 1$ and label the site $1$ such that $\ave{\nb_1^{k/2}}_\Omega=\mQ_\Omega^{k/2}$.
From the inequality~\eqref{Ineq_nb_i_k_1}, for arbitrary $k_1 \le k$ and $i\in \Lambda$, we obtain 
$\ave{\nb_i^{k_1/2}}_\Omega \le \mQ_\Omega^{k_1/2} $. 

For the decomposition of the ground state of 
\begin{align}
	\ket{\Omega_N} = \sum_{m=0}^\infty a_m \ket{m}_1 \ket{\chi_m}_{\Lambda_1} , \quad \Lambda_1= \Lambda\setminus \{1\}. 
\end{align}
We define the reference quantum state $\ket{\Phi_N}$ as 
\begin{align}
	\ket{\Phi_N}=  \ket{0}_1 \ket{\Omega_{N,\Lambda_1}}_{\Lambda_1} ,
\end{align}
where $\Omega_{N,\Lambda_1}$ is the ground state of the subset Hamiltonian $H_{\Lambda_1}$. 

By decomposing the Hamiltonian as 
\begin{align}
	H= \widehat{H_1} + H_{\Lambda_1} ,
\end{align}
we have 
\begin{align}
	\label{ineq_H_tilde_rho_Omega_ree}
	&\bra{\Phi_N} H \ket{\Phi_N}- \bra{\Omega_N} H \ket{\Omega_N} \notag \\
	&=E_{0,\Lambda_1}(N) -\sum_{m=0}^\infty |a_m|^2 \bra{\chi_m} H_{\Lambda_1} \ket{\chi_m}_{\Lambda_1}
	- \bra{\Omega_N}  \widehat{H_1} \ket{\Omega_N} \ge 0,
\end{align}
where we use $\bra{\Phi_N} H_{\Lambda_1} \ket{\Phi_N} =\bra{\Omega_{N,\Lambda_1}}H_{\Lambda_1}\ket{\Omega_{N,\Lambda_1}}  =E_{0,\Lambda_1}(N)$ 
and $\bra{\Phi_N} \widehat{H_1}\ket{\Phi_N}=0$.

From the inequality~\eqref{lem:subset_ground_energy_main_ineq_m}, we obtain 
\begin{align}
	\label{chi_m_ave_H_Lamnda_1}
	\bra{\chi_m} H_{\Lambda_1} \ket{\chi_m}_{\Lambda_1} \ge E_{0,\Lambda_1}(N-m)\ge E_{0,\Lambda_1}(N) - m \delta_{\bar{N}^\ast},
\end{align}
where we let 
\begin{align}
	\bar{N}^\ast = \frac{N}{|\Lambda|-1}=\frac{|\Lambda|}{|\Lambda|-1} \cdot \frac{N}{|\Lambda|} \le  \frac{2N}{|\Lambda|} \for |\Lambda| \ge 2. 
\end{align}
Using the inequality~\eqref{chi_m_ave_H_Lamnda_1}, we immediately obtain 
\begin{align}
	\label{a_m_Lambda/1}
	\sum_{m=0}^\infty |a_m|^2   \bra{\chi_m} H_{\Lambda_1} \ket{\chi_m}_{\Lambda_1} 
	&\ge E_{0,\Lambda_1}(N)  -\delta_{\bar{N}^\ast}  \sum_{m=0}^\infty |a_m|^2 m  \notag \\
	&= E_{0,\Lambda_1}(N)  -\delta_{\bar{N}^\ast}  \bra{\Omega_N} \nb_1 \ket{\Omega_N} \notag \\
	&\ge E_{0,\Lambda_1}(N)  -\delta_{\bar{N}^\ast} \mQ_\Omega^{(k-1)/2} ,
\end{align}
where we use the condition $\bra{\Omega_N}\nb_1 \ket{\Omega_N} \le \bra{\Omega_N}\nb^{(k-1)/2}_1 \ket{\Omega_N} \le \mQ_\Omega^{(k-1)/2}$ from the assumption of $k\ge 3$. 
By applying the above inequality to~\eqref{ineq_H_tilde_rho_Omega_ree}, we reach the inequality of
\begin{align}
	\label{ineq_H_tilde_rho_Omega__2}
	\delta_{\bar{N}^\ast} \mQ_\Omega^{(k-1)/2} - \bra{\Omega_N}  \widehat{H_1} \ket{\Omega_N} \ge \bra{\Phi_N} H \ket{\Phi_N}- \bra{\Omega_N} H \ket{\Omega_N} \ge 0.
\end{align}

Moreover, by employing the same analyses to derive the inequality~\eqref{Omega_over_line_upp_fin}, we can derive 
\begin{align}
	\label{Omega_N_over_line_upp_fin}
	- \bra{\Omega_N}  \widehat{H_1} \ket{\Omega_N} \le - U_1  \mQ_\Omega^{k/2} + \br{ \bar{J}_{1,k} \mQ_\Omega^{k/2} + 2 \bar{\mJ} \mQ_\Omega^{(k-1)/2}} ,
\end{align}
which reduces the inequality~\eqref{ineq_H_tilde_rho_Omega__2} to 
\begin{align}
	& - (U_1-\bar{J}_{1,k})  \mQ_\Omega^{k/2} +\br{\delta_{\bar{N}^\ast} +  2 \bar{\mJ} }\mQ_\Omega^{(k-1)/2}  \ge 0 \notag \\
	&\longrightarrow   \mQ_\Omega \le \br{ \frac{\delta_{\bar{N}^\ast}+2 \bar{\mJ}}{U_1-\bar{J}_{1,k}}}^2 
	\le \br{ \frac{\delta_{2N/|\Lambda|}+2 \bar{\mJ}}{U_1-\bar{J}_{1,k}}}^2.
\end{align}
This completes the proof. $\square$

{~}

\hrulefill{\bf [ End of Proof of Proposition~\ref{Prop:Moment_upper:bound_conserve}]}

{~}

In the case where the total boson number is conserved, we can also prove a similar statement to Corollary~\ref{corol:large_q_quantum_energy_unconserve}:
\begin{corol} \label{corol:large_q_quantum_energy}
	We adopt the same setup as in Corollary~\ref{corol:large_q_quantum_energy_unconserve}.
	When the total boson number is conserved, the energy $\mE_\mQ$ in Eq.~\eqref{definition_mE_q}, i.e., $\mE_\mQ= \inf_{\ket{\Psi_\mQ}} \br{ \bra{\Psi_\mQ} H \ket{\Psi_\mQ}}$, satisfies 
	\begin{align}
		\label{mainineq:corol:large_q_quantum_energy}
		\mE_\mQ- E_{0,\Lambda}(N)   \ge (U_1 -\bar{J}_{1,k}) \mQ^{k/2} - \br{2 \bar{\mJ} +  \delta_{2N/|\Lambda|} }  \mQ^{(k-1)/2}  .
	\end{align}
\end{corol}

\textit{Proof of Corollary~\ref{corol:large_q_quantum_energy}.}
Let us assume $\bra{\Psi_\mQ} \nb^{k/2}_1 \ket{\Psi_\mQ}=\mQ^{k/2}$ without loss of generality.
Then, by decomposing 
\begin{align}
	\ket{\Psi_\mQ}=\sum_m a'_m \ket{m}_1 \ket{\chi'_m}_{\Lambda_1} , 
\end{align}
we obtain 
\begin{align}
	\label{lower_Psi_q_H_psi_q}
	&\bra{\Psi_\mQ} H \ket{\Psi_\mQ}=\bra{\Psi_\mQ} \widehat{H_1} + H_{\Lambda_1} \ket{\Psi_\mQ} \notag \\
	&\ge   \bra{\Psi_\mQ} \widehat{H_1}  \ket{\Psi_\mQ} + \sum_{m} |a'_m|^2 \bra{\chi'_m} H_{\Lambda_1} \ket{\chi'_m}_{\Lambda_1} \notag\\
	&\ge \bra{\Psi_\mQ} \widehat{H_1}  \ket{\Psi_\mQ}+ \sum_{m}|a'_m|^2 E_{0,\Lambda_1}(N-m)
	\ge \bra{\Psi_\mQ} \widehat{H_1}  \ket{\Psi_\mQ}+ E_{0,\Lambda_1}(N)  - \delta_{2N/|\Lambda|} \sum_{m}|a'_m|^2 m ,
\end{align}
where $E_{0,\Lambda_1}(N-m)$ has been defined as the ground energy of $H_{\Lambda_1}$ with $N-m$ bosons, which satisfies the inequality~\eqref{chi_m_ave_H_Lamnda_1}.

We can derive the same inequality as~\eqref{Omega_N_over_line_upp_fin} for $\bra{\Psi_\mQ} \widehat{H_1}  \ket{\Psi_\mQ}$, which yields the lower bound of 
\begin{align}
	\bra{\Psi_\mQ} \widehat{H_1}  \ket{\Psi_\mQ} \ge U_1  \mQ^{k/2} - \br{ \bar{J}_{1,k} \mQ^{k/2} +  2 \bar{\mJ} \mQ^{(k-1)/2}} ,
\end{align}
which reduces the inequality~\eqref{lower_Psi_q_H_psi_q} to the desired inequality: 
\begin{align}
	\label{lower_Psi_q_H_psi_q_2}
	\bra{\Psi_\mQ} H \ket{\Psi_\mQ}
	&\ge  (U_1 -\bar{J}_{1,k}) \mQ^{k/2} -  2 \bar{\mJ} \mQ^{(k-1)/2} + E_{0,\Lambda_1}(N)  - \delta_{2N/|\Lambda|} \sum_{m}|a'_m|^2 m \notag \\
	&\ge (U_1 -\bar{J}_{1,k}) \mQ^{k/2} - \br{2 \bar{\mJ} + \delta_{2N/|\Lambda|}}  \mQ^{(k-1)/2} + E_{0,\Lambda}(N) ,  
\end{align}
where, in the last inequality, we use Lemma~\ref{lem:subset_ground_energy} to get $E_{0,\Lambda_1}(N) \ge E_{0,\Lambda}(N)$ and the inequality of 
\begin{align}
	\sum_{m}|a'_m|^2 m 
	=  \bra{\Psi_\mQ}\nb_1 \ket{\Psi_\mQ} \le  \bra{\Psi_\mQ}\nb_1^{(k-1)/2} \ket{\Psi_\mQ} .
\end{align}
Note that we have assumed $k\ge 3$ [$(k-1)/2\ge 1$]. 
This completes the proof. $\square$

{~}

\hrulefill{\bf [ End of Proof of Corollary~\ref{corol:large_q_quantum_energy}]}

{~}

\section{Boson number distribution for Bose-Hubbard class} \label{sec:Boson number distribution for Bose-Hubbard class}

\subsection{Main theorem}

Based on the results in the previous section, we derive concentration bounds for the boson number distribution on an arbitrary site. 
In this section, for the case where the boson number is conserved, we simply denote the ground state and the ground energy by $\ket{\Omega}$ and $E_0$ by omitting $N$ dependence. 
Similarly, the overall flow of the proof is illustrated in  Supplementary Figure~\ref{fig_supp_BH_boson_number}.

We prove the following theorem (see~\ref{sec:Proof_boson_dist_preliminary} and \ref{sec:Proof_boson_dist} for the proof):
\begin{theorem} \label{main_thm/Boson number distribution/BH}
	Let us denote the quantity $\check{\mJ}$ which gives the lower bound of 
	\begin{align}
		\label{assum_main_thm/Boson number distribution/BH}
		\mE_\mQ- E_0  \ge  (U_i -\bar{J}_{i,k}) \mQ^{k/2} -\check{\mJ}  \mQ^{(k-1)/2} \for  \forall i\in \Lambda,
	\end{align}
	where $\mE_\mQ$  has been defined in Eq.~\eqref{definition_mE_q}. 
	Then, for an arbitrary site $i\in \Lambda$, the boson number distribution decays exponentially as follows:
	\begin{align}
		\label{main_thm/Boson number distribution/BH/ineq}
		\bra{\Omega}  \Pi_{i,\ge x} \ket{\Omega} \le  \br{ \frac{1-\sqrt{1-16\zeta_{i,0}^2}}{4\zeta_{i,0}}}^{2(x-M_{i,0})/k}  
	\end{align} 
	with 
	\begin{align}
		\zeta_{i,0}:=\frac{\bar{J}_{i,k}}{U_i- \mathfrak{u}_i -\bar{J}_{i,k}} ,
	\end{align} 
	and 
	\begin{align}
		\label{Definitiion_M_i_0}
		M_{i,0}:= \max \brr{2^{2/k}\br{\frac{\check{\mJ}}{U_i -\bar{J}_{i,k}}}^{2}, \br{ \frac{\check{\mJ} \bar{J}_{i,k} + 2 \bar{\mJ} (U_i- \mathfrak{u}_i -\bar{J}_{i,k})}{\mathfrak{u}_i\bar{J}_{i,k}} }^2},
	\end{align} 
	where $\mathfrak{u}_i $ is an arbitrary positive constant. 
	We remind that the projection $\Pi_{i,\ge x}$ was defined in Eq.~\eqref{Pi_X_I_definition}. 
\end{theorem}

{\bf Remark.} 
We note that the decay rate satisfies 
\begin{align}
	\frac{1-\sqrt{1-16\zeta_{i,0}^2}}{4\zeta_{i,0}} < 1 \for \zeta_{i,0}<\frac{1}{4} . 
\end{align} 
From Assumption~\ref{assump:Repulsive condition} of $U_i >5\bar{J}_{i,k}$, we can find $\mathfrak{u}_i>0$ such that $\zeta_{i,0}<1/4$\footnote{For example, by defining $\Delta U_i$ by $U_i =5\bar{J}_{i,k}+ \Delta U_i$, we can choose $\mathfrak{u}_i=\Delta U_i/2>0$ and obtain 
	\begin{align}
		\zeta_{i,0} = \frac{\bar{J}_{i,k}}{4\bar{J}_{i,k}+ \Delta U_i/2} < \frac{1}{4}.  
	\end{align} 
}.

Regarding the parameter $\check{\mJ}$, from Corollaries~\ref{corol:large_q_quantum_energy_unconserve} and~\ref{corol:large_q_quantum_energy}, 
we have already derived 
\begin{align}
	\check{\mJ} =\begin{cases}  
		2 \bar{\mJ} &\for \textrm{Boson number is not conserved}  ,\\
		2 \bar{\mJ} +  \delta_{2N/|\Lambda|}  &\for \textrm{Boson number is conserved} .  
	\end{cases}
\end{align}
The condition~\eqref{assum_main_thm/Boson number distribution/BH} also implies an upper bound for the $(k/2)$th order moment.
By choosing $\ket{\Psi_Q}=\ket{\Omega}$ in Eq.~\eqref{definition_mE_q}, which gives $\mE_\mQ= E_0 $, we have   
\begin{align}
	0 \ge  (U_i -\bar{J}_{i,k}) \bra{\Omega} \nb_i^{k/2} \ket{\Omega} -\check{\mJ}\bra{\Omega} \nb_i^{k/2} \ket{\Omega}^{(k-1)/k} 
	\longrightarrow 
	\bra{\Omega} \nb_i^{k/2} \ket{\Omega}^{1/k} \le \frac{\check{\mJ}}{U_i -\bar{J}_{i,k}},
\end{align}
and hence 
\begin{align}
	\label{upper_bound_nb_i_k/2_thm}
	\bra{\Omega} \nb_i^{k/2} \ket{\Omega} \le \br{\frac{\check{\mJ}}{U_i -\bar{J}_{i,k}}}^k.
\end{align}

The parameter $M_{i,0}$ cannot be well-defined in the case where $\bar{J}_{i,k}=0$.
In this case, we can prove the following alternative corollary:
\begin{corol} \label{main_corol/Boson number distribution/BH}
	Let us adopt the same setup as in Theorem~\ref{main_thm/Boson number distribution/BH} and assume $\bar{J}_{i,k}=0$. 
	Then, for an arbitrary site $i\in \Lambda$, the boson number distribution decays exponentially as follows:
	\begin{align}
		\label{main_thm/Boson number distribution/BH/ineq_k-1}
		\bra{\Omega}  \Pi_{i,\ge x} \ket{\Omega} \le e^{-2(x-M_{i,0})/k}  
	\end{align} 
	with
	\begin{align}
		\label{Definitiion_M_i_0_k-1}
		M_{i,0}:=  \max \brr{2^{2/k}\br{\frac{\check{\mJ}}{U_i}}^{2}, \br{ \frac{16\bar{\mJ}+\check{\mJ}}{U_i}}^2}.
	\end{align} 
	We prove the statement after the proof of Theorem~\ref{main_thm/Boson number distribution/BH}. 
\end{corol}

\subsection{Preliminaries}\label{sec:Proof_boson_dist_preliminary} 
Without loss of generality, we consider the boson number distribution at the site $1$. 
We first decompose the ground state $\ket{\Omega}$ as  
\begin{align}
	\label{decomp_Omega_dist}
	&\ket{\Omega} = \sum_{m=0}^{\bar{m}} \Pi_{1, I_m} \ket{\Omega} = \sum_{m=0}^{\bar{m}} p_m^{1/2} \ket{\omega_m} ,  \notag \\
	&I_0=[0,M) ,\quad  I_m=[M+(m-1)k,M+mk)  \ (1\le m\le \bar{m}), \quad  I_{\bar{m}}=[M+(\bar{m}-1)k,\infty) 
\end{align}
with 
\begin{align}
	p_m:=\bra{\Omega}  \Pi_{1, I_m} \ket{\Omega} , \quad   \ket{\omega_m}:= p_m^{-1/2}\Pi_{1, I_m} \ket{\Omega} ,
\end{align}
where we defined $ \Pi_{1, I_m}$ using Eq.~\eqref{def_Pi_X_q} as
\begin{align}
	\Pi_{1, I_m} = \sum_{N\in I_m} \Pi_{1,N}.
\end{align}
Now, the integer $M$ is a control parameter which will be chosen afterward (see Lemma~\ref{lem:choice_of_M}).

The Hamiltonian~\eqref{Boson_most_general} can change the local boson number up to $k$, and hence we have 
\begin{align}
	\Pi_{1, I_m} H \Pi_{1, I_{m'}} = 0 \for |m-m'|> 1 . 
\end{align}
Under the above notation, the target probability of $\bra{\Omega}  \Pi_{1,\ge x} \ket{\Omega}$ is upper-bounded by 
\begin{align}
	\label{Desired_prob_distribution}
	\bra{\Omega}  \Pi_{1,\ge x} \ket{\Omega} \le p_{\bar{m}} , 
\end{align}
where we choose $\bar{m}$ as 
\begin{align}
	\label{Desired_prob_distribution__2}
	\bar{m}=\floor{1+\frac{x-M}{k}}. 
\end{align}

We here define $\mQ_m$ as follows: 
\begin{align}
	\label{mQ_definition}
	\max_{i \in \Lambda} \bra{\omega_m} \nb_i^{k/2}  \ket{\omega_m}=\mQ^{k/2}_m ,
\end{align}
which also implies
\begin{align}
	\label{tilde_mQ_definition}
	\mQ^{k/2}_m \ge  (M+mk-k)^{k/2} \for m\ge 1
\end{align}
from $\bra{\omega_m} \nb_1^{k/2}  \ket{\omega_m} \ge [M+(m-1)k]^{k/2}$. 
Moreover, from the condition~\eqref{assum_main_thm/Boson number distribution/BH}, we immediately obtain 
\begin{align}
	\bra{\omega_m} H \ket{\omega_m}- E_0 \ge (U_i -\bar{J}_{i,k}) \mQ_m^{k/2} -\check{\mJ}  \mQ_m^{(k-1)/2}.
\end{align}

For the choice of the integer $M$, we prove the following lemma, which will be used for the proof of Theorem~\ref{main_thm/Boson number distribution/BH}:
\begin{lemma} \label{lem:choice_of_M}
	By choosing the integer $M$ such that 
	\begin{align}
		\label{M_0_chioice_lem}
		M= \ceil{M_0},\quad M_0= \max \brr{2^{2/k}\br{\frac{\check{\mJ}}{U_1 -\bar{J}_{1,k}}}^{2}, \br{ \frac{\check{\mJ} \bar{J}_{1,k} + 2 \bar{\mJ} (U_1- \mathfrak{u}_1 -\bar{J}_{1,k})}{\mathfrak{u}_1\bar{J}_{1,k}} }^2},
	\end{align}
	we have 
	\begin{align}
		&p_0\ge \frac{1}{2} , \quad \mQ_0^{k/2} \le 2\br{\frac{\check{\mJ}}{U_i -\bar{J}_{i,k}}}^k  , 
		\label{p_0_ge_1/2_mQ_0} 
	\end{align}
	and 
	\begin{align}
		\label{U_1-bar_J_1_k_tilde_mQ}
		(U_1 -\bar{J}_{1,k})\mQ_m^{k/2} -\check{\mJ}  \mQ_m^{(k-1)/2}  
		\ge  \frac{U_1- \mathfrak{u}_1 -\bar{J}_{1,k}}{\bar{J}_{1,k}} \br{ \bar{J}_{1,k} \mQ_m^{k/2} +  2 \bar{\mJ} \mQ_m^{(k-1)/2} } 
	\end{align}
	for $m\ge 1$, 
	where $\mathfrak{u}_1$ is an arbitrary positive parameter. 
\end{lemma}

\textit{Proof of Lemma~\ref{lem:choice_of_M}.}
For the first inequality of $p_0$, we use the Markov inequality as follows:
\begin{align}
	p_0=  \bra{\Omega} \Pi_{1, I_0} \ket{\Omega} = 1- \sum_{N\ge M} \norm{ \Pi_{1,N} \ket{\Omega}}^2 
	&\ge 1- \frac{\bra{\Omega}\nb_1^{k/2}\ket{\Omega}}{M^{k/2}} \notag \\
	&\ge 1- \frac{1}{M^{k/2}} \br{\frac{\check{\mJ}}{U_1 -\bar{J}_{1,k}}}^k ,
\end{align}
where we use the inequality~\eqref{upper_bound_nb_i_k/2_thm}. 
Therefore, $p_0 \ge 1/2$ is derived from the choice of $M_0$ in~\eqref{M_0_chioice_lem}. 

For the second inequality in~\eqref{p_0_ge_1/2_mQ_0} for $\mQ_0$, we first obtain 
\begin{align}
	\bra{\Omega} \nb_i^{k/2} \ket{\Omega} \ge p_0 \bra{\omega_0} \nb_i^{k/2} \ket{\omega_0} 
	\ge  \frac{ \mQ_0^{k/2} }{2} ,
\end{align}
where we use $p_0\ge 1/2$ and the definition~\eqref{mQ_definition}. 
By applying the above inequality to~\eqref{upper_bound_nb_i_k/2_thm}, we obtain  
\begin{align}
	\frac{ \mQ_0^{k/2} }{2} \le \br{\frac{\check{\mJ}}{U_i -\bar{J}_{i,k}}}^k,
\end{align}
which reduces to the second inequality~\eqref{p_0_ge_1/2_mQ_0}. 

The third inequality~\eqref{U_1-bar_J_1_k_tilde_mQ} is proven as follows:
\begin{align}
	&(U_1 -\bar{J}_{1,k})\mQ_m^{k/2} -\check{\mJ}  \mQ_m^{(k-1)/2}  \notag \\
	&= \frac{U_1- \mathfrak{u}_1 -\bar{J}_{1,k}}{\bar{J}_{1,k}} \br{\bar{J}_{1,k} \mQ_m^{k/2} +  2 \bar{\mJ} \mQ_m^{(k-1)/2} }
	+\mathfrak{u}_1 \mQ_m^{k/2} -\check{\mJ}  \mQ_m^{(k-1)/2} -  \frac{2 \bar{\mJ} (U_1- \mathfrak{u}_1 -\bar{J}_{1,k})}{\bar{J}_{1,k}}  \mQ_m^{(k-1)/2} \notag \\
	&= \frac{U_1- \mathfrak{u}_1 -\bar{J}_{1,k}}{\bar{J}_{1,k}}   \br{\bar{J}_{1,k} \mQ_m^{k/2} +  2 \bar{\mJ} \mQ_m^{(k-1)/2} } 
	+ \mathfrak{u}_1 \mQ_m^{(k-1)/2} \brr{\mQ_m^{1/2} -   \frac{\check{\mJ} \bar{J}_{1,k} + 2 \bar{\mJ} (U_1- \mathfrak{u}_1 -\bar{J}_{1,k})}{\mathfrak{u}_1\bar{J}_{1,k}}  }\notag \\
	&\ge \frac{U_1- \mathfrak{u}_1 -\bar{J}_{1,k}}{\bar{J}_{1,k}}  \br{\bar{J}_{1,k} \mQ_m^{k/2} +  2 \bar{\mJ} \mQ_m^{(k-1)/2} } ,
\end{align} 
where we use $\mQ_m^{1/2}\ge M^{1/2} \ge M_0^{1/2}\ge \brr{\check{\mJ} \bar{J}_{1,k} + 2 \bar{\mJ} (U_1- \mathfrak{u}_1 -\bar{J}_{1,k})}/(\mathfrak{u}_1\bar{J}_{1,k})$ from the definition in~\eqref{tilde_mQ_definition}.

This completes the proof. $\square$

{~}

\hrulefill{\bf [ End of Proof of Lemma~\ref{lem:choice_of_M}]}

{~}

In particular, for the case of $\bar{J}_{1,k}=0$, we prove the following corollary: 
\begin{corol} \label{corol:choice_of_M}
	By choosing the integer $M$ such that 
	\begin{align}
		\label{M_0_chioice_lem_bar_J=0}
		M= \ceil{M_0},\quad M_0= \max \brr{2^{2/k}\br{\frac{\check{\mJ}}{U_1}}^{2}, \br{ \frac{16\bar{\mJ}+\check{\mJ}}{U_1}}^2},
	\end{align}
	We obtain the same inequalities as in~\eqref{p_0_ge_1/2_mQ_0} and the upper bound of 
	\begin{align}
		\label{U_1-bar_J_1_k_tilde_mQ_J=0}
		U_1 \mQ_m^{k/2} -\check{\mJ}  \mQ_m^{(k-1)/2} 
		\ge 16 \bar{\mJ} \mQ_m^{(k-1)/2}. 
	\end{align} 
\end{corol}

\textit{Proof of Corollary~\ref{corol:choice_of_M}.}
The inequalities in \eqref{p_0_ge_1/2_mQ_0} holds by simply setting $\bar{J}_{1,k}=0$. Hence, we only have to reconsider the inequality~\eqref{U_1-bar_J_1_k_tilde_mQ}.
We calculate
\begin{align}
	\frac{(U_1 -\bar{J}_{1,k})\mQ_m^{k/2} -\check{\mJ}  \mQ_m^{(k-1)/2}}{ \bar{J}_{1,k} \mQ_m^{k/2} +  2 \bar{\mJ} \mQ_m^{(k-1)/2}}  
	=\frac{U_1\mQ_m^{k/2} -\check{\mJ}  \mQ_m^{(k-1)/2}}{ 2 \bar{\mJ} \mQ_m^{(k-1)/2}} 
	&= \frac{U_1\mQ_m^{1/2} -\check{\mJ}}{ 2 \bar{\mJ}} \ge  \frac{U_1M^{1/2} -\check{\mJ}}{ 2 \bar{\mJ}} \ge 8,
\end{align} 
which is ensured for $M>(16\bar{\mJ}+\check{\mJ})^2/U_1^2$. 
This completes the proof. $\square$

\subsection{Proof of Theorem~\ref{main_thm/Boson number distribution/BH}}
\label{sec:Proof_boson_dist}

Based on the decomposition~\eqref{decomp_Omega_dist}, we consider the energy contribution of the state $\ket{\omega_m}$ to the average $\bra{\Omega} H\ket{\Omega} $, which is characterized by the parameter $p_m$.
In detail, we let 
\begin{align}
	\ket{\Omega^{\neq m}} =\frac{1}{\sqrt{1- p_m} } \sum_{s\neq m}^\infty p_s^{1/2} \ket{\omega_s} ,
\end{align}
where the state $\ket{\Omega^{\neq m}} $ is normalized since $\sum_{s\neq m}^\infty p_s = 1- p_m$. 
We aim to separate the contribution by $s=m$ as follows: 
\begin{align}
	E_0= (1- p_m ) \bra{\Omega^{\neq m}} H\ket{\Omega^{\neq m} } 
	+p_m\Delta E_m  \notag \\
	\longrightarrow \bra{\Omega^{\neq m}} H\ket{\Omega^{\neq m} } =\frac{E_0  -p_m \Delta E_m}{1 -p_m} ,
\end{align}
where $\Delta E_m$ depends on $\{p_s\}_{s=0}^m$ and will be calculated below.

Because $\ket{\Omega^{\neq m}}$ should satisfy the condition of 
\begin{align}
	\bra{\Omega^{\neq m}} H\ket{\Omega^{\neq m} } \ge E_0 ,
\end{align}
the coefficient $|a_m|$ need to satisfy
\begin{align}
	\label{cond_Delta_E_n_E_0_N}
	\bra{\Omega^{\neq m}} H\ket{\Omega^{\neq m} } =\frac{E_0  -p_m \Delta E_m}{1 -p_m} \ge E_0. 
\end{align}
The inequality implies 
\begin{align}
	\label{Delta_E_m_lower}
	\Delta E_m \le  E_0.
\end{align}
as long as $p_m\neq 0,1$

Let us first consider the case of $p_m=0$. In this case, the ground energy is simply given by
\begin{align}
	\bra{\Omega}H\ket{\Omega} = \bra{\Omega^{<m}}H\ket{\Omega^{<m}} + \bra{\Omega^{>m}}H\ket{\Omega^{>m}}  .
\end{align}
For $m>0$, because of $p_0\ge 1/2$ as in~\eqref{p_0_ge_1/2_mQ_0}, we can ensure
$\ket{\Omega^{>m}}\neq \ket{\Omega}$ and $\bra{\Omega^{>m}}H\ket{\Omega^{>m}}>E_0$. 
Therefore, we obtain $p_{m+1}=p_{m+2}=\cdots =p_{\bar{m}}=0$, which trivially yields the main inequality~\eqref{main_thm/Boson number distribution/BH/ineq} from~\eqref{Desired_prob_distribution}. 
On the other hand, the case of $p_m=1$ is prohibited because of $p_0\ge 1/2$.
We thus need to consider the inequality~\eqref{Delta_E_m_lower} for the non-trivial cases. 

In the following, we consider the parameter $\Delta E_m$, which is calculated as 
\begin{align}
	\label{Delta_E_m_upp}
	&p_m \Delta E_m  \notag\\
	&= p_m \bra{\omega_m} H \ket{\omega_m} 
	- \sqrt{p_m p_{m+1}} \br{ \bra{\omega_{m+1}}\widehat{H_{0,1}}(\mathbf{b},\mathbf{b}^\dagger) \ket{\omega_m} +{\rm c.c.} }
	- \sqrt{p_m p_{m-1}} \br{ \bra{\omega_{m-1}} \widehat{H_{0,1}}(\mathbf{b},\mathbf{b}^\dagger) \ket{\omega_m} +{\rm c.c.} } \notag \\
	&\ge p_m \bra{\omega_m} H \ket{\omega_m} 
	-2 \sqrt{p_m p_{m+1}} \br{\bra{\omega_{m+1}} \overline{H_{0,1}}(\mathbf{b},\mathbf{b}^\dagger) \ket{\omega_{m+1}}  \bra{\omega_{m}} \overline{H_{0,1}}(\mathbf{b},\mathbf{b}^\dagger) \ket{\omega_{m}} }^{1/2} \notag \\
	&\quad \quad\quad\quad\quad\quad\quad - 2\sqrt{p_m p_{m-1}}\br{\bra{\omega_{m-1}} \overline{H_{0,1}}(\mathbf{b},\mathbf{b}^\dagger) \ket{\omega_{m-1}}  \bra{\omega_{m}} \overline{H_{0,1}}(\mathbf{b},\mathbf{b}^\dagger) \ket{\omega_{m}} }^{1/2} ,
\end{align}
where we use the Cauchy-Schwarz inequality, $\Pi_{1, I_m} H_0(\mathbf{b},\mathbf{b}^\dagger) \Pi_{1, I_{m'}} = \Pi_{1, I_m} \widehat{H_{0,1}}(\mathbf{b},\mathbf{b}^\dagger) \Pi_{1, I_{m'}}$ for $m\neq m'$, and the definition~\eqref{overline_H_0_i_vec} for $\overline{H_{0,1}}(\mathbf{b},\mathbf{b}^\dagger)$. 
We obtain a similar inequality to \eqref{Omega_over_line_upp_fin} as 
\begin{align}
	\label{omega_m_m-1_H_ave}
	\bra{\omega_m} \overline{H_{0,1}}(\mathbf{b},\mathbf{b}^\dagger) \ket{\omega_m}  \le \bar{J}_{1,k} \mQ_m^{k/2} +  2 \bar{\mJ} \mQ_m^{(k-1)/2} = : T^2_m 
	\for \forall m, 
\end{align}
where $ \mQ_m$ has been defined by Eq.~\eqref{mQ_definition}. 
Also, by using the condition~\eqref{assum_main_thm/Boson number distribution/BH}, we have 
\begin{align}
	\label{omega_m_H_ave}
	\bra{\omega_m} H \ket{\omega_m}  
	&\ge  (U_1 -\bar{J}_{1,k}) \mQ_m^{k/2} -\check{\mJ}  \mQ_m^{(k-1)/2}+E_0 \notag \\
	&\ge \frac{1}{\zeta_0} T^2_m + E_0 ,  \quad \zeta_0:=\frac{\bar{J}_{1,k}}{U_1- \mathfrak{u}_1 -\bar{J}_{1,k}}, 
\end{align}
where the parameter $\mathfrak{u}_1$ is defined in Lemma~\ref{lem:choice_of_M}, and we use the inequality~\eqref{U_1-bar_J_1_k_tilde_mQ} for $\mQ_m$.

By applying the inequalities~\eqref{omega_m_m-1_H_ave} and~\eqref{omega_m_H_ave} to \eqref{Delta_E_m_upp}, we obtain 
\begin{align}
	p_m \Delta E_m\ge 
	&p_m \brr{\frac{1}{\zeta_0} T^2_m + E_0} -2 \sqrt{p_m  p_{m+1} } T_mT_{m+1}
	-2 \sqrt{p_mp_{m-1}} T_m T_{m-1} .
\end{align}
From the inequality~\eqref{Delta_E_m_lower}, i.e., $p_m E_0 \ge p_m \Delta E_m $, we can derive 
\begin{align}
	\label{sqrt_p_m_recursive}
	\sqrt{p_m} \le  2\zeta_0 \br{ \sqrt{p_{m+1}} \frac{T_{m+1}}{T_m} 
		+ \sqrt{p_{m-1}}\frac{T_{m-1}}{T_m}  } .
\end{align}
Here, the quantity $T_m$ depends on $\mQ_m$, which cannot be controlled in general.
Hence, the coefficients $2\zeta_0 T_{m+1}/T_m$ and $2\zeta_0 T_{m-1}/T_m$ can be arbitrarily larger than $1$.
At first glance, this prohibits us from deriving a meaningful inequality for $\sqrt{p_m}$. 
With a careful estimation, we can prove the following lemma:

\begin{lemma} \label{lem:number_sequance_solve}
	Let $\{x_m\}_{m=0}^{\bar{m}}$ be an arbitrary set of positive numbers.  
	We then consider a number sequence $\{a_m\}_{m=0}^{\bar{m}}$ such that 
	\begin{align}
		a_m \le \zeta \frac{x_{m-1}}{x_m} a_{m-1} + \zeta \frac{x_{m+1}}{x_m} a_{m+1},\quad 
		\zeta \le \frac{1}{2},
	\end{align}
	which holds for $m\ge 1$.
	We then obtain the upper bound of 
	\begin{align}
		a_m \le \br{\frac{1-\sqrt{1-4\zeta^2}}{2\zeta}}^m \frac{x_0a_0}{x_m} 
	\end{align}
	for $\forall m\in [0,\bar{m}]$. 
\end{lemma}

\textit{Proof of Lemma~\ref{lem:number_sequance_solve}.}
In the proof, we let $a_{\bar m+1}:=0$.
We first prove 
\begin{align}
	\label{a_m_inequality_m-1}
	a_m \le z_\zeta \zeta \frac{x_{m-1}}{x_m} a_{m-1} ,\quad z_\zeta=\frac{1-\sqrt{1-4\zeta^2}}{2\zeta^2}. 
\end{align}
Note that under the assumption of $\zeta\le 1/2$, we have $1\le z_\zeta \le 2$. 
For the proof, we use induction. For $m=\bar{m}$, we trivially obtain 
\begin{align}
	a_m \le \zeta \frac{x_{m-1}}{x_m} a_{m-1} \le z_\zeta \zeta \frac{x_{m-1}}{x_m} a_{m-1} 
\end{align}
because of $z_\zeta  \ge 1$ and $a_{\bar m+1}=0$. 
We assume the inequality~\eqref{a_m_inequality_m-1} for $m\in [m_0,\bar{m}]$ and prove the case of $m=m_0-1$. 
We have 
\begin{align}
	\label{a_m_0-1_zeta_up}
	a_{m_0-1} \le \zeta \frac{x_{m_0-2}}{x_{m_0-1}} a_{m_0-2} + \zeta \frac{x_{m_0}}{x_{m_0-1}} a_{m_0}.
\end{align}
From the inequality~\eqref{a_m_inequality_m-1} for $m=m_0$, we have 
\begin{align}
	\zeta \frac{x_{m_0}}{x_{m_0-1}} a_{m_0}\le 
	\zeta \frac{x_{m_0}}{x_{m_0-1}}\cdot z_\zeta \zeta  \frac{x_{m_0-1}}{x_{m_0}} a_{m_0-1}=z_\zeta \zeta^2 a_{m_0-1}  ,
\end{align}
which reduces the inequality~\eqref{a_m_0-1_zeta_up} to 
\begin{align}
	\label{a_m_0-1_zeta_up__2}
	a_{m_0-1} \le \frac{\zeta}{1-z_\zeta \zeta^2} \cdot \frac{x_{m_0-2}}{x_{m_0-1}} a_{m_0-2}= z_\zeta \zeta \frac{x_{m_0-2}}{x_{m_0-1}} a_{m_0-2}, 
\end{align}
where we use the form of $z_\zeta=(1-\sqrt{1-4\zeta^2})/(2\zeta^2)$. 

Using the inequality~\eqref{a_m_inequality_m-1}, we obtain for $\forall m$
\begin{align}
	a_m \le (z_\zeta \zeta)^m \frac{x_0}{x_m} a_0  .
\end{align}
This completes the proof.  $\square$

{~}

\hrulefill{\bf [ End of Proof of Lemma~\ref{lem:number_sequance_solve}]}

{~}

We apply Lemma~\ref{lem:number_sequance_solve} to the inequality~\eqref{sqrt_p_m_recursive} with 
\begin{align}
	a_m \to \sqrt{p_m},\quad x_m \to T_m   , \quad \zeta \to 2\zeta_0 ,
\end{align}
and obtain 
\begin{align}
	\label{upper_bound_p_m_fin0}
	&\sqrt{p_m} \le  \br{ \frac{1-\sqrt{1-16\zeta_0^2}}{4\zeta_0}}^m \frac{T_0}{T_m} \sqrt{p_0} \notag \\
	&\longrightarrow p_m \le  \br{ \frac{1-\sqrt{1-16\zeta_0^2}}{4\zeta_0}}^{2m} \frac{T^2_0}{T^2_m} . 
\end{align}

We finally estimate the parameters $T_0$ and $T_m$. For $T_0$, we obtain the upper bound of
\begin{align}
	T_0^2 
	&= \bar{J}_{1,k} \mQ_0^{k/2} +  2 \bar{\mJ} \mQ_0^{(k-1)/2} \notag \\
	&\le \bar{J}_{1,k} \bar{\mQ}_0^{k/2} +  2 \bar{\mJ} \bar{\mQ}_0^{(k-1)/2}  ,\quad 
	\bar{Q}_0^{k/2}:= 2\br{\frac{\check{\mJ}}{U_i -\bar{J}_{i,k}}}^k ,
\end{align}
where we use the upper bound~\eqref{p_0_ge_1/2_mQ_0}. 
Also, we have $\mQ_m \ge M+mk-k$ from~\eqref{tilde_mQ_definition}, and hence 
\begin{align}
	T_m^2 = \bar{J}_{1,k} \mQ_m^{k/2} +  2 \bar{\mJ} \mQ_m^{(k-1)/2} 
	&\ge  
	\bar{J}_{1,k} (M+mk-k)^{k/2} + 2 \bar{\mJ} (M+mk-k)^{(k-1)/2}  \notag \\
	&\ge  
	\bar{J}_{1,k} \bar{Q}_0^{k/2}+ 2 \bar{\mJ} \bar{Q}_0^{(k-1)/2} ,
\end{align}
where we use $(M+mk-k)^{k/2} \ge M_0^{k/2} \ge \bar{Q}_0^{k/2}$ from the definition~\eqref{M_0_chioice_lem}. 
We thus obtain $T_m^2 \ge T_0^2$, which reduces the inequality~\eqref{upper_bound_p_m_fin0} to 
\begin{align}
	\label{p_bar_m_final_upp_bound_1}
	p_{\bar{m}} \le  \br{ \frac{1-\sqrt{1-16\zeta_0^2}}{4\zeta_0}}^{2\bar{m}} 
\end{align} 
by choosing $m=\bar{m}$, where $\bar{m}$ was defined in Eq.~\eqref{Desired_prob_distribution__2}.  
By using 
\begin{align}
	\floor{1+\frac{x-\ceil{x_0}}{k}} \ge \frac{x-x_0}{k} 
\end{align}
for any $x\in \mathbb{Z}$ and $k\in \mathbb{N}$, we obtain 
\begin{align}
	\bar{m}=\floor{1+\frac{x-\ceil{M_0}}{k}} \ge \frac{x-M_0}{k} ,
\end{align}
which reduces inequality~\eqref{p_bar_m_final_upp_bound_1} to the desired form~\eqref{main_thm/Boson number distribution/BH/ineq}.
This completes the proof of Theorem~\ref{main_thm/Boson number distribution/BH}. $\square$

{~}

The proof of Corollary~\ref{main_corol/Boson number distribution/BH} is exactly the same since the proof does not rely on whether or not $\bar{J}_{1,k}=0$.
As the only difference, we adopt the $M_0$ in Corollary~\ref{corol:choice_of_M} instead of Lemma~\ref{lem:choice_of_M}. 
Under the choice of Eq.~\eqref{M_0_chioice_lem_bar_J=0}, we have $\zeta_0=1/8$ because of the inequality~\eqref{U_1-bar_J_1_k_tilde_mQ_J=0}. 
Then, we obtain 
\begin{align}
	\frac{1-\sqrt{1-16\zeta_0^2}}{4\zeta_0}= 2 - \sqrt{3} = 0.267949 \cdots \le e^{-1},  
\end{align} 
which yields the inequality~\eqref{main_thm/Boson number distribution/BH/ineq_k-1}.
This completes the proof of Corollary~\ref{main_corol/Boson number distribution/BH}. $\square$

\clearpage

\section{Boson number distribution in $\phi4$-type models} \label{sec:Boson number distribution in phi4 model}

\subsection{Main theorem}

We here consider the $\phi4$-type which is given by from Eqs.~\eqref{phi_4_type_model} and \eqref{mF_phi_explicit}:
\begin{align}
	H& = \sum_{i\in \Lambda} \mu_i \pi_i^2 + \mF(\boldsymbol{\phi}) \notag \\
	&=\sum_{i\in \Lambda} \mu_i \pi_i^2 +\sum_{k_1=1}^{k/2} \sum_{i_1,i_2,\ldots, i_{2k_1} \in \Lambda} f_{i_1,i_2,\ldots, i_{2k_1} } \phi_{i_1}\phi_{i_2} \cdots \phi_{i_{2k_1}} .  
	\label{phi_4_type_model_again}
\end{align}
In this case, Assumption~\ref{assump:Repulsive condition} no longer holds, and we need a qualitatively different approach.
For the convenience of readers, we show the basic parameters $\bar{\mu}$ in Eq.~\eqref{def_bar_mu_phi} and $\bar{f}$ in Eq.~\eqref{Definition/of_bar_f} again: 
\begin{align}
	\label{def_bar_mu_phi__re}
	\bar{\mu} := \max_{i\in \Lambda} (|\mu_i|) ,
\end{align}
and 
\begin{align}
	\label{Definition/of_bar_f__re}
	\bar{f}= \max_{i\in \Lambda} \br{ 
		\sum_{k_1=1}^{k/2} \sum_{\substack{ i_1,i_2,\ldots, i_{2k_1} \in \Lambda \\ \{i_1,i_2,\ldots, i_{2k_1}\} \ni i}} 
		\abs{ f_{i_1,i_2,\ldots, i_{2k_1} } } } ,
\end{align}
respectively.

In the subsequent subsections, we aim to prove the following theorem:  
\begin{theorem} \label{thm_boson_dist_phi4}
	For an arbitrary site $i$, the boson number distribution satisfies the following sub-exponential concentration bound:
	\begin{align}
		\label{main_ineq_corol_boson_dist_phi4}
		&\bra{\Omega} \Pi_{i, >x} \ket{\Omega} \le 4e^k \exp \br{-\frac{k x^{1/k}}{8e \tilde{C}}},
	\end{align}
	with $\tilde{C}$ defined in Eq.~\eqref{prop:main_inequality_pi_moment_rewrite} below, 
	where the projection $\Pi_{i, >x}$ has been defined in Eq.~\eqref{Pi_X_I_definition}. 
\end{theorem}
The overall proof structure of Theorem 2 is summarized in Supplementary Figure~\ref{fig_supp_phi4_boson_number}.

\begin{figure}
	\begin{center}
		\includegraphics[width=\textwidth]{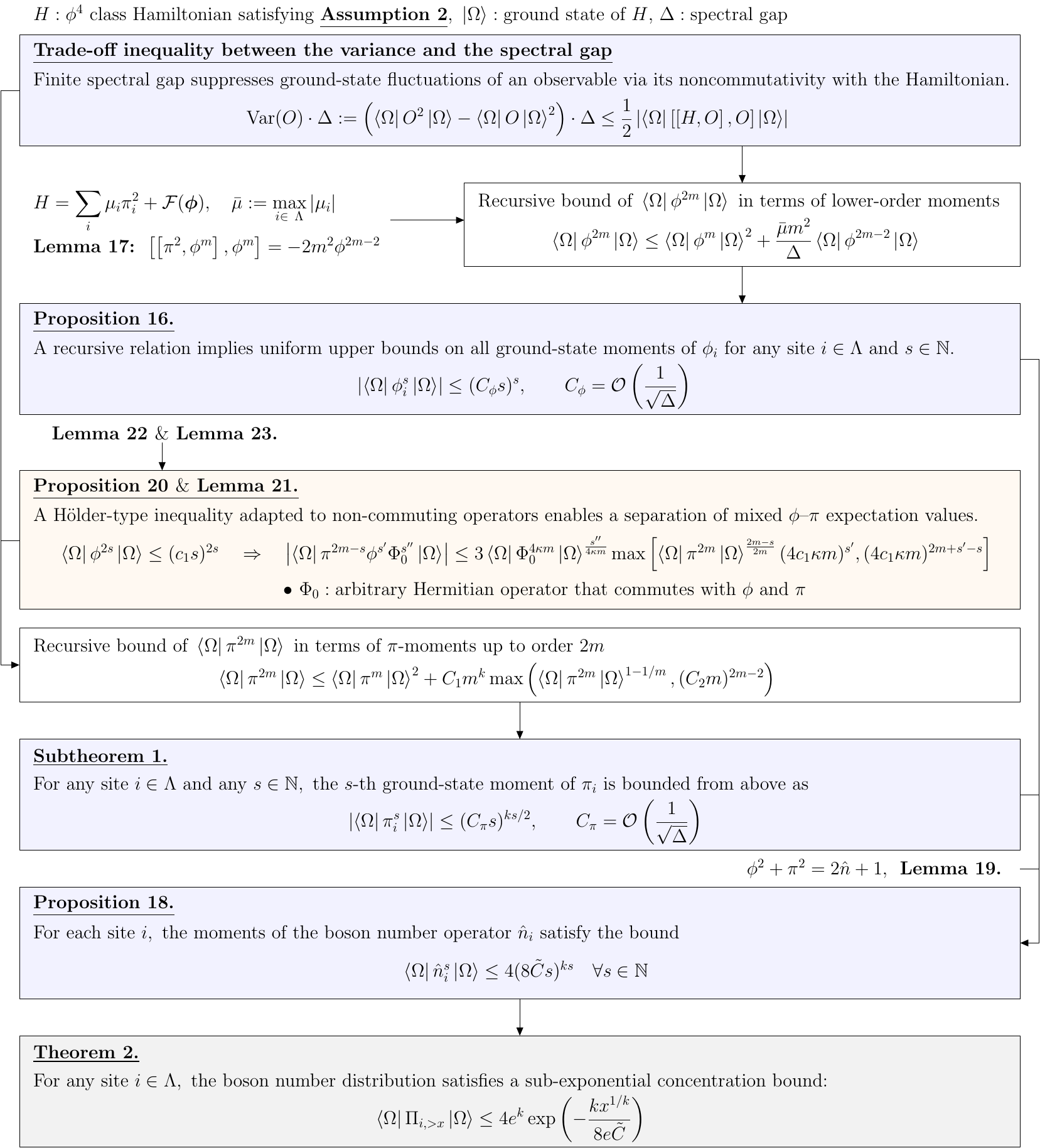}
	\end{center}
	\caption{This figure outlines the proof of Theorem 2, which establishes the subexponential decay of the boson-number distribution for $\phi4$ class Hamiltonians.}
	\label{fig_supp_phi4_boson_number}
\end{figure}

{\bf Remark.}
The primary difference between Theorem~\ref{main_thm/Boson number distribution/BH} for the Bose-Hubbard classes and this theorem is the dependence on the spectral gap $\Delta$.
In the former case, we need no assumption on the spectral gap; instead, we need a stronger condition of repulsive interactions (Assumption~\ref{assump:Repulsive condition}).

In the $\phi4$ cases, we only need the existence of the spectral gap. We utilize Assumption~\ref{assump:Parity symmetry} of the parity symmetry to ensure $\abs{\bra{\Omega} \phi_i \ket{\Omega}}=0$, which leads to the explicit definition of $\tilde{C}$ as follows: 
\begin{align}
	\label{tilde_C_explicit_from}
	\tilde{C} 
	&= k^2 \br{\frac{\bar{f}'}{\bar{\mu}}}^{1/k}\brr{ 2 \max_{i\in \Lambda} \abs{\bra{\Omega} \phi_i \ket{\Omega}} + 4 \br{ \frac{\bar{\mu}}{\Delta}}^{1/2} +1  } \notag \\
	&= k^2 \br{\frac{\bar{f}'}{\bar{\mu}}}^{1/k}\brr{ 4 \br{ \frac{\bar{\mu}}{\Delta}}^{1/2} +1  }, \quad \bar{f}'=\max(\bar{f} ,\bar{\mu}/2) .
\end{align} 
From $\tilde{C}\propto \Delta^{-1/2}$, we can see that the decay of the probability depends on the spectral gap as follows:
\begin{align}
	&\bra{\Omega} \Pi_{i, >x} \ket{\Omega} \le e^{-\Omega(x^{1/k} \Delta^{1/2})} = \exp \brrr{-\Omega\brr{ \br{\frac{x}{\Delta^{-k/2}}}^{1/k} }} \notag \\
	&\longrightarrow \bra{\Omega} \Pi_{i, >x} \ket{\Omega} \le \exp \brrr{-\Omega\brr{ \br{\frac{x}{\Delta^{-2}}}^{1/4} }} \for k=4.
\end{align}
The optimality of the gap dependence is still an open question. 
On the other hand, the numerical calculations suggest that sub-exponential decay will not improve to exponential decay, while the optimal $k$ dependence of the sub-exponential form is unclear (see Supplementary Figure~\ref{fig:phi4_concentration}).

\begin{figure}
	\centering
	\includegraphics[scale=0.5]{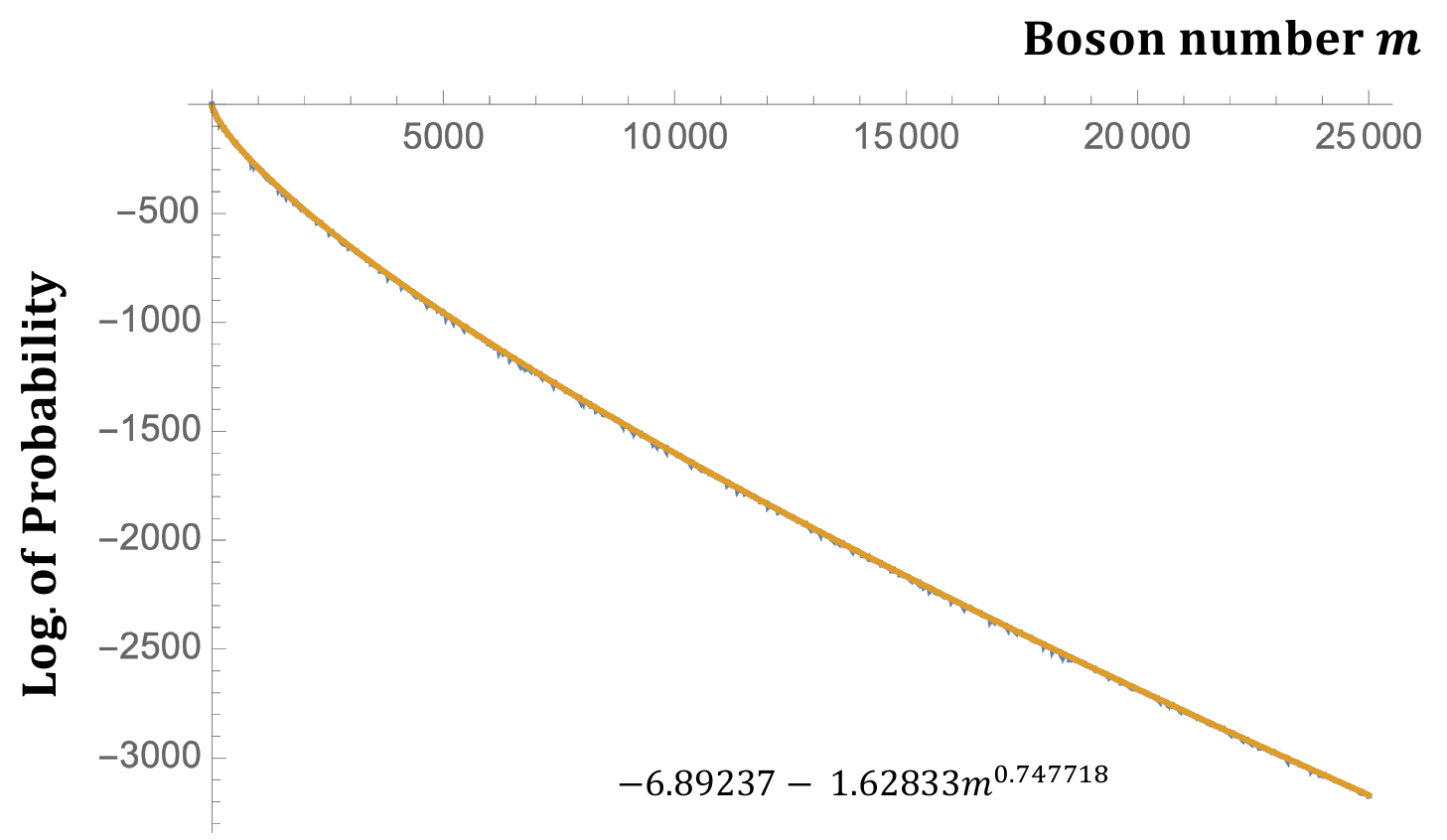}
	\caption{We consider the single-site $\phi^4$ model with Hamiltonian $H = \pi^2 + \phi^2 + \phi^4$. The probability $\bra{\Omega} \Pi_{=m} \ket{\Omega}$ $(0 \leq m \leq N)$ is calculated, where the total boson number cutoff is set to $N=25000$ and numerical precision is ensured up to 5000 digits.
		The orange line shows a subexponential fit given by $-6.89237 - 1.62833\times m^{0.747718}$, which indicates a subexponential decay of the form $e^{-3m/4}$.
	}
	\label{fig:phi4_concentration}
\end{figure}

\subsection{Useful relations}

\subsubsection{Generalized commutator relation}

If two operators $\mA$ and $\mB$ satisfy $[\mA,\mB]=1$, we obtain 
\begin{align}
	\label{generalized/Commutator_ineq}
	\brr{\mA^m, \mB^n} = \sum_{k=1}^{\min(m,n)} C_{k,m,n} \mB^{n-k}  \mA^{m-k} 
	=-  \sum_{k=1}^{\min(m,n)} (-1)^{k} C_{k,m,n}\mA^{m-k}  \mB^{n-k}  
\end{align}
with 
\begin{align}
	\label{Def_C_k_m_n}
	C_{k,m,n}:= k! \binom{m}{k} \binom{n}{k} .
\end{align}
This expression is immediately derived by parametrizing $\mA\to x_A \mA$ and $\mB\to x_B \mB$ with the expression of 
\begin{align}
	e^{x_A \mA} e^{x_B \mB} =  e^{x_B\mB} e^{x_A\mA}e^{x_Ax_B[\mA,\mB]} , 
\end{align}
where we use the Zassenhaus formula as $e^{x_A \mA+x_B \mB}= e^{x_A \mA} e^{x_B \mB} e^{-x_Ax_B[\mA,\mB]/2}= e^{x_B \mB}  e^{x_A \mA} e^{-x_Ax_B[\mB,\mA]/2}$. 
By comparing the terms with $x_A^mx_B^n$, we derive Eq.~\eqref{generalized/Commutator_ineq}.

By using these relations, for $\phi$ and $\pi$ with $[\phi,\pi]=i$ (or $[-i \phi,\pi]=1$), we obtain 
\begin{align}
	\label{phi_m_pi_n_commutator}
	\brr{\phi^m, \pi^n}= \sum_{k=1}^{\min(m,n)}i^k C_{k,m,n} \pi^{n-k}  \phi^{m-k}  =- \sum_{k=1}^{\min(m,n)} (-i)^{k}  C_{k,m,n} \phi^{m-k} \pi^{n-k} .
\end{align}

\subsubsection{Tradeoff relation between the variance and the spectral gap} 

As a key analytical tool, we use the following trade-off inequality, which connects the variance and the spectral gap~\cite{Kuwahara_2013,Kuwahara_2016_asymptotic,Kuwahara_2017}: 
\begin{align}
	\label{trade_off_relation}
	\var(O)\cdot \Delta \le \frac{1}{2} \abs{\bra{\Omega} \brr{\brr{H,O},O} \ket{\Omega}}, 
\end{align}
where the proof is elementary (for example, see Appendix A in Ref.~\cite{Kuwahara_2016_asymptotic},
and we use the above upper bound without proof.

\subsection{Moment of operator $\phi$}

We here consider the moment of the $\phi$ operator as $\abs{\bra{\Omega} \phi^{s} \ket{\Omega}} $, where we omit the lattice index $i$ for simplicity.
The following proposition holds for an arbitrary $\phi_i$ ($i\in \Lambda$): 
\begin{prop} \label{prop:phi_distribution}
	Under the existence of the spectral gap $\Delta$, the moment of $\phi$ is bounded from above by 
	\begin{align}
		\label{prop:phi_distribution/main_ineq}
		\abs{\bra{\Omega} \phi^{s} \ket{\Omega}} &\le  \brr {\br{\abs{\bra{\Omega} \phi \ket{\Omega}} + 2 \br{ \frac{\bar{\mu}}{\Delta}}^{1/2} }s}^{s} ,
	\end{align}
	for arbitrary $s\in \mathbb{N}$, where the parameter $\bar{\mu}$ has been defined in Eq.~\eqref{def_bar_mu_phi__re}, i.e., $\bar{\mu} := \max_{i\in \Lambda} (|\mu_i|)$.  
\end{prop}

{\bf Remark.} Under the assumption of the parity symmetry~\eqref{mF_phi_explicit}, we can always let 
\begin{align}
	\abs{\bra{\Omega} \phi \ket{\Omega}} =0 ,
\end{align}
while under the breakdown of the parity symmetry, we need to estimate the upper bound of $\abs{\bra{\Omega} \phi \ket{\Omega}}$.
It is possible by employing a similar analysis to the proof of Proposition~\ref{Prop:Moment_upper:bound_non-conserve}.
However, for this purpose, we need an additional condition similar to Assumption~\ref{assump:Repulsive condition} for the Bose-Hubbard cases, and the unconditional proof\footnote{We mean by the unconditional proof that any non-critical ground states with the spectral gap satisfy a boson-number concentration bound as well as the entanglement area law without any additional conditions.} of the entanglement area law is, in general, impossible.

\subsubsection{Proof of Proposition~\ref{prop:phi_distribution}}

We aim to upper bound the variance of $\phi^m$ for general $m$.
By applying the trade-off relation~\eqref{trade_off_relation}, we need to consider $\bra{\Omega} \brr{\brr{H,\phi^m},\phi^m} \ket{\Omega}$. Here, only the term of $ \sum_{i\in \Lambda} \mu_i \pi_i^2 $ in the Hamiltonian contributes to the commutator. 
Therefore, from $\max_{i\in\Lambda}(|\mu_i|)=\bar{\mu}$ in Eq.~\eqref{def_bar_mu_phi__re}, we obtain 
\begin{align}
	\label{trade_off_relation_phi^2}
	\Delta  \br{\bra{\Omega} \phi^{2m} \ket{\Omega}  - \bra{\Omega}\phi^m\ket{\Omega}^2 } \le \frac{\bar{\mu}}{2} \abs{\bra{\Omega} \brr{\brr{\pi^2,\phi^m},\phi^m} \ket{\Omega}}. 
\end{align}
To estimate the RHS of the above inequality, we consider the following lemma:
\begin{lemma} \label{Lem:double_commutator_phi^2}
	For an arbitrary power of $\phi$, the double commutator $\brr{\brr{\pi^2,\phi^m},\phi^m}$ is given as follows:
	\begin{align}
		\label{Lem:double_commutator_phi^2_main}
		\brr{\brr{\pi^2,\phi^m},\phi^m} = -2 m^2\phi^{2m-2}  . 
	\end{align}
\end{lemma}

\textit{Proof of Lemma~\ref{Lem:double_commutator_phi^2}.} 
We begin with the commutator $ \brr{\pi^2,\phi^m}$.  
Because of $[\pi,\phi]=-i$, we have 
\begin{align}
	\brr{\pi^2,\phi^m}=\pi [\pi,\phi^m] +[\pi,\phi^m]  \pi = -im \br{\pi \phi^{m-1}+\phi^{m-1}  \pi }.
\end{align}
In the same way, we can also derive
\begin{align}
	\brr{\br{\pi \phi^{m-1}+\phi^{m-1}  \pi }, \phi^m}= -2 i m \phi^{2m-2} .
\end{align}
By combining the above two equations, we prove Eq.~\eqref{Lem:double_commutator_phi^2_main}. $\square$

{~}

\hrulefill{\bf [ End of Proof of Lemma~\ref{Lem:double_commutator_phi^2}]}

{~}

By applying Lemma~\ref{Lem:double_commutator_phi^2} to the inequality~\eqref{trade_off_relation_phi^2}, we obtain 
\begin{align}
	\label{trade_off_relation_phi^2_2}
	\Delta \cdot \br{\bra{\Omega} \phi^{2m} \ket{\Omega}  - \bra{\Omega}\phi^m\ket{\Omega}^2 }  \le \frac{\bar{\mu}}{2} \abs{\bra{\Omega} \brr{\brr{\pi^2,\phi^m},\phi^m} \ket{\Omega}} \le \bar{\mu} m^2 \bra{\Omega} \phi^{2m-2} \ket{\Omega} ,
\end{align}
which gives 
\begin{align}
	\label{trade_off_relation_phi^2_2_fin}
	\bra{\Omega} \phi^{2m} \ket{\Omega} \le \bra{\Omega} \phi^m \ket{\Omega}^2 + \frac{\bar{\mu} m^2}{\Delta} \bra{\Omega} \phi^{2m-2} \ket{\Omega} .
\end{align}

We here rewrite the target inequality~\eqref{prop:phi_distribution/main_ineq} as follows: 
\begin{align}
	\label{moment_upp_bound_assume}
	\abs{\bra{\Omega} \phi^s \ket{\Omega}} \le \br{C  s}^{s} , \quad C =\abs{\bra{\Omega} \phi \ket{\Omega}} 
	+ 2 \br{ \frac{\bar{\mu}}{\Delta}}^{1/2} . 
\end{align}
For the proof, we use induction.  
For $s=1$, the inequality trivially holds because of $C  \ge \abs{\bra{\Omega} \phi \ket{\Omega}}$.
Also, for $s=2$, the inequality~\eqref{trade_off_relation_phi^2_2_fin} gives 
\begin{align}
	\bra{\Omega} \phi^2 \ket{\Omega} \le \bra{\Omega} \phi \ket{\Omega}^2 + \frac{\bar{\mu}}{\Delta} \le C^2 + \frac{C^2}{4} = \frac{5C^2}{4} \le(2C)^2 ,
\end{align}
where we use $\abs{\bra{\Omega} \phi \ket{\Omega}}\le C$ and $(\bar{\mu}/\Delta)^{1/2} \le C/2$. 

We assume the inequality~\eqref{moment_upp_bound_assume} up to $s=2m-2$ and consider the cases of $s=2m-1$ and $2m$, where $m\ge 2$
We first consider $s=2m$ and obtain 
\begin{align}
	\label{moment_2m_th_induction}
	\bra{\Omega} \phi^{2m} \ket{\Omega}& \le \bra{\Omega} \phi^m \ket{\Omega}^2 + \frac{\bar{\mu}m^2}{\Delta} \bra{\Omega} \phi^{2m-2} \ket{\Omega} \notag \\
	&\le \br{Cm}^{2m} +   \frac{\bar{\mu}m^2}{\Delta} \br{2Cm}^{2m-2} \notag \\
	&\le \br{2Cm}^{2m} \br{2^{-2m} +  \frac{\bar{\mu}}{4C^2\Delta} } \le \br{2Cm}^{2m},
\end{align}
where we use the inequality of 
\begin{align}
	2^{-2m} +  \frac{1}{4C^2} \cdot\frac{\bar{\mu}}{\Delta} \le \frac{1}{16}+  \frac{1}{16}= \frac{1}{8}<1 .
\end{align}
Note that $2^{-2m} \le 1/8$ because of $m\ge 2$. 
For the case of $s=2m-1$, we utilize the inequality~\eqref{moment_2m_th_induction} to derive
\begin{align}
	\label{2m-1_phi_Omega_ave}
	\abs{\bra{\Omega} \phi^{2m-1} \ket{\Omega}} &\le \bra{\Omega} \phi^{2m} \ket{\Omega}^{1-1/(2m)} 
	\le \brr{\frac{1}{8}  \br{2Cm}^{2m}}^{1-1/(2m)}  \le \br{\frac{1}{8}}^{3/4}   \br{2Cm}^{2m-1}  \notag \\
	&\le \br{\frac{1}{8}}^{3/4}  \brr{C(2m-1)}^{2m-1} \br{1+\frac{1}{2m-1}}^{2m-1}
	\le e \br{\frac{1}{8}}^{3/4} \brr{C(2m-1)}^{2m-1}  \le   \brr{C(2m-1)}^{2m-1} .
\end{align}
We thus prove the inequality~\eqref{moment_upp_bound_assume} for all $s\in \mathbb{N}$. 

This completes the proof of Proposition~\ref{prop:phi_distribution}. $\square$

\subsection{Moment of operator $\pi$}

Using the upper bound on the moment function $|\bra{\Omega} \phi^{s} \ket{\Omega}|$, we derive an upper bound for $|\bra{\Omega} \pi^{s} \ket{\Omega}|$.
The major difficulty for the $\phi4$-type Hamiltonian lies in this point. 
We aim to prove the following subtheorem (see~\ref{Sec:Proof_Distribution of pi operator} for the proof):

\begin{subtheorem} \label{subthm:pi_distribution}
	Under the existence of the spectral gap $\Delta$, the moment of $\pi_{i}$ is bounded from above by 
	\begin{align}
		\label{prop:main_inequality_pi_moment}
		\bra{\Omega} \pi_i^s\ket{\Omega}
		\le \brr{\frac{\bar{f}'}{\bar{\mu}} \br{\check{c}_1k^2 s}^k }^{s/2} ,
	\end{align}
	for arbitrary $s\in \mathbb{N}$ and $i\in \Lambda$,
	where $\bar{f}'=\max(\bar{f} ,\bar{\mu}/2)$ and $\check{c}_1$ is defined as 
	\begin{align}
		\label{def:tilde_c_1}
		\check{c}_1=2 \max_{i\in \Lambda} \abs{\bra{\Omega} \phi_i \ket{\Omega}} + 4 \br{ \frac{\bar{\mu}}{\Delta}}^{1/2} +1. 
	\end{align}
	We recall that the parameters $\bar{\mu}$ and $\bar{f}$ were defined in Eqs.~\eqref{def_bar_mu_phi__re} and \eqref{Definition/of_bar_f__re}, respectively.
\end{subtheorem}

{\bf Remark.} As in Proposition~\ref{prop:phi_distribution}, we can set $\abs{\bra{\Omega} \phi \ket{\Omega}} =0$ under the parity symmetry~\eqref{mF_phi_explicit}. 
We rewrite the inequality~\eqref{prop:main_inequality_pi_moment} as 
\begin{align}
	\label{prop:main_inequality_pi_moment_rewrite}
	\bra{\Omega} \pi_i^s\ket{\Omega}
	\le \br{\tilde{C} s}^{ks/2} , \quad \tilde{C} = \check{c}_1k^2 \br{\frac{\bar{f}'}{\bar{\mu}}}^{1/k}   . 
\end{align}
Using the Markov inequality, we obtain the probability distribution of $\pi_i$ as 
\begin{align}
	\label{prop:main_inequality_prob_bound}
	\bra{\Omega} \Pi_{|\pi_i|>x} \ket{\Omega} 
	&\le \min_{s\in\mathbb{N}}\brr{ \br{\frac{\tilde{C} s}{x^{2/k}}}^{ks/2} } \notag \\
	&\le \exp \br{-\frac{k}{2} \floor{\frac{x^{2/k}}{e\tilde{C}}}} \le e^{k/2} e^{-x^{2/k}/(2e\tilde{C}/k)} ,
\end{align}
where we choose $s$ as $s=\floor {x^{k/2}/(e\tilde{C})}$.
This inequality gives subexponential decay instead of exponential decay, 
where $ \Pi_{|\pi_i|>x} $ denotes the projection onto the eigenspace of $\pi_i$ whose absolute eigenvalues are larger than $x$.

By comparing the moment bound~\eqref{prop:main_inequality_pi_moment_rewrite} with that for $\phi$, i.e.,  
\begin{align}
	\abs{\bra{\Omega} \phi^{s} \ket{\Omega}} \le  \brr {\br{\abs{\bra{\Omega} \phi \ket{\Omega}} + 2 \br{ \frac{\bar{\mu}}{\Delta}}^{1/2} }s}^{s}, 
\end{align} 
we can also ensure the following looser upper bound as 
\begin{align}
	\label{phi_s_upper_bound_re}
	\abs{\bra{\Omega} \phi^{s} \ket{\Omega}} \le   \br{\tilde{C} s}^{ks/2} .
\end{align}
Therefore, for both of $\abs{\bra{\Omega} \phi^{s} \ket{\Omega}} $ and $\abs{\bra{\Omega} \pi^{s} \ket{\Omega}} $, the upper bound $ \br{\tilde{C} s}^{ks/2} $ holds.

\subsection{Moment of the number operator $\nb$: concluding the proof of Theorem~\ref{thm_boson_dist_phi4}}

By combining Proposition~\ref{prop:phi_distribution} and Subtheorem~\ref{subthm:pi_distribution}, we finally prove the main theorem on the boson number distribution for the $\phi4$ class.
For this purpose, we first prove the following proposition on the moment upper bound (see~\ref{sec_Proof of Proposition_prop_boson_dist_phi4} for the proof): 
\begin{prop} \label{prop_boson_dist_phi4}
	For an arbitrary site $i$, the moment function of the boson number operator $\nb_i$ satisfies the following bound:
	\begin{align}
		\label{main_ineq_thm_boson_dist_phi4}
		\bra{\Omega} \nb^s \ket{\Omega} \le 4\br{8\tilde{C} s}^{ks} ,
	\end{align}
	where $s$ is an arbitrary positive integer, and the parameter $\tilde{C}$ is defined in Eq.~\eqref{prop:main_inequality_pi_moment_rewrite}. 
\end{prop}

Based on the above proposition, we immediately prove Theorem~\ref{thm_boson_dist_phi4} regarding the boson number distribution:

\subsubsection{Proof of Theorem~\ref{thm_boson_dist_phi4}.}

As in the inequality~\eqref{prop:main_inequality_prob_bound}, we use the Markov inequality as follows: 
\begin{align}
	\label{prop:main_inequality_prob_bound_nb}
	\bra{\Omega} \Pi_{i, >x} \ket{\Omega} 
	&\le \min_{s\in\mathbb{N}}\brr{ 4\br{\frac{8\tilde{C} s}{x^{1/k}}}^{ks}  } .
\end{align}
By choosing $s$ as 
\begin{align}
	s= \floor{\frac{x^{1/k}}{8e \tilde{C}} } \ge \frac{x^{1/k}}{8e \tilde{C}}-1,
\end{align}
we reduce the inequality~\eqref{prop:main_inequality_prob_bound_nb} to the desired form: 
\begin{align}
	\bra{\Omega} \Pi_{i, >x} \ket{\Omega} 
	&\le  4\exp \br{- k \floor{\frac{x^{1/k}}{8e \tilde{C}} }} \le  4e^k \exp  \br{-\frac{k x^{1/k}}{8e \tilde{C}}} .
\end{align}
This completes the proof of Theorem~\ref{thm_boson_dist_phi4}. $\square$

\subsubsection{Proof of Proposition~\ref{prop_boson_dist_phi4}} \label{sec_Proof of Proposition_prop_boson_dist_phi4}
For simplicity of the notations, we omit the site index $i$. 
From the definitions of $\phi$ and $\pi$ in Eq.~\eqref{phi_pi_definition}, we have 
\begin{align}
	\phi^2 + \pi^2 = 2\nb+1 ,
\end{align}
and hence 
\begin{align}
	\label{phi2pi2_2n+1}
	\bra{\Omega} (\phi^2 + \pi^2)^s \ket{\Omega} = \bra{\Omega} (2\nb+1)^s \ket{\Omega}  ,
\end{align}
which allows us to estimate the higher-order moments of the boson number from
the moments of  $\bra{\Omega}\phi^s\ket{\Omega}$ and $\bra{\Omega}\pi^s\ket{\Omega}$.

In the following, we generally decompose
\begin{align}
	\label{phi2+pi2_decomp}
	(\phi^2 + \pi^2)^s = \sum_{\substack{m_1,m_2\le 2s \\ m_1+m_2 \le 2s}} \lambda^{(s)}_{m_1,m_2}  \phi^{m_1}\pi^{m_2} ,
\end{align}
where we can find such an expression using the commutation relation~\eqref{phi_m_pi_n_commutator}.
Unfortunately, it is a challenging task to find the explicit form of $\lambda^{(s)}_{m_1,m_2}$, and hence, we upper-bound the absolute value.  
For this purpose, we prove the following lemma:
\begin{lemma} \label{lem:coefficient/lamnda_s_m1m2} 
	For arbitrary $s$, $m_1$ and $m_2$ such that $m_1,m_2\le 2s$ and $m_1+m_2 \le 2s$, we obtain the upper bound of 
	\begin{align}
		\label{lem:coefficient/lamnda_s_m1m2/main_ineq} 
		\abs{ \lambda^{(s)}_{m_1,m_2} }\le 4^s s^{2s-m_1-m_2} .
	\end{align}
\end{lemma}

{~}

\textit{Proof of Lemma~\ref{lem:coefficient/lamnda_s_m1m2}.}
For the proof, we use the induction method.
For $s=1$, the inequality trivially holds since only the terms of $\lambda^{(s)}_{2,0}$ and $\lambda^{(s)}_{0,2}$ are non-zero and equal to $1$. 
For $s=2$, we have 
\begin{align}
	(\phi^2 + \pi^2)^{2}
	= \phi^4 + \pi^4 + \phi^2 \pi^2 +  \pi^2 \phi^2
	= \phi^4 + \pi^4 +2 \phi^2 \pi^2 - 4 i   \phi \pi -2 ,
\end{align}
where we use Eq.~\eqref{pi2_phi_m1_commutator} below with $m_1=2$. 
The above equation proves the inequality~\eqref{lem:coefficient/lamnda_s_m1m2/main_ineq} for $\lambda^{(2)}_{m_1,m_2}$.

We assume the inequality for general $s$ with $s\ge 2$ and consider the case of $s+1$.
Using the decomposition~\eqref{phi2+pi2_decomp}, we have 
\begin{align}
	\label{phi2+pi2_decomp_s;1}
	(\phi^2 + \pi^2)^{s+1}
	&= (\phi^2 + \pi^2) \sum_{\substack{m_1,m_2\le 2s \\ m_1+m_2 \le 2s}} \lambda^{(s)}_{m_1,m_2}  \phi^{m_1}\pi^{m_2} \notag \\
	&= \sum_{\substack{m_1,m_2\le 2s \\ m_1+m_2 \le 2s}} \lambda^{(s)}_{m_1,m_2} 
	\brr{  \phi^{m_1+2}\pi^{m_2} +  \phi^{m_1}\pi^{m_2+2} -2i m_1 \phi^{m_1-1}\pi^{m_2+1} - m_1(m_1-1)\phi^{m_1-2}\pi^{m_2} }  \notag \\
	&= \sum_{\substack{m_1,m_2\le 2s+2 \\ m_1+m_2 \le 2s+2}} \lambda^{(s+1)}_{m_1,m_2}  \phi^{m_1}\pi^{m_2} , 
\end{align}
where we use the commutation relation~\eqref{phi_m_pi_n_commutator} with $m\to m_1$ and $n\to 2$ as follows 
\begin{align}
	\pi^2 \phi^{m_1} = \phi^{m_1} \pi^2 -  [\phi^{m_1}, \pi^2] 
	&= \phi^{m_1} \pi^2 +\sum_{k=1}^{2} (-i)^{k}  k! \binom{2}{k} \binom{m_1}{k} \phi^{m_1-k} \pi^{2-k}  \notag \\
	&=  \phi^{m_1} \pi^2 - 2 i m_1\phi^{m_1-1} \pi - m_1(m_1-1) \phi^{m_1-2}   .
	\label{pi2_phi_m1_commutator}
\end{align}
From the equation, we obtain 
\begin{align}
	\lambda^{(s+1)}_{m_1,m_2} = \lambda^{(s)}_{m_1-2,m_2} +  \lambda^{(s)}_{m_1,m_2-2} - 2i (m_1+1)  \lambda^{(s)}_{m_1+1,m_2-1} - (m_1+1)(m_1+2)  \lambda^{(s)}_{m_1+2,m_2} .
\end{align}
where we let $\lambda^{(s)}_{m_1,m_2} = 0$ if the condition $m_1,m_2\le 2s$ or $m_1+m_2 \le 2s$ is not satisfied. 

Using the assumption for $\lambda^{(s)}_{m_1,m_2}$ and the inequality~\eqref{lem:coefficient/lamnda_s_m1m2/main_ineq}, we derive 
\begin{align}
	\abs{\lambda^{(s+1)}_{m_1,m_2}} 
	&\le  4^s \br{2 s^{2s-m_1-m_2+2} +2(m_1+2) s^{2s-m_1-m_2} + (m_1+2)^2 s^{2s-m_1-m_2-2} }\notag \\
	&\le  4^{s+1}  (s+1)^{2s-m_1-m_2+2} \brr{\frac{1}{2} + \frac{(m_1+1)}{2(s+1)^2} + \frac{(m_1+2)^2}{4(s+1)^4} } 
	\le  4^{s+1}  (s+1)^{2s-m_1-m_2+2} ,
\end{align}
where, in the last inequality, we use $m_1+2 \le 2s+2 =2(s+1)$ (from $m_1+m_2\le 2s$) and $s\ge 2$.
We thus prove the inequality~\eqref{lem:coefficient/lamnda_s_m1m2/main_ineq} in the case of $s+1$.
This completes the proof. $\square$

{~}

\hrulefill{\bf [ End of Proof of Lemma~\ref{lem:coefficient/lamnda_s_m1m2}]}

{~}

Using the decomposition~\eqref{phi2+pi2_decomp} with the Cauchy-Schwarz inequality, we obtain 
\begin{align}
	\label{phi2+pi2_decomp_cal}
	\bra{\Omega} (\phi^2 + \pi^2)^s \ket{\Omega} 
	&\le \sum_{s_0=0}^s  \sum_{m_1+m_2 = 2s_0} \abs{\lambda^{(s)}_{m_1,m_2}} 
	\sqrt{\bra{\Omega}  \phi^{2m_1} \ket{\Omega}   \bra{\Omega}  \pi^{2m_2} \ket{\Omega}}  \notag \\
	&\le \sum_{s_0=0}^s  \sum_{m_1+m_2 = 2s_0} 4^s s^{2s-m_1-m_2} \cdot 
	\br{2\tilde{C} m_1}^{km_1/2}  \br{2\tilde{C} m_2}^{km_2/2}  \notag \\
	&\le \sum_{s_0=0}^s  \sum_{m_1+m_2 = 2s_0} 4^s s^{2s-2s_0} \cdot 
	\br{4\tilde{C} s}^{ks_0}  \notag \\
	&\le (2s+1)4^s\br{4\tilde{C} s}^{ks}\sum_{s_0=0}^s    \brr{\frac{s^2}{ (4\tilde{C} s)^{k}}}^{s-s_0}    
	\le (2s+1)4^s\br{4\tilde{C} s}^{ks} \frac{1}{1-\frac{s^2}{ (4\tilde{C} s)^{k}}} ,
\end{align}
where we use Lemma~\ref{lem:coefficient/lamnda_s_m1m2} and the inequalities~\eqref{prop:main_inequality_pi_moment_rewrite} and \eqref{phi_s_upper_bound_re}. 
Finally, from the upper bound of 
\begin{align}
	\frac{s^2}{ (4\tilde{C} s)^{k}} \le \frac{1}{ (4\tilde{C})^{k}} \le \frac{1}{4} ,
\end{align} 
we reduce the inequality~\eqref{phi2+pi2_decomp_cal} to 
\begin{align}
	\bra{\Omega} (\phi^2 + \pi^2)^s \ket{\Omega}  \le  2 (2s+1)4^s \br{4\tilde{C} s}^{ks} \le  2 (2s+1) \br{8\tilde{C} s}^{ks} ,
\end{align} 
where we use $k\ge 2$ and $\tilde{C} \ge 1$ [see Eq.~\eqref{prop:main_inequality_pi_moment_rewrite}]. 

From Eq.~\eqref{phi2pi2_2n+1}, we have $\bra{\Omega} (\phi^2 + \pi^2)^s \ket{\Omega}=\bra{\Omega} (2\nb+1)^s \ket{\Omega} \ge \frac{2s+1}{2} \bra{\Omega} \nb^s \ket{\Omega}$, and hence 
we prove the inequality~\eqref{main_ineq_thm_boson_dist_phi4}. 
This completes the proof of Proposition~\ref{prop_boson_dist_phi4}. $\square$

\subsection{Proof of Subtheorem~\ref{subthm:pi_distribution}} \label{Sec:Proof_Distribution of pi operator}

\subsubsection{Key proposition}

For the proof, we have to treat the expectation as 
\begin{align}
	\abs{\bra{\Omega}\pi^{2m-s} \phi^{s'} \ket{\Omega}} . 
\end{align}
To estimate it, we need to separate the average with respect to $\pi^{2m}$ using the known moment bound~\eqref{prop:phi_distribution/main_ineq} for $\abs{\bra{\Omega} \phi^{s} \ket{\Omega}}$.  
The following proposition plays a key role in our analyses (see~\ref{sec:Proof of Proposition_prop:ave_phi_s_pi_2m-s} for the proof):

\begin{prop} \label{prop:ave_phi_s_pi_2m-s}
	Let $\ave{O}$ be the expectation value with respect to a given quantum state. 
	We assume that the following upper bound for $\ave{\phi^{2s}}$ holds for positive $c_1$:
	\begin{align}
		\label{prop:ave_phi_s_pi_2m-s/assump}
		&\ave{\phi^{2s}} \le   (c_1 s)^{2s}  \quad (c_1\ge 1) \for \forall s \in\mathbb{N}. 
	\end{align}
	Then, for $\forall m\ge 1$, $\forall s \in [1,2m]$ and $\forall s',s''$ such that $s'+s''\le \kappa s$ ($\kappa\ge 1$), we obtain the upper bound of
	\begin{align}
		\label{main_ineq_phi_s_pi_2m_0-s}
		\abs{\ave{\pi^{2m-s} \phi^{s'} \Phi_0^{s''}}} \le 3 \ave{\Phi_0^{4\kappa m}}^{\frac{s''}{4\kappa m}} \max\brr{ \ave{\pi^{2m}}^{1-\frac{s}{2m}} \br{4c_1\kappa m}^{s'},  \br{4c_1\kappa m}^{2m+s'-s}}  ,
	\end{align}
	where $\Phi_0$ is an arbitrary Hermitian operator that commutes with $\phi$ and $\pi$. 
\end{prop}

{\bf Remark.} 
The proposition has a similar manner to the H\"older inequality.
The difficulty here is that the H\"older inequality does not hold for non-commuting operators. 
We can only utilize the Cauchy-Schwarz inequality as 
\begin{align}
	\abs{ \ave{\pi^{2m-s} \phi^{s'} }} \le  \sqrt{\ave{\phi^{2s'}} \ave{ \pi^{4m-2s} }} .
\end{align}
In the above inequality, the RHS includes $\ave{ \pi^{4m-2s}}$, which may be a higher-order moment than $\ave{\pi^{2m}}$ if $4m-2s\ge 2m$.
This prohibits us from utilizing the inequality in upper-bounding $\bra{\Omega}\pi^s \ket{\Omega}$.

In the proof of Subtheorem~\ref{subthm:pi_distribution}, Proposition~\ref{prop:ave_phi_s_pi_2m-s} is invoked at the critical bottleneck where the trade-off inequality reduces the variance of $\pi^m$ to expectation values of double commutators involving $\pi^m$ and interaction terms that are polynomial in $\phi$.
At this stage, a naive application of the Cauchy-Schwarz inequality would necessarily introduce higher $\pi$-moments and thus break the inductive scheme.
Proposition~\ref{prop:ave_phi_s_pi_2m-s} prevents this escalation by systematically converting all such mixed expectations back into expressions controlled by the top moment $\langle \pi^{2m}\rangle$, up to explicit and manageable polynomial factors in $m$.
This mechanism closes the induction on $m$ and leads directly to the subexponential moment bounds, and hence the tail estimates, for the $\pi$-operator claimed in Subtheorem~\ref{subthm:pi_distribution}.

Based on the above proposition, we can prove the following lemma:
\begin{lemma} \label{double_commutatro_pi_m}
	Under the setup of Proposition~\ref{prop:ave_phi_s_pi_2m-s}, 
	for arbitrary positive $l_1, l_2 \in \mathbb{N}$, the double commutator $ \brr{\brr{\phi^{l_1} ,\pi^m},\pi^m}\Phi_0^{l_2}$ satisfies the norm inequality as 
	\begin{align}
		\label{main:ave_phi_l_pi_m_ineq/fin}
		\abs{ \ave{ \brr{\brr{\phi^{l_1} ,\pi^m},\pi^m}\Phi_0^{l_2}}} \le 3 \ave{\Phi_0^{4\kappa_l m}}^{\frac{l_2}{4\kappa_l m}}  (2l_1 m)^2  G_{1,m}^{l_1-2} G_{2,m}^{2m-2} ,
	\end{align}
	where $l=l_1+l_2$, and $G_{1,m}, G_{2,m}$ are defined as 
	\begin{align}
		\label{G_1mG_2m:definition}
		G_{1,m}=4c_1\kappa_l m, \AND G_{2,m}=\max\br{\ave{\pi^{2m}}^{1/(2m)}, 4c_1\kappa_l m} 
	\end{align}
	with the choice of $\kappa_l = \max[1, (l-2)/2]$. 
	We recall that $[\Phi_0,\phi]=[\Phi_0,\pi]=0$. 
\end{lemma}

\textit{Proof of Lemma~\ref{double_commutatro_pi_m}.}
For the commutator $\brr{\phi^{l_1} ,\pi^m} $ for $\forall l \in \mathbb{N}$, we obtain from $[\phi,\pi]=i$ 
\begin{align}
	\brr{\phi^{l_1} ,\pi^m} =\sum_{s=0}^{l_1-1} \phi^ {s} \brr{\phi ,\pi^m} \phi^ {l_1-1-s}= 
	i m \sum_{s=0}^{l_1-1} \phi^ {s} \pi^{m-1} \phi^ {l_1-1-s} ,
\end{align}
By using the relation of Eq.~\eqref{phi_m_pi_n_commutator}, we reduce the above inequality to the form of 
\begin{align}
	\brr{\phi^{l_1} ,\pi^m} =i m \sum_{s=0}^{l_1-1}  \sum_{k=0}^{\min(s,m-1)}i^k C_{k,s,m-1} \pi^{m-1-k} \phi^ {l_1-1-k} . 
\end{align}

In the same way, we obtain 
\begin{align}
	\label{phi_1_l_pi_m_Eq_1}
	\brr{\brr{\phi^{l_1} ,\pi^m},\pi^m}
	& =i m \sum_{s=0}^{l_1-1}  \sum_{k=0}^{\min(s,m-1)}i^k C_{k,s,m-1} \pi^{m-1-k} \brr{\phi^ {l_1-1-k} ,\pi^m} \notag \\
	&=-m^2 \sum_{s=0}^{l_1-1}  \sum_{k=0}^{\min(s,m-1)}i^k C_{k,s,m-1} \pi^{m-1-k} \sum_{s'=0}^{l_1-2-k}   \sum_{k'=0}^{\min(s',m-1)}i^{k'} C_{k',s',m-1}\pi^{m-1-k'} \phi^ {l_1-2-k-k'}  \notag \\
	&= -m^2 \sum_{s=0}^{l_1-1}  \sum_{k=0}^{\min(s,m-1)} \sum_{s'=0}^{l_1-2-k}   \sum_{k'=0}^{\min(s',m-1)} i^{k+k'}  C_{k,s,m-1}C_{k',s',m-1}\pi^{2m-2-k-k'}  \phi^ {l_1-2-k-k'} .
\end{align}
In particular, for $l_1=1$ and $l_1=2$, we have 
\begin{align}
	\brr{\brr{\phi ,\pi^m},\pi^m}= 0 ,\quad 
	\brr{\brr{\phi^2 ,\pi^m},\pi^m}= -m^2\pi^{2m-2}  ,
\end{align}
respectively. 
Also, for $m=1$, we have 
\begin{align}
	\brr{\brr{\phi^{l_1} ,\pi},\pi} = -\sum_{s=0}^{l_1-1}  \sum_{s'=0}^{l_1-2}  \phi^ {l_1-2} =- l_1(l_1-1)\phi^ {l_1-2}  .
\end{align}
Therefore, in the following, we consider $l_1\ge 3$ and $m\ge 2$ as the non-trivial regimes.

We then use Proposition~\ref{prop:ave_phi_s_pi_2m-s} to obtain an upper bound of
\begin{align}
	\abs{\ave{ \pi^{2m-2-k-k'}  \phi^ {l_1-2-k-k'} \Phi_0^{l_2}} } . 
\end{align}
We here apply the inequality~\eqref{main_ineq_phi_s_pi_2m_0-s} with the choice of 
\begin{align}
	s \to 2+k+k' ,\quad s'\to l_1-2-k-k' ,\quad s'' \to l_2 , \quad  \kappa\to \kappa_l:=\max\br{ 1, \frac{l-2}{2}} ,
\end{align}
where $\kappa$ was defined by a constant satisfying $s'+s''\le \kappa s$ ($\kappa\ge 1$). Note that $l_1+l_2=l$.  
We then obtain 
\begin{align}
	\label{reduce_2m-2_l-2-k}
	&\abs{\ave{\pi^{2m-2-k-k'}  \phi^ {l_1-2-k-k'}  \Phi_0^{l_2} } } \notag \\
	&\le 3\ave{\Phi_0^{4\kappa_l m}}^{\frac{l_2}{4\kappa_l m}} \max\brr{ \ave{\pi^{2m}}^{1-\frac{2+k+k'}{2m}} \br{4c_1\kappa_l m}^{l_1-2-k-k' },  \br{4c_1\kappa_l m}^{2m+l_1-2(2+k+k')}}  \notag \\
	&=  3\ave{\Phi_0^{4\kappa_l m}}^{\frac{l_2}{4\kappa_l m}}\br{4c_1\kappa_l m}^{l_1-2-k-k' } \max\brr{ \ave{\pi^{2m}}^{1-\frac{2+k+k'}{2m}},  \br{4c_1\kappa_l m}^{2m-(2+k+k')}} \notag \\
	&=3 \ave{\Phi_0^{4\kappa_l m}}^{\frac{l_2}{4\kappa_l m}} G_{1,m}^{l_1-2-k-k'} G_{2,m}^{2m-(2+k+k')}  ,
\end{align}
where we use the definitions for $G_{1,m}$ and $G_{2,m}$ in Eq.~\eqref{G_1mG_2m:definition}.

By applying the inequality~\eqref{reduce_2m-2_l-2-k} to Eq.~\eqref{phi_1_l_pi_m_Eq_1}, we derive
\begin{align}
	\label{ave_phi_l_pi_m_ineq}
	&\abs{\ave{\brr{\brr{\phi^{l_1} ,\pi^m},\pi^m}\Phi_0^{l_2} }} \notag \\
	&\le 
	3m^2 \ave{\Phi_0^{4\kappa m}}^{\frac{l_2}{4\kappa m}} \sum_{s=0}^{l_1-1}  \sum_{k=0}^{\min(s,m-1)} \sum_{s'=0}^{l_1-2-k}   \sum_{k'=0}^{\min(s',m-1)} C_{k,s,m-1}C_{k',s',m-1}G_{1,m}^{l_1-2-k-k'} G_{2,m}^{2m-(2+k+k')}
	\notag \\
	&\le 3m^2\ave{\Phi_0^{4\kappa m}}^{\frac{l_2}{4\kappa m}} G_{1,m}^{l_1-2} G_{2,m}^{2m-2} \br{ \sum_{s=0}^{l_1-1}  \sum_{k=0}^{\min(s,m-1)}  C_{k,s,m-1} (G_{1,m}G_{2,m})^{-k}}^2 ,
\end{align}
where we use that $C_{k,m,n}$ monotonically increases with $m$ and $n$ from the definition~\eqref{Def_C_k_m_n}, i.e., $C_{k,m,n}:= k! \binom{m}{k} \binom{n}{k} $.
We calculate the summation as 
\begin{align}
	\label{ave_phi_l_pi_m_summation}
	\sum_{s=0}^{l_1-1}  \sum_{k=0}^{\min(s,m-1)}  C_{k,s,m-1} (G_{1,m}G_{2,m})^{-k} 
	&=  \sum_{s=0}^{l_1-1}  \sum_{k=0}^{\min(s,m-1)} k! \binom{s}{k} \binom{m-1}{k}(G_{1,m}G_{2,m})^{-k}  \notag \\
	&\le  \sum_{s=0}^{l_1-1}  \sum_{k=0}^{\min(s,m-1)}  \binom{s}{k} m^k \br{\frac{1}{16c_1^2\kappa_l^2 m^2}}^{k} \notag \\
	&\le \sum_{s=0}^{l_1-1}  \br{1+\frac{1}{16c_1^2\kappa_l^2 m}}^{s} \notag \\
	&=16c_1^2\kappa_l^2 m \brr{\br{1+\frac{1}{16c_1^2\kappa_l^2 m}}^{l_1}-1 } \le 2l_1 ,
\end{align}
where, in the last inequality, we use the following upper bound from $c_1\ge 1$, $\kappa_l=\max(1, l/2-1)$, and $l\ge 3$: 
\begin{align}
	\br{1+\frac{1}{16c_1^2\kappa_l^2 m}}^{l_1} -1
	\le  \frac{2l_1}{16c_1^2\kappa_l^2 m}.
\end{align}
Note that $(1+1/y)^x -1 \le 2x/y$ for $x\ge 0$ as long as $y\ge x$. 
By combining the inequalities~\eqref{ave_phi_l_pi_m_ineq} and~\eqref{ave_phi_l_pi_m_summation}, we reach the main inequality~\eqref{main:ave_phi_l_pi_m_ineq/fin}.
This completes the proof of Lemma~\ref{double_commutatro_pi_m}.   $\square$

{~}

\hrulefill{\bf [ End of Proof of Lemma~\ref{double_commutatro_pi_m}]}

{~}

\subsubsection{Proof of Subtheorem~\ref{subthm:pi_distribution}}
In the proof, we treat the moment of the $\pi_1$ operator at the site $1$.
We then analyze the variance of $\pi_1^m$ using the trade-off inequality~\eqref{trade_off_relation} as follows: 
\begin{align}
	\label{moment:pi_1^m_upper_bound}
	\var(\pi_1^m) \cdot \Delta 
	&\le \frac{1}{2} \abs{\bra{\Omega} \brr{\brr{H,\pi_1^m},\pi_1^m} \ket{\Omega}} \notag\\
	&\le \frac{1}{2} \sum_{k_1=1}^{k/2} \sum_{\substack{ i_1,i_2,\ldots, i_{2k_1} \in \Lambda \\ \{i_1,i_2,\ldots, i_{2k_1} \} \ni 1}} |f_{i_1,i_2,\ldots, i_{2k_1} }|  \abs{\bra{\Omega}\brr{\brr{ \phi_{i_1}\phi_{i_2} \cdots \phi_{i_{2k_1}} ,\pi_1^m},\pi_1^m}\ket{\Omega}} .
\end{align}

For simplicity, we consider the quantity of 
\begin{align}
	\label{Omega_l_1__l_j_exp}
	\abs{\bra{\Omega}\brr{\brr{ \phi_1^{l_1} \phi_2^{t_2} \cdots \phi_j^{t_j} ,\pi_1^m},\pi_1^m}\ket{\Omega}}
	= \abs{\bra{\Omega}\brr{\brr{ \phi_1^{l_1} ,\pi_1^m},\pi_1^m}  \phi_2^{t_2}\phi_3^{t_3} \cdots \phi_j^{t_j} \ket{\Omega}}
\end{align}
for an arbitrary positive integer $j$, where $l_1+t_2+\cdots + t_j=2k_1$ is assumed. 
By denoting $\Phi_0$ as 
\begin{align}
	\Phi_0 =\phi_2^{t_2}\phi_3^{t_3} \cdots \phi_j^{t_j} ,
\end{align}
we rewrite Eq.~\eqref{Omega_l_1__l_j_exp} as  
\begin{align}
	\bra{\Omega}\brr{\brr{ \phi_1^{l_1} \phi_2^{t_2} \cdots \phi_j^{t_j} ,\pi_1^m},\pi_1^m}\ket{\Omega}
	= \bra{\Omega}\brr{\brr{ \phi_1^{l_1} ,\pi_1^m},\pi_1^m} \Phi_0\ket{\Omega} .
	\label{Omega_phi_1tom_estimation}
\end{align}
We aim to estimate the upper bound of the above expectation value. 

In order to apply Lemma~\ref{double_commutatro_pi_m} to the estimation of Eq.~\eqref{Omega_phi_1tom_estimation}, 
we use Proposition~\ref{prop:phi_distribution} to obtain
\begin{align}
	\label{Omega_phi_2k_1_uppe}
	\abs{\bra{\Omega} \phi_i^{2k_1} \ket{\Omega}} \le \br{\check{c}_1k_1}^{2k_1} \for \forall i\in\Lambda
\end{align}
with 
\begin{align}
	\label{check_c_1_2_max_def}
	\check{c}_1=2 \max_{i\in \Lambda} \abs{\bra{\Omega} \phi_i \ket{\Omega}} + 4 \br{ \frac{\bar{\mu}}{\Delta}}^{1/2} +1, 
\end{align}
where we add $+1$ in Eq.~\eqref{check_c_1_2_max_def} to ensure $\check{c}_1\ge 1$.  
Now, we set $l_2=1$ and $l=l_1+l_2 \le 2k_1$ in Lemma~\ref{double_commutatro_pi_m}, which makes $\kappa_l\le (l-2)/2 \le k_1$. 
Then, the inequality~\eqref{main:ave_phi_l_pi_m_ineq/fin} in Lemma~\ref{double_commutatro_pi_m} yields
\begin{align}
	\label{main:used_phi_l_pi_m_ineq/fin}
	\bra{\Omega}\brr{\brr{ \phi_1^{l_1} \phi_2^{t_2} \cdots \phi_j^{t_j} ,\pi_1^m},\pi_1^m}\ket{\Omega}
	\le 3 \bra{\Omega} \Phi_0^{4k_1 m} \ket{\Omega}^{1/(4k_1 m)}  (4k_1 m)^2 \tilde{G}_{1,m}^{l_1-2} \tilde{G}_{2,m}^{2m-2}
\end{align}
where we set $c_1\to \check{c}_1$, $\kappa_l\to k_1$ in Eq.~\eqref{G_1mG_2m:definition} and define
\begin{align} 
	\label{G_1mG_2m:definition__re}
	\tilde{G}_{1,m}=4\check{c}_1k_1 m, \AND \tilde{G}_{2,m}=\max\br{\bra{\Omega} \pi_1^{2m} \ket{\Omega}^{1/(2m)}, 4\check{c}_1 k_1 m}.
\end{align}
Note that the RHS of \eqref{main:ave_phi_l_pi_m_ineq/fin} monotonically increases with $\kappa_l$.

Furthermore, using the H\"older inequality, we obtain 
\begin{align}
	\label{upp_Omega_Phi_0_4sm}
	\bra{\Omega} \Phi_0^{4k_1 m} \ket{\Omega} 
	&= \bra{\Omega}  \phi_2^{4t_2 k_1 m} \phi_3^{4t_3 k_1 m}  \cdots  \phi_j^{4t_j k_1 m} \ket{\Omega} \notag \\
	&\le \bra{\Omega}  \phi_2^{4(2k_1- l_1) k_1 m} \ket{\Omega}^{\frac{t_2}{2k_1- l_1}} \cdot \bra{\Omega}  \phi_3^{4(2k_1- l_1) k_1 m} \ket{\Omega}^{\frac{t_3}{2k_1- l_1}} \cdots \bra{\Omega}  \phi_j^{4(2k_1- l_1) k_1 m} \ket{\Omega}^{\frac{t_j}{2k_1- l_1}}  \notag \\
	&\le \brr{2 \check{c}_1 (2k_1- l_1) k_1 m}^{4t_2 k_1 m } \cdot \brr{2 \check{c}_1 (2k_1- l_1) k_1 m}^{4t_3 k_1 m } \cdots \brr{2 \check{c}_1 (2k_1- l_1) k_1 m}^{4t_j k_1 m } \notag\\
	&\le \br{4 \check{c}_1 k_1^2 m}^{4k_1 m (2k_1- l_1)},
\end{align}
where we use the inequality~\eqref{Omega_phi_2k_1_uppe} and $t_2+t_3+\cdots+ t_j = 2k_1- l_1 \le 2k_1$. 
By applying the inequality~\eqref{upp_Omega_Phi_0_4sm} to~\eqref{main:used_phi_l_pi_m_ineq/fin}, we have 
\begin{align}
	\label{upp_phi_1_phi_2_ldots_phi_j}
	|\bra{\Omega}\brr{\brr{ \phi_1^{l_1} \phi_2^{t_2} \cdots \phi_j^{t_j} ,\pi_1^m},\pi_1^m}\ket{\Omega}|
	&\le 3 \br{4 \check{c}_1 k_1^2 m}^{2k_1- l_1} (4k_1 m)^2 \tilde{G}_{1,m}^{l_1-2} \tilde{G}_{2,m}^{2m-2} \notag \\
	&\le \frac{3\Delta}{16\bar{\mu}} \br{4 \check{c}_1 k_1^2 m}^{2k_1}  \max\br{\bra{\Omega} \pi_1^{2m} \ket{\Omega}^{1- 1/m}, \br{4\check{c}_1 k_1^2 m}^{2m-2}} ,
\end{align}
where we use $\check{c}_1\ge  4 \br{ \bar{\mu}/\Delta}^{1/2}$ from Eq.~\eqref{def:tilde_c_1} to derive
\begin{align}
	(4k_1 m)^2 \tilde{G}_{1,m}^{-2} =\frac{1}{\check{c}_1^2} \le \frac{\Delta}{16 \bar{\mu}} .
\end{align}

From the upper bound~\eqref{upp_phi_1_phi_2_ldots_phi_j}, we upper-bound 
$|\bra{\Omega} \phi_{i_1}\phi_{i_2} \cdots \phi_{i_{2k_1}}\ket{\Omega}|$ as 
\begin{align}
	\label{upp_phi_1_phi_2_ldots_phi_j_general}
	|\bra{\Omega} \brr{\brr{  \phi_{i_1}\phi_{i_2} \cdots \phi_{i_{2k_1}},\pi_1^m},\pi_1^m}\ket{\Omega}|
	\le \frac{3\Delta}{16\bar{\mu}} \br{4 \check{c}_1 k_1^2 m}^{2k_1}  \max\br{\bra{\Omega} \pi_1^{2m} \ket{\Omega}^{1- 1/m}, \br{4\check{c}_1 k_1^2 m}^{2m-2}} .
\end{align}
By applying it to the inequality~\eqref{moment:pi_1^m_upper_bound}, we have 
\begin{align}
	\label{upp_Ham_pi^m_general}
	\var(\pi_1^m) \cdot \Delta 
	&\le \frac{3\Delta}{16\bar{\mu}} \sum_{k_1=1}^{k/2} \sum_{\substack{ i_1,i_2,\ldots, i_{2k_1} \in \Lambda \\ \{i_1,i_2,\ldots, i_{2k_1} \} \ni 1}} 
	\br{4 \check{c}_1 k_1^2 m}^{2k_1} \max\br{\bra{\Omega} \pi_1^{2m} \ket{\Omega}^{1- 1/m}, \br{4\check{c}_1 k_1^2 m}^{2m-2}}    |f_{i_1,i_2,\ldots, i_{2k_1}}| \notag \\
	&\le \frac{3\bar{f}\Delta}{16\bar{\mu}} \cdot \frac{(\check{c}_1 k^2 m)^2}{(\check{c}_1 k^2 m)^2-1}  \br{\check{c}_1 k^2 m}^{k}  \max\br{\bra{\Omega} \pi_1^{2m} \ket{\Omega}^{1- 1/m}, \br{\check{c}_1 k^2 m}^{2m-2}}   \notag \\
	&\le  \frac{\bar{f}\Delta}{\bar{\mu}} \br{\check{c}_1 k^2 m}^{k}  \max\br{\bra{\Omega} \pi_1^{2m} \ket{\Omega}^{1- 1/m}, \br{\check{c}_1 k^2 m}^{2m-2}}  ,
\end{align}
where we use the definition~\eqref{Definition/of_bar_f__re} for $\bar{f}$, which we reshow as follows:
\begin{align}
	\bar{f}= \max_{i\in \Lambda} \br{ 
		\sum_{k_1=1}^{k/2} \sum_{\substack{ i_1,i_2,\ldots, i_{2k_1} \in \Lambda \\ \{i_1,i_2,\ldots, i_{2k_1}\} \ni i}} 
		\abs{ f_{i_1,i_2,\ldots, i_{2k_1} } } }. 
\end{align}

From the inequality~\eqref{upp_Ham_pi^m_general}, we derive 
\begin{align}
	\label{moment:pi_1^m_upper_bound_simple}
	\bra{\Omega} \pi_1^{2m} \ket{\Omega}
	\le \bra{\Omega} \pi_1^{m} \ket{\Omega}^2+ \frac{\bar{f}}{\bar{\mu}} \br{\check{c}_1 k^2 m}^{k}  \max\br{\bra{\Omega} \pi_1^{2m} \ket{\Omega}^{1- 1/m}, \br{\check{c}_1 k^2 m}^{2m-2}}  .
\end{align}
As in the proof of Proposition~\ref{prop:phi_distribution}, we use the induction method to derive the target inequality~\eqref{prop:main_inequality_pi_moment}, which we show again as follows: 
\begin{align}
	\label{prop:main_inequality_pi_moment_re}
	\bra{\Omega} \pi_1^s\ket{\Omega}
	\le \brr{\frac{\bar{f}'}{\bar{\mu}} \br{\check{c}_1k^2 s}^k }^{s/2} 
\end{align}
with $\bar{f}'=\max(\bar{f} ,\bar{\mu}/2)$.
We begin with $s=1$. The Hamiltonian is invariant under $\pi_i\to -\pi_i$ for $\forall i\in\Lambda$, and hence $\bra{\Omega} \pi_1\ket{\Omega}=0$.
For $s=2$, from the inequality~\eqref{moment:pi_1^m_upper_bound_simple} with $\bra{\Omega} \pi_1\ket{\Omega}=0$ and $m=1$, we obtain 
the inequality~\eqref{prop:main_inequality_pi_moment} as follows:
\begin{align}
	\bra{\Omega} \pi_1^2 \ket{\Omega}
	\le \frac{\bar{f} \br{\check{c}_1 k^2}^{k}}{\bar{\mu}}  \le  \frac{\bar{f}'(2\check{c}_1k^2 )^k}{\bar{\mu}}  .  
\end{align}

We assume the inequality up to $s=2m-2$ and consider the cases of $s=2m-1, 2m$.
We begin with the case of $s=2m$.
In the inequality~\eqref{moment:pi_1^m_upper_bound_simple}, for the case where $\bra{\Omega} \pi_1^{2m} \ket{\Omega}^{1- 1/m} \le \br{\check{c}_1 k^2 m}^{2m-2}$, we have 
\begin{align}
	&\bra{\Omega} \pi_1^{2m} \ket{\Omega}^{1- 1/m} \le  \br{\check{c}_1 k^2 m}^{2m-2}  \notag\\
	&\quad \longrightarrow \quad \bra{\Omega} \pi_1^{2m} \ket{\Omega} \le  \br{\check{c}_1 k^2 m}^{2m}\le 
	\brr{\frac{\bar{f}'}{\bar{\mu}} \br{\check{c}_1k^2 \cdot 2m}^k }^{m} ,
\end{align}
which immediately yields the main inequality~\eqref{prop:main_inequality_pi_moment_re}, where we use $2\bar{f}'\ge \bar{\mu}$ and $k\ge 2$.
We therefore consider the case of $\bra{\Omega} \pi_1^{2m} \ket{\Omega}^{1- 1/m} > \br{\check{c}_1 k^2 m}^{2m-2}$, and the inequality~\eqref{moment:pi_1^m_upper_bound_simple} reduces to
\begin{align}
	\bra{\Omega} \pi_1^{2m} \ket{\Omega} \le \bra{\Omega} \pi_1^{m} \ket{\Omega}^2 + \frac{\bar{f}' \br{\check{c}_1 k^2 m}^{k}}{\bar{\mu}} \bra{\Omega} \pi_1^{2m} \ket{\Omega}^{1- 1/m} .
\end{align}
We rewrite it as
\begin{align}
	&\bra{\Omega} \pi_1^{2m} \ket{\Omega}^{1/m}  
	\le \frac{ \bra{\Omega} \pi_1^{m} \ket{\Omega}^2}{\bra{\Omega} \pi_1^{2m} \ket{\Omega}^{1-1/m}} + \frac{\bar{f}' \br{\check{c}_1 k^2 m}^{k}}{\bar{\mu}}
	\le  \bra{\Omega} \pi_1^{m} \ket{\Omega}^{2/m} + \frac{\bar{f}' \br{\check{c}_1 k^2 m}^{k}}{\bar{\mu}}  \notag \\
	\longrightarrow &\bra{\Omega} \pi_1^{2m} \ket{\Omega} \le \br{ \bra{\Omega} \pi_1^{m} \ket{\Omega}^{2/m} + \frac{\bar{f}' \br{\check{c}_1 k^2 m}^{k}}{\bar{\mu}}}^m,
\end{align}
where we use $\bra{\Omega} \pi_1^{m} \ket{\Omega}\le \bra{\Omega} \pi_1^{2m} \ket{\Omega}^{1/2}$.
Using the inequality~\eqref{prop:main_inequality_pi_moment_re} for $\bra{\Omega} \pi_1^{m} \ket{\Omega}$, we have 
\begin{align}
	\label{Omega_ave_pi_1^2m}
	\bra{\Omega} \pi_1^{2m} \ket{\Omega} 
	&\le \br{ \frac{\bar{f}'\br{\check{c}_1k^2 m}^k}{\bar{\mu}}  + \frac{\bar{f}' \br{\check{c}_1 k^2 m}^{k}}{\bar{\mu}}}^m \notag \\
	&= \brr{ 2^{-k+1} \cdot \frac{\bar{f}'}{\bar{\mu}} \br{2\check{c}_1 k^2 m}^{k}}^m 
	\le \brr{  \frac{\bar{f}'}{\bar{\mu}} \br{2\check{c}_1 k^2 m}^{k}}^m ,
\end{align}
where we use $k\ge 2$ in the last inequality. 

Finally, we consider the case of $s=2m-1$, which is upper-bounded in the similar way to~\eqref{2m-1_phi_Omega_ave} as follows:
\begin{align}
	\abs{\bra{\Omega} \pi_1^{2m-1} \ket{\Omega}} \le \bra{\Omega} \pi_1^{2m} \ket{\Omega}^{1-1/(2m)} 
	&\le \brr{ 2^{-k+1} \cdot \frac{\bar{f}'}{\bar{\mu}} \br{2\check{c}_1 k^2 m}^{k}}^{m-1/2}  \notag \\
	&\le \brr{ 2^{-k+1} \cdot \br{\frac{2m}{2m-1}}^{k}}^{m-1/2}  \brrr{\frac{\bar{f}'}{\bar{\mu}} \brr{\check{c}_1 k^2 \br{2m-1}}^{k}}^{m-1/2}  \notag \\
	&\le \brrr{\frac{\bar{f}'}{\bar{\mu}} \brr{\check{c}_1 k^2 \br{2m-1}}^{k}}^{m-1/2} ,
\end{align}
where, in the second inequality, we use the upper bound~\eqref{Omega_ave_pi_1^2m}, and in the last inequality, we use 
\begin{align}
	2^{-k+1} \cdot \br{\frac{2m}{2m-1}}^{k} \le \frac{8}{9} \for k\ge 2,\quad m\ge 2.
\end{align}
This completes the proof of Subtheorem~\ref{subthm:pi_distribution}. $\square$

\newpage
\subsection{Proof of Proposition~\ref{prop:ave_phi_s_pi_2m-s}} \label{sec:Proof of Proposition_prop:ave_phi_s_pi_2m-s}
\subsubsection{Strategy and roadmap.}

We briefly outline the structure of the proof of Proposition~\ref{prop:ave_phi_s_pi_2m-s}
before presenting the detailed arguments.
The proposition provides an upper bound on mixed expectation values of the form  $\langle \pi^{2m-s}\,\phi^{s'}\,\Phi_0^{s''}\rangle$, under the moment assumption on $\phi$ stated explicitly in the proposition.
The proof proceeds by induction on the moment order $m$ and is designed to control such
expressions in terms of the top moment $\langle \pi^{2m}\rangle$,
despite the fact that $\pi$ and $\phi$ do not commute.

The argument is based on a recursive reduction of the power of $\pi$ appearing inside the
expectation value.
The starting point is a Cauchy-Schwarz estimate that extracts one factor of
$\langle \pi^{2m}\rangle$, leaving an expectation in which powers of $\pi$ are interlaced
with powers of $\phi$.
At this stage, $\pi$ must be moved past $\phi$.
Because of non-commutativity, commuting $\pi$ through $\phi$ produces additional terms
arising from commutators.
Iterating this procedure naturally leads to a decomposition into a main contribution,
in which the $\pi$-power is reduced in a direct way, and a collection of error terms
generated by commutators.

To organize this structure, we introduce a recursive family of quantities.
At recursion depth $p$, the main term is written as
$L_p = |\langle \pi^{2m-s_p}\,\phi^{s'_p}\,\Phi_0^{s''_p}\rangle|$,
where the exponent $s_p$ measures how much of the original $\pi$-power has been reduced.
All additional terms generated by commutators at the previous step are grouped into a second
quantity $K_p$, which consists of a finite linear combination of expectation values of the form
$\langle \pi^{2m-s_p-k}\,\phi^{s'_p-k}\,\Phi_0^{s''_p}\rangle$ with $k\ge 1$.
Thus, $L_p$ represents the principal contribution at scale $p$, while $K_p$ collects all
error terms that arise purely from the non-commutativity of $\pi$ and $\phi$.
The exponents $s_p,s'_p,s''_p$ are chosen to grow dyadically with $p$, reflecting the repeated
use of Cauchy-Schwarz and ensuring that the remaining power of $\pi$ decreases rapidly.

With these definitions, one obtains a universal recursive inequality of the form
$L_p \leq \langle \pi^{2m}\rangle^{1/2}\,\bigl(L_{p+1}+K_{p+1}\bigr)^{1/2}$,
valid as long as the remaining power of $\pi$ is positive.
This recursion is iterated until the $\pi$-power becomes sufficiently small, at which point a direct Cauchy-Schwarz estimate closes the bound.
The induction hypothesis on $m$ is used to control the expectations appearing in $K_p$,
which always involve strictly smaller total powers of $\pi$.

To compare the resulting bounds obtained at different recursion depths,
we introduce a reference quantity $W_{m,p}$.
The quantity $W_{m,p}$ is constructed so as to capture the maximal size that an expression
of the form $L_p$ or $K_p$ can attain under the induction hypothesis.
It combines the dependence on the top moment $\langle \pi^{2m}\rangle$,
explicit polynomial factors in $m$ originating from the assumed bounds on $\phi$,
and the contribution of $\Phi_0$ into a single expression of fixed functional form.

Lemma~\ref{lamm:upp_W_m_s} establishes that the reference quantities $W_{m,p}$
are uniformly controlled by the base quantity $W_{m,0}$ for all recursion depths $p$.
This result shows that the dyadic recursion does not amplify constants and allows all
bounds obtained at intermediate scales to be compared directly with the final target bound.

Lemma~\ref{lamm:upp_K_m} provides a quantitative estimate on the commutator contributions
$K_p$.
It shows that each $K_p$ is suppressed relative to $W_{m,p}$ by an explicit factor,
so that commutator-generated terms never dominate the recursive scheme.
This estimate is the key ingredient that prevents the accumulation of errors
arising from non-commutativity.

Together, Lemmas~\ref{lamm:upp_W_m_s} and \ref{lamm:upp_K_m} imply that both the main terms
$L_p$ and the commutator terms $K_p$ remain uniformly controlled throughout the recursion.
As a consequence, all branches of the recursive argument are bounded by a fixed multiple
of $W_{m,0}$, allowing the induction on $m$ to close and yielding the bound stated in
Proposition~\ref{prop:ave_phi_s_pi_2m-s}.

\subsubsection{Proof.}
Throughout the proof, we often use the inequality of 
\begin{align}
	\ave{O^{2m}} \le \ave{O^{2m'}}^{m/m'} \for m'\ge m ,
	\label{moment_relation_basci}
\end{align}
which gives $\ave{O^{2m}}^{1/(2m)} \le \ave{O^{2m'}}^{1/(2m')}$.

We prove the statement by induction.
For $m=1$, the inequality is trivially satisfied from 
\begin{align}
	\abs{\ave{\pi^{2-s} \phi^{s'} \Phi_0^{s''}}} &\le 
	\begin{cases}
		\ave{\pi^{2}}^{1/2}  \ave{\phi^{2s'} \Phi_0^{2s''}}^{1/2}&\for s=1, \\
		\ave{\phi^{2s'} \Phi_0^{2s''}}^{1/2} &\for s=2 ,
	\end{cases} \notag \\
	&\le 
	\begin{cases}
		\ave{\pi^{2m}}^{1-\frac{s}{2m}} \br{2 c_1\kappa m}^{s'} \ave{\Phi_0^{4\kappa m}}^{\frac{s''}{4\kappa m}} &\for s=1, \\
		\br{2 c_1\kappa m}^{s'}  \ave{\Phi_0^{4\kappa m}}^{\frac{s''}{4\kappa m}}  &\for s=2 ,
	\end{cases} 
\end{align}
where, in the first inequality, we use the Cauchy-Schwarz inequality, and in the second inequality, 
we use the H\"older inequality with $\tilde{s}=s'+s''\le \kappa s \le 2\kappa m$ (note that $s\le 2m$):
\begin{align}
	\label{ave_phi_2s'_2s''}
	\ave{\phi^{2s'} \Phi_0^{2s''}}
	&\le  \ave{\phi^{2\tilde{s}}}^{\frac{s'}{\tilde{s}}} \ave{\Phi_0^{2\tilde{s}}}^{\frac{s''}{\tilde{s}}}  \le \br{2 c_1\kappa m}^{2s'} \ave{\Phi_0^{4\kappa m}}^{\frac{s''}{2\kappa m}} .
\end{align}
Note that we have assume $\ave{\phi^{2s}} \le (c_1s)^{2s}$ and the inequality~\eqref{moment_relation_basci} was used for $\ave{\Phi_0^{2\tilde{s}}}^{\frac{s''}{\tilde{s}}}$.

We then assume the target inequality for $m\le m_0-1$:
\begin{align}
	\label{induction_phis_pi_2m-s}
	\abs{\ave{\pi^{2m-s} \phi^{s'} \Phi_0^{s''}}} \le \zeta \ave{\Phi_0^{4\kappa m}}^{\frac{s''}{4\kappa m}} \max\brr{ \ave{\pi^{2m}}^{1-\frac{s}{2m}} \br{4c_1\kappa m}^{s'},  \br{4c_1\kappa m}^{2m+s'-s}}  ,
\end{align}
where $\zeta$ is proven to be chosen as $\zeta=3$ afterward [see the discussion below the inequality~\eqref{uppe_L0_in_general_0}]. 
We aim to prove the case of $m=m_0\ge2$ under the assumption of~\eqref{induction_phis_pi_2m-s}.

For this purpose, we take the following two steps. In the first step, we will prove the inequality of 
\begin{align}
	\label{induction_phis_pi_2m-s_fist/step}
	\abs{\ave{\pi^{2m_0-1 - s} \phi^{s'} \Phi_0^{s''}}} \le  \zeta \ave{\Phi_0^{4\kappa m_0}}^{\frac{s''}{4\kappa m_0}} \max\brr{ \ave{\pi^{2m_0}}^{1-\frac{s+1}{2m_0}} \br{4c_1\kappa m_0}^{s'},  \br{4c_1\kappa m_0}^{2m_0+s'-(s+1)}} 
\end{align}
for $s \in[1,2m_0-1]$ and $s'+s'' \le \kappa s$. 
This inequality can be proved based on the assumption of~\eqref{induction_phis_pi_2m-s} for $m\le m_0-1$. 

In order to reach the target inequality (i.e., the inequality~\eqref{induction_phis_pi_2m-s} with $m=m_0$), we utilize the above inequality~\eqref{induction_phis_pi_2m-s_fist/step} and derive 
\begin{align}
	\label{induction_phis_pi_2m-s_second/one}
	\abs{\ave{\pi^{2m_0-s} \phi^{s'} \Phi_0^{s''}}} \le  \zeta \ave{\Phi_0^{4\kappa m_0}}^{\frac{s''}{4\kappa m_0}} \max\brr{ \ave{\pi^{2m_0}}^{1-\frac{s}{2m_0}} \br{4c_1\kappa m_0}^{s'},  \br{4c_1\kappa m_0}^{2m_0+s'-s}} 
\end{align}
for $s \in[1,2m_0]$ and $s'+s'' \le \kappa s$.

In the following, we aim to prove the inequality~\eqref{induction_phis_pi_2m-s_fist/step}, but the same analyses will be applied to the proof of~\eqref{induction_phis_pi_2m-s_second/one}.
We first consider the case of $s+1\ge m_0$, which gives $2[2m_0-(s+1)] \le 2m_0$. 
In this case, we immediately obtain from the Cauchy-Schwarz inequality 
\begin{align}
	\label{proof_case_s_ge_2m_0-s/fin} 
	\abs{ \ave{\pi^{2m_0-(s+1)} \phi^{s'} \Phi_0^{s''}}} 
	&\le \ave{\pi^{2[2m_0-(s+1)]}}^{1/2} \ave{\phi^{2s'}\Phi_0^{2s''}}^{1/2} \notag \\
	&\le \ave{\pi^{2m_0}}^{\frac{2m_0-(s+1)}{2m_0}} \brr{ \br{2 c_1\kappa m_0}^{2s'} \ave{\Phi_0^{4\kappa m_0}}^{\frac{s''}{2\kappa m_0}}}^{1/2} ,
\end{align}
where we use the inequality~\eqref{ave_phi_2s'_2s''} in the last inequality. 
This reduces to the inequality~\eqref{induction_phis_pi_2m-s_fist/step}.

We next consider the case of $s+1< m_0$. 
We start with the Cauchy-Schwarz inequality as 
\begin{align}
	\label{first/step_upp_bound_s_s'}
	\abs{ \ave{\pi^{2m_0-(s+1)} \phi^{s'}  \Phi_0^{s''}}} \le
	\ave{\pi^{2m_0}}^{1/2} \ave{\phi^{s'} \pi^{2m_0-2-2s} \phi^{s'}  \Phi_0^{2s''}  }^{1/2} \for s+1< m_0.
\end{align} 
We then apply the relation~\eqref{phi_m_pi_n_commutator} to $\phi^{s'} \pi^{2m_0-2s-2}$ and obtain  
\begin{align}
	\label{phi_s_pi_2m0-2s_phis}
	&\ave{\phi^{s'} \pi^{2m_0-2s-2} \phi^{s'}  \Phi_0^{2 s''}} \notag \\
	&=\ave{ \pi^{2m_0-2s-2}  \phi^{2s'}  \Phi_0^{2s''}  } + \sum_{k=1}^{\min(2m_0-2s-2,s)} i^k C_{k,2m_0-2s-2,s'}\ave{ \pi^{2m_0-2s-2-k}  \phi^{2s'-k}  \Phi_0^{2s''}  } .
\end{align}
We here define 
\begin{align}
	\label{s_p_s'_p_def}
	s_p:= 2^p (s+1),\quad  s'_p:= 2^p s', \AND s''_p= 2^p s''. 
\end{align}
Then, by defining $L_p$ and $K_p$ as 
\begin{align}
	\label{K_p_L_p_definition}
	&L_p:= \abs{ \ave{ \pi^{2m_0-s_p} \phi^{s'_p}\Phi_0^{s''_p} } },\notag \\
	&K_p:= \sum_{k=1}^{\min(2m_0-s_p,s'_{p-1})} i^k C_{k,2m_0-s_p,s'_{p-1}}\ave{ \pi^{2m_0-s_p-k}  \phi^{s'_p-k} \Phi_0^{s''_p} } ,
\end{align}
we reduce the inequality~\eqref{first/step_upp_bound_s_s'} to 
\begin{align}
	\label{recursive_L_p_K_p_p=0}
	\abs{ \ave{\pi^{2m_0-\tilde{s}} \phi^{s'_0} \Phi_0^{s''_0}}} = L_0
	&\le \ave{\pi^{2m_0}}^{1/2} \br{L_1 +K_1}^{1/2}   .
\end{align}
By generalizing the above inequality, we can also derive 
\begin{align}
	\label{recursive_L_p_K_p}
	L_p\le \ave{\pi^{2m_0}}^{1/2} \br{L_{p+1} +K_{p+1}}^{1/2} \for s_p < m_0.  
\end{align}

In the following, we define $W_{m,p}$ as follows:
\begin{align}
	\label{def_W_m0_s}
	W_{m,p} :=\ave{\Phi_0^{4\kappa m}}^{\frac{s''_p}{4\kappa m}} \max\brr{ \ave{\pi^{2m}}^{1-\frac{s_p}{2m}} \br{4c_1\kappa m}^{s'_p},  \br{4c_1\kappa m}^{2m+s'_p-s_p}} . 
\end{align}
Note that the $W_{m,0}$ corresponds to the RHS of the target inequality~\eqref{induction_phis_pi_2m-s_fist/step} from the definition of~\eqref{s_p_s'_p_def}. 

To analyze the recursive relation~\eqref{recursive_L_p_K_p}, we define $p_0$ as the minimum non-negative integer that satisfies
\begin{align}
	\label{p_0_definition}
	L_p > K_p \for p\le p_0, \quad L_{p_0+1} \le K_{p_0+1}, 
\end{align}
where, in particular for $p_0=0$, we do not require $L_0 > K_0$ and only the condition $L_{p+1} \le K_{p+1}$ is satisfied for all $p$. 
We first consider the case where $p_0=0$. In this case, the inequality~\eqref{recursive_L_p_K_p_p=0} gives  
\begin{align}
	\label{L_0_upper_bound_for_p0=0}
	L_0\le \ave{\pi^{2m_0}}^{1/2} \br{L_1 +K_1}^{1/2} \le 2^{1/2}\ave{\pi^{2m_0}}^{1/2} K_1^{1/2} \for p_0=0 . 
\end{align}

When $p_0\ge 1$, we prove the following inequality for arbitrary $p_1 < p_0$:  
\begin{align}
	\label{L_0_upper_bound_for_p1}
	L_0 \le a_{p_1} \ave{\pi^{2m_0}}^{1-1/2^{p_1+1}} \br{L_{p_1+1}}^{1/2^{p_1+1}}  ,\quad a_{p_1} = \prod_{s=0}^{p_1} \br{1+\frac{1}{2^{s+1}}} .
\end{align}
Note that $L_{p_1} > K_{p_1}$ $(p_1 < p_0)$ holds from the definition~\eqref{p_0_definition} for $p_0$. 
We rely on the induction method. First, for $p_1=0$, the inequality~\eqref{L_0_upper_bound_for_p1} is derived from~\eqref{recursive_L_p_K_p_p=0} as follows:
\begin{align}
	L_0\le \ave{\pi^{2m_0}}^{1/2} \br{L_1 +K_1}^{1/2} \le \br{1+\frac{1}{2}} \ave{\pi^{2m_0}}^{1/2} L_1^{1/2} ,
\end{align}
where we use the following general inequality for $x\ge y$ with $\alpha=1/2$: 
\begin{align}
	\label{relation_x+y^alpha}
	\br{x+ y}^\alpha =x^\alpha \br{1+ \frac{y}{x}}^\alpha \le x^\alpha \br{1+ \alpha \cdot \frac{y}{x}} \le (1+\alpha) x^\alpha\quad (0<\alpha <1). 
\end{align}
By assuming the inequality~\eqref{L_0_upper_bound_for_p1} up to $p=p_1-1$ ($p_1< p_0$), we obtain 
\begin{align}
	L_0& \le a_{p_1-1} \ave{\pi^{2m_0}}^{1-1/2^{p_1}} \br{L_{p_1}}^{1/2^{p_1}}   \notag \\
	&\le a_{p_1-1} \ave{\pi^{2m_0}}^{1-1/2^{p_1}} \br{\ave{\pi^{2m_0}}^{1/2} \br{L_{p_1+1} +K_{p_1+1}}^{1/2}}^{1/2^{p_1}} \notag \\
	&=a_{p_1-1} \ave{\pi^{2m_0}}^{1-1/2^{p_1+1}}  \br{L_{p_1+1} +K_{p_1+1}}^{1/2^{p_1+1}} \notag \\
	&\le a_{p_1-1} \ave{\pi^{2m_0}}^{1-1/2^{p_1+1}} \br{1+ \frac{1}{2^{p_1+1}}}  \br{L_{p_1+1}}^{1/2^{p_1+1}} 
	= a_{p_1} \ave{\pi^{2m_0}}^{1-1/2^{p_1+1}}\br{L_{p_1+1}}^{1/2^{p_1+1}}  ,
\end{align}
where, in the third inequality, we use the relation~\eqref{relation_x+y^alpha} with the condition of $L_{p_1+1} > K_{p_1+1}$ for $p_1<p_0$ (or $p_1+1\le p_0$). 
This completes the proof of the inequality~\eqref{L_0_upper_bound_for_p1}. 

In the following, we define the integer $\bar{p}$ ($\ge0$) as 
\begin{align}
	s_{\bar{p}}=2^{\bar{p}} (s+1) <  m_0 \le 2^{\bar{p}+1} (s+1)=s_{\bar{p}+1}  < 2m_0.
\end{align}
We then separate the cases of $\bar{p}<p_0$ and $\bar{p}\ge p_0$.

For $\bar{p}<p_0$, we use the inequality~\eqref{L_0_upper_bound_for_p1} with $p_1=\bar{p}<p_0$, we obtain 
\begin{align}
	L_0 \le a_{\bar{p}} \ave{\pi^{2m_0}}^{1-1/2^{\bar{p}+1}} \br{L_{\bar{p}+1}}^{1/2^{\bar{p}+1}}  .
	\label{uppe_L0_p1<p_0}
\end{align}
Then, by using the inequality~\eqref{proof_case_s_ge_2m_0-s/fin} and $m_0 \le s_{\bar{p}+1} < 2m_0$, we obtain 
\begin{align}
	L_{\bar{p}+1}  = \abs{ \ave{ \pi^{2m_0-s_{\bar{p}+1}} \phi^{s'_{\bar{p}+1}}\Phi_0^{s''_{\bar{p}+1}}}} \le W_{m_0,\bar{p}+1} ,
\end{align}
which reduces the inequality~\eqref{uppe_L0_p1<p_0} to
\begin{align}
	L_0 \le a_{\bar{p}} \ave{\pi^{2m_0}}^{1-1/2^{\bar{p}+1}} \br{W_{m_0,\bar{p}+1}}^{1/2^{\bar{p}+1}}  .
	\label{uppe_L0_bar_p<p_0} 
\end{align}

For $\bar{p}\ge p_0$, we use the inequality~\eqref{L_0_upper_bound_for_p1} with $p_1=p_0-1$, we obtain 
\begin{align}
	\label{L_0_upper_bound_for_p1=p_0-1}
	L_0 \le a_{p_0-1} \ave{\pi^{2m_0}}^{1-1/2^{p_0}} \br{L_{p_0}}^{1/2^{p_0}}   .
\end{align}
Because of $s_{p_0+1} < 2m_0$ from $s_{p_0} \le s_{\bar{p}} <m_0$, the inequality~\eqref{recursive_L_p_K_p} gives
\begin{align}
	\br{L_{p_0}}^{1/2^{p_0}}   \le \ave{\pi^{2m_0}}^{1/2^{p_0+1}} \brr{ \br{L_{p_0+1} + K_{p_0+1}}^{1/2} }^{1/2^{p_0}} .
\end{align}
By applying the above inequality to~\eqref{L_0_upper_bound_for_p1=p_0-1}, we derive 
\begin{align}
	L_0 &\le a_{p_0-1} \ave{\pi^{2m_0}}^{1-1/2^{p_0}}\cdot \ave{\pi^{2m_0}}^{1/2^{p_0+1}} \brr{ \br{L_{p_0+1} + K_{p_0+1}}^{1/2} }^{1/2^{p_0}} \notag \\
	&\le 2^{1/2^{p_0+1}} a_{p_0-1} \ave{\pi^{2m_0}}^{1-1/2^{p_0+1}} \br{K_{p_0+1}}^{1/2^{p_0+1}},
	\label{uppe_L0_bar_p_ge_p_0} 
\end{align}
where, in the second inequality, we use $L_{p_0+1} \le K_{p_0+1}$ from the definition~\eqref{p_0_definition} of $p_0$.

Therefore, by combining the inequalities~\eqref{L_0_upper_bound_for_p0=0},~\eqref{uppe_L0_bar_p<p_0} and \eqref{uppe_L0_bar_p_ge_p_0}, we obtain 
\begin{align}
	L_0 &\le 3\max\brr{\ave{\pi^{2m_0}}^{1/2} K_1^{1/2}, \ave{\pi^{2m_0}}^{1-1/2^{\bar{p}+1}} \br{W_{m_0,\bar{p}+1}}^{1/2^{\bar{p}+1}} ,\ave{\pi^{2m_0}}^{1-1/2^{p_0+1}} \br{K_{p_0+1}}^{1/2^{p_0+1}}} ,
	\label{uppe_L0_in_general_0} 
\end{align}
where we use the inequality of 
\begin{align}
	2^{1/2^{p+1}} a_{p-1} \le 2.38423\cdots  < 3\for  \forall p\in \mathbb{N}.
\end{align}

To further reduce the upper bound to the desired form as in~\eqref{main_ineq_phi_s_pi_2m_0-s}, we prove the following two lemmas: 

\begin{lemma} \label{lamm:upp_W_m_s}
	For an arbitrary $s_p$ such that $s_p\le 2m_0$, we prove the following upper bound:
	\begin{align}
		\ave{\pi^{2m_0}}^{1-1/2^p} \br{W_{m_0,p}}^{1/2^p} \le W_{m_0,0}  .
	\end{align}
	We recall that $s_p:= 2^p (s+1)$ and $s'_p:= 2^p s'$ as in Eq.~\eqref{s_p_s'_p_def}. 
\end{lemma}

\begin{lemma} \label{lamm:upp_K_m}
	The quantity $K_p$ in Eq.~\eqref{K_p_L_p_definition} is upper-bounded as follows:
	\begin{align}
		\label{lamm:upp_K_m_main_ineq}
		K_p 
		&\le \zeta W_{m_0,p}  \br{e^{1/(8c_1^2 \kappa)}-1} \le \frac{\zeta}{7c_1^2\kappa} W_{m_0,p} .
	\end{align}
\end{lemma}
\noindent We recall that $\zeta$ is the control parameter that was adopted in the inequality~\eqref{induction_phis_pi_2m-s}. 

By applying Lemmas~\ref{lamm:upp_W_m_s} and \ref{lamm:upp_K_m} to the inequality~\eqref{uppe_L0_in_general_0}, we obtain 
\begin{align}
	L_0 &\le 3 W_{m_0,0}  \max\brr{\br{\frac{\zeta}{7c_1^2\kappa}}^{1/2}, 1, \br{\frac{\zeta}{7c_1^2\kappa}}^{1/2^{p_0+1}} } .
	\label{uppe_L0_in_general_0_again} 
\end{align}
Because of the assumption of $c_1\ge 1$ and $\kappa\ge 1$, we have $\zeta/(7c_1^2\kappa)\le \zeta / 7 \le 1$ for $\zeta\le 7$.
Therefore, by choosing $\zeta = 3$, we obtain 
\begin{align}
	\max\brr{\br{\frac{\zeta}{7c_1^2\kappa}}^{1/2}, 1, \br{\frac{\zeta}{7c_1^2\kappa}}^{1/2^{p_0+1}} } =1,
\end{align}
which reduces~\eqref{uppe_L0_in_general_0_again} to the target inequality~\eqref{induction_phis_pi_2m-s_fist/step}.

To derive the second inequality~\eqref{induction_phis_pi_2m-s_second/one}.
We follow the same analytical processes. 
For the case of $s\ge m_0$, using the Cauchy-Schwarz inequality as in~\eqref{proof_case_s_ge_2m_0-s/fin}, 
we obtain the inequality~\eqref{induction_phis_pi_2m-s_second/one}.

For the case of $s<m_0$, 
the inequality~\eqref{first/step_upp_bound_s_s'} is replaced by 
\begin{align}
	\label{first/step_upp_bound_s_s'__2}
	\abs{ \ave{\pi^{2m_0-s} \phi^{s'}  \Phi_0^{s''}}} \le
	\ave{\pi^{2m_0}}^{1/2} \ave{\phi^{s'} \pi^{2m_0-2s} \phi^{s'}  \Phi_0^{2s''}  }^{1/2} \for s< m_0.
\end{align} 
Then, by redefining 
\begin{align}
	s_p:= 2^p s,\quad  s'_p:= 2^p s', \AND s''_p= 2^p s'',  
\end{align}
we can derive the same inequality as~\eqref{recursive_L_p_K_p} by redefining $L_p$ and $K_p$ using the above $s_p$ in Eq.~\eqref{K_p_L_p_definition}: 
\begin{align}
	L_p\le \ave{\pi^{2m_0}}^{1/2} \br{L_{p+1} +K_{p+1}}^{1/2} \for s_p < m_0.  
\end{align}
The remaining parts are the same in the proof for the first inequality~\eqref{induction_phis_pi_2m-s_fist/step}.

One difference stems from the proof of Lemma~\ref{lamm:upp_K_m}. 
There, we analyze $K_p$ as 
\begin{align}
	&K_p:= \sum_{k=1}^{\min(2m_0-s_p,s'_{p-1})} i^k C_{k,2m_0-s_p,s'_{p-1}}\ave{ \pi^{2m_0-s_p-k}  \phi^{s'_p-k}\Phi_0^{s''_p} } ,
\end{align}
and treat the expectation $\ave{ \pi^{2m_0-s_p-k}  \phi^{s'_p-k}\Phi_0^{s''_p} } $. 
To estimate it, we rely on the proved inequality~\eqref{induction_phis_pi_2m-s_fist/step} instead of the inequality~\eqref{induction_phis_pi_2m-s},
By writing 
\begin{align}
	\label{sm_0-sp-k_cannot_be_estiated}
	\ave{ \pi^{2m_0-s_p-k}  \phi^{s'_p-k} \Phi_0^{s''_p}} = \ave{ \pi^{2m_0-1-(s_p+k-1)}  \phi^{(s'_p-k)} \Phi_0^{s''_p}} ,
\end{align}
we have $(s'_p-k) + s''_p\le  \kappa\br{s_p+k-1}$ because of
\begin{align}
	\label{kappa_upper_bound_ineq}
	(s'_p-k) + s''_p &=2^p( s'+s'') - k \le 2^p \kappa s - k  \notag \\
	&\le \kappa \brr{2^ps  + k-1}  \notag \\
	&=  \kappa\br{s_p+k-1}
\end{align}
for $k\ge 1$. 
This allows us to use the inequality~\eqref{induction_phis_pi_2m-s_fist/step} with 
\begin{align}
	s \to s_p+k-1, \quad s' \to s'_p-k,\quad s'' \to s''_p ,
\end{align}
which yields the same inequality as~\eqref{upp_pi_2m0-2s+2s'-k_2s+2s'-k}
\begin{align}
	\ave{ \pi^{2m_0-s_p-k}  \phi^{s'_p-k} \Phi_0^{s''_p}}
	\le  \zeta \ave{\Phi_0^{4\kappa m_0}}^{\frac{s''_p}{4\kappa m_0}} \max\brr{ \ave{\pi^{2m_0}}^{1-\frac{s_p+k}{2m_0}} \br{4c_1\kappa m_0}^{s'_p-k},  \br{4c_1\kappa m_0}^{2m_0+s'_p-s_p - 2k}}  .
\end{align}
Hence, we can apply the same analyses going forward. 

The reason why we cannot directly use the inequality~\eqref{induction_phis_pi_2m-s} for~\eqref{sm_0-sp-k_cannot_be_estiated} is that the condition~\eqref{kappa_upper_bound_ineq} cannot be ensured. If we rewrite 
$$\ave{ \pi^{2m_0-s_p-k}  \phi^{s'_p-k} \Phi_0^{s''_p}} 
=\ave{\pi^{2(m_0-1)-(s_p+k-2)}  \phi^{(s'_p-k)} \Phi_0^{s''_p}}$$ and apply~\eqref{induction_phis_pi_2m-s} with $m\to m_0-1$, $s \to s_p+k-2$, $s' \to s'_p-k$, and $s'' \to s''_p$, the following condition must be satisfied:
\begin{align}
	s'_p-k + s''_p \le  \kappa(s_p+k-2) . \label{must_be_satisfied!}
\end{align}
However, in the case where $s'_p+ s''_p=\kappa  s_p$, $k=1$ and $\kappa=2$, the LHS is equal to $2  s_p -1$, which is larger than the RHS, i.e., $\kappa(s_p+k-2)=2 s_p-2$. 
We thus cannot ensure the condition~\eqref{must_be_satisfied!}. 

By obtaining the inequalities~\eqref{induction_phis_pi_2m-s_fist/step} and \eqref{induction_phis_pi_2m-s_second/one}, 
we reach the main inequality~\eqref{induction_phis_pi_2m-s}  with $\zeta=3$. 
This completes the proof of Proposition~\ref{prop:ave_phi_s_pi_2m-s}. $\square$

\subsubsection{Proof of Lemma~\ref{lamm:upp_W_m_s}}

From the definition of $W_{m,p}  $ in Eq.~\eqref{def_W_m0_s}, we aim to upper-bound 
\begin{align}
	&\ave{\pi^{2m_0}}^{1-1/2^p} \br{W_{m_0,p}}^{1/2^p}  \notag \\
	&= \ave{\pi^{2m_0}}^{1-1/2^p}\brrr{\ave{\Phi_0^{4\kappa m_0}}^{\frac{s''_p}{4\kappa m_0}} \max\brr{ \ave{\pi^{2m_0}}^{1-\frac{s_p}{2m_0}} (4c_1 \kappa m_0)^{s'_p},  (4c_1 \kappa m_0)^{2m_0+s'_p-s_p}} }^{1/2^p} \notag \\
	&=  \ave{\pi^{2m_0}}^{1-1/2^p} \ave{\Phi_0^{4\kappa m_0}}^{\frac{s''}{4\kappa m_0}} \max\brr{ \ave{\pi^{2m_0}}^{\frac{1}{2^p}-\frac{s+1}{2m_0}} (4c_1 \kappa m_0)^{s'},  (4c_1 \kappa m_0)^{2m_0/2^p+s'-(s+1)}} ,
	\label{ave_pi_2m0_1-1/2^p_lemma}
\end{align}
where, in the second equality, we use $s_p:= 2^p (s+1)$, $s'_p:= 2^p s'$ and $s''_p=2^p s''$ from the definition~\eqref{s_p_s'_p_def}.
In the case of $\ave{\pi^{2m_0}} \ge (4c_1 \kappa m_0)^{2m_0}$, we have 
\begin{align}
	\label{lamm:upp_W_m_s_pr1}
	&\ave{\pi^{2m_0}}^{1-1/2^p} \max\brr{ \ave{\pi^{2m_0}}^{\frac{1}{2^p}-\frac{s+1}{2m_0}} (4c_1 \kappa m_0)^{s'},  (4c_1 \kappa m_0)^{2m_0/2^p+s'-(s+1)}} \notag \\
	&=  \ave{\pi^{2m_0}}^{1-1/2^p} \ave{\pi^{2m_0}}^{\frac{1}{2^p}-\frac{s+1}{2m_0}} (4c_1 \kappa m_0)^{s'}  \notag \\
	&= \ave{\pi^{2m_0}}^{1-\frac{s+1}{2m_0}} (4c_1 \kappa m_0)^{s'}  .
\end{align} 
On the other hand, in the case of $\ave{\pi^{2m_0}} < (4c_1 \kappa m_0)^{2m_0}$, we obtain 
\begin{align}
	\label{lamm:upp_W_m_s_pr2}
	&\ave{\pi^{2m_0}}^{1-1/2^p} \max\brr{ \ave{\pi^{2m_0}}^{\frac{1}{2^p}-\frac{s+1}{2m_0}} (4c_1 \kappa m_0)^{s'},  (4c_1 \kappa m_0)^{2m_0/2^p+s'-(s+1)}} \notag \\
	&< \brr{(4c_1 \kappa m_0)^{2m_0}}^{1-1/2^p} \cdot  (4c_1 \kappa m_0)^{2m_0/2^p+s'-(s+1)} \notag \\
	&=(4c_1 \kappa m_0)^{2m_0+s'-(s+1)} .
\end{align} 
By combining the upper bounds~\eqref{lamm:upp_W_m_s_pr1} and \eqref{lamm:upp_W_m_s_pr2} with the inequality~\eqref{ave_pi_2m0_1-1/2^p_lemma}, we reach the upper bound of 
\begin{align}
	\ave{\pi^{2m_0}}^{1-1/2^p} \br{W_{m_0,p}}^{1/2^p}  
	&\le  \ave{\Phi_0^{4\kappa m_0}}^{\frac{s''}{4\kappa m_0}} \max\brr{\ave{\pi^{2m_0}}^{1-\frac{s+1}{2m_0}} (4c_1 \kappa m_0)^{s'}, (4c_1 \kappa m_0)^{2m_0+s'-(s+1)} } \notag \\
	&= W_{m_0,0} . 
\end{align}
This completes the proof. $\square$

\subsubsection{Proof of Lemma~\ref{lamm:upp_K_m}}

We first show the definition~\eqref{K_p_L_p_definition} for $K_p$ again: 
\begin{align}
	\label{K_p_L_p_definition_again_show}
	&K_p:= \sum_{k=1}^{\min(2m_0-s_p,s'_{p-1})} i^k C_{k,2m_0-s_p,s'_{p-1}}\ave{ \pi^{2m_0-s_p-k}  \phi^{s'_p-k}\Phi_0^{s''_p} } .
\end{align}
By writing 
\begin{align}
	\label{pi_2m0-sp-k_phi_s'p-k}
	\ave{ \pi^{2m_0-s_p-k}  \phi^{s'_p-k} \Phi_0^{s''_p}} = \ave{ \pi^{2(m_0-1)-(s_p+k-2)}  \phi^{(s'_p-k)} \Phi_0^{s''_p}} ,
\end{align}
we have 
\begin{align}
	(s'_p-k) + s''_p &=2^p( s'+s'') - k \le 2^p \kappa s - k  \notag \\
	&\le \kappa \brr{2^p(s+1) + k-2  }  \notag \\
	&=  \kappa\br{s_p+k-2}
\end{align}
for $k\ge 1$ and $p\ge0$, where we use $s'+s'' \le \kappa s$ for the condition of the inequality~\eqref{induction_phis_pi_2m-s_fist/step}.   Note that $s_p=2^p(s+1)$. 

Hence, we can apply the inequality~\eqref{induction_phis_pi_2m-s} with 
\begin{align}
	m\to m_0-1, \quad s \to s_p+k-2, \quad s' \to s'_p-k,\quad s'' \to s''_p ,
\end{align}
we upper-bound Eq.~\eqref{pi_2m0-sp-k_phi_s'p-k} by
\begin{align}
	\label{upp_pi_2m0-2s+2s'-k_2s+2s'-k}
	&\abs{\ave{ \pi^{2m_0-s_p-k}  \phi^{s'_p-k} \Phi_0^{s''_p}}} \notag \\
	&\le 
	\zeta \ave{\Phi_0^{4\kappa m_0}}^{\frac{s''}{4\kappa m_0}} \max\brr{ \ave{ \pi^{2m_0-2}}^{\frac{2m_0-s_p-k}{2m_0-2}}   \brr{4c_1\kappa(m_0-1)}^{s'_p-k} , 
		\brr{4c_1\kappa(m_0-1)}^{2m_0+s'_p-s_p-2k}}
	\notag \\
	&\le 
	\zeta\ave{\Phi_0^{4\kappa m_0}}^{\frac{s''}{4\kappa m_0}} \max\brr{ \ave{ \pi^{2m_0}}^{\frac{2m_0-s_p-k}{2m_0}}  \br{4c_1\kappa m_0}^{s'_p-k} , 
		\br{4c_1\kappa m_0}^{2m_0+s'_p-s_p-2k}}  ,
\end{align}
where, in the second inequality, we use the inequality~\eqref{moment_relation_basci} to get $\ave{ \pi^{2m_0-2}}^{1/(2m_0-2)}\le \ave{ \pi^{2m_0}}^{1/(2m_0)}$.
Then, in the case of $\ave{ \pi^{2m_0}} \le \br{4c_1\kappa m_0}^{2m_0}$, we have 
\begin{align}
	\max\brr{ \ave{ \pi^{2m_0}}^{\frac{2m_0-s_p-k}{2m_0}}  \br{4c_1\kappa m_0}^{s'_p-k} , 
		\br{4c_1\kappa m_0}^{2m_0+s'_p-s_p-2k}}  =  \br{4c_1\kappa m_0}^{2m_0+s'_p-s_p-2k} .
\end{align}
On the other hand, in the case of $\ave{ \pi^{2m_0}} > \br{4c_1\kappa m_0}^{2m_0}$, the following upper bound holds:
\begin{align}
	\max\brr{ \ave{ \pi^{2m_0}}^{\frac{2m_0-s_p-k}{2m_0}}  \br{4c_1\kappa m_0}^{s'_p-k} , 
		\br{4c_1\kappa m_0}^{2m_0+s'_p-s_p-2k}}    
	&= \ave{\pi^{2m_0}}^{\frac{2m_0-s_p}{2m_0}} \frac{ \br{4c_1\kappa m_0}^{s'_p-k} }{\ave{\pi^{2m_0}}^{k/(2m_0)}} \notag \\
	&\le  \ave{\pi^{2m_0}}^{\frac{2m_0-s_p}{2m_0}}\br{4c_1\kappa m_0}^{s'_p-2k} . 
\end{align}
By applying the above two inequalities to~\eqref{upp_pi_2m0-2s+2s'-k_2s+2s'-k}, we obtain 
\begin{align}
	\label{upp_pi_2m0-2s+2s'-k_2s+2s'-k_/2}
	\abs{\ave{ \pi^{2m_0-s_p-k}  \phi^{s'_p-k}  \Phi_0^{s''_p}}}
	&\le\zeta (4c_1\kappa m_0)^{-2k}\ave{\Phi_0^{4\kappa m_0}}^{\frac{s''}{4\kappa m_0}} \max\brr{\ave{\pi^{2m_0}}^{\frac{2m_0-s_p}{2m_0}}\br{4c_1\kappa m_0}^{s'_p}, \br{4c_1\kappa m_0}^{2m_0+s'_p-s_p}} \notag \\
	&=\zeta (4c_1\kappa m_0)^{-2k}W_{m_0,p},
\end{align}
where we use the definition~\eqref{def_W_m0_s} for $W_{m_0,p}$. 

We use the inequality~\eqref{upp_pi_2m0-2s+2s'-k_2s+2s'-k_/2} to upper-bound the quantity $K_p$ in Eq.~\eqref{K_p_L_p_definition_again_show} in the following way:
\begin{align}
	\label{K_p_upper_bound_sum}
	K_p 
	&\le \zeta W_{m_0,p} \sum_{k=1}^{\min(2m_0-s_p,s'_{p-1})} C_{k,2m_0-s_p,s'_{p-1}} (4c_1\kappa m_0)^{-2k}.
\end{align}
For the summation, we calculate 
\begin{align}
	\label{K_p_upper_bound_sum_estimation}
	\sum_{k=1}^{\min(2m_0-s_p,s'_{p-1})} C_{k,2m_0-s_p,s'_{p-1}} (4c_1\kappa m_0)^{-2k}
	&=  \sum_{k=1}^{\min(2m_0-s_p,s'_{p-1})} k! \binom{2m_0-s_p}{k} \binom{s'_{p-1}}{k}(4c_1\kappa m_0)^{-2k} \notag\\
	&\le  \sum_{k=1}^{s'_{p-1}} \binom{s'_{p-1}}{k} \br{\frac{2m_0}{16c_1^2\kappa^2 m_0^2}}^{k} \notag \\
	&= \br{ 1+ \frac{1}{8c_1^2\kappa^2 m_0}}^{s'_{p-1}} -1 \le e^{1/(8c_1^2 \kappa)}-1,
\end{align}
where, in the last inequality, we use $s'_{p-1} \le s'_{p-1} + s''_{p-1} \le \kappa s_{p-1} \le  \kappa m_0 $ from $s_p \le 2m_0$.
By applying the upper bound~\eqref{K_p_upper_bound_sum_estimation} to \eqref{K_p_upper_bound_sum}, we obtain 
\begin{align}
	\label{K_p_upper_bound_fin}
	K_p 
	&\le \zeta W_{m_0,p}  \brr{e^{1/(8c_1^2 \kappa)}-1} \le  \frac{8(e^{1/8}-1)}{8c_1^2  \kappa}  \zeta W_{m_0,p} \le  \frac{\zeta}{7c_1^2\kappa} W_{m_0,p},
\end{align}
where we use $e^{1/x} -1 \le x_0(e^{1/{x_0}}-1)/x $ for $x\ge x_0$ with $8c_1^2 \kappa\ge 8$ from $c_1\ge 1$ and $\kappa\ge 1$. 
We thus prove the main inequality~\eqref{lamm:upp_K_m_main_ineq}.
This completes the proof. $\square$

{~} \\

{~} \\

\part{Entanglement area law in bosonic systems with long-range interactions}

\section{1D entanglement area law of interacting bosons} 

\subsection{Setup and assumptions}  \label{1D:Setup and assumptions}

We consider a one-dimensional chain located on $\mathbb{Z}$\footnote{Without loss of generality, we can make Hamiltonian on a finite set $\Lambda$ defined on $\mathbb{Z}$ by adding zero operators.}; that is, each of the site $i$ is characterized by the position $x\in \mathbb{Z}$.   
We separate the total system into $L=(-\infty, 0]$ and $R=[1,\infty)$ and consider the entanglement entropy for the ground state $\ket{\Omega}$ under the assumption of the spectral gap $\Delta$.  
In the following, we also assume the boson number distribution in the form of 
\begin{align}
	\label{Assum_boson_concentration}
	\norm{\Pi_{i,> N}  \ket{\Omega}} \le \mc e^{-\mb N^{1/\ma}} \for \forall i\in \Lambda,
\end{align}
where $\{\ma,\mc\}$ are $\orderof{1}$ constants, and $\mb$ depends on spectral gap as 
\begin{align}
	\label{Delta_dependence_mb}
	\mb \propto \Delta^{2\upsilon/(\ma k)} 
\end{align}
with $\upsilon$ an $\orderof{1}$ constant.

We consider a general boson model with $k$-body interactions, as in  
\begin{align}
	\label{general_k-local_op_boson}
	H = \sum_{Z: |Z|\le k} h_Z , 
\end{align} 
where $h_Z$ consists of the boson number operator and satisfies 
\begin{align}
	\label{ineq_h_Z_operator}
	\norm{h_Z\Pi_{\Lambda,\le N}  } \le  J_Z N^{k/2} \for N\ge 1,
\end{align}
with
\begin{align}
	\label{ineq_J_Z_operator_bar_J_1}
	\max_{i,i'} \br{\sum_{Z:Z\ni \{i,i'\}} J_Z } \le g \bar{J}(\dist_{i,i'}) .
\end{align}
We assume that $\bar{J}(0)=1$ and $\bar{J}(\dist_{i,i'})$ decays faster than $\dist_{i,i'}^{-2}$, which satisfies
\begin{align}
	\label{cond_r^2_1_decay}
	\br{r^2+1} \bar{J}(r) \le \frac{1}{r^{\bar{\alpha}}+1}\quad  (\bar{\alpha}>0).
\end{align} 
Note that we recover the condition~\eqref{Boson_most_general_k-local_g_ext} by letting $i'=i$. 

For an arbitrary operator $O$, we define the Schmidt rank ${\rm SR} (O)$ as the minimum integer such that
\begin{align}
	O = \sum_{m=1}^{{\rm SR} (O)} O_{L,m} \otimes O_{R,m},
\end{align}
where $O_{L,m}$ and $O_{R,m}$ are supported on the subsystems $L$ and $R$, respectively.  
Also, the entanglement entropy $S_L(\Omega)$ of the ground state $\ket{\Omega}$ is defined by
\begin{align}
	S_L(\Omega) :=- \sum_{j=1}^{\infty} \lambda^2_j \log(\lambda^2_j) 
\end{align}
with the Schmidt decomposition of the ground state as 
\begin{align}
	\ket{\Omega} =\sum_{j=1}^{\infty} \lambda_j \ket{L_j} \otimes \ket{R_j},
\end{align}
where each of the states $\{\ket{L_j}\}_j$ (resp. $\{\ket{R_j}\}_j$) is supported on $L$ (resp. $R$).

\subsection{Main statement}
The main statement here is the following theorem regarding the entanglement area law: 

\begin{theorem} \label{thm:main_theorem_area_law}
	Under the setup in \ref{1D:Setup and assumptions}, the entanglement entropy for an arbitrary partition is upper-bounded as follows:
	\begin{align}
		\label{thm:main_theorem_area_law/main_ineq}
		S_L(\Omega)  \le  C_0 \Delta^{-(1+2/\bar{\alpha})(\upsilon+1)} \brr{\log\br{1 /\Delta}}^{4+3/\bar{\alpha} +\chi(1+2/\bar{\alpha}) } ,
	\end{align}
	where $\chi = k\ma/2$, $\upsilon$ and $\bar{\alpha}$ are defined in Eqs.~\eqref{Delta_dependence_mb} and \eqref{cond_r^2_1_decay}, respectively, and $C_0$ is a constant that depends only on system details.
	In particular, in the case of $\upsilon=0$, we obtain 
	\begin{align}
		\label{thm:main_theorem_area_law/main_ineq_upsilon=0}
		S_L(\Omega)   \le  C_0 \Delta^{-1-2/\bar{\alpha}} \brr{\log\br{1 /\Delta}}^{3+3/\bar{\alpha} +\chi(1+2/\bar{\alpha}) } \log\log(1/\Delta).
	\end{align}
	
	Moreover, there exists a matrix product state (MPS) $\ket{{\rm M}_{\mD}}$ that approximates the ground state in the sense of 
	\begin{align}
		\label{MPS_approx_Area_law}
		\norm{\tr_{X^\co} \br{\ket{\Omega}\bra{\Omega} -\ket{{\rm M}_{\mD}}\bra{{\rm M}_{\mD}}}}_1 \le  \delta |X|
	\end{align}
	for an arbitrary subset $X\subseteq \Lambda$, where $\mD$ is the bond dimension and chosen as  
	\begin{align}
		\mD=\exp\brrr{ C_1 \Delta^{-(1+2/\bar{\alpha})(\upsilon+1)} \brr{\log\br{1 /\Delta}}^{4+3/\bar{\alpha} +\chi(1+2/\bar{\alpha}) } +C_2 \frac{\log^{\chi/2+5/2}\brr{1/(\delta \Delta)}}{\Delta^{(\upsilon+1)/2}}} .
	\end{align}
\end{theorem}

{\bf Remark.}
For the Bose-Hubbard classes, from Theorem~\ref{main_thm/Boson number distribution/BH}, the exponential decay of the boson number distribution~\eqref{main_thm/Boson number distribution/BH/ineq} gives $\ma=1$ and $\mb,\mc =\orderof{1}$ in~\eqref{Assum_boson_concentration}, and hence $\upsilon=0$ and $\chi=k/2$. 
We thus prove the entanglement area law in the form of 
\begin{align}
	\label{thm:main_theorem_area_law/main_ineq_BH}
	S_L(\Omega)  \le  C_0 \Delta^{-1-2/\bar{\alpha}} \brr{\log\br{1 /\Delta}}^{3+3/\bar{\alpha} +k(1+2/\bar{\alpha})/2} \log\log(1/\Delta),
\end{align}

On the other hand, for the $\phi4$ classes, the inequality~\eqref{main_ineq_corol_boson_dist_phi4} in Theorem~\ref{thm_boson_dist_phi4} gives 
\begin{align}
	&\norm{ \Pi_{i, >x} \ket{\Omega} } =\sqrt{ \bra{\Omega} \Pi_{i, >x} \ket{\Omega} } \le  2e^{k/2} \exp  \br{-\frac{k x^{1/k}}{16e \tilde{C}}} ,
\end{align}
which yields $\ma=k$, $\mb=k/(16e \tilde{C}) \propto \Delta^{1/2}$ (i.e., $2\upsilon/(\ma k)=1/2$) from Eq.~\eqref{tilde_C_explicit_from}, and $\mc=2e^{k/2}$. 
We hence obtain $\chi=k^2/2$ and $\upsilon=k^2/4$, which reduces the inequality~\eqref{thm:main_theorem_area_law/main_ineq} to
\begin{align}
	\label{thm:main_theorem_area_law/main_ineq_phi4}
	S_L(\Omega)  \le  C_0 \Delta^{-(1+2/\bar{\alpha})(k^2/4+1)} \brr{\log\br{1 /\Delta}}^{4+3/\bar{\alpha} +k^2(1+2/\bar{\alpha})/2} .
\end{align}

Also, regarding the MPS approximation, a particularly important case is $\Delta=\orderof{1}$ and $\delta=1/\poly(n)$, where the sufficient bond dimension is given by 
\begin{align}
	\mD \propto \exp\brr{ \log^{\chi/2+5/2}\br{n}} , 
\end{align}
which is a quasi-polynomial form with respect to the system size.

\subsection{Brief Outline of the proof strategy}

In this section, we provide a high-level overview of the proof of the area law for one-dimensional systems with long-range boson-boson interactions (See Supplementary Fig.~\ref{fig_supp_area_law}). The main approach we follow is based on the approximate ground state projector (AGSP) technique, initially developed in Refs.~\cite{PhysRevB.85.195145,arad2013area}
for systems with short-range interactions. This method was later extended to long-range interactions in Ref.~\cite{Kuwahara2020arealaw}.
In this work, we further extend the method to incorporate unbounded bosonic Hamiltonians.

\begin{figure}
	\begin{center}
		\includegraphics[width=\textwidth]{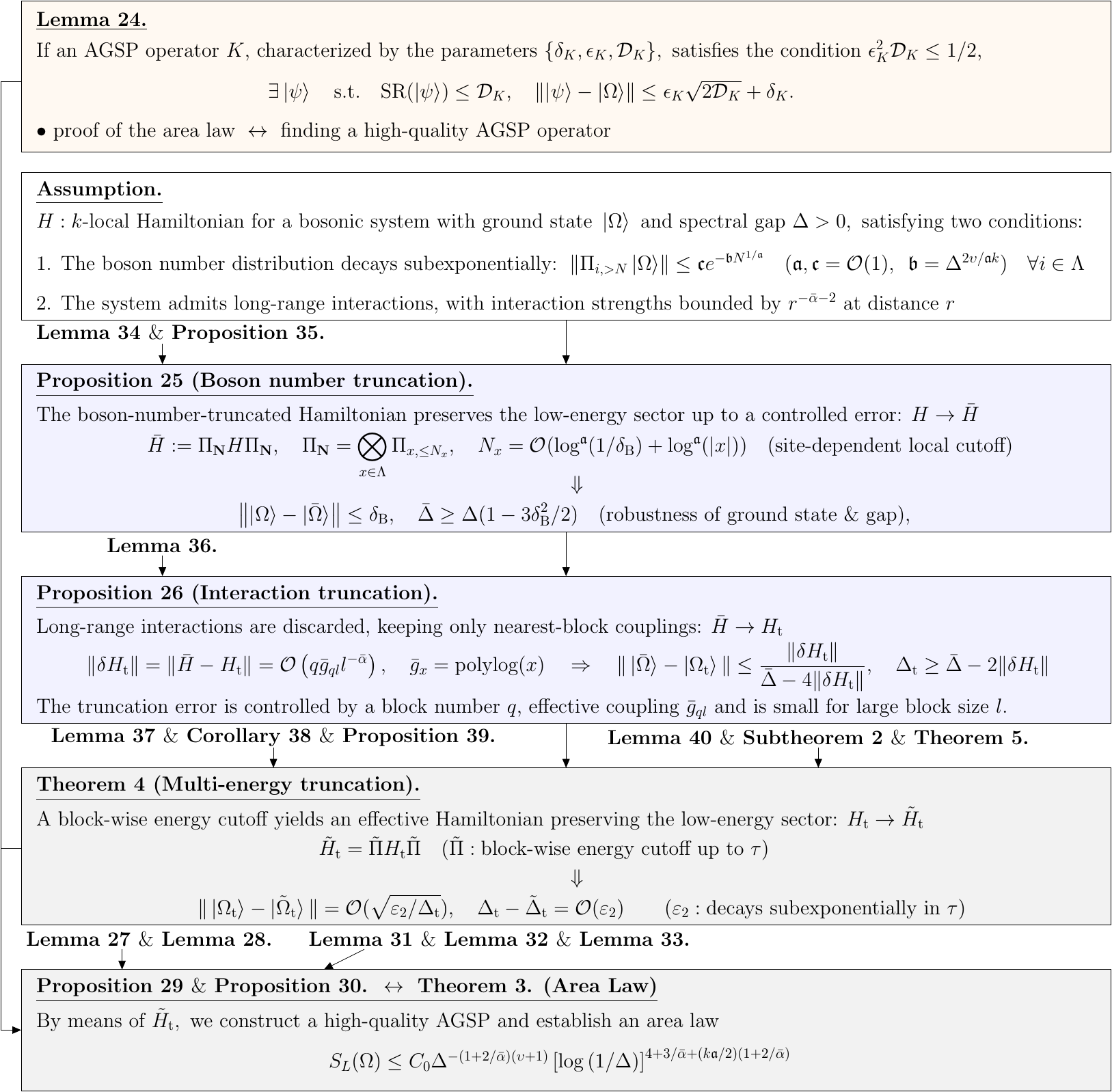}
	\end{center}
	\caption{This figure outlines the proof of Theorem 3, culminating in an area law for bosonic systems.}
	\label{fig_supp_area_law}
\end{figure}

The AGSP operator, denoted by $K$, approximates the ground state projector $\ket{\Omega}\bra{\Omega}$. Typically,  the Schmidt rank of the AGSP $K$ and the precision of the approximation have a trade-off relationship (see Sec .~\ref {sec:Approximate Ground State Projection (AGSP)}).
The key advantage is that if a suitable AGSP with low Schmidt rank and high precision is found, a low-entangled state that closely approximates the ground state can be obtained. This quantitative result is detailed in Lemma~\ref{prop:Overlap between the ground state and low-entangled state}. Therefore, the goal of the area law proof is to construct such a high-quality AGSP operator.

We begin by imposing a boson-number cut-off on each site, so as to render the local Hilbert space finite-dimensional everywhere. Importantly, the cut-off at site $x$ is chosen to grow (at most) poly-logarithmically with the distance from the boundary. This ensures that sites closer to the boundary have stricter cut-offs, while those farther away allow more bosons [see Eq.~\eqref{Prop:eff:boson_number_truncation_proj_choice_Nx}].
Through this truncation, we can approximately preserve both the ground state and the spectral gap. Importantly, a uniform cutoff cannot be applied since the error grows with system size. After truncation, the Hamiltonian remains unbounded in regions sufficiently far from the boundary. As we will demonstrate in Theorem~\ref{Effective Hamiltonian_multi_truncation} for energy truncation, this step yields a clear difference from the bounded Hamiltonian. The resulting effective Hamiltonian by boson-number truncations, $\bar{H}$, retains the important properties of the original ground state (see Proposition~\ref{Prop:eff:boson_number_truncation}).

Next, we apply an interaction truncation near the boundary. If all long-range interactions across the entire region were truncated, the modified Hamiltonian $H_\tc$ would differ significantly from the original $\bar{H}$, with $\|\bar{H} - H_\tc\| \sim \mathcal{O}(|\Lambda|)$. This would lead to substantial changes in the ground state. Therefore, we restrict the truncation of long-range interactions to the vicinity of the boundary between subsystems $L$ and $R$ (see Supplementary Figure~\ref{fig:Area_law_Ham_truncate}). Around the boundary, we divide the neighboring region to the boundary into $q$ blocks $\{B_s\}_{s=0}^{q+1}$, each of length $l$, such that only adjacent blocks interact. The remaining two blocks $B_0$ and $B_{q+1}$ are extended to the left and right ends, respectively.  
The error $\|\bar{H} - H_\tc\|$ can be effectively controlled by adjusting the number of blocks $q$ and their length $l$ (Proposition~\ref{lemm:error_int_truncation}).

The third truncation involves applying an energy cutoff within the blocks mentioned above (see also Fig.~\ref{fig:effective_Ham} below). In previous studies of effective Hamiltonians~\cite{arad2013area,Arad_2016},
the energy cutoff was applied only to the edge blocks ($B_0$ and $B_{q+1}$). For systems with long-range interactions, it is crucial to perform the energy cutoff to all blocks to obtain a better entanglement area law~\cite{Kuwahara2020arealaw}.
This process transforms the Hamiltonian as $H_\tc \to \tilde{H}_\tc$. 
In total, the original Hamiltonian is transformed as 
\begin{align}
	H \xrightarrow{\textrm{Boson number truncation}} \bar{H} \xrightarrow{\textrm{Interaction truncation}} H_\tc \xrightarrow{\textrm{Multi-energy truncation}}   \tilde{H}_\tc . 
\end{align}
While the preservation of the ground state in bounded Hamiltonians has been well studied~\cite{Arad_2016},
unbounded Hamiltonians like $H_\tc$ pose additional challenges, such as divergence in imaginary time evolution, which was a key technique in Ref.~\cite{Arad_2016}.
To address this, we have to avoid the divergence by carefully treating the unbounded Hamiltonian.
We here employ an alternative method developed in Ref.~\cite{Anshu_2021},
which was originally used to treat Hamiltonians lacking strict $k$-locality. Consequently, we establish a modified theorem for the preservation of the ground state in $\tilde{H}_\tc$ (Theorem~\ref{Effective Hamiltonian_multi_truncation}  in~\ref{Sec:Effective Hamiltonian with Multi-energy Cut-off}).

With the effective Hamiltonian $\tilde{H}_\tc$ in hand, we can now construct the AGSP operator.
Following Ref.~\cite{arad2013area},
we utilize Chebyshev polynomials, where the accuracy of the AGSP is roughly given by $e^{-m \sqrt{\Delta/\norm{\tilde{H}_\tc}}}$ (Lemma~\ref{lemm:Chebyshev_AGSP}), where $m$ is the polynomial degree and $\Delta$ is the spectral gap. The Schmidt rank of the polynomial of $\tilde{H}_\tc$ is bounded as shown in Lemma~\ref{thm:Schmit_rank_Ham_power_0}. Then, by optimizing all parameters, i.e., the boson-number truncations, the number of blocks $q+2$, the block length $l$, and the degree of the Chebyshev polynomial, we construct a suitable AGSP operator to meet the desired property in Lemma~\ref{prop:Overlap between the ground state and low-entangled state}. This leads to Proposition~\ref{prop1:truncate_gs_overlap}, where we obtain a low-Schmidt-rank state with a large overlap with the ground state.

Finally, by applying a sequence of AGSP operators, we derive an upper bound on the entanglement entropy (Lemma~\ref{prop:entropy_and_AGSP}) and estimate the required Schmidt rank to approximate the ground state with sufficient accuracy. This is summarized in Proposition~\ref{prop0:overlap_AGSP_entropy_bound}, completing the proof of the main theorem.

\section{Proof of Theorem~\ref{thm:main_theorem_area_law}.}

We utilize basic statements in Ref.~\cite{Kuwahara2020arealaw}
without proofs, which treats the bounded Hamiltonians, such as a spin or fermion model, with power-law decaying interactions. 
The primary difference here is that the Hilbert space dimension and the interaction energy are unbounded.

\subsection{Approximate Ground State Projection (AGSP)} \label{sec:Approximate Ground State Projection (AGSP)}

We introduce the projection operator onto the ground state. Constructing the exact ground-state projection operator is generally challenging, so we consider an approximate one:
\begin{align}
	K\ket{\Gs}  \simeq \ket{\Gs} \quad \text{and} \quad \| K (1 - \ket{\Gs}\bra{\Gs}) \| \simeq 0,
	\label{AGSP:formal_def}
\end{align}
where $(1 - \ket{\Gs}\bra{\Gs})$ is the projection operator onto the excited states' space. We assume $K$ is Hermitian (i.e., $K=K^\dagger$). 

Next, we characterize the approximate ground state projection (AGSP) operator by three parameters: $\{\delta_K, \epsilon_K, \mD_K\}$. Let $\ket{\Gs_K}$ be the quantum state invariant under $K$ such that:
\begin{align}
	K \ket{\Gs_K} = \ket{\Gs_K}. \label{def_of_Gs_K}
\end{align}

The parameters are defined by the following inequalities:
\begin{align}
	\|\ket{\Gs}  - \ket{\Gs_K} \|  \le \delta_K, \quad \| K (1- \ket{\Gs_K}\bra{\Gs_K}) \| \le \epsilon_K, \quad \text{and} \quad \text{SR}(K) \le \mD_K, \label{Def_AGSP_error_chap2}
\end{align}
where $\text{SR}(K)$ is the Schmidt rank of $K$ between the subsystems $L$ and $R$.
The second inequality in~\eqref{Def_AGSP_error_chap2} implies that for any state $\ket{\psi_\bot}$ orthogonal to $\ket{\Gs_K}$ (i.e., $\langle \psi_\bot \ket{\Gs_K}=0$), we have:
\begin{align}
	\| K \ket{\psi_\bot} \| = \| K (1- \ket{\Gs_K}\bra{\Gs_K}) \ket{\psi_\bot} \| \le \epsilon_K. \label{Def_AGSP_error_second}
\end{align}

Note that $\ket{\Gs_K}$ is an approximate ground state when $\delta_K \simeq 0$. If $\delta_K=\epsilon_K=0$, $K$ is the exact ground state projector, i.e., $K = \ket{\Gs}\bra{\Gs}$. In the standard AGSP definition~\cite{PhysRevB.85.195145,arad2013area,Kuwahara_2017},
the parameter $\delta_K$ is typically not considered. However, in long-range interacting systems, the error $\|\ket{\Gs} - \ket{\Gs_K}\|$ can be significant, requiring careful consideration of $\delta_K$.

\begin{lemma}[Supplementary Proposition 2 in Ref.~\cite{Kuwahara2020arealaw}]  \label{prop:Overlap between the ground state and low-entangled state}
	Let $K$ be an AGSP operator for $\ket{\Gs}$ with the parameters ($\delta_{K}, \epsilon_{K}, \mD_{K}$). 
	If the following inequality holds 
	\begin{align}
		\epsilon_{K}^2 \mD_{K} \le  \frac{1}{2}, \label{cond:Boot_strapping_lemma}
	\end{align}
	there exists a quantum state  $\ket{\psi}$ with ${\rm SR}(\ket{\psi})\le \mD_{K}$ such that
	\begin{align}
		\left \| \ket{\psi} - \ket{\Gs} \right\|\le \epsilon_{K}\sqrt{2 \mD_{K}   } + \delta_{K}.
		\label{ineq:AGSP_bootstrap_norm_distance}
	\end{align}
\end{lemma}

\subsection{Effective Hamiltonian by boson-number truncations}

We here adopt the boson-number truncation of $\bar{\Pi}$ and consider the effective Hamiltonian $\bar{H}$ in the form of 
\begin{align}
	\bar{H}:=\bar{\Pi} H \bar{\Pi} = \sum_{Z} \bar{h}_Z
\end{align}
with the ground state $\ket{\bar{\Omega}}$ and the spectral gap $\bar{\Delta}$. 

Regarding the above effective Hamiltonian, we will prove the following proposition (see~\ref{Proof of Proposition_Prop:eff:boson_number_truncation} for the proof): 
\begin{prop} \label{Prop:eff:boson_number_truncation}
	Let us adopt $\Pi_{\mathbf{N}}$ as the projection $\bar{\Pi}$, which is given by 
	\begin{align}
		\label{Prop:eff:boson_number_truncation_proj}
		\bar{\Pi}= \Pi_{\mathbf{N}} := \bigotimes_{x\in \Lambda}  \Pi_{x,\le N_x} , 
	\end{align}
	with 
	\begin{align}
		\label{Prop:eff:boson_number_truncation_proj_choice_Nx}
		N_x = \mb^{-\ma} \brr{ \log \br{\epsilon_\bo^{-1}} + \log \br{{|x|^{3} +1}}}^{\ma} \le 
		2^{\ma }\mb^{-\ma} \brr{ \log^{\ma} \br{\epsilon_\bo^{-1}} + \log^{\ma} \br{{|x|^{3} +1}}}.
	\end{align} 
	Then, the Hamiltonian $\bar{H}= \Pi_{\mathbf{N}} H \Pi_{\mathbf{N}}$ preserves the ground state and the spectral gap as follows:
	\begin{align}
		\label{norm_Omega_minus_bar_Omega}
		\norm{ \ket{\Omega } - \ket{\bar{\Omega} }} \le \delta_\bo, 
		\quad \bar{\Delta} \ge  \Delta \br{1-\frac{3}{2} \delta_\bo^2}, 
	\end{align}
	with $\epsilon_\bo$ given by
	\begin{align}
		\label{epsilon_0_choice_delta_0}
		\epsilon_\bo = w_0 \delta_\bo^2  \Delta \log^{-\ma k/2} \br{\delta_\bo^{-1}} ,
	\end{align} 
	where $w_0$ is an $\orderof{1}$ constant and we choose $\delta_\bo$ such that $\delta_\bo \le 1/\sqrt{6}$.  
	Note that $\ket{\bar{\Omega}}$ and $\bar{\Delta}$ are the ground state and the spectral gap of $\bar{H}$. 
\end{prop}

Under the projection~\eqref{Prop:eff:boson_number_truncation_proj},  
each of the sites has a Hilbert space dimension $d_x$ that depends on $x$, which is upper-bounded by $N_{x}$ as follows:
\begin{align}
	\label{def:d_x_local_dimension}
	&d_x \le  2^{\ma }\mb^{-\ma} \brr{ \log^{\ma} \br{\epsilon_\bo^{-1}} + \log^{\ma} \br{{x^3 +1}}}
	\le  d_0+ d_1 \log^{\ma} (x+1) ,  \notag \\
	&d_0= 2^{\ma }\mb^{-\ma} \log^{\ma} \br{\epsilon_\bo^{-1}} ,\quad d_1= 6^{\ma}\mb^{-\ma} ,
\end{align} 
where we use $\log^{\ma} \br{x^3 +1} \le\log^{\ma} \br{x +1}^3=3^\ma \log^{\ma} \br{x +1}$. 
Here, we can write $d_1=\orderof{\mb^{-\ma}}$, which only depends on the system details, while the dimension $d_0$ also depends on the error $\epsilon_\bo$ for the boson-number truncations.
Hence, we need to derive the area law so that the $(d_0,d_1)$ dependences are correctly taken into account (see~\ref{Sec:Proof of Proposition_prop1_truncate_gs_overlap}). 

Moreover, we rewrite the condition~\eqref{ineq_h_Z_operator} as 
\begin{align}
	\label{ineq_bar_h_Z_operator}
	\norm{\bar{h}_Z} \le  J_Z N_x^{k/2} \for Z\subseteq [-x,x]  . 
\end{align}
Then, using the upper bound of~\eqref{ineq_J_Z_operator_bar_J_1}, we have 
\begin{align}
	\label{ineq_h_Z_operator_barg_1_2}
	\max_{i,i'} \br{\sum_{Z:Z\ni \{i,i'\}, Z\subset [-x,x]} \norm{\bar{h}_Z} } \le g N_x^{k/2}  \bar{J}(\dist_{i,i'}) .
\end{align}
Because of 
\begin{align}
	\label{def_bar_g_x}
	g N_x^{k/2} 
	&\le g\mb^{-\ma k/2} \brr{ \log \br{\epsilon_\bo^{-1}} + \log \br{{|x|^{3} +1}}}^{\ma k/2} \notag \\
	& \le  g_0+ g_1 \log^{\chi} (x+1)=:\bar{g}_x , 
\end{align} 
with
\begin{align}
	\label{def:g_x_local_energy}
	g_0 =  g \mb^{-\ma k/2} 2^{\ma k/2} \log^{\ma k/2} \br{\epsilon_\bo^{-1}},\quad g_1 =  g \mb^{-\ma k/2} 6^{\ma k/2} ,\quad \chi=\frac{\ma k}{2} , 
\end{align} 
we can reduce the inequality~\eqref{ineq_h_Z_operator_barg_1_2} to 
\begin{align}
	\label{general_k-local_op_unbounded_cond}
	\sum_{Z:Z\ni \{i,i'\}, Z\subset [-x,x] } \|\bar{h}_Z\| \le \bar{g}_x \bar{J}(\dist_{i,i'})  ,
\end{align}  
which characterizes the interaction strength of the effective Hamiltonian  
$
\bar{H} = \sum_{Z: |Z|\le k} \bar{h}_Z  .
$

\subsection{Interaction-truncated Hamiltonian}

\begin{figure}
	\centering
	\includegraphics[scale=0.5]{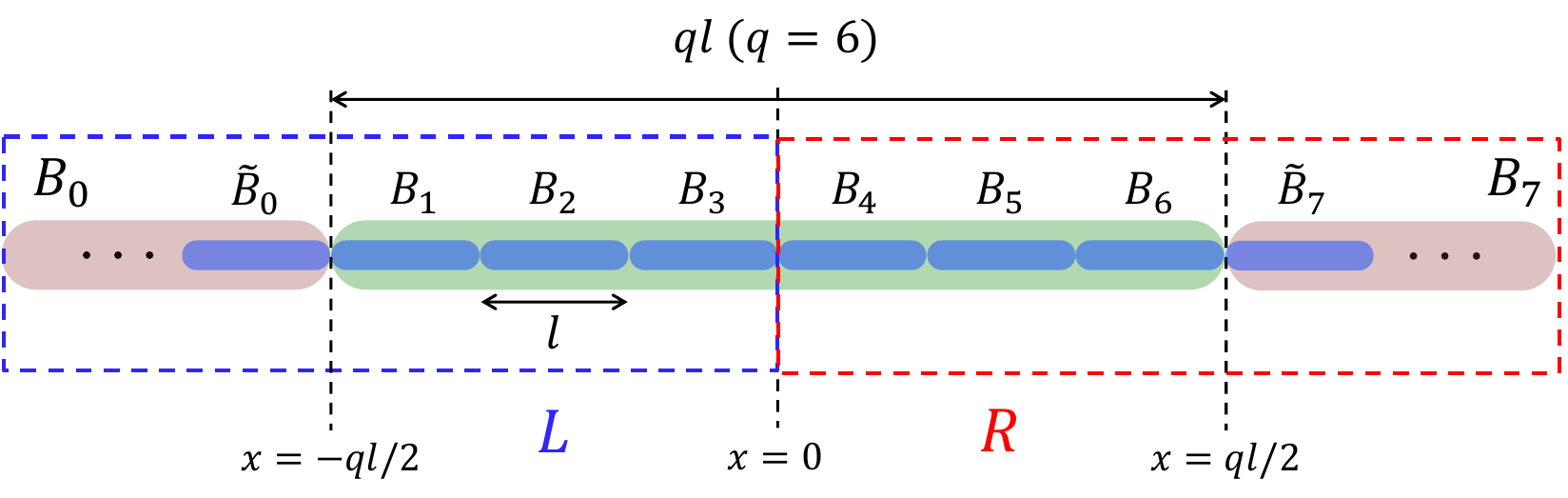}
	\caption{Interaction truncation in the Hamiltonian. The system is decomposed into $(q+2)$ blocks ($q=6$ shown above). 
		Blocks $\{B_s\}_{s=1}^q$ each have length $l$, and edge blocks $B_0$ and $B_{q+1}$ extend to the system’s left and right ends. 
		We truncate all interactions between separated blocks, so the truncated Hamiltonian $H_\tc$ [Eq.~\eqref{def:truncated_Hamiltonian}] remains close to the original Hamiltonian $H$, as shown in Lemma~\ref{lemm:error_int_truncation}.}
	\label{fig:Area_law_Ham_truncate}
\end{figure}

We decompose the system into blocks $B_0$, $\{B_s\}_{s=1}^q$, and $B_{q+1}$, such that $\bigcup_{s=0}^{q+1} B_s = \Lambda$, where $q$ is an even integer with the condition $q \geq 4$ and $|B_s|=l$ for $1 \leq s \leq q$. The subsets $L$ and $R$ are expressed as:
\begin{align}
	L = \bigcup_{s=0}^{q/2} B_s,\quad R = \bigcup_{s=q/2+1}^{q+1} B_s. \label{Block_definition_LR}
\end{align}
Following Ref.~\cite{Kuwahara2020arealaw}, we also define the interaction operator $V_{X,Y}(\Lambda_0)$ between two subsystems $X\subset \Lambda$ and $Y\subset \Lambda$ as follows. 
\begin{align}
	V_{X,Y}(\Lambda_0) :=\sum_{\substack{Z: Z\subset \Lambda_0 \\ Z\cap X\neq \emptyset , Z\cap Y\neq \emptyset} } \bar{h}_Z  ,
	\label{Def:V_X_Y}
\end{align} 
where the subset $\Lambda_0\subset \Lambda $ and  $X\sqcup Y \subset \Lambda_0 $ are arbitrarily chosen.

After truncating all interactions between non-adjacent blocks, only interactions between adjacent blocks remain:
\begin{align}
	H_\tc = \sum_{s=0}^{q+1} h_s + \sum_{s=0}^q h_{s,s+1}, \label{def:truncated_Hamiltonian}
\end{align}
where $h_{s,s+1} := V_{B_s,B_{s+1}}(B_s \sqcup B_{s+1})$ (with $X = B_s$, $Y = B_{s+1}$, and $\Lambda_0 = B_s \sqcup B_{s+1}$ in the definition of $V_{X,Y}(\Lambda_0)$ in Eq.~\eqref{Def:V_X_Y}), and $h_s$ collects all terms supported on $B_s$. 
In particular, for $s=0$ and $s=q$, we let 
\begin{align}
	h_{0,1}:= V_{\tilde{B}_0,B_1}(\tilde{B}_0 \sqcup B_{1}),
	\quad h_{q,q+1}:= V_{B_q,\tilde{B}_{q+1}}(B_q \sqcup \tilde{B}_{q+1} ), 
\end{align}
where $\tilde{B}_0=[-ql/2-l, -ql/2)$ and $\tilde{B}_{q+1}=[ql/2, ql/2+l)$ (see Supplementary Fig.~\ref{fig:Area_law_Ham_truncate}).

In  Lemma~\ref{lem:ar_V_X_Y_upper} in \ref{sec:Error estimation of interaction truncation}, we will prove the upper bound of 
\begin{align}
	\label{bar_V_X_Y_uppe_fin_again}
	\overline{V}_{X,Y}:= \sum_{Z:Z\cap X\neq \emptyset, Z\cap Y\neq \emptyset} \|\bar{h}_Z\|
	\le 2 \eta_1  \eta_2 \bar{g}_{|x|+r}  \br{r^2+1} \bar{J}(r) ,
\end{align}
where $X= (-\infty,x]$, $Y= [y,\infty)$ ($y-x=r$), and $\eta_1$ and $\eta_2$ are $\orderof{1}$ constants.
By using the inequality~\eqref{bar_V_X_Y_uppe_fin_again} with $r=1$ and $|x| \le ql/2$, we have:
\begin{align}
	\| h_{s,s+1} \|
	&\le \sum_{\substack{Z:Z\subset B_s \sqcup B_{s+1} \\ Z\cap B_s \neq \emptyset,\ Z \cap B_{s+1}\neq \emptyset} } \norm{h_Z}\le4 \eta_1  \eta_2 \bar{g}_{ql} \bar{J}(1) 
	=:c_0\bar{g}_{ql}   , \label{truncated_Hamiltonian_block_interaction}
\end{align}
where we use $\bar{g}_{ql/2+1}\le \bar{g}_{ql}$ from $ql\ge 2$ and the monotonic increase of $\bar{g}_x$

As notations, we define $\ket{\Gs_\tc}$ and $\Delta_\tc$ as the ground state and the spectral gap of the truncated Hamiltonian $H_\tc$, respectively. 
We also denote the ground energy of $H_\tc$ by $E_{\tc,0}$:
$
H_\tc \ket{\Omega_\tc}=E_{\tc,0} \ket{\Omega_\tc} .
$
Under the inequality~\eqref{bar_V_X_Y_uppe_fin_again}, we can prove the following proposition (see~\ref{sec:Error estimation of interaction truncation} for the proof):

\begin{prop} \label{lemm:error_int_truncation}
	The norm distance between $\bar{H}$ and $H_\tc$ is bounded from above by
	\begin{align}
		\norm{\delta H_\tc}=\norm{\bar{H}-H_\tc}  \le    4 \eta_1  \eta_2 q \bar{g}_{ql}\br{l^2+1} \bar{J}(l)  , \label{lemma:truncate__ineq}
	\end{align}
	where we define $\delta H_\tc := \bar{H}-H_\tc$.
	Also, the spectral gap $\Delta_\tc$ of $H_\tc$  is bounded from below by
	\begin{align}
		\Delta_\tc \ge \bar{\Delta} - 2 \norm{\delta H_\tc} \ge \bar{\Delta} -  8 \eta_1  \eta_2 q \bar{g}_{ql}  \br{l^2+1} \bar{J}(l)  . \label{lemma:truncate_ineq_gap}
	\end{align}
	
	Under the assumption of $4\norm{\delta H_\tc} < \bar{\Delta}$, the ground state $\ket{\bar{\Gs}}$ have an overlap with that of the truncated Hamiltonian $\ket{\Gs_\tc}$ as follows:
	\begin{align}
		\| \ket{\bar{\Gs}}-\ket{\Gs_\tc} \| \le \frac{ \norm{\delta H_\tc}}{\bar{\Delta} - 4\norm{\delta H_\tc}} . \label{overlap_Gs_t_Gs}
	\end{align}
	Also, for an arbitrary quantum state $\ket{\phi}$, the norm distance between $\ket{\bar{\Gs}}$ and $\ket{\phi}$ is bounded from above by
	\begin{align}
		\| \ket{\bar{\Gs}}-\ket{\phi} \|\le  \| \ket{\Gs_\tc}-\ket{\phi} \| +\frac{ \norm{\delta H_\tc}}{\bar{\Delta} -4 \norm{\delta H_\tc}}.
		\label{overlap_Gs_phi_norm}
	\end{align}
\end{prop}

\subsection{Effective Hamiltonian with Multi-energy Cut-off} \label{Sec:Effective Hamiltonian with Multi-energy Cut-off}

\begin{figure}
	\centering
	\includegraphics[scale=0.5]{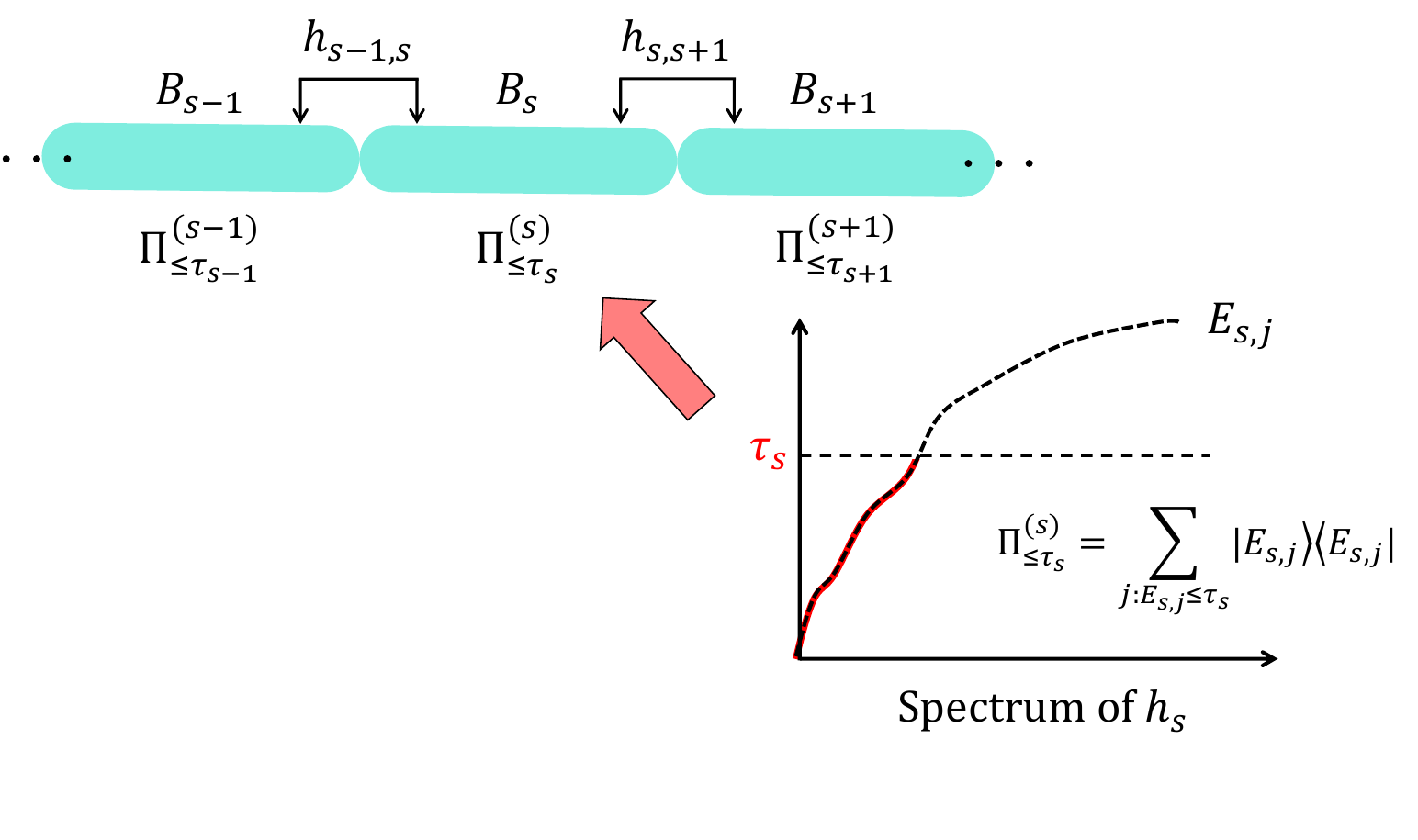}
	\caption{Schematic of the effective Hamiltonian $\tilde{H}_\tc$. We modify the energy spectrum in each $\{h_s\}_{s=0}^{q+1}$ so that energies above $\tau_s$ are constant, while $\{h_{s,s+1}\}_{s=0}^q$ remains the same as the original Hamiltonian. The low-energy spectrum is approximately preserved. The accuracy improves subexponentially with the cut-off energy $\tau$ (Theorem~\ref{Effective Hamiltonian_multi_truncation}).}
	\label{fig:effective_Ham}
\end{figure}

{~}\\
In constructing the AGSP operator~\eqref{AGSP:formal_def}, we require an effective Hamiltonian $\tilde{H}_\tc$ with a small norm that retains the low-energy properties of the original Hamiltonian $H_\tc$. To achieve this, we apply the energy cut-off to the Hamiltonian $H_\tc$ in Eq.~\eqref{def:truncated_Hamiltonian}, which plays a crucial role in Refs.~\cite{arad2013area,Arad2017,Arad_2016}.
We adopt here the multi-energy cutoff used in Ref.~\cite{Kuwahara2020arealaw} for the long-range area law.  

For each block Hamiltonian $\{h_s\}_{s=0}^{q+1}$, we adopt the following spectral decomposition 
\begin{align}
	h_s &= \sum_{E_{s,j}} E_{s,j} \ket{E_{s,j}}\bra{E_{s,j}}  ,\label{truncation_effective_Hamiltonian_h_s}
\end{align}
where $\{E_{s,j}, \ket{E_{s,j}}\}_j$ are the eigenvalues and eigenstates of $h_s$, respectively. 
We define the projection operator onto the eigenspace of $h_s$ as 
\begin{align}
	\Pi^{(s)}_{I}=\sum_{E_{s,j} \in I} \ket{E_{s,j}}\bra{E_{s,j}}  
	\label{projection_to_spectrum}
\end{align}
for $I \subset \mathbb{R}$. Especially for $\Pi^{(s)}_{(-\infty,E)}$ and $\Pi^{(s)}_{(-\infty,E]}$, we denote them by $\Pi^{(s)}_{< E}$ and $\Pi^{(s)}_{\le E}$, respectively. 
In the same way, we define $\Pi^{(s)}_{> E}$ and $\Pi^{(s)}_{\ge E}$.

Using the projections $\{\Pi^{(s)}_{\le \tau_s}\}_{s=0}^{q+1}$, we define the total projection $\tilde{\Pi}$ as 
\begin{align}
	\tilde{\Pi}=\bigotimes_{s=0}^{q+1} \Pi^{(s)}_{\le \tau_s} ,
	\label{projection_to_spectrum_total}
\end{align}
with
\begin{align}
	\tau_s = E_{s,0} + \tau \for s=0,1,2\ldots, q+1  \label{tau_s_tau} .
\end{align}
By the projection $\Pi^{(s)}_{\le \tau_s}$, we have 
\begin{align}
	\Pi^{(s)}_{\le \tau_s} h_s \Pi^{(s)}_{\le \tau_s}&= \sum_{E_{s,j}\le \tau_s} E_{s,j} \ket{E_{s,j}}\bra{E_{s,j}} .
\end{align}

By using the above notations, we describe the effective Hamiltonian $\tilde{H}_\tc$ as 
\begin{align}
	\tilde{H}_\tc =\tilde{\Pi} H_\tc \tilde{\Pi} . 
	\label{explicit_tilde_H_eff}
\end{align}
We note that the above definition differs slightly from the original one.
In Ref.~\cite{Kuwahara2020arealaw}, the following expression was considered: 
\begin{align}
	\tilde{H}'_\tc =\sum_{s=0}^{q+1} \tilde{h}_s + \sum_{s=0}^q h_{s,s+1} ,\quad  \tilde{h}_s := h_s \Pi^{(s)}_{\le \tau_s}+ \tau_s \Pi^{(s)}_{> \tau_s} 
	\label{explicit_tilde_H_eff'}
\end{align}
for $s=0,1,2\ldots, q+1$.

One advantage of utilizing the definition~\eqref{explicit_tilde_H_eff} is that we can utilize Lemma~\ref{lemma:effective_global}.
Here, we do not need spectral analyses of $\tilde{H}_\tc$, which is quite challenging in our setup with unbounded Hamiltonians\footnote{As will be mentioned in Lemma~\ref{lemma:commutator_unbounded}, the Hamiltonian $H_\tc$ does not converge the imaginary time evolution as $e^{-\beta H_\tc} O e^{\beta H_\tc}$. Hence, we have to avoid using the imaginary time evolution as in \ref{sec:Key Proposition and difficult point}.
	However, the spectral analyses of $\tilde{H}_\tc$ in Supplementary Note 4. F of Ref.~\cite{Kuwahara2020arealaw}, which fully utilizes the imaginary time evolution, is difficult to extend to our setup.}
As shown below, we can prove the preservation of the ground state and the spectral gap by choosing the truncation parameter $\tau$ sufficiently large (see~\ref{sec:Effective Hamiltonian theory_proof} for the proof).

\begin{theorem} \label{Effective Hamiltonian_multi_truncation} 
	Let us define $\varepsilon_1$ and $\varepsilon_2$ as 
	\begin{align}
		\label{def_epsilon_1_eff_H}
		\varepsilon_1=  2q \mE_{\tau  - 4 c_0\bar{g}_{ql}  -T_0}  ,\quad 
		\varepsilon_2= \frac{\varepsilon_1}{1-\varepsilon_1}  2q \br{ \tau  +2 c_0\bar{g}_{ql} } ,
	\end{align}
	where $\mE_y$ ($y\ge 0$) is a sub-exponentially decaying function defined in Eq.~\eqref{def:varepsilon_y}, and $T_0$ is defined by $T_{m=0}$ using $T_m$ in Eq.~\eqref{assump_kappa_form}.
	Then, we obtain
	\begin{align}
		\label{thm:effective_global/main1_eff}
		\norm{ \ket{\Omega_\tc } - \ket{\tilde{\Omega}_\tc}} \le \sqrt{\frac{2\varepsilon_2}{\Delta_\tc}},
	\end{align} 
	and the spectral gap $\tilde{\Delta}$ is lower-bounded by
	\begin{align}
		\label{thm:effective_global_gap_eff}
		\tilde{\Delta}_\tc \ge \Delta_\tc - 3 \varepsilon_2    .
	\end{align} 
\end{theorem}

{\bf Remark.} 
In Ref.~\cite{PhysRevA.108.042422}, short-range bosonic area laws in specific models were considered. 
Therein, the theorem on the effective Hamiltonian in Ref.~\cite{arad2013area}, which was derived for bounded Hamiltonians, was utilized\footnote{More precisely,  Appendix D 1 of Ref.~\cite{PhysRevA.108.042422} considers the Hilbert space truncation only around the boundary up to a finite distance. At this stage, interaction strength is unbounded beyond a finite distance from the boundary.}.
However, the unboundedness of the total Hamiltonian makes a troublesome problem, as will be shown in \ref{Sec:Multi-commutator bound}. 
The error parameters $\varepsilon_1$ and $\varepsilon_2$ give a subexponential decay with respect to $\tau$, which sets a main difference in bounded Hamiltonian cases that give the exponential error decay. 

For the convenience of readers, we show the explicit form of $\mE_{y}$ in Eq.~\eqref{def:varepsilon_y}:
\begin{align}
	\label{mE_Y_again}
	\mE_{y}=\mu_1 \exp\br{- \frac{y}{8e \tilde{T}_{y/T_0}}}+  \mu_2  \exp\brr{-\frac{1}{2}\br{\frac{y}{4e\tilde{c}_2g_1}}^{1/(1+\chi)}},
\end{align}
with
\begin{align}
	&\tilde{T}_{y/T_0}:= (2 c_0 \tilde{c}_3 +\tilde{c}_1)  \bar{g}_{y/T_0+ql } ,\quad T_0= (2 c_0 \tilde{c}_3 +\tilde{c}_1)  \bar{g}_{ql } ,
	\notag \\
	&\mu_1=1+e \int_{0}^\infty (z+2) \exp\brr{- \frac{1}{8e} \cdot \frac{z}{1 + \log^\chi(z+3)}}dz ,\quad \mu_2:= 1+e \int_{0}^\infty (z+2) \exp\brr{-\frac{1}{2}\br{\frac{ 2 c_0 \tilde{c}_3 +\tilde{c}_1}{4e\tilde{c}_2}  z}^{1/(1+\chi)}}  dz, \notag \\
	&c_0=4 \eta_1  \eta_2 \bar{J}(1),\quad   \tilde{c}_1=\frac{2^{\chi +\bar{\alpha}+3} 4k\eta_1}{1- 2^{-\bar{\alpha}}} , \quad 
	\tilde{c}_2= \tilde{c}_1  \brr{ \frac{2\chi  (2+\bar{\alpha})}{\bar{\alpha}}}^{\chi } ,\quad 
	\tilde{c}_3 = 2^{\bar{\alpha}}+ \frac{2(2+\bar{\alpha})}{\bar{\alpha}} , 
\end{align}
where we refer to Eq.~\eqref{def:varepsilon_y} for $\mE_{y}$, Eq.~\eqref{definition_tilde_T_x} for $\tilde{T}_{y/T_0}$, 
Eq.~\eqref{assump_kappa_form} for $T_0$, Eqs.~\eqref{defin:mu_1} and \eqref{defin:mu_2} for $\mu_1,\mu_2$, 
Eq.~\eqref{truncated_Hamiltonian_block_interaction} for $c_0$, 
Eq.~\eqref{def::tilde_c_1c_2} for $\tilde{c}_1,\tilde{c}_2$, and Eq.~\eqref{defi:tilde_c_3} for $\tilde{c}_3$.
We recall that the parameter $\eta_p$ is defined by the inequality~\eqref{def:eta_parameter}, and the parameter $\bar{g}_x$ has been defined as 
$\bar{g}_x = g_0+ g_1 \log^{\chi } (x+1)$ in Eq.~\eqref{def_bar_g_x}. 

In the following, we assume 
\begin{align}
	\label{tau_c_tau_condition}
	\tau \le  \bar{g}_{ql} ql .
\end{align}
Then, by letting $y= \tau - 4c_0 \bar{g}_{ql} -T_0$, we have 
\begin{align}
	\frac{y}{T_0} = \frac{\tau - 4c_0 \bar{g}_{ql} -T_0 }{T_0}  \le \frac{1}{2 c_0 \tilde{c}_3 +\tilde{c}_1}  ql  \le ql
\end{align}
from $\tilde{c}_1\ge 1$, and hence, we upper-bound $\bar{g}_{y/T_0+ql }$ by $\bar{g}_{2ql }$, which also yields an upper bound for the parameter $\mE_{y}$ as follows:
\begin{align}
	\label{mE_tau_insert_again}
	\mE_{\tau - 4c_0 \bar{g}_{ql} -T_0}
	&\le\mu_1 \exp\brr{- \frac{\tau - 4c_0 \bar{g}_{ql} -T_0}{8e (2 c_0 \tilde{c}_3 +\tilde{c}_1)  \bar{g}_{2ql } }}+  \mu_2  \exp\brr{-\frac{1}{2}\br{\frac{\tau - 4c_0 \bar{g}_{ql} -T_0}{4e\tilde{c}_2g_1}}^{1/(1+\chi)}} \notag \\
	&\le \mu_1 \exp\brr{- \frac{\tau }{16e (2 c_0 \tilde{c}_3 +\tilde{c}_1)  \bar{g}_{2ql } }}+  \mu_2  \exp\brr{-\frac{1}{2}\br{\frac{\tau }{8e\tilde{c}_2g_1}}^{1/(1+\chi)}}. 
\end{align}
where, in the second inequality, we impose the condition $\tau\ge 2(4c_0 \bar{g}_{ql} +T_0)$. 

Using the inequality~\eqref{mE_tau_insert_again}, we consider the condition for $\tau$ to satisfy $\varepsilon_2 \le \varepsilon_\ast \Delta_\tc$.
For this purpose, we first consider the condition for $\varepsilon_1 \le \varepsilon_{1,\ast}$, which is satisfied by choosing $\tau$ as 
\begin{align}
	\label{cond_for_epsilon_1}
	\tau \ge 16e (2 c_0 \tilde{c}_3 +\tilde{c}_1)  \bar{g}_{2ql }  \log\br{\frac{2 \mu_1}{\varepsilon_{1,\ast}}}+ 
	8e\tilde{c}_2g_1 \brr{2\log\br{\frac{2 \mu_2}{\varepsilon_{1,\ast}}}}^{1+\chi}  \for \varepsilon_1 \le \varepsilon_{1,\ast} . 
\end{align}
Also, from Eq.~\eqref{def_epsilon_1_eff_H}, the main condition $\varepsilon_2 \le \varepsilon_\ast \Delta_\tc$ implies  
\begin{align}
	\label{cond_for_epsilon_1'}
	\frac{\varepsilon_1}{1-\varepsilon_1}  2q \br{ \tau  +2 c_0\bar{g}_{ql} } \le \varepsilon_\ast \Delta_\tc
	&\longrightarrow \varepsilon_1 \le \frac{\varepsilon_\ast \Delta_\tc }{\varepsilon_\ast\Delta_\tc+ 2q (\tau  +2 c_0\bar{g}_{ql})}\notag \\
	&\longrightarrow \varepsilon_1 \le \frac{\varepsilon_\ast \Delta_\tc }{\varepsilon_\ast\Delta_\tc+ 2q ( \bar{g}_{ql} ql   +2 c_0\bar{g}_{ql})}  \for \tau \le  \bar{g}_{ql} ql  ,
\end{align}
Therefore, by combining~\eqref{cond_for_epsilon_1} and~\eqref{cond_for_epsilon_1'}, we have 
\begin{align}
	\label{cond_for_epsilon_2}
	&\tau \ge 16e (2 c_0 \tilde{c}_3 +\tilde{c}_1)  \bar{g}_{2ql }  \log\brr{2 \mu_1 \br{1+ \frac{2q ( \bar{g}_{ql} ql   +2 c_0\bar{g}_{ql})}{\varepsilon_\ast \Delta_\tc}}} + 
	2^{4+\chi}e\tilde{c}_2g_1 \log^{1+\chi}\brr{2 \mu_2 \br{1+ \frac{2q ( \bar{g}_{ql} ql   +2 c_0\bar{g}_{ql})}{\varepsilon_\ast \Delta_\tc}}}  \notag \\
	&\longrightarrow \tau \ge \msC_1\bar{g}_{2ql }  \log \br{\frac{\msC_2   q^2l  \bar{g}_{ql} }{\varepsilon_\ast\Delta_\tc}}
	+ \msC_3 g_1   \log^{1+\chi} \br{\frac{\msC_4   q^2l  \bar{g}_{ql} }{\varepsilon_\ast\Delta_\tc}} \for \varepsilon_2 \le \varepsilon_\ast \Delta_\tc ,
\end{align}
where $\{\msC_1,\msC_2,\msC_3,\msC_4\}$ are constants of $\orderof{1}$.

Here, the condition~\eqref{cond_for_epsilon_2} includes the condition~\eqref{cond_for_epsilon_1}, and hence we have to choose $\tau$ so that the conditions~\eqref{tau_c_tau_condition} and~\eqref{cond_for_epsilon_2} are satisfied:
\begin{align}
	\label{cond_for_epsilon_summary}
	\msC_1\bar{g}_{2ql }  \log \br{\frac{\msC_2   q^2l  \bar{g}_{ql} }{\varepsilon_\ast\Delta_\tc}}
	+ \msC_3 g_1   \log^{1+\chi} \br{\frac{\msC_4   q^2l  \bar{g}_{ql} }{\varepsilon_\ast\Delta_\tc}} 
	\le   \tau \le  \bar{g}_{ql} ql   
\end{align}
to ensure $\varepsilon_2 \le \varepsilon_\ast \Delta_\tc$.

\subsection{AGSP construction}

We here construct the AGSP based on the Chebyshev polynomials~\cite{arad2013area}. 
We use the following lemma, derived from Supplementary Note 2. H of Ref.~\cite{Kuwahara2020arealaw}:
\begin{lemma} \label{lemm:Chebyshev_AGSP}
	Let $T_m(x)$ be the Chebyshev polynomial as 
	\begin{align}
		T_m(x):=\frac{\left(x+\sqrt{x^2-1}\right)^m + \left(x-\sqrt{x^2-1}\right)^m}{2}. \label{AGSP_Chebyshev_polynomial}
	\end{align}
	Then, the polynomial of 
	\begin{align}
		\label{lemm:Chebyshev_AGSP_construction_AGSP}
		K(m,x) = \frac{T_m \Bigl [ \frac{2x- (\|\tilde{H}_\tc-\tilde{E}_{\tc,0}\| +\tilde{\Delta}_\tc) }{\|\tilde{H}_\tc-\tilde{E}_{\tc,0}\| -\tilde{\Delta}_\tc}\Bigr] }{T_m \Bigl [-\frac{\|\tilde{H}_\tc-\tilde{E}_{\tc,0}\| +\tilde{\Delta}_\tc}{\|\tilde{H}_\tc-\tilde{E}_{\tc,0}\| -\tilde{\Delta}_\tc}\Bigr] } 
	\end{align}
	satisfies the inequality of 
	\begin{align}
		\label{lemm:Chebyshev_AGSP_main}
		\norm{ K(m, \tilde{H}_\tc-\tilde{E}_{\tc,0}) \br{1-\ket{\tilde{\Omega}_\tc}\bra{\tilde{\Omega}_\tc}} } \le 2e^{-2m \sqrt{\tilde{\Delta}_\tc/\|\tilde{H}_\tc-\tilde{E}_{\tc,0}\| }},
	\end{align}
	where $\ket{\tilde{\Omega}_\tc}$ and $\tilde{E}_{\tc,0}$ are the ground state and the ground energy of $\tilde{H}_\tc$, respectively. 
\end{lemma}

{\bf Remark.}
By applying the same analyses to derive the inequality~\eqref{epsilon_H_tc_upp2}, we can derive 
\begin{align}
	\label{norm_bound_tilde_H_tc}
	\norm{ \tilde{H}_\tc-\tilde{E}_{\tc,0} } \le\norm{( \tilde{H}_\tc -\tilde{E}_{\tc,0}) \tilde{\Pi} }  \le 2q \br{ \tau  +2 c_0\bar{g}_{ql} } .
\end{align}
Using the upper bound~\eqref{norm_bound_tilde_H_tc}, we can reduce the inequality~\eqref{lemm:Chebyshev_AGSP_main} to 
\begin{align}
	\label{lemm:Chebyshev_AGSP_main_explicit}
	\norm{ K(m,\tilde{H}_\tc) \br{1-\ket{\tilde{\Omega}_\tc}\bra{\tilde{\Omega}_\tc}} } \le 2e^{-2m \sqrt{\tilde{\Delta}_\tc/ \brr{2q \br{ \tau  +2 c_0\bar{g}_{ql} }  } }}.
\end{align}

\subsection{Schmidt rank of the polynomials of the effective Hamiltonian}

We here consider the Schmidt rank of the power of the truncated Hamiltonian ${\rm SR} (\tilde{H}_\tc^m)$.
For any partition $L \sqcup R$ and the projections $P_L$ and $P_R$, the combined projection $P_L \otimes P_R$ does not change the Schmidt rank between $L$ and $R$. 
On the other hand, we have to distinguish the Schmidt ranks of $(P_L \otimes P_R H P_L \otimes P_R)^m$ and $P_L \otimes P_R H^m P_L \otimes P_R$.
They usually have a different Schmidt rank.
Nevertheless, by carefully following  Supplementary Lemma~8 and Proposition~4 of Ref.~\cite{Kuwahara2020arealaw}, we can recover the following lemma for the effective Hamiltonian $\tilde{\Pi} H_\tc \tilde{\Pi}$ under the choice of $\tilde{\Pi}$ as in Eq.~\eqref{projection_to_spectrum_total}: 
\begin{lemma} [Supplementary Lemma~8 and Proposition~4 in Ref.~\cite{Kuwahara2020arealaw}] \label{thm:Schmit_rank_Ham_power_0}
	The Schmidt rank of the power of the truncated Hamiltonian ${\rm SR} (\tilde{H}_\tc^m)$ is bounded from above by
	\begin{align}
		{\rm SR} (\tilde{H}_\tc^m) \le  \min\brrr{  \brr{2+(2d_{ql}l)^{k}}^{m}, d_{ql}^{ql}(q+m+1)^{q+1}  [ e(q+1)^2 (2d_{ql}l)^{k}]^{m/(q+1)} }
		, \label{ineq:Schmit_rank_Ham_power_0}
	\end{align}
	where $d_x$ is defined as the Hilbert space dimension on the site $x$ as in Eq.~\eqref{def:d_x_local_dimension}. 
\end{lemma}

\textit{Proof of Lemma~\ref{thm:Schmit_rank_Ham_power_0}.}
The proof is exactly the same as that in  Supplementary Lemma~8 and Proposition~4 of Ref.~\cite{Kuwahara2020arealaw}, which is based on Ref.~\cite{arad2013area}.
The only point to consider is the Schmidt rank of the operators $\{h_{s,s+1}\}_{s=0}^{q+1}$. 
These operators are supported on $\tilde{B}_0 \cup \br{ \bigcup_{s=1}^q B_s} \cup \tilde{B}_{q+1}=[-(q/2+1)l,(q/2+1)l]$ (see Supplementary Fig.~\ref{fig:Area_law_Ham_truncate}).
From Eq.~\eqref{def:d_x_local_dimension}, the local Hilbert space is now given by $d_{(q/2+1)l} \le d_{ql}$.
We thus prove the inequality~\eqref{ineq:Schmit_rank_Ham_power_0}. $\square$

\subsection{Quantum state with a small Schmidt rank and a large overlap with the ground state}

We have the ingredients to find a good AGSP operator in the sense that the condition~\eqref{cond:Boot_strapping_lemma} in Lemma~\ref{prop:Overlap between the ground state and low-entangled state} is satisfied. 
Using it, we can prove the existence of a quantum state that has a small Schmidt rank and a large overlap with the ground state: 

\begin{prop}  \label{prop1:truncate_gs_overlap}
	There exists a quantum state $\ket{\phi}$ such that 
	\begin{align}
		\|\ket{\Gs} -\ket{\phi} \| \le \frac{1}{2}
	\end{align}
	with
	\begin{align}
		\label{main:ineq:prop1:truncate_gs_overlap}
		\log \brr{{\rm SR}(\ket{\phi})}\le  c_\ast  \Delta^{-(1+2/\bar{\alpha})(\upsilon+1)} \brr{\log\br{1 /\Delta}}^{4+3/\bar{\alpha} +\chi(1+2/\bar{\alpha}) }  ,
	\end{align}
	where $c_\ast$ is a constant that depends only on system details.
	In particular, for $\upsilon=0$, the upper bound is improved to 
	\begin{align}
		\label{main:ineq:prop1:truncate_gs_overlap_upsilon=0}
		\log \brr{{\rm SR}(\ket{\phi})}\le  c_\ast  \Delta^{-1-2/\bar{\alpha}} \brr{\log\br{1 /\Delta}}^{3+3/\bar{\alpha} +\chi(1+2/\bar{\alpha}) }  \log\log(1/\Delta).
	\end{align}
\end{prop}

\subsubsection{Guide to the parameter trade-offs for the proof of Proposition~\ref{prop1:truncate_gs_overlap}}

We briefly explain the parameter choices in the proof of Proposition~\ref{prop1:truncate_gs_overlap} from a scaling perspective.
The construction involves several truncations followed by an AGSP based on a polynomial of degree $m$, and the parameters are chosen so that each approximation error is reduced to a constant level, while the Schmidt rank of the resulting state grows only moderately.

The boson-number truncation is controlled by $\delta_{\bo}$.
Fixing $\delta_{\bo}=\mathcal{O}(1)$ ensures that the ground state is preserved up to a constant error,
$\norm{\ket{\Omega}-\ket{\bar{\Omega}}}=\mathcal{O}(1)$,
and that the spectral gap remains stable, $\bar{\Delta}=\Theta(\Delta)$.
At the same time, this choice induces effective local dimensions and interaction strengths that grow at most polylogarithmically in the distance to the boundary.
Choosing $\delta_{\bo}$ parametrically smaller would strictly worsen these quantities and, as a consequence, deteriorate the final Schmidt-rank scaling.

The interaction truncation is governed by the block length $l$ and the number of blocks $q$.
The discarded interaction norm scales as
$\norm{\delta H_{\tc}}=\mathcal{O}\br{q\,\bar g_{ql}\,l^{-\bar{\alpha}}}$,
where the effective interaction strength on the relevant scale satisfies
$\bar g_{ql}=\mathcal{O}\br{\Delta^{-\upsilon}\log^\chi(ql)}$.
Thus, taking
\begin{align}
	l=\mathcal{O}\br{\br{\frac{q\log^\chi(q/\Delta)}{\Delta^{\upsilon+1}}}^{1/\bar{\alpha}}}
\end{align}
is sufficient to ensure $\norm{\delta H_{\tc}}=\mathcal{O}(\Delta)$ and hence $\Delta_{\tc}=\Theta(\Delta)$.
While increasing $l$ further would indeed suppress the interaction truncation error, it would simultaneously enlarge the effective operator support and worsen the Schmidt-rank estimates.
The above choice, therefore, represents a minimal block length that achieves the required accuracy without unnecessarily inflating entanglement.

The multi-energy cut-off introduces the parameter $\tau$.
The corresponding approximation error decays subexponentially as
$\varepsilon_2=\mathcal{E}_{\tau-\mathcal{O}(\bar g_{ql})}$,
so choosing
$\tau=\mathcal{O}\br{\Delta^{-\upsilon}\log^{1+\chi}(q/\Delta)}$
is sufficient to make $\varepsilon_2=\mathcal{O}(\Delta)$.
This guarantees that both the overlap
$\norm{\ket{\Omega_{\tc}}-\ket{\tilde{\Omega}_{\tc}}}$
and the effective gap $\tilde{\Delta}_{\tc}$ remain stable, while keeping the spectral radius
$\norm{\tilde H_{\tc}-\tilde E_{\tc,0}}=\mathcal{O}\br{q(\tau+\bar g_{ql})}$
within a range that can still be resolved efficiently by the Chebyshev filter.

Finally, the remaining parameters $q$ and the polynomial degree $m$ are fixed by balancing the AGSP suppression against the growth of the Schmidt rank.
From the Chebyshev construction, the AGSP error satisfies
\begin{align}
	\epsilon_K  = \mathcal{O}\br{\exp\brr{- m \sqrt{\frac{\Delta^{\upsilon+1}}{q \log^{1+\chi}\br{q/\Delta}}}}},
\end{align}
which decays exponentially in $m$ for fixed $q$.
On the other hand, Lemma~\ref{thm:Schmit_rank_Ham_power_0} yields the Schmidt-rank bound
\begin{align}
	\log\mathcal{D}_K = \mathcal{O}\br{\br{q l}\, \log d_{ql} + \frac{m}{q}\, \log\br{d_{ql} l}}.
\end{align}
Using the previously fixed choice of $l$ and the corresponding bounds on the effective local dimension,
the above expression reduces to
\begin{align}
	\log\mathcal{D}_K = \mathcal{O}\br{q \br{\frac{q \log^\chi(q/\Delta)}{\Delta^{\upsilon+1}}}^{1/\bar{\alpha}} \log(q/\Delta) + \frac{m}{q}\log(q/\Delta)}.
\end{align}
Here the dependence on the local Hilbert space enters only through the worst-case block dimension $d_{ql}$,
which grows at most polylogarithmically in $q/\Delta$, as established in the boson-number truncation step.

Thus, after fixing $l$, the Schmidt-rank growth separates into a contribution that is independent of $m$
and a contribution that grows linearly in $m$ but is suppressed by a factor $1/q$.
The bootstrapping argument requires that the AGSP error be chosen so that
$\epsilon_K \sqrt{\mathcal{D}_K}$ and $\epsilon_K^2 \mathcal{D}_K$
are bounded by fixed constants, for instance by $1/2$ and $1/8$, respectively.
To handle these constraints simultaneously, we impose the sufficient condition
$\epsilon_K \mathcal{D}_K \le \epsilon_K^{1/2}$,
which guarantees both bounds once $\epsilon_K$ is sufficiently small
and does not affect the leading asymptotic scaling.

Intuitively, the parameters $q$ and $m$ play complementary roles in the AGSP construction.
Increasing $q$ suppresses the $m$-dependent contribution to $\log\mathcal{D}_K$ by a factor $1/q$,
but simultaneously enlarges the $m$-independent contribution.
The choice of $q$ therefore balances the growth of these two terms.
Once $q$ is fixed, the remaining freedom lies in the polynomial degree $m$,
which controls the strength of the exponential suppression provided by the Chebyshev filter.
This condition fixes the scale of $q$ by requiring that the exponential decay of $\epsilon_K$
dominates the $m$-independent contribution to $\log\mathcal{D}_K$,
leading to the choice $q = \mathcal{O}\br{\Delta^{-(\upsilon+1)} \log^{3+\chi}\br{1/\Delta}}$.
For this choice of $q$, the polynomial degree $m$ is then taken sufficiently large so that
the exponential suppression in $\epsilon_K$ compensates for the remaining
$m$-dependent term $\frac{m}{q}\log(q/\Delta)$ in $\log\mathcal{D}_K$.
This determines the scaling of $m$ and yields the Schmidt-rank bound stated in
Proposition~\ref{prop1:truncate_gs_overlap}.

\subsubsection{Proof of Proposition~\ref{prop1:truncate_gs_overlap}} \label{Sec:Proof of Proposition_prop1_truncate_gs_overlap}
We begin with choosing the parameter $\delta_\bo$ in Proposition~\ref{Prop:eff:boson_number_truncation} as 
\begin{align}
	\delta_\bo=\frac{1}{8} ,
\end{align}
which gives 
\begin{align}
	\label{error_bar_Omega_cal}
	&\norm{ \ket{\Omega } - \ket{\bar{\Omega} }} \le \frac{1}{8},  \quad \bar{\Delta} \ge \frac{3}{4}\Delta ,\quad  \epsilon_\bo = \frac{w_0  \Delta }{64 \log^{\ma k/2} \br{8}  }  .
\end{align}
Under the above choice, the parameters $d_0$ and $g_0$ in Eqs.~\eqref{def:d_x_local_dimension} and \eqref{def:g_x_local_energy}, respectively, are now given by 
\begin{align}
	\label{d_0_g_0_propto}
	&d_0\propto \Delta^{-2\upsilon/k} \log^\ma (1/\Delta),\quad d_1\propto \Delta^{-2\upsilon/k} , \notag\\
	&g_0\propto \Delta^{-\upsilon} \log^{\chi} (1/\Delta) , \quad g_1\propto \Delta^{-\upsilon}  ,
\end{align}
where we use the relation~\eqref{Delta_dependence_mb} for $\mb$, i.e., $\mb \propto \Delta^{2\upsilon/(\ma k)} $.

We next choose the number of blocks $q$ such that 
\begin{align}
	\norm{\delta H_\tc} \le \frac{\Delta}{16}
	\quad  \longrightarrow \quad  \Delta_\tc \ge \bar{\Delta} - \frac{\Delta}{8} \ge \frac{5}{8} \Delta  ,\label{condition_delta_H_t}
\end{align}
where the first inequality yields the second one because of the inequality \eqref{lemma:truncate_ineq_gap}, i.e., $\Delta_\tc\ge \bar \Delta - 2\norm{\delta H_\tc}$.
Using Proposition~\ref{lemm:error_int_truncation} and the inequality~\eqref{cond_r^2_1_decay}, the above condition is satisfied 
\begin{align}
	& 4 \eta_1  \eta_2 q \bar{g}_{ql}\br{l^2+1} \bar{J}(l) \le   \frac{4 \eta_1  \eta_2 q  \bar{g}_{ql}}{l^{\bar{\alpha}}+1} \le \frac{\Delta}{16}  \notag \\
	& \longrightarrow l^{\bar{\alpha}}+1 \ge (64 \eta_1  \eta_2) \frac{q  \bar{g}_{ql}}{\Delta} = (64 \eta_1  \eta_2) \frac{q\brr{g_0 + g_1\log^{\chi}(ql+1)}}{\Delta} \notag \\
	& \longrightarrow l = w_1 \brr{\frac{q\log^\chi(q/\Delta)}{\Delta^{\upsilon+1}}}^{1/\bar{\alpha}}  , 
	\label{choice_of_the_length_l}
\end{align}
where $w_1$ is a constant of $\orderof{1}$ and we use~\eqref{d_0_g_0_propto}. 
Under this choice, we also obtain 
\begin{align}
	\label{bar_g_ql_upper_bound}
	\bar{g}_{ql} \le \frac{\Delta}{64 \eta_1  \eta_2} \frac{l^{\bar{\alpha}}+1}{q}\le  w_2 \Delta^{-\upsilon} \log^\chi(q/\Delta) 
\end{align}
with $w_2=\orderof{1}$. 
This choice of $q$ gives the following inequality from the inequality~\eqref{overlap_Gs_phi_norm}: 
\begin{align}
	&\norm{ \ket{\bar{\Gs}}-\ket{\phi} } \le  \norm{ \ket{\Gs_\tc}-\ket{\phi} }+\frac{1}{8}  \notag \\
	&\longrightarrow \norm{ \ket{\Gs}-\ket{\phi} }\le  \| \ket{\Gs_\tc}-\ket{\phi} \| +  \norm{ \ket{\Gs}-\ket{\bar{\Gs}}}  + \frac{1}{8} \le 
	\| \ket{\Gs_\tc}-\ket{\phi} \| + \frac{1}{4} 
	\label{Gs_phi_distance_ineq_4/15}
\end{align}
for an arbitrary quantum state $\ket{\phi}$, where we use the inequality~\eqref{error_bar_Omega_cal}. 

Next, from Lemma~\ref{prop:Overlap between the ground state and low-entangled state}, if an operator $K$ satisfies the AGSP condition such that 
\begin{align}
	&\epsilon_{K}^2 \mD_{K} \le  \frac{1}{2}, \label{boot_strapping_cond_prop5} 
\end{align}
we can find a quantum state $\ket{\psi}$ satisfying
\begin{align}
	&\| \ket{\Gs}-\ket{\psi} \|  \le \epsilon_{K} \sqrt{2 \mD_{K}} + \delta_{K} \quad {\rm with} \quad {\rm SR}(\ket{\psi}) \le \mD_{K},
\end{align}
where the parameters $\{\delta_{K}, \epsilon_{K},\mD_{K}\}$ are defined in Eq.~\eqref{Def_AGSP_error_chap2}. 
Hence, we aim to prove the existence of the AGSP operator such that 
\begin{align}
	&\epsilon_{K} \sqrt{2 \mD_{K}} + \delta_{K} \le  \frac{1}{2} , \label{trace_distance_cond_prop5}\\
	& \log(\mD_{K}) \le c_\ast    \Delta^{-(1+2/\bar{\alpha})(\upsilon+1)}  \brr{\log\br{1 /\Delta}}^{4+3/\bar{\alpha} +\chi(1+2/\bar{\alpha})}  , \label{Schmidt_rank_cond_prop5}
\end{align}
which proves the Proposition~\ref{prop1:truncate_gs_overlap} by replacing $\ket{\phi}$ with $\ket{\psi}$  in the inequality~\eqref{Gs_phi_distance_ineq_4/15}.

In the construction of the AGSP operator, we utilize Eq.~\eqref{lemm:Chebyshev_AGSP_construction_AGSP}, which is based on the Chebyshev polynomials with respect to the effective Hamiltonian $\tilde{H}_\tc$ from the truncated Hamiltonian $H_\tc$. 
Here, the error is estimated by the inequality~\eqref{lemm:Chebyshev_AGSP_main_explicit}.
We adopt the operator $K(m,\tilde{H}_\tc - \tilde{E}_{\tc,0})$ as the AGSP operator $K$ for the ground state $\ket{\Gs}$, and in the following, we aim to estimate the AGSP parameters $(\delta_{K}, \epsilon_{K}, \mD_{K})$ as defined in \eqref{Def_AGSP_error_chap2}.

First, we choose such that $\delta_{K}\le 3/8$.
Here, $\delta_{K}$ is estimated as 
\begin{align}
	&\delta_{K}= \| \ket{\Gs}-\ket{\tilde{\Gs}_\tc} \|  \le  \norm{ \ket{\Gs}- \ket{\Gs_\tc}}+   \norm{ \ket{\Gs_\tc} -\ket{\tilde{\Gs}_\tc} } \le 
	\frac{1}{4}  +  \norm{ \ket{\Gs_\tc} -\ket{\tilde{\Gs}_\tc} }  ,
\end{align}
where, in the second inequality, we use~\eqref{Gs_phi_distance_ineq_4/15} with $\ket{\phi} \to \ket{\Gs_\tc}$.
For the norm of $ \norm{ \ket{\Gs_\tc} -\ket{\tilde{\Gs}_\tc} } $, we obtain from the inequality~\eqref{thm:effective_global/main1_eff} in Theorem~\ref{Effective Hamiltonian_multi_truncation}: 
\begin{align}
	&\norm{ \ket{\Omega_\tc } - \ket{\tilde{\Omega}_\tc}} \le \sqrt{\frac{2\varepsilon_2}{\Delta_\tc}} .
\end{align} 
Hence, the condition $\delta_{K}\le 3/8$ is ensured by 
\begin{align}
	\sqrt{\frac{2\varepsilon_2}{\Delta_\tc}} \le \frac{1}{8}
	\longrightarrow  \varepsilon_2\le \frac{\Delta_\tc}{128}
	&\longrightarrow \tau \ge \msC'_1\bar{g}_{2ql }  \log \br{\frac{\msC'_2   q^2l  \bar{g}_{ql} }{\Delta}}
	+ \msC'_3  g_1  \log^{1+\chi} \br{\frac{\msC'_4   q^2l  \bar{g}_{ql} }{\Delta}} \notag \\
	&\longrightarrow \tau \ge w_3  \Delta^{-\upsilon}\log^{1+\chi}\br{q/\Delta}
	\label{upper_bound_tau_w3}
\end{align} 
for appropriate choices of $\orderof{1}$ constants $\{\msC'_1,\msC'_2,\msC'_3,\msC'_4\}$ and $w_3$, where, in the third inequality, we use the inequalities~\eqref{cond_for_epsilon_summary} with $\epsilon_\ast \to 1/128$ and $\Delta_\tc \ge 5\Delta/8$ as in \eqref{condition_delta_H_t}, and in the last inequality, we use the upper bound~\eqref{bar_g_ql_upper_bound}. 
Under the above choice, we can also ensure
\begin{align}
	\label{thm:effective_global_gap_eff_apply}
	\tilde{\Delta}_\tc \ge \Delta_\tc - 3 \varepsilon_2  \ge 0.9 \Delta_\tc \ge \frac{1}{2} \Delta ,
\end{align} 
where we use the upper bound~\eqref{thm:effective_global_gap_eff}.

Second, from the inequality~\eqref{lemm:Chebyshev_AGSP_main_explicit}, we obtain
\begin{align}
	\epsilon_{K}  \le 2e^{-2m \sqrt{\tilde{\Delta}_\tc/ \brr{2q \br{ \tau  +2 c_0\bar{g}_{ql} }  } }}
	&\le 2\exp \left(-2m \sqrt{\frac{\Delta}{4q \br{ \tau  +2 c_0\bar{g}_{ql} }}} \right) \notag \\
	&\le \exp \left(-w_4 m \sqrt{\frac{\Delta^{\upsilon+1}}{q \log^{1+\chi}\br{q /\Delta}}} \right) 
	, \label{error_k_m_poly_AGSP_prop_area_1}
\end{align}
where we use the upper bounds~\eqref{upper_bound_tau_w3} and~\eqref{thm:effective_global_gap_eff_apply}, and $w_4$ is an $\orderof{1}$ constant. 
Third, from Lemma~\ref{thm:Schmit_rank_Ham_power_0}, we obtain
\begin{align}
	\mD_{K}={\rm SR} [K(m,\tilde{H}_\tc - \tilde{E}_{\tc,0})] 
	&\le   d_{ql}^{ql}(q+m+1)^{q+1}  [ e(q+1)^2 (2d_{ql}l)^{k}]^{m/(q+1)} \notag \\
	&\le  d_{ql}^{2ql}[ e(q+1)^2 (2d_{ql}l)^{k}]^{m/(q+1)} \notag \\
	&\le \exp\brr{ w_5 \log(q/\Delta) q  \brr{\frac{q\log^\chi(q/\Delta)}{\Delta^{\upsilon+1}}}^{1/\bar{\alpha}} 
		+ \frac{w_6 m}{q} \log (q/\Delta) }   
	\label{D_K_tc_Schmidt_rank}
\end{align}
by assuming $(q+m+1)^{q+1} \le d_{ql}^{ql}$, where we use $d_0\propto \Delta^{-2\upsilon/k} \log^\ma (1/\Delta)$, $d_1\propto \Delta^{-2\upsilon/k}$ in \eqref{d_0_g_0_propto} and 
\begin{align}
	\log(d_{ql} l) = \log(d_{ql}) + \log(l) &\le    \log\brr{d_0+ d_1 \log^\ma(ql+1)} +   \log\brrr{w_1 \brr{\frac{q\log^\chi(q/\Delta)}{\Delta^{\upsilon+1}}}^{1/\bar{\alpha}}}  \notag \\
	&\le {\rm const.}  \log (q/\Delta) .
\end{align}
Note that we ensure the condition $(q+m+1)^{q+1} \le d_{ql}^{ql}$ as long as $\log(m) \lesssim l \propto (q/\Delta^{\upsilon+1})^{1/\bar{\alpha}}$, which can be indeed satisfied under the choice of Eq.~\eqref{choice_of_m}.

Then, for the inequality~\eqref{trace_distance_cond_prop5} to be satisfied, we need to choose $m$ and $q$ such that
\begin{align}
	\epsilon_{K} \sqrt{2\mD_{K}}\le \frac{1}{8} .   \label{proof_two_cond_prop5_reduce}
\end{align}
In the following, we choose $m$ and $q$ as such $\epsilon_{K} \mD_{K} \le \epsilon_{K}^{1/2}$, which reduces the conditions~\eqref{boot_strapping_cond_prop5} and \eqref{proof_two_cond_prop5_reduce} to
\begin{align}
	&\epsilon_{K}^2 \mD_{K}\le \epsilon_{K}^{3/2} \le  \frac{1}{2}, \quad \quad 
	\epsilon_{K} \sqrt{2\mD_{K} }  \le \sqrt{2} \epsilon_{K}^{3/4} \le \frac{1}{8}  .  \label{proof_two_cond_prop5}
\end{align}
From the inequalities~\eqref{error_k_m_poly_AGSP_prop_area_1} and \eqref{D_K_tc_Schmidt_rank}, 
the condition $\epsilon_{K} \mD_{K} \le \epsilon_{K}^{1/2}$ is satisfied for 
\begin{align}
	&w_5 \log(q/\Delta)q  \brr{\frac{q\log^\chi(q/\Delta)}{\Delta^{\upsilon+1}}}^{1/\bar{\alpha}}  \le \frac{w_4 m }{4} \sqrt{\frac{\Delta^{\upsilon+1}}{q \log^{1+\chi}\br{q /\Delta}}}  
	,\notag \\
	&\frac{w_6 m}{q} \log (q/\Delta)   \le \frac{w_4 m }{4} \sqrt{\frac{\Delta^{\upsilon+1}}{q \log^{1+\chi}\br{q /\Delta}}}  . \label{condition_m_l_AGSP}
\end{align}

The first inequality in \eqref{condition_m_l_AGSP} gives the lower bound of $m$ as follows:
\begin{align}
	&m \ge \frac{4w_5}{w_4} \Delta^{-(1/2+1/\bar{\alpha})(\upsilon+1)} q^{3/2 +1/\bar{\alpha} } \brr{\log\br{q /\Delta}}^{\chi/\bar{\alpha}+(1+\chi)/2+1} \notag \\
	&\longrightarrow m \ge w_7  \Delta^{-(1/2+1/\bar{\alpha})(\upsilon+1)} q^{3/2 +1/\bar{\alpha} }\brr{\log\br{q /\Delta}}^{\chi/\bar{\alpha}+(1+\chi)/2+1},
	\label{condition_m_l_AGSP_first_one}
\end{align} 
where $w_7=4w_5/w_4=\orderof{1}$. 
The second one in~\eqref{condition_m_l_AGSP} implies 
\begin{align}
	&\frac{q^{1/2}}{\log^{3/2+\chi/2} (q/\Delta) }
	\ge \frac{4w_6}{w_4} \frac{1}{\sqrt{\Delta^{\upsilon+1}}} \notag \\
	&\longrightarrow  \frac{q}{\log^{3+\chi} (q/\Delta) }
	\ge \br{\frac{4w_6}{w_4} }^2 \frac{1}{\Delta^{\upsilon+1}}\longrightarrow  q  \ge \frac{w_8}{\Delta^{\upsilon+1}} \log^{3+\chi} (1/\Delta).
\end{align} 
We thus choose 
$
q = \ceil{w_8 (1/\Delta^{\upsilon+1}) \log^{3+\chi} (1/\Delta)}, 
$
and then the parameters $\epsilon_{K}$ decays exponentially with respect to $m$. 
Therefore, by choosing the following constant $w_9$ appropriately, the choice of 
\begin{align}
	\label{choice_of_m}
	&m= w_9 \Delta^{-(2+2/\bar{\alpha})(\upsilon+1)} \brr{\log\br{1 /\Delta}}^{3/\bar{\alpha}+6+2\chi(\bar{\alpha}+1)/\bar{\alpha}}  
\end{align} 
satisfies~\eqref{proof_two_cond_prop5} and \eqref{condition_m_l_AGSP_first_one}.

Finally, under the above choices, we can upper-bound the Schmidt rank $\mD_{K}$ from the inequality~\eqref{D_K_tc_Schmidt_rank} as 
\begin{align}
	\log (\mD_{K}) \le w_{10}  \Delta^{-(1+2/\bar{\alpha})(\upsilon+1)}  \brr{\log\br{1 /\Delta}}^{4+3/\bar{\alpha} +\chi(1+2/\bar{\alpha})}  ,
\end{align}
where $w_{10}$ is a constant of $\orderof{1}$.
We thus obtain the inequality~\eqref{Schmidt_rank_cond_prop5}. 

For $\upsilon=0$, we have $d_0\propto \log^\ma (1/\Delta)$, $d_1\propto 1$, and hence $\log(d_{ql}) \lesssim \log\log(1/\Delta)$.
This point improves the inequality~\eqref{D_K_tc_Schmidt_rank} to 
\begin{align}
	\mD_{K}={\rm SR} [K(m,\tilde{H}_\tc)] 
	\le \exp\brr{ w_5 \log\log(1/\Delta) q  \brr{\frac{q\log^\chi(q/\Delta)}{\Delta^{\upsilon+1}}}^{1/\bar{\alpha}} 
		+ \frac{w_6 m}{q} \log (q/\Delta) }   .
\end{align}
We then follow the same calculation and eventually prove the upper bound~\eqref{main:ineq:prop1:truncate_gs_overlap_upsilon=0}.

This completes the proof of Proposition~\ref{prop1:truncate_gs_overlap}.   $\square$

\subsection{Completing the proof }

We now have all the ingredients to prove the main theorem. Based on the quantum state constructed in Proposition~\ref{prop1:truncate_gs_overlap}, we approximate the ground state with an arbitrary accuracy while controlling the Schmidt rank, which allows us to prove an upper bound for the entanglement entropy.

\begin{prop} \label{prop0:overlap_AGSP_entropy_bound}
	Let $\ket{\phi}$ be an arbitrary quantum state such that 
	\begin{align}
		\|\ket{\Gs} - \ket{\phi} \| \le \frac{1}{2} 
		\label{def_phi_Prop_theorem_main}
	\end{align}
	with $\mD_\phi := {\rm SR}(\ket{\phi})$. 
	Then, there exists a quantum state $\ket{\psi}$ approximating the ground state $\ket{\Gs}$ such that
	\begin{align}
		\|\ket{\Gs} - \ket{\psi} \| \le \delta,
	\end{align}
	and the Schmidt rank of $\ket{\psi}$ satisfies
	\begin{align}
		\log [{\rm SR}(\ket{\psi})] \leq \log (\mD_\phi) + \tilde{c}_\ast \frac{\log^{\chi/2+5/2}\brr{1/(\delta \Delta)}}{\Delta^{(\upsilon+1)/2}}.
		\label{schmidt_rank_error_delta_SR}
	\end{align}
	Furthermore, the entanglement entropy $S(\ket{\Gs})$ is bounded from above by
	\begin{align}
		S(\ket{\Gs}) \le\log (\mD_\phi) +  \tilde{c}'_\ast \frac{\log^{\chi/2+5/2}\br{1/\Delta}}{\Delta^{(\upsilon+1)/2}} . 
		\label{area_law_phi_Gs_d}
	\end{align}
	Here, $\tilde{c}_\ast$ and $\tilde{c}'_\ast$ are $\orderof{1}$ constants. 
\end{prop}

By applying Proposition~\ref{prop1:truncate_gs_overlap} to Proposition~\ref{prop0:overlap_AGSP_entropy_bound}, we immediately prove Theorem~\ref{thm:main_theorem_area_law}. 
Here, $\log (\mD_\phi)$ corresponds to the RHS of the inequality~\eqref{main:ineq:prop1:truncate_gs_overlap}, which is larger than  $\tilde{c}'_\ast \log^{\chi/2+5/2}\br{1/\Delta}/\sqrt{\Delta}$ in the inequality~\eqref{area_law_phi_Gs_d} for $\Delta \ll 1$. 
We thus prove the first main inequality~\eqref{thm:main_theorem_area_law/main_ineq}.  

To prove the second main inequality~\eqref{MPS_approx_Area_law}, we follow the same approach as in Ref.~\cite{Kuwahara2020arealaw}.
The proof relies on the following two lemmas.
The first lemma connects the MPS approximation and the truncation error of the Schmidt rank: 
\begin{lemma}[Lemma~1 in Ref.~\cite{PhysRevB.73.094423}] \label{lemma_Verstraete_Cirac}
	We here label the total system $\Lambda$ as $\Lambda=\{1,2,\ldots,n\}$.
	Then, let $\ket{\psi}$ be an arbitrary quantum state, decomposed as follows between the subsets $\{1,2,\ldots, x\}$ and $\{x+1,x+2,\ldots,n\}$:
	\begin{align}
		\label{psi_def_schmidt}
		\ket{\psi} = \sum_{m=1}^\infty \mu_m^{(x)} \ket{\psi_{\leq x,m}} \otimes \ket{\psi_{> x,m}},
	\end{align}
	where $\{\mu_m^{(x)}\}_{m=1}^\infty$ are the Schmidt coefficients in descending order.
	Then, there exists an MPS approximation $\ket{{\rm M}_{\psi,\mD}}$ with bond dimension $\mD$\footnote{To distinguish the notation of the spatial dimension $D$, we adopt the cursive notation $\mD$ for the bond dimension.}, approximating $\ket{\psi}$ such that:
	\begin{align}
		\label{main_lemma_Verstraete_Cirac_0}
		\| \ket{\psi} - \ket{{\rm M}_{\psi,\mD}} \| \leq 2 \sum_{x=1}^{n-1} \delta_{x,\mD}, \quad \delta_{x,\mD} := \sum_{m > \mD} |\mu_m^{(x)}|^2.
	\end{align}
	Moreover, the inequality is generalized as follows:
	\begin{align}
		\label{main_lemma_Verstraete_Cirac}
		\norm{ \tr_{X^\co} \br{ \ket{\psi}\bra{\psi} - \ket{{\rm M}_{\psi,\mD}} \bra{{\rm M}_{\psi,\mD}} } }_1 \leq 2 \sum_{x=x_0-1}^{x_1} \delta_{x,\mD},
	\end{align}
	where the subset $X=[x_0,x_1]$ ($x_0 \le x_1$) can arbitrarily chosen.
\end{lemma}

From the above lemma, we aim to derive the upper bound on the error $\delta_{x,\mD}$ by the Schmidt rank truncation for the ground state $\ket{\Omega}$.
For this purpose, we utilize the Eckart-Young theorem:
\begin{lemma}[Eckart-Young theorem~\cite{Eckart1936}] \label{lemma:Eckart-Young theorem}
	Given a normalized state $\ket{\psi}$ as in Eq.~\eqref{psi_def_schmidt}, for any quantum state $\ket{\psi'}$, the following inequality holds:
	\begin{align}
		\sum_{m > {\rm SR}(\ket{\psi'})} |\mu_m^{(i)}|^2 \leq \| \ket{\psi} - \ket{\psi'}\|^2,
		\label{main_lemma:Eckart-Young theorem}
	\end{align}
	where ${\rm SR}(\ket{\psi'})$ is defined for the decomposition of $\{1,2,\ldots, i\}$ and $\{i+1,i+2,\ldots,n\}$.
\end{lemma}

To apply the Eckart-Young theorem, we use the inequality~\eqref{schmidt_rank_error_delta_SR}, which ensures the existence of a quantum state $\ket{\psi_\mD}$ such that 
\begin{align}
	\|\ket{\Gs} - \ket{\psi_\mD} \| \le \delta,
\end{align}
and the Schmidt rank of $\ket{\psi_\mD}$ satisfies
\begin{align}
	\log(\mD) \leq \log (\mD_\phi) + \tilde{c}_\ast \frac{\log^{\chi/2+5/2}\brr{1/(\delta \Delta)}}{\Delta^{(\upsilon+1)/2}}.
\end{align}
This implies that by choosing 
\begin{align}
	\mD_0= \exp\brrr{ C_1 \Delta^{-(1+2/\bar{\alpha})(\upsilon+1)} \brr{\log\br{1 /\Delta}}^{4+3/\bar{\alpha} +\chi(1+2/\bar{\alpha}) } +C_2 \frac{\log^{\chi/2+5/2}\brr{1/(\delta \Delta)}}{\Delta^{(\upsilon+1)/2}} }, 
\end{align}
we can ensure $\delta_{x,\mD_0} \le \delta^2$ for an arbitrary bi-partition of the ground state. 
By applying it to the inequality~\eqref{main_lemma_Verstraete_Cirac}, we prove the second main inequality~\eqref{MPS_approx_Area_law}, where we use $|X|=x_1-x_0+1$ from $X=[x_0,x_1]$.

This completes the proof. $\square$

\subsubsection{Preliminary lemma}

We first introduce the following proposition that relates the AGSP operators and the entanglement entropy:
\begin{lemma}[Supplementary Proposition~3 in Ref.~\cite{Kuwahara2020arealaw}] \label{prop:entropy_and_AGSP}
	Let $\{K_p\}_{p=1}^\infty$ be a set of AGSP operators.
	such that the errors $\epsilon_{K_p}$ and $\delta_{K_p}$ decrease with the index $p$ and goes to zero in the limit of $p\to \infty$, i.e., $\epsilon_{K_\infty}=0$, $\delta_{K_\infty}=0$. That is, the operator $K_\infty$ is the exact ground-state projector.
	Also, we set $\ket{\psi_\mD}$ be an arbitrary quantum state with 
	\begin{align}
		\| \ket{\psi_\mD} -\ket{\Gs}\|= \nu_0 \quad {\rm and} \quad  {\rm SR}(\ket{\psi_\mD})=\mD. 
		\label{def:nu_0_psiD}
	\end{align}
	Then, we prove for each of $\{K_p\}_{p=1}^{\infty}$
	\begin{align}
		\left \| \frac{ K_p e^{-i\theta_p}\ket{\psi_\mD} }{\| K_p \ket{\psi_\mD}\|} - \ket{\Gs} \right\|  \le \Gamma_{K_p} 
		\label{norm_distance_AGSP_pth}
	\end{align}
	with $\theta_p \in \mathbb{R}$ appropriately chosen, where $\{\Gamma_{K_p}\}_{p=1}^\infty$ are defined as 
	\begin{align}
		\Gamma_{K_p} := \frac{\epsilon_{K_p}}{1- \nu_0- \delta_{K_p}} + \delta_{K_p}. \label{Def:gamma_p_bar}
	\end{align}
	Moreover, under the condition $\Gamma_{K_p}\le 1$ for all $p$, the entanglement entropy $S(\ket{\Gs})$ is bounded from above by
	\begin{align}
		S(\ket{\Gs}) &\le \log (\mD) -  \sum_{p=0}^\infty \Gamma_{K_p}^2  \log  \frac{\Gamma_{K_p}^2 }{3\mD_{K_{p+1}}} , \label{Basic_inequality_for_entropy_bound_fin}
	\end{align}
	where we set $\Gamma_{K_0}:=1$.
\end{lemma}

\subsubsection{Proof of Proposition~\ref{prop0:overlap_AGSP_entropy_bound}}
The proof is based on Lemma~\ref{prop:entropy_and_AGSP}. 
We set $q=2$ and construct the AGSP operator for $\ket{\Gs}$ by using $K(m,\tilde{H}_\tc)$ as in the proof of  Proposition~\ref{prop1:truncate_gs_overlap}. 
Then, the AGSP parameters $\{\delta_K, \epsilon_K, D_K \}$ depends on the parameters $\epsilon_\bo$, $l$, $m$ and $\tau$. 
We adopt the state $\ket{\phi}$ in Proposition~\ref{prop1:truncate_gs_overlap} as the quantum state $\ket{\psi_\mD}$ in Eq.~\eqref{def:nu_0_psiD}. Then, we can ensure 
\begin{align}
	\nu_0 \le \frac{1}{2},
\end{align}
which reduces the inequality~\eqref{Basic_inequality_for_entropy_bound_fin} to 
\begin{align}
	S(\ket{\Gs}) &\le \log (\mD_\phi) -  \sum_{p=0}^\infty \Gamma_{K_p}^2  \log  \frac{\Gamma_{K_p}^2 }{3\mD_{K_{p+1}}} , \label{ground_state_entanglement_Gs}
\end{align}
where we denote $\mD_\phi={\rm SR}(\ket{\phi})$. 

In the following, we consider a set of AGSP $\{K_p\}_{p=1}^\infty$ such that 
\begin{align}
	\label{Gamma_K_p_condition}
	\Gamma_{K_p} \le \frac{1}{p},
\end{align}
and estimating the Schmidt rank $\mD_{K_p}$ to achieve it. 
The condition~\eqref{Gamma_K_p_condition} is satisfied when 
\begin{align}
	\delta_{K_p} \le \frac{1}{3p}, \quad \frac{\epsilon_{K_p}}{1 - \nu_0 - \delta_{K_p}} \le \frac{2}{3p} \quad \text{or} \quad \epsilon_{K_p} \le \frac{1}{9p},
	\label{choice_of_delta_epsilon_s}
\end{align}
where the second inequality is derived from
\begin{align}
	\frac{\epsilon_{K_p}}{1 - \nu_0 - \delta_{K_p}} \le \frac{\epsilon_{K_p}}{\frac{1}{2} - \delta_{K_p}} \le \frac{\epsilon_{K_p}}{\frac{1}{2} - \frac{1}{3}} = 6\epsilon_{K_p} \le \frac{2}{3p}.
	\label{choice_of_delta_epsilon_s_2}
\end{align}
Here, we use $\nu_0 \le \frac{1}{2}$.

We begin with the parameter $\epsilon_\bo$ in Proposition~\ref{Prop:eff:boson_number_truncation} and choose it as 
\begin{align}
	\label{epsilon_0_choice_delta_0_rre}
	\epsilon_\bo = w_0 \delta_\bo^2  \Delta \log^{-\ma k/2} \br{\delta_\bo^{-1}} ,
\end{align} 
which gives 
\begin{align}
	\label{bar_Delta_ge_3/4_Delta}
	\norm{ \ket{\Omega } - \ket{\bar{\Omega} }} \le \delta_\bo, 
	\quad \bar{\Delta} \ge \Delta \br{1 - \frac{3\delta_\bo^2}{2}} 
\end{align}
from the inequality~\eqref{norm_Omega_minus_bar_Omega}. 
Then, for 
\begin{align}
	\delta_\bo= \frac{1}{9p}, 
\end{align}
we have from Eq.~\eqref{epsilon_0_choice_delta_0}
\begin{align}
	\epsilon_\bo = \frac{w_0}{81p^2\log^{\ma k/2} \br{9p}}  \Delta  ,\quad \bar{\Delta}\ge \Delta \br{1 - \frac{1}{54p^2}} \ge 
	\frac{3}{4} \Delta.
\end{align} 
The above choice provides the parameters $d_0,d_1$ and $g_0,g_1$ in a similar way to~\eqref{d_0_g_0_propto} as follows: 
\begin{align}
	\label{d_0_g_0_propto2}
	&d_0\propto \Delta^{-2\upsilon/k} \log^\ma (p/\Delta),\quad d_1\propto \Delta^{-2\upsilon/k} , \notag\\
	&g_0\propto \Delta^{-\upsilon} \log^{\chi} (p/\Delta) , \quad g_1\propto \Delta^{-\upsilon}  ,
\end{align}
where we use Eqs.~\eqref{def:d_x_local_dimension} and \eqref{def:g_x_local_energy}.

Second, from the inequality~\eqref{lemma:truncate__ineq} with $q=2$, we get the upper bound of $\norm{\delta H_\tc}$ as:
\begin{align}
	\norm{\delta H_\tc} \le 8\eta_1\eta_2 \bar{g}_{2l} (l^2+1) \bar{J}(l) \le \frac{8\eta_1\eta_2 \bar{g}_{2l}}{l^{\bar{\alpha}}+1} ,
	\label{lemma:truncate__ineq_prop_area_law}
\end{align}
where we use the condition~\eqref{cond_r^2_1_decay}.
From the inequalities~\eqref{lemma:truncate_ineq_gap} and~\eqref{bar_Delta_ge_3/4_Delta}, we obtain
\begin{align}
	\label{cond_Delta_tc_bar}
	\Delta_\tc \ge \bar{\Delta} - 2\norm{\delta H_\tc} \ge \frac{3}{4}\Delta- 2\norm{\delta H_\tc}, 
\end{align}
We then choose the parameter $l$ so that the following condition is satisfied [see also~\eqref{overlap_Gs_t_Gs}]:
\begin{align}
	\label{The condition_for_Gs_Gs_tc}
	\| \ket{\bar{\Gs}}-\ket{\Gs_\tc} \| \le\frac{ \norm{\delta H_\tc}}{\bar{\Delta} -4 \norm{\delta H_\tc}} \le\frac{\Delta/\bar{\Delta}}{12p} \longrightarrow \norm{\delta H_\tc} \le \frac{\Delta}{12p+4(\Delta/\bar{\Delta})} \le \frac{\Delta}{12p}.
\end{align}
By applying the above to~\eqref{cond_Delta_tc_bar}, we get the lower bound of 
\begin{align}
	\label{lower_bound_Delta_tc_7/12}
	\Delta_\tc \geq  \frac{3}{4}\Delta - \frac{\Delta}{6p} \ge \frac{7}{12}\Delta 
	\for p\ge 1.
\end{align} 
To satisfy the above condition~\eqref{The condition_for_Gs_Gs_tc}, we have to satisfy the lower bound of $l$ from the inequality~\eqref{lemma:truncate__ineq_prop_area_law} as follows:
\begin{align}
	& \frac{8\eta_1\eta_2 \bar{g}_{2l}}{l^{\bar{\alpha}}+1} \leq \frac{\Delta}{12p} \longrightarrow \frac{l^{\bar{\alpha}}}{\bar{g}_{2l}} \ge (8\eta_1\eta_2) \frac{12p}{\Delta}   \longrightarrow  l \ge \tilde{w}_1 \brr{ \frac{p}{\Delta^{\upsilon+1}}\log^{\chi} (p/\Delta)}^{1/\bar{\alpha}} ,
	\label{cond_l_Ht_0}
\end{align}
where we use $\bar{g}_{2l}=g_0 + g_1 \log^\chi(2l+1) \lesssim \Delta^{-\upsilon}  \log^{\chi} (p/\Delta) + \Delta^{-\upsilon}  \log^\chi(2l+1)$ from~\eqref{d_0_g_0_propto2}. 
Under the inequality~\eqref{cond_l_Ht_0} for $l$, we also obtain 
\begin{align}
	\label{bar_g_2l_upper_bound}
	\bar{g}_{2l} \le \tilde{w}_2 \Delta^{-\upsilon} \log^\chi(p/\Delta) , \quad  \tilde{w}_2 =\orderof{1} .
\end{align}
Note that a similar inequality holds for $\bar{g}_{2ql}= \bar{g}_{4l}$, which will be utilized in~\eqref{eq:varepsilon_1_varepsilon_2_cond}. 

Furthermore, from the inequality~\eqref{overlap_Gs_phi_norm}, for an arbitrary quantum state $\ket{\phi}$, we have
\begin{align}
	\norm{ \ket{\bar{\Gs}}-\ket{\phi} } &\le  \| \ket{\Gs_\tc}-\ket{\phi} \| +\frac{ \norm{\delta H_\tc}}{\bar{\Delta} -4 \norm{\delta H_\tc}}\leq \|\ket{\Gs_\tc} - \ket{\phi}\| + \frac{1}{9p} ,
	\label{Gs_ket_phi_distance_last_prop}
\end{align}
where we use the inequality~\eqref{The condition_for_Gs_Gs_tc} and $(\Delta/\bar{\Delta})/(12p) \le 1/(9p)$ in the second inequality.
By combining the above inequality with~\eqref{bar_Delta_ge_3/4_Delta} [$\delta_\bo=1/(9p)$], we have 
\begin{align}
	\label{Omega_distance_Omega_tc}
	\norm{ \ket{\Gs}-\ket{\phi} }  \le \norm{ \ket{\Omega } - \ket{\bar{\Omega} }}+ \|\ket{\Gs_\tc} - \ket{\phi}\| + \frac{1}{9p} \le \|\ket{\Gs_\tc} - \ket{\phi}\| + \frac{2}{9p} .
\end{align}

Third, for the construction of the AGSP operator, we use the effective Hamiltonian from the truncated Hamiltonian $H_\tc$ with $q=2$. 
Then, from the inequality~\eqref{thm:effective_global/main1_eff} in Theorem~\ref{Effective Hamiltonian_multi_truncation}, the condition 
\begin{align}
	\label{cond_tilde_Gs_Tc_Gs_tc}
	&\|\ket{\tilde{\Gs}_\tc} - \ket{\Gs_\tc}\| \le \frac{1}{9p}
\end{align}
is satisfied by
\begin{align}
	\label{eq:varepsilon_1_varepsilon_2_cond}
	\sqrt{\frac{2\varepsilon_2}{\Delta_\tc}}  \le \frac{1}{9p} 
	&\longrightarrow   \varepsilon_2 \le \frac{\Delta_\tc}{162p^2}  \notag \\
	&\longrightarrow \tau \ge \msC''_1\bar{g}_{4l}  \log \br{\frac{\msC''_2 l  \bar{g}_{2l} p^2}{\Delta}}
	+ \msC''_3 g_1   \log^{1+\chi} \br{\frac{\msC''_4l \bar{g}_{2l} p^2}{\Delta}} \for \varepsilon_2 \le \frac{\Delta_\tc}{162p^2} 
\end{align}
with $\{\msC''_1,\msC''_2,\msC''_3,\msC''_4\}$ constants of $\orderof{1}$, 
where we use the inequality~\eqref{cond_for_epsilon_summary} to derive the upper bound for $\tau$. 
Using the obtained upper bounds~\eqref{cond_l_Ht_0} and \eqref{bar_g_2l_upper_bound}, we choose $\tau$ in the form of 
\begin{align}
	\label{choice_of_tau_K_p}
	\tau = \tilde{w}_3  \Delta^{-\upsilon} \log^{1+\chi} (p/\Delta) ,\quad \tilde{w}_3=\orderof{1} 
\end{align}
in order to satisfy the condition~\eqref{cond_tilde_Gs_Tc_Gs_tc}. 
Then, by combining the inequalities~\eqref{Omega_distance_Omega_tc} and \eqref{cond_tilde_Gs_Tc_Gs_tc} with $\ket{\phi}=\ket{\tilde{\Gs}_\tc}$, we obtain 
\begin{align}
	\label{Omega_distance_tilde_Omega_tc_2}
	\norm{ \ket{\Gs}-\ket{\tilde{\Gs}_\tc} }  \le \frac{1}{3p},
\end{align}
which yields the first desired condition for $\delta_{K_p}$, i.e., $\delta_{K_p} \le 1/(3p)$. 

We next consider the second condition for $\epsilon_{K_p}$ of $\epsilon_{K_p} \le 1/(9p)$. 
The remaining control parameter is $m$, which is the degree of the Chebyshev polynomial. 
First, the condition~\eqref{eq:varepsilon_1_varepsilon_2_cond} implies 
\begin{align}
	\label{thm:effective_global_gap_eff_apply_Kp}
	\tilde{\Delta}_\tc \ge  \Delta_\tc - 3 \varepsilon_2  \ge \frac{53}{54} \Delta_\tc \ge \frac{1}{2} \Delta .
\end{align} 
from the inequality~\eqref{thm:effective_global_gap_eff}, where we use $\Delta_\tc \geq 7\Delta/12$ as from inequality~\eqref{lower_bound_Delta_tc_7/12}.  
Second, we obtain a similar inequality to \eqref{error_k_m_poly_AGSP_prop_area_1} as
\begin{align}
	\epsilon_{K_p} \le 2e^{-2m \sqrt{\tilde{\Delta}_\tc/ \brr{4\br{ \tau  +2 c_0\bar{g}_{2l} }  } }} 
	\le 2\exp\left(-m \sqrt{\frac{\Delta^{\upsilon+1}}{2\brr{\tilde{w}_3  \log^{\chi+1}(p/\Delta) + 2c_0 \tilde{w}_2 \log^\chi(p/\Delta)}}}\right),
	\label{error_k_m_poly_AGSP_prop_area_2}
\end{align}
where we use the inequality~\eqref{bar_g_2l_upper_bound} and Eq.~\eqref{choice_of_tau_K_p} for $\bar{g}_{2l}$ and $\tau$, respectively. 
Therefore, the condition $\epsilon_{K_p} \le 1/(9p)$ is satisfied by choosing 
\begin{align}
	m =\ceil{ \sqrt{\frac{2\brr{\tilde{w}_3  \log^{\chi+1}(p/\Delta) + 2c_0 \tilde{w}_2 \log^\chi(p/\Delta)}}{\Delta^{\upsilon+1}}} \log(18p)} \le  \tilde{w}_4\Delta^{-(\upsilon+1)/2}  \log^{\chi/2+3/2} (p/\Delta) ,
\end{align}
where $ \tilde{w}_4=\orderof{1}$. 

We finally estimate the Schmidt rank $\mD_{K_p}$ and calculate $\log(3\mD_{K_p})$ that appears in the upper bound~\eqref{Basic_inequality_for_entropy_bound_fin}. 
From Lemma~\ref{thm:Schmit_rank_Ham_power_0}, we have
\begin{align}
	\mD_{K_p} \le \brr{2+(2d_{2l} l)^{k}}^{m}  \le  \exp\brr{\tilde{w}_5m   \log(p/\Delta) } \le \exp\brr{ \tilde{w}_4\tilde{w}_5 \Delta^{-(\upsilon+1)/2}  \log^{\chi/2+5/2} (p/\Delta)}  , 
	\label{the_Schmidt_rank_prop_area_law}
\end{align}
where $d_{2l} \lesssim \Delta^{-2\upsilon/k} \log^\ma (p/\Delta)$ is derived from the inequality \eqref{d_0_g_0_propto2} and the condition~\eqref{cond_l_Ht_0} for $l$. 
We thus upper-bound $\log(\mD_{K_p})$ by
\begin{align}
	\label{upper_bound_log_3D_K}
	\log(\mD_{K_p}) \le \tilde{w}_6 \frac{\log^{\chi/2+5/2}(3p/\Delta)}{\Delta^{(\upsilon+1)/2}} , \quad \tilde{w}_6=\orderof{1} . 
\end{align}

Using the upper bound~\eqref{upper_bound_log_3D_K}, we immediately derive the first main inequality~\eqref{schmidt_rank_error_delta_SR} as follows. 
From the inequality~\eqref{norm_distance_AGSP_pth} with \eqref{Gamma_K_p_condition} and \eqref{choice_of_delta_epsilon_s}, the quantum state 
$$
\ket{\psi_p}:= \frac{K_p e^{-i\theta_p}\ket{\phi}}{\| K_p \ket{\phi}\|}
$$ 
satisfies 
\begin{align}
	\norm{\ket{\psi_p} - \ket{\Gs} } \le \Gamma_{K_p} \le \frac{1}{p}
\end{align}
under an appropriate choice of the phase factor $\theta_p$. Also, the Schmidt rank of $\ket{\psi_p}$ is upper-bounded by 
\begin{align}
	\log \brr{ {\rm SR} (\ket{\psi_p}) }\le \log (\mD_\phi) + \log(\mD_{K_p})\le \log (\mD_\phi) +  \tilde{w}_6 \frac{\log^{\chi/2+5/2}(3p/\Delta)}{\Delta^{(\upsilon+1)/2}}  ,
\end{align}
which yields the inequality~\eqref{schmidt_rank_error_delta_SR} by choosing $p=\ceil{1/\delta}$.

Lastly, by combining the upper bounds~\eqref{upper_bound_log_3D_K} and $\Gamma_{K_p}\le 1/p$ with the inequality~\eqref{ground_state_entanglement_Gs}, we obtain
\begin{align}
	S(\ket{\Gs}) &\le \log (\mD_\phi) + \log (3\mD_{K_1}) +  \sum_{p=1}^\infty \frac{1}{p^2} \brrr{ \log (p^2)+  \tilde{w}_6 \frac{\log^{\chi/2+5/2}\brr{3(p+1)/\Delta}}{\Delta^{(\upsilon+1)/2}}} \notag \\ 
	&\le \log (\mD_\phi) +   \tilde{w}_7 \frac{\log^{\chi/2+5/2}\br{1/\Delta}}{\Delta^{(\upsilon+1)/2}},
	\label{ground_state_entanglement_Gs_applicatio_1}
\end{align}
where we let $\Gamma_0=1$ in the case of $p=0$ and $\tilde{w}_7$ is an $\orderof{1}$ constant. 
This completes the second main inequality~\eqref{area_law_phi_Gs_d}, completing the proof of Proposition~\ref{prop0:overlap_AGSP_entropy_bound}. $\square$

\section{Proof of Proposition~\ref{Prop:eff:boson_number_truncation}: effective Hamiltonian by boson-number truncation}
\label{Proof of Proposition_Prop:eff:boson_number_truncation}

\subsection{Effective Hamiltonian by arbitrary projection}

For an arbitrary projection $\Pi$, we prove the following proposition~\cite{Generalized_area}, which is similar to Supplemental Lemma~4 of Ref.~\cite{Kuwahara2020arealaw}:
\begin{lemma} \label{lemma:effective_global}
	Let us set the ground energy to be equal to zero, i.e., $H\ket{\Omega}=0$. 
	We consider the Hilbert space spanned by $\Pi$ and consider 
	the effective Hamiltonian $\tilde{H}$ in this restricted Hilbert space:
	\begin{align}
		\tilde{H} :=  \Pi  H  \Pi  ,
	\end{align}
	Then, the ground state $\ket{\tilde{\Omega}}$ of $\tilde{H}$, which satisfies $\Pi \ket{\tilde{\Omega}}=\ket{\tilde{\Omega}}$, is close to the original ground state $\ket{\Omega}$ in the sense that
	\begin{align}
		\label{lemma:effective_global/main1}
		\norm{ \ket{\Omega } - \ket{\tilde{\Omega}}} \le \sqrt{\frac{2\epsilon_H}{ \Delta }}   ,\quad 
		\epsilon_H:=\frac{\bra{\Omega} \Pi H \Pi \ket{\Omega}}{\norm{\Pi \ket{\Omega}}^2}  ,
	\end{align} 
	and the spectral gap $\tilde{\Delta}$ is lower-bounded by
	\begin{align}
		\label{lemma:effective_global_gap}
		\tilde{\Delta} \ge  \Delta-  3\epsilon_H   .
	\end{align} 
\end{lemma}

{\bf Remark.} A straightforward estimation of $\epsilon_H$ reads 
\begin{align}
	\epsilon_H= \frac{\bra{\Omega} (1-\Pi) H(1- \Pi )\ket{\Omega}}{\norm{\Pi \ket{\Omega}}^2}  \le \frac{\norm{H}\cdot  \norm{(1- \Pi )\ket{\Omega}}^2}{\norm{\Pi \ket{\Omega}}^2} .
\end{align} 
However, in the thermodynamic limit ($|\Lambda|\to\infty$), the RHS of the inequality diverges to infinity, and hence we have to estimate $\epsilon_H$ more carefully (see Proposition~\ref{prop:Gs_preservation}).

\subsubsection{Proof of Lemma~\ref{lemma:effective_global}}

We first expand $\ket{\tilde{\Omega}}$ as 
\begin{align}
	&\ket{\tilde{\Omega}} =e^{i\theta} \sqrt{1-\epsilon_{\tilde{\Omega}}}  \ket{\Omega} +\sqrt{\epsilon_{\tilde{\Omega}}}  \ket{\psi_{\bot}} , 
	\quad \bra{\Omega} \psi_\bot \rangle=0 ,
\end{align}
where $e^{i\theta}$ characterizes a phase factor and can be chosen to be equal to $1$ (i.e., $\theta=0$).  
We then derive
\begin{align}
	\label{zeta_2_^2_upp}
	\bra{\tilde{\Omega}} H\ket{\tilde{\Omega}}  =   \epsilon_{\tilde{\Omega}} \bra{\psi_\bot} H \ket{\psi_{\bot}} \ge \epsilon_{\tilde{\Omega}} \Delta ,
\end{align}
where we use $\bra{\Omega} H \ket{\Omega} =0$. 
Because of $\Pi\ket{\tilde{\Omega}}= \ket{\tilde{\Omega}}$ and 
\begin{align}
	\bra{\tilde{\Omega}} H\ket{\tilde{\Omega}}=  \bra{\tilde{\Omega}} \Pi H  \Pi\ket{\tilde{\Omega}}=\bra{\tilde{\Omega}} \tilde{H}\ket{\tilde{\Omega}}  \le \frac{\bra{\Omega} \Pi  \tilde{H} \Pi \ket{\Omega}}{\norm{\Pi \ket{\Omega}}^2}
	=\frac{\bra{\Omega} \Pi  H \Pi \ket{\Omega}}{\norm{\Pi \ket{\Omega}}^2} =\epsilon_H , 
\end{align}
we can reduce the inequality~\eqref{zeta_2_^2_upp} to
\begin{align}
	\epsilon_{\tilde{\Omega}} \le\frac{\epsilon_H}{ \Delta } .
\end{align}
Then, we consider 
\begin{align}
	\label{tildeOmega_Omega_distance}
	\norm{ \ket{\Omega}-\ket{\tilde{\Omega}} }^2= 
	2- 2 {\rm Re} (\langle \tilde{\Omega} \ket{\Omega} )  = 2 -2\sqrt{1-\epsilon_{\tilde{\Omega}}}\le 2 \epsilon_{\tilde{\Omega}} \le  \frac{2\epsilon_H}{ \Delta }  ,
\end{align}
where we use $2-2\sqrt{1-x} \le 2x$ for $0\le x\le 1$. 
Note that for $\epsilon_H /\Delta \ge 1$ the above upper bound is included in the trivial bound as $\norm{ \ket{\Omega}-\ket{\tilde{\Omega}} }^2 \le 2$. 

We next consider the second main inequality~\eqref{lemma:effective_global_gap}. 
We let $\ket{\tilde{E}_1}$ be the first excited state for $\tilde{H}$ such that $\tilde{H}\ket{\tilde{E}_1}=\tilde{E}_1\ket{\tilde{E}_1}$.
Then, using $\langle  \tilde{E}_1 \ket{\tilde{\Omega}}=0$, we obtain
\begin{align}
	| \langle  \tilde{E}_1 \ket{\Omega}| = \abs{ \bra{ \tilde{E}_1 }  ( \ket{\Omega} - \ket{\tilde{\Omega}})} \le 
	\sqrt{ \frac{2\epsilon_H}{\Delta} },
\end{align}
where we use the upper bound~\eqref{tildeOmega_Omega_distance} in the last inequality. 
Therefore, we have 
\begin{align}
	\tilde{E}_1= \bra{ \tilde{E}_1 } \tilde{H} \ket{\tilde{E}_1} =\bra{ \tilde{E}_1 }  \Pi H  \Pi \ket{\tilde{E}_1} = \bra{ \tilde{E}_1 } H\ket{\tilde{E}_1} \ge 
	(1-| \langle  \tilde{E}_1 \ket{\Omega}| ^2) \Delta \ge 
	\br{1-   \frac{2\epsilon_H}{\Delta}  } \Delta ,
\end{align}
where we use $ \Pi  \ket{\tilde{E}_1} =  \ket{\tilde{E}_1}$. 
On the other hand, the ground energy $\tilde{E}_0$ is smaller than 
\begin{align}
	\frac{\bra{\Omega} \Pi \tilde{H} \Pi \ket{\Omega}}{\norm{\Pi \ket{\Omega}}^2}  = \frac{\bra{\Omega} \Pi H \Pi \ket{\Omega}}{\norm{\Pi \ket{\Omega}}^2} =
	\epsilon_H .
\end{align}
By combining the above inequalities, we obtain the lower bound of the spectral gap:
\begin{align}
	\tilde{\Delta} = \tilde{E}_1 - \tilde{E}_0 \ge \Delta-  3\epsilon_H . 
\end{align}
This completes the proof. $\square$

{~}

\hrulefill{\bf [ End of Proof of Lemma~\ref{lemma:effective_global}]}

{~}

\subsection{Projection of boson-number truncation}

Here,  in Lemma~\ref{lemma:effective_global}, we choose the projection $\Pi$ as the boson-number truncation operator. 
We briefly show the setup of Proposition~\ref{Prop:eff:boson_number_truncation} again.
Under the assumption of 
\begin{align}
	\label{Assum_boson_concentration/re}
	\norm{\Pi_{i,> N}  \ket{\Omega}} \le \mc e^{-\mb N^{1/\ma}} \for \forall i\in \Lambda,
\end{align}
we will determine an appropriate truncation boson number.

\begin{figure}[tt]
	\centering
	\includegraphics[clip, scale=0.4]{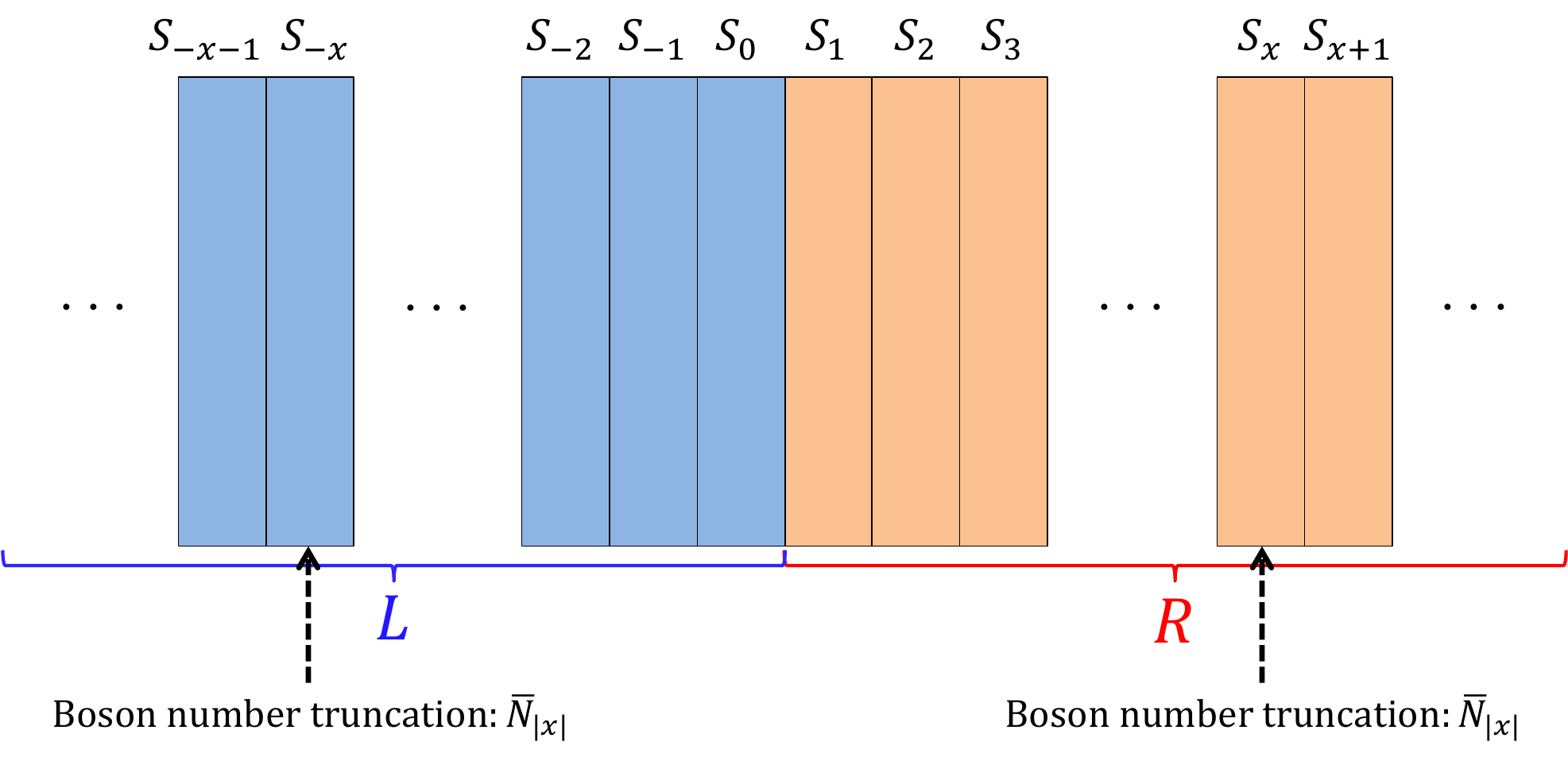}
	\caption{Schematic picture of boson-number truncation. 
		To apply the method to high-dimensional systems in the future, we consider the two-dimensional lattice here. 
		The truncation of the boson numbers poly-logarithmically increases as the site separates from the boundary between $L$ and $R$. 
	}
	\label{fig:Boson_truncation.pdf}
\end{figure}

To make the discussion more general, we consider the high-dimensional systems. 
We slice the total system to $\{S_x\}_{x=-\infty}^\infty$ such that 
\begin{align}
	\Lambda= \bigotimes_{x=-\infty}^{\infty} S_x , 
\end{align}
where $x=0$ is the boundary between $L$ and $R$ (see Supplementary Fig.~\ref{fig:Boson_truncation.pdf}):
\begin{align}
	L= \bigotimes_{x=-\infty}^{0} S_x ,\quad  R= \bigotimes_{x=1}^{\infty} S_x  . 
\end{align}

Let us define the truncation number as $N_i$ ($i\in S_x$) as 
\begin{align}
	\label{definition_of_m_|x|}
	N_i = 
	\mb^{-\ma} \brr{ \log \br{\epsilon_\bo^{-1}} + \log \br{{|x|^{3D} +1}}}^{\ma} \le 
	2^{\ma }\mb^{-\ma} \brr{ \log^{\ma} \br{\epsilon_\bo^{-1}} + \log^{\ma} \br{{|x|^{3D} +1}}}.
\end{align}
We then obtain from~\eqref{Assum_boson_concentration/re} 
\begin{align}
	\label{specific_choice_N_i}
	\norm{\Pi_{i,> N_i} \ket{\Omega}} \le \frac{\mc\epsilon_\bo}{x^{3D}+1} \for \forall i\in S_x.
\end{align}
Note that the above setup is the same as in Proposition~\ref{Prop:eff:boson_number_truncation} in the one-dimensional cases.

We prove the following proposition (see~\ref{sec:Proof of Proposition_prop:Gs_preservation} for the proof):
\begin{prop} \label{prop:Gs_preservation}
	Let us adopt the interaction decay of high-dimensional Hamiltonians as 
	\begin{align}
		\label{ineq_h_Z_operator_re}
		\norm{h_Z\Pi_{\Lambda,\le N}  } \le  J_Z N^{k/2}
	\end{align}
	with
	\begin{align}
		\label{ineq_h_Z_operator_barg_1_re}
		\max_{i,i'} \br{\sum_{Z:Z\ni \{i,i'\}} J_Z } \le \frac{\bar{J}_1}{\dist_{i,i'}^\alpha+1} .
	\end{align}
	
	Under the projection of $\Pi_{\mathbf{N}}$, the parameter $\epsilon_H$ in Lemma~\ref{lemma:effective_global} is upper-bounded as follows:
	\begin{align}
		\label{prop:Gs_preservation_main2}
		\epsilon_H \le  \frac{\gamma^2 k\bar{J}_1\mc \epsilon_\bo |S_0|^2 2^{D+3}}{1-\br{4 \gamma \mc \epsilon_\bo |S_0|}^2}
		\brr{  \frac{3f_0 \mathfrak{w}_D(\alpha-D)}{\alpha-D-1} +f_0 \log^{\ma k/2} \br{\epsilon_\bo^{-1}}},
	\end{align}
	where we define $f_0=\br{2^{\ma+1}/\mb^\ma}^{k/2}$, and a constant $\mathfrak{w}_D$ is defined by
	\begin{align}
		\mathfrak{w}_D:= \sup_{x\in \mathbb{Z}}\brr{ \frac{\log^{\ma k/2} \br{ |x|^{3D} +1}}{|x|+1} }. 
	\end{align}
\end{prop}

{\bf Remark.}
From the statement, we roughly obtain
\begin{align}
	\epsilon_H = \epsilon_\bo \log^{\ma k/2} \br{\epsilon_\bo^{-1}} \orderof{ |\partial L|^2  } ,
\end{align}
where we use $S_0=\partial L$.

\subsection{Proof of Proposition~\ref{Prop:eff:boson_number_truncation}}

Based on Proposition~\ref{prop:Gs_preservation}, we here prove Proposition~\ref{Prop:eff:boson_number_truncation}.
For the convenience of readers, we show it again.

{\bf Proposition~\ref{Prop:eff:boson_number_truncation}.}
\textit{
	Let us adopt $\Pi_{\mathbf{N}}$ as the projection $\bar{\Pi}$, which is given by 
	\begin{align}
		\bar{\Pi}= \Pi_{\mathbf{N}} := \bigotimes_{x\in \Lambda}  \Pi_{x,\le N_x} , 
	\end{align}
	with 
	\begin{align}
		N_x = \mb^{-\ma} \brr{ \log \br{\epsilon_\bo^{-1}} + \log \br{{|x|^{3} +1}}}^{\ma} \le 
		2^{\ma }\mb^{-\ma} \brr{ \log^{\ma} \br{\epsilon_\bo^{-1}} + \log^{\ma} \br{{|x|^{3} +1}}}.
	\end{align} 
	Then, the Hamiltonian $\bar{H}= \Pi_{\mathbf{N}} H \Pi_{\mathbf{N}}$ preserves the ground state and the spectral gap as follows:
	\begin{align}
		\label{norm_Omega_minus_bar_Omega_re}
		\norm{ \ket{\Omega } - \ket{\bar{\Omega} }} \le \delta_\bo, 
		\quad \bar{\Delta} \ge  \Delta \br{1-\frac{3}{2} \delta_\bo^2} , 
	\end{align}
	with $\epsilon_\bo$ given by
	\begin{align}
		\label{epsilon_0_choice_delta_0_re}
		\epsilon_\bo = w_0 \delta_\bo^2  \Delta \log^{-\ma k/2} \br{\delta_\bo^{-1}} ,
	\end{align} 
	where $w_0$ is an $\orderof{1}$ constant and we choose $\delta_\bo$ such that $\delta_\bo \le 1/\sqrt{6}$.  
	Note that $\ket{\bar{\Omega}}$ and $\bar{\Delta}$ are the ground state and the spectral gap of $\bar{H}$. 
}

In 1D setup, we reduce the inequality~\eqref{prop:Gs_preservation_main2} to 
\begin{align}
	\label{prop:Gs_preservation_main2_1D}
	\epsilon_H \le  \frac{16\gamma^2 k \bar{J}_1\mc \epsilon_\bo}{1-\br{4 \gamma \mc \epsilon_\bo}^2} \brr{  \frac{3f_0 \mathfrak{w}_1(\alpha-1)}{\alpha-2} +f_0 \log^{\ma k/2} \br{\epsilon_\bo^{-1}}} =: \bar{\epsilon}_H.
\end{align}
For the proof, we set 
\begin{align}
	\sqrt{\frac{2 \bar{\epsilon}_H}{\Delta}} \le \delta_\bo\longrightarrow 
	\bar{\epsilon}_H \le \frac{\Delta}{2} \delta_\bo^2 .
\end{align} 
From the inequality~\eqref{lemma:effective_global/main1}, this choice yields the first main inequality in~\eqref{norm_Omega_minus_bar_Omega_re}. 
Because the parameters $\{\gamma, \mc,\bar{J}_1,f_0, \mathfrak{w}_1, \alpha,\ma, k \}$ are $\orderof{1}$ constants, the above inequality is satisfied by choosing as in~\eqref{epsilon_0_choice_delta_0_re}.

Under the same choice, we also obtain from the inequality~\eqref{lemma:effective_global_gap}
\begin{align}
	\bar{\Delta} \ge  \Delta-  3\epsilon_H  \ge \Delta \br{1-\frac{3}{2} \delta_\bo^2},
\end{align} 
which gives the desired lower bound for $\bar{\Delta}$ in~\eqref{norm_Omega_minus_bar_Omega_re}.
This completes the proof of Proposition~\ref{Prop:eff:boson_number_truncation}. $\square$

\subsection{Proof of Proposition~\ref{prop:Gs_preservation}} \label{sec:Proof of Proposition_prop:Gs_preservation}

We start from the lower bound of $\norm{\Pi_{\mathbf{N}} \ket{\Omega}}^2=\bra{\Omega}\Pi_{\mathbf{N}}\ket{\Omega}$, which is also given by
\begin{align}
	\label{Omega_exp_Pi_vec_N}
	\bra{\Omega}\Pi_{\mathbf{N}}\ket{\Omega}= 1- \bra{\Omega}\Pi_{\mathbf{N}}^\co \ket{\Omega} =1-\norm{ \Pi^\co_{\mathbf{N}} \ket{\Omega} } ^2,
\end{align}
where $\Pi_{\mathbf{N}}+ \Pi_{\mathbf{N}}^\co=1$. 

We adopt a set of subset $X_1,X_2,\ldots, X_{|\Lambda|}$ such that $X_{s+1} \supset X_s$ and $|X_{s+1}-X_s|=1$.
By denoting 
\begin{align}
	\Pi_s= \bigotimes_{i\in X_s} \Pi_{i,\le N_i} ,
\end{align}
we obtain 
\begin{align}
	\label{Pi_vec_q_omega_decomp}
	\Pi_{\mathbf{N}} \ket{\Omega} 
	&= \Pi_{|\Lambda|} \ket{\Omega} =\ket{\Omega} + \sum_{s=1}^{|\Lambda|} \br{  \Pi_{s} -\Pi_{s-1}}  \ket{\Omega} ,
\end{align}
where we let $X_0=\emptyset$. 
From the above equation, we obtain 
\begin{align}
	\label{norm_Pi_vec_q_ket_Omega}
	\norm{ \Pi^\co_{\mathbf{N}} \ket{\Omega} } 
	=\norm{\ket{\Omega}- \Pi_{\mathbf{N}} \ket{\Omega} } 
	\le \sum_{s=1}^{|\Lambda|} \norm{\br{  \Pi_{s} -\Pi_{s-1}}  \ket{\Omega}} 
	\le \sum_{i\in \Lambda} \norm{\br{ \Pi_{i,\le N_i}  -1} \ket{\Omega}} 
	=  \sum_{i\in \Lambda} \norm{\Pi_{i,> N_i}\ket{\Omega}} .
\end{align}

Using the inequalities~\eqref{Assum_boson_concentration/re} and \eqref{specific_choice_N_i}, we obtain 
\begin{align}
	\sum_{i\in \Lambda}  \norm{\Pi_{i,> N_i} \ket{\Omega}} \le \mc  \sum_{i\in \Lambda}  e^{-\mb N_i^{1/\ma}}
	\le\mc \epsilon_\bo \sum_{x=-\infty}^\infty  \sum_{i\in S_x} \frac{1}{|x|^{3D}+1} 
	\le 4 \gamma \mc \epsilon_\bo |S_0|  ,
\end{align}
where we use $|S_x| + |S_{-x}| \le \sum_{i\in S_0} |\partial i[x]| \le \gamma (x^{D-1}+1) |S_0|$ for $x\ge 0$ [see~\eqref{def:parameter_gamma} for the definition of $\gamma$] to derive 
\begin{align}
	\sum_{x=-\infty}^\infty  \sum_{i\in S_x} \frac{1}{|x|^{3D}+1} \le 
	\gamma |S_0| \sum_{x=0}^\infty \frac{x^{D-1}+1 }{x^{3D}+1} \le 4 \gamma |S_0|  .
\end{align}
Note that $\sum_{x=0}^\infty \frac{x^{D-1}+1 }{x^{3D}+1}$ is maximized for $D=1$ and $\sum_{x=0}^\infty \frac{2}{x^{3}+1} \approx 3.37301 < 4$. 
We thus prove $\norm{ \Pi^\co_{\mathbf{N}} \ket{\Omega} } \le  4 \gamma \mc \epsilon_\bo |S_0|$, which lower-bound Eq.~\eqref{Omega_exp_Pi_vec_N} by
\begin{align}
	\label{prop:Gs_preservation_main1}
	\bra{\Omega}\Pi_{\mathbf{N}}\ket{\Omega}\ge  1- \br{4 \gamma \mc \epsilon_\bo |S_0|}^2. 
\end{align}

We then aim to calculate $\epsilon_H$, i.e., $\epsilon_H:= \bra{\Omega} \Pi_{\mathbf{N}}(H-E_0)\Pi_{\mathbf{N}}\ket{\Omega}/\bra{\Omega} \Pi_{\mathbf{N}}\ket{\Omega}$. 
By letting 
\begin{align}
	\Pi_{\mathbf{N}_Z}:= \bigotimes_{i\in Z} \Pi_{i,\le N_i} \AND \Pi_{\mathbf{N}_Z}^\co:=1- \Pi_{\mathbf{N}_Z},
\end{align}
we have from $\Pi_{\mathbf{N}}=\Pi_{\mathbf{N}_Z} \Pi_{\mathbf{N}_{Z^\co}}$
\begin{align}
	\bra{\Omega} \Pi_{\mathbf{N}} H \Pi_{\mathbf{N}} \ket{\Omega} 
	&= \sum_{Z:Z\subset \Lambda} \bra{\Omega} \Pi_{\mathbf{N}_{Z^\co}} \Pi_{\mathbf{N}_Z}  h_Z  \Pi_{\mathbf{N}}   \ket{\Omega}  \notag \\
	&=  \sum_{Z:Z\subset \Lambda} 
	\br{ \bra{\Omega} \Pi_{\mathbf{N}_{Z^\co}} h_Z \Pi_{\mathbf{N}}\ket{\Omega}- \bra{\Omega}  \Pi_{\mathbf{N}_{Z^\co}} \Pi^\co_{\mathbf{N}_Z}  h_Z  \Pi_{\mathbf{N}} \ket{\Omega} } \notag \\
	&= \sum_{Z:Z\subset \Lambda} 
	\br{ \bra{\Omega}h_Z \Pi_{\mathbf{N}}   \ket{\Omega} - \bra{\Omega}\Pi^\co_{\mathbf{N}_Z} h_Z   \Pi_{\mathbf{N}} \ket{\Omega} } \notag \\
	&=  \bra{\Omega} H\Pi_{\mathbf{N}}  \ket{\Omega}- \sum_{Z:Z\subset \Lambda} 
	\bra{\Omega}   \Pi^\co_{\mathbf{N}_Z} h_Z  \Pi_{\mathbf{N}}  \ket{\Omega}  \notag \\
	&=   \bra{\Omega} \Pi_{\mathbf{N}} \ket{\Omega}E_0 - \sum_{Z:Z\subset \Lambda} 
	\bra{\Omega} \Pi^\co_{\mathbf{N}_Z} h_Z \Pi_{\mathbf{N}}  \ket{\Omega}  ,
\end{align}
where we use $\Pi_{\mathbf{N}} \Pi_{\mathbf{N}_{Z^\co}}=\Pi_{\mathbf{N}} $ and $[\Pi_{\mathbf{N}_{Z^\co}},h_Z]= [\Pi_{\mathbf{N}_{Z^\co}},  \Pi^\co_{\mathbf{N}_Z} ]=0$. 
From the above equation, we obtain 
\begin{align}
	\abs{\bra{\Omega} \Pi_{\mathbf{N}} (H-E_0) \Pi_{\mathbf{N}} \ket{\Omega}} 
	\le \sum_{Z:Z\subset \Lambda} 
	\norm{\Pi^\co_{\mathbf{N}_Z} \ket{\Omega}} \cdot \norm{h_Z \Pi_{\mathbf{N}} } .
	\label{Omega_Pi_vec_q_Omega}
\end{align}

Using the inequality~\eqref{ineq_h_Z_operator_re}, i.e., $\norm{h_Z\Pi_{\Lambda,\le N}  } \le  J_Z N^{k/2}$, we have 
\begin{align}
	\norm{h_Z\Pi_{\mathbf{N}_Z} } \le  J_Z \brr{\max_{i\in Z} (N_i)}^{k/2} ,
\end{align}
and 
\begin{align}
	\norm{\Pi^\co_{\mathbf{N}_Z} \ket{\Omega}}   \le |Z| \max_{i\in Z} \br{ \mc e^{-\mb N_i^{1/\ma}}} \le k \max_{i\in Z} \br{ \mc e^{-\mb N_i^{1/\ma}}},
\end{align}
where we use a similar inequality to~\eqref{norm_Pi_vec_q_ket_Omega}. 
By applying the above two inequalities to the RHS of the inequality~\eqref{Omega_Pi_vec_q_Omega}. we upper-bound  
\begin{align}
	\label{upp_sum_Z_Zsubset/Lambda}
	& \sum_{Z:Z\subset \Lambda} 
	\norm{h_Z\Pi_{\mathbf{N}}}\cdot \norm{\Pi^\co_{\mathbf{N}_Z} \ket{\Omega}}   \notag \\
	&\le
	\sum_{x_1=-\infty}^{\infty} \sum_{x_2: |x_2| \ge |x_1|}\sum_{i_1 \in S_{x_1}} \sum_{i_2 \in S_{x_2}} \sum_{Z:Z\ni \{i_1,i_2\}}  
	k J_Z N_{x}^{k/2} \cdot \frac{\mc \epsilon_\bo}{|x_1|^{3D}+1} \notag \\
	&\le\sum_{x_1=-\infty}^{\infty} \sum_{x_2: |x_2| \ge |x_1|} \frac{k \bar{J}_1|S_{x_1}|\cdot |S_{x_2}| }{(|x_2|-|x_1|)^\alpha+1}\brrr{2^{\ma }\mb^{-\ma} \brr{ \log^{\ma} \br{\epsilon_\bo^{-1}} + \log^{\ma} \br{{|x_2|^{3D} +1}}} }^{k/2} \cdot  \frac{\mc \epsilon_\bo}{|x_1|^{3D}+1} ,
\end{align}
where we use the definition of~$N_{x}$ in Eq.~\eqref{definition_of_m_|x|}, which also gives the inequality~\eqref{specific_choice_N_i}, and the upper bound~\eqref{ineq_h_Z_operator_barg_1_re} with $\dist_{i_1,i_2} \le |x_2|-|x_1|$.

We obtain the upper bound of 
\begin{align}
	&\brrr{2^{\ma }\mb^{-\ma} \brr{ \log^{\ma} \br{\epsilon_\bo^{-1}} + \log^{\ma} \br{{|x_2|^{3D} +1}}} }^{k/2} \notag \\
	&\le 2^{k/2}2^{\ma k/2}\mb^{-\ma k/2}  \brr{ \log^{\ma k/2} \br{{|x_2|^{3D} +1}}  +  \log^{\ma k/2}\br{\epsilon_\bo^{-1}}}  \notag \\
	&\le \br{2^{\ma+1}/\mb^\ma}^{k/2} \brr{ \mathfrak{w}_D (|x_2|+1) +\log^{\ma k/2} \br{\epsilon_\bo^{-1}} }
	=f_0 \mathfrak{w}_D (|x_2|+1) +f_0 \log^{\ma k/2} \br{\epsilon_\bo^{-1}}
\end{align}
with $f_0=\br{2^{\ma+1}/\mb^\ma}^{k/2}$ and  
\begin{align}
	\mathfrak{w}_D:= \sup_{x \in \mathbb{Z}}\brr{ \frac{\log^{\ma k/2} \br{ |x|^{3D} +1}}{|x|+1} }.
\end{align}
We therefore reduce the RHS of the inequality~\eqref{upp_sum_Z_Zsubset/Lambda} to
\begin{align}
	\label{sum_Z_Z_subset_Lamna_norm}
	& \sum_{Z:Z\subset \Lambda} 
	\norm{h_Z\Pi_{\mathbf{N}}}\cdot \norm{\Pi^\co_{\mathbf{N}_Z} \ket{\Omega}}   \notag \\
	&\le  \gamma^2 k\bar{J}_1\mc \epsilon_\bo |S_0|^2  \sum_{x_1=0}^{\infty} \sum_{r=0}^\infty
	\brr{f_0 \mathfrak{w}_D (x_1+r+1) +f_0 \log^{\ma k/2} \br{\epsilon_\bo^{-1}}} \cdot  \frac{(x_1^{D-1}+1) \brr{\br{x_1+r}^{D-1}+1}  }{\br{|x_1|^{3D}+1}\br{r^\alpha+1} } ,
\end{align}
where we let $|x_2|=|x_1| +r$ and use $|S_{x}|+|S_{-x}| \le \gamma (|x|^{D-1}+1)|S_0|$ for $\forall x>0$ from~\eqref{def:parameter_gamma}. 
In the following, we separately estimate 
\begin{align}
	\label{eta_D_term_estimation0}
	\sum_{x_1=0}^{\infty} \sum_{r=0}^\infty f_0 \mathfrak{w}_D (x_1+r+1)  \frac{(x_1^{D-1}+1) \brr{\br{x_1+r}^{D-1}+1}  }{\br{|x_1|^{3D}+1}\br{r^\alpha+1} } ,
\end{align}
and 
\begin{align}
	\label{no_eta_D_term_estimation0}
	\sum_{x_1=0}^{\infty} \sum_{r=0}^\infty
	\brr{f_0 \log^{\ma k/2} \br{\epsilon_\bo^{-1}}} \cdot  \frac{(x_1^{D-1}+1) \brr{\br{x_1+r}^{D-1}+1}  }{\br{|x_1|^{3D}+1}\br{r^\alpha+1} } .
\end{align}

We begin with the quantity~\eqref{eta_D_term_estimation0}. 
Using $(s^a+1)(s^b+1)\le 2(s^{a+b}+1)$\footnote{The inequality derive from $2(s^{a+b}+1)-(s^a+1)(s^b+1)=s^{a+b}-s^a-s^b+1=(s^{a}-1)(s^{b}-1)$.} for $s\in \mathbb{N}$ and $a,b\ge0$, we have 
\begin{align}
	\label{eta_D_term_estimation}
	&\sum_{x_1=0}^{\infty} \sum_{r=0}^\infty
	f_0 \mathfrak{w}_D (x_1+r+1) \frac{(x_1^{D-1}+1) \brr{\br{x_1+r}^{D-1}+1}  }{\br{|x_1|^{3D}+1}\br{r^\alpha+1} } \notag \\
	&\le  2f_0\mathfrak{w}_D\sum_{x_1=0}^{\infty} \sum_{r=0}^\infty
	\frac{(x_1^{D-1}+1) \brr{\br{x_1+r}^{D}+1}  }{\br{|x_1|^{3D}+1}\br{r^\alpha+1} } \notag \\
	&\le 2f_0 \mathfrak{w}_D 2^D \sum_{x_1=0}^{\infty}  \frac{(x_1^{D-1}+1) (x_1^D+1)}{x_1^{3D}+1}
	\sum_{r=0}^\infty
	\frac{r^D+1}{r^\alpha+1}   ,
\end{align}
where we use $(x_1+r)^D+1\le \max\brr{(2x_1)^D+1,(2r)^D+1}\le 2^D \br{x_1^D+1} (r^D+1)$.
For the summations in the RHS of~\eqref{eta_D_term_estimation}, we can derive 
\begin{align}
	\sum_{x_1=0}^{\infty}  \frac{(x_1^{D-1}+1) (x_1^D+1)}{x_1^{3D}+1}
	\le 2\sum_{x_1=0}^{\infty}  \frac{x_1^{2D-1}+1}{x_1^{3D}+1} \le 2\sum_{x_1=0}^{\infty}  \frac{x_1+1}{x_1^{3}+1} \approx 5.59629 < 6,
\end{align}
and 
\begin{align}
	\sum_{r=0}^\infty  \frac{r^D+1}{r^\alpha+1}  
	= 2+ \sum_{r=2}^\infty  \frac{r^D+1}{r^\alpha+1} \le 2+  2\sum_{r=2}^\infty r^{D-\alpha} \le 2+2 \int_1^\infty z^{D-\alpha} dz = 2+\frac{2}{\alpha-D-1}
	=\frac{2(\alpha-D)}{\alpha-D-1},
\end{align}
where we use $\alpha>D+1$ and $D\ge 1$. 
By applying them to the inequality~\eqref{eta_D_term_estimation}, we obtain 
\begin{align}
	\label{eta_D_term_estimation_fin}
	\textrm{\eqref{eta_D_term_estimation0}}
	\le  \frac{24 f_0 \mathfrak{w}_D 2^{D}(\alpha-D)}{\alpha-D-1}.
\end{align}

In the same way, for the summation~\eqref{no_eta_D_term_estimation0}, we can derive 
\begin{align}
	\label{no_eta_D_term_estimation_fin}
	\textrm{\eqref{no_eta_D_term_estimation0}}
	&=\sum_{x_1=0}^{\infty} \sum_{r=0}^\infty
	\brr{f_0 \log^{\ma k/2} \br{\epsilon_\bo^{-1}}}   \frac{(x_1^{D-1}+1) \brr{\br{x_1+r}^{D-1}+1}  }{\br{|x_1|^{3D}+1}\br{r^\alpha+1} }  \notag \\
	&\le 2^{D-1} \brr{f_0 \log^{\ma k/2} \br{\epsilon_\bo^{-1}}}  \sum_{x_1=0}^{\infty}  \frac{(x_1^{D-1}+1) (x_1^{D-1}+1)}{x_1^{3D}+1}
	\sum_{r=0}^\infty\frac{r^{D-1}+1}{r^\alpha+1}   \notag \\
	&\le  2^{D-1} \brr{f_0 \log^{\ma k/2} \br{\epsilon_\bo^{-1}}}  \times (6.74601\cdots) \times (2.07666\cdots) \le 2^{D+3} \brr{f_0 \log^{\ma k/2} \br{\epsilon_\bo^{-1}}} ,
\end{align}
where we use $\alpha>D+1$.  

By applying the inequality~\eqref{eta_D_term_estimation_fin} and \eqref{no_eta_D_term_estimation_fin} to the RHS of~\eqref{sum_Z_Z_subset_Lamna_norm}, we prove 
\begin{align}
	& \sum_{Z:Z\subset \Lambda} 
	\norm{h_Z\Pi_{\mathbf{N}}}\cdot \norm{\Pi^\co_{\mathbf{N}_Z} \ket{\Omega}} \le   \gamma^2k \bar{J}_1\mc \epsilon_\bo |S_0|^2 \cdot 
	2^{D+3}  \brr{  \frac{3f_0 \mathfrak{w}_D(\alpha-D)}{\alpha-D-1} +f_0 \log^{\ma k/2} \br{\epsilon_\bo^{-1}}}.
\end{align}
By applying the above inequality to~\eqref{Omega_Pi_vec_q_Omega} and combining it with~\eqref{prop:Gs_preservation_main1}, 
we prove the main inequality~\eqref{prop:Gs_preservation_main2}. 
This completes the proof. $\square$

\section{Proof of Proposition~\ref{lemm:error_int_truncation}: Error estimation of interaction truncation}
\label{sec:Error estimation of interaction truncation}

The proofs of the inequalities~\eqref{lemma:truncate_ineq_gap},~\eqref{overlap_Gs_t_Gs} and \eqref{overlap_Gs_phi_norm} immediately follows from the proof of Supplemental Lemma~4 of Ref.~\cite{Kuwahara2020arealaw}.
We only have to prove the inequality~\eqref{lemma:truncate__ineq}.
For this purpose, we use the following lemma:

\begin{lemma} \label{lem:ar_V_X_Y_upper}
	Let us define the subsets $X$ and $Y$ as 
	\begin{align}
		X= (-\infty,x] ,\quad Y= [y,\infty),\quad y-x=r . 
	\end{align}
	Then, for an arbitrary site $i\in \Lambda[r_0]$, we obtain 
	\begin{align}
		\label{bar_V_X_Y_uppe_fin}
		\overline{V}_{X,Y}:= \sum_{Z:Z\cap X\neq \emptyset, Z\cap Y\neq \emptyset} \norm{\bar{h}_Z}
		\le 2 \eta_1  \eta_2 \bar{g}_{|x|+r}  \br{r^2+1} \bar{J}(r) ,
	\end{align}
	where we define the parameter $\eta_p$ ($\ge 1$) by  
	\begin{align}
		\sum_{y=y_0}^\infty  \bar{g}_{r+y}  (y^{p-1}+1) \bar{J}(y) \le  \eta_p \bar{g}_{r+y_0}  (y_0^p+1) \bar{J}(y_0) \for p < \alpha . 
		\label{def:eta_parameter}
	\end{align}
\end{lemma}

For convenience, we here relabel
\begin{align}
	B_0 \setminus \tilde{B}_0 \to B_{-1}, \quad  \tilde{B}_0\to B_0, \quad 
	\tilde{B}_{q+1} \to B_{q+1} ,\quad 
	B_{q+1} \setminus \tilde{B}_{q+1} \to B_{q+2} .
\end{align}
We then define 
\begin{align}
	&X_s := \bigcup_{j\ge s+2} B_{j}  ,  \quad 
	\Lambda_s=\bigcup_{j\ge s} B_{j} ,
\end{align}
where $s \in [-1,q]$. 
From the definition of the truncated Hamiltonian~\eqref{def:truncated_Hamiltonian}, we obtain 
\begin{align}
	\|\bar{H}-H_\tc\| 
	\le  \sum_{s=-1}^q \overline{V}_{B_s ,X_s} . \label{ineq_H_minus_H_t}
\end{align}
Here, $B_s \subset [-\infty, -ql/2+s l]$ for $s\in [-1,q]$ from the definition, and hence by applying Lemma~\ref{lem:ar_V_X_Y_upper} with $x=ql/2+l$ and $r=l$, we obtain the desired inequality of
\begin{align}
	\|\bar{H}-H_\tc\| \le   2 \eta_1  \eta_2 (q+2) \bar{g}_{ql/2+2l}  \br{l^2+1} \bar{J}(l) \le 
	4 \eta_1  \eta_2 q \bar{g}_{ql}  \br{l^2+1} \bar{J}(l) ,
\end{align}
where we use the monotonic increase of $\bar{g}_x$ and $q\ge 4$ (which gives $q/2+2\le q$). 
This completes the proof. $\square$

\subsubsection{Proof of Lemma~\ref{lem:ar_V_X_Y_upper}.}

\begin{figure}
	\centering
	\includegraphics[scale=0.6]{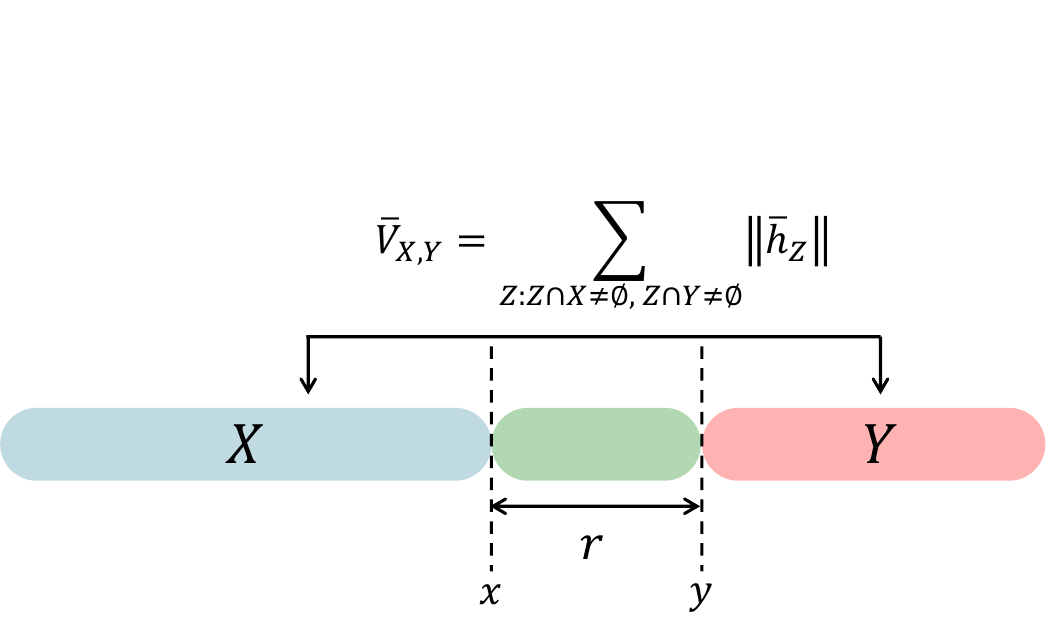}
	\caption{Block-block interaction $\overline{V}_{X,Y}$. We denote the right-end site of $X$ and the left-end site of $Y$ as $x$ and $y$, respectively. 
		The interaction $\overline{V}_{X,Y}$ picks up all the interaction norms of $\norm{\bar{h}_Z}$ that act on both $X$ and $Y$.  
	}\label{fig:Block_interaction}
\end{figure}

As long as $\diam(Z) =s$ with $Z\ni x$, we have $Z\subset [-|x|-s,|x|+s]$, and hence 
\begin{align}
	\sum_{Z:Z\ni x, \diam (Z)= s} \norm{\bar{h}_Z} &\le 
	\sum_{i:\dist_{x,i}=s}\ \sum_{Z:Z\ni \{i,x\}, Z\subset [-|x|-s,|x|+s]} \norm{\bar{h}_Z} \notag \\
	&\le
	2 \bar{g}_{|x|+s} \bar{J}(s) ,
\end{align}
where we use $|\partial x[s]| \le 2$ in one dimension.  
In general, for $x' \in \Lambda$ and $Y\subseteq \Lambda$ such that $\dist_{x',Y} = r'$, 
we have 
\begin{align}
	\sum_{Z:Z\cap \{x'\} \neq \emptyset, Z\cap Y\neq \emptyset} \norm{\bar{h}_Z} 
	=\sum_{Z:Z\ni x', \diam (Z)\ge r'} \norm{\bar{h}_Z} 
	&\le 2 \sum_{s=r'}^\infty  \bar{g}_{|x'|+s} \bar{J}(s)\notag\\
	&\le 2 \eta_1 \bar{g}_{|x'|+r'}  \br{r'+1} \bar{J}(r') ,
\end{align}
where we use the inequality~\eqref{def:eta_parameter} with $p=1$.
Therefore, from $\dist_{X,Y}=r$, we obtain 
\begin{align}
	\label{bar_V_X_Y_uppe_1}
	\sum_{Z:Z\cap X\neq \emptyset, Z\cap Y\neq \emptyset} \norm{\bar{h}_Z}
	& \le \sum_{x'=-\infty}^x \sum_{Z:Z\cap \{x'\}\neq \emptyset, Z\cap Y\neq \emptyset} \norm{\bar{h}_Z}  \notag \\
	&\le 2 \eta_1 \sum_{x'=-\infty}^x \bar{g}_{|x'|+x-x'+r}  \brr{(x-x'+r)+1} \bar{J}(x-x'+r),
\end{align}
where we use $r'=\dist_{x',Y} = \dist_{x,Y} + (x-x') = r + x-x'$. 
By replacing $x'$ with $x-s$, we reduce the above inequality to 
\begin{align}
	\label{bar_V_X_Y_uppe_2}
	2\eta_1 \sum_{x'=-\infty}^x  \bar{g}_{|x'|+x-x'+r}    \brr{(x-x'+r)+1} \bar{J}(x-x'+r) 
	&\le 2 \eta_1 \sum_{s=0}^\infty  \bar{g}_{|x|+r+s}  \brr{(r+s)+1} \bar{J}(r+s) \notag \\
	&=2\eta_1 \sum_{s=r}^\infty  \bar{g}_{|x|+s}  \br{s+1} \bar{J}(s)  \notag \\
	&\le 2\eta_1 \eta_{2} \bar{g}_{|x|+r}  \br{r^2+1} \bar{J}(r).
\end{align}
By combining the inequalities~\eqref{bar_V_X_Y_uppe_1} and \eqref{bar_V_X_Y_uppe_2}, we prove the main inequality~\eqref{bar_V_X_Y_uppe_fin}. 
This completes the proof. $\square$

\section{Proof of Theorem~\ref{Effective Hamiltonian_multi_truncation}: Effective Hamiltonian theory} \label{sec:Effective Hamiltonian theory_proof}
In this section, we prove Theorem~\ref{Effective Hamiltonian_multi_truncation}, which ensures the preservation of the ground state and the spectral gap by the multi-energy cut-off.
The main proof will be shown in \ref{sec:Proof of the Main Theorem_eff_ham}.

\subsection{Notations} \label{sec:preliminaries_eff_proof}

We first remind several notations. 
The projection operator onto the eigenspace of $h_s$ is defined as 
\begin{align}
	\Pi^{(s)}_{I}=\sum_{E_{s,j} \in I} \ket{E_{s,j}}\bra{E_{s,j}}  
	\label{projection_to_spectrum/again}
\end{align}
for $I \subset \mathbb{R}$. Especially for $\Pi^{(s)}_{(-\infty,E)}$ and $\Pi^{(s)}_{(-\infty,E]}$, we denote them by $\Pi^{(s)}_{< E}$ and $\Pi^{(s)}_{\le E}$, respectively. 
In the same way, we define $\Pi^{(s)}_{> E}$ and $\Pi^{(s)}_{\ge E}$.
By using the above notations, we define the projection $\tilde{\Pi}$ as 
\begin{align}
	\tilde{\Pi}= \bigotimes_{s=0}^{q+1}\Pi^{(s)}_{\le \tau_s}.
	\label{projection_to_spectrum_tilde_Pi}
\end{align}
The effective Hamiltonian $\tilde{H}_\tc$ is 
\begin{align}
	\tilde{H}_\tc =\tilde{\Pi} H_\tc \tilde{\Pi} 
	\label{explicit_tilde_H_eff_again}
\end{align}
for $s=0,1,2\ldots, q+1$, where we choose the cut-off energies $\{\tau_s\}_{s=0}^{q+1}$ as 
\begin{align}
	\tau_s= E_{s,0} + \tau \for s=0,1,2\ldots, q+1.
\end{align}
We denote the ground state of $\tilde{H}_\tc $ by $\ket{\tilde{\Omega}_\tc}$. 

For the total Hamiltonian $H_\tc$, we define $\{ E_{\tc,j}, \ket{E_{\tc,j}}\}_{j}$ are the eigenvalues and the eigenstates of $H_\tc$, respectively.
Using them, we define $\Pi_{\tc,I}$ as 
\begin{align}
	&\Pi_{\tc, I}=\sum_{E_{\tc,j} \in I} \ket{E_{\tc,j}}\bra{E_{\tc,j}}. \label{projector_total_energy}
\end{align}

\subsection{Multi-commutator bound} \label{Sec:Multi-commutator bound}
For the proof of the main theorem, we consider the norm of $\norm{\Pi_{\tc,\ge E+E'}  O_s  \Pi_{\tc,\le E}}$ that plays a key role (see the proof below), where $O_s$ commute with $h_s$: $[O_s,h_s]=0$.   
A standard approach in Ref.~\cite{Arad_2016} utilizes the imaginary time evolution as 
\begin{align}
	\|\Pi_{\tc,\ge E+E'}  O_s  \Pi_{\tc,\le E}\| &= 
	\left \|\Pi_{\tc,\ge E+E'} e^{-\beta H_\tc} e^{\beta H_\tc} O_s e^{-\beta H_\tc} e^{\beta H_\tc}\Pi _{\le E} \right\|
	\le  e^{-\beta E'}\norm{ e^{\beta H_\tc}  O_s   e^{-\beta H_\tc}}  \label{upper_bound_norm_Pi_E'_O_s_Pi_E} .
\end{align}
Here, the Hamiltonian $H_\tc$ has an unbounded norm as a site becomes distant from the boundary [see the inequality~\eqref{general_k-local_op_unbounded_cond}].  
As shown in the following Lemma~\ref{lemma:commutator_unbounded}, the multi-commutator $\ad_{H_\tc}^m (O_s)$ has a norm that scales as $m^{m(1+\chi)}$ (see~\ref{sec:Proof of Lemma_lemma:commutator_unbounded} for the proof): 

\begin{lemma} \label{lemma:commutator_unbounded}
	Let us define $O_{Z_0}$ as an arbitrary operator with unit norm, where $Z_0$ satisfies 
	\begin{align}
		Z_0 \subseteq [-\ell,\ell] ,\quad |Z_0|\le k . 
	\end{align} 
	Then, for an arbitrary $k$-local Hamiltonian $\bar{H}$ satisfying the conditions~\eqref{general_k-local_op_unbounded_cond}, we have
	\begin{align}
		\label{communtaro_bound_long_range_main}
		\norm{ \ad_{\bar{H}}^m  (O_{Z_0} )} \le 
		2^{\bar{\alpha}} \br{\tilde{c}_1 m  \bar{g}_{m+\ell}}^m +
		2\br{\frac{2+\bar{\alpha}}{\bar{\alpha}}}
		\br{ \tilde{c}_2g_1 m^{\chi +1}}^m ,
	\end{align}
	with $\tilde{c}_1$ and $\tilde{c}_2$ defined as 
	\begin{align}
		\label{def::tilde_c_1c_2}
		\tilde{c}_1=\frac{2^{\chi +\bar{\alpha}+3} 4k\eta_1}{1- 2^{-\bar{\alpha}}} , \quad 
		\tilde{c}_2= \tilde{c}_1  \brr{ \frac{2\chi  (2+\bar{\alpha})}{\bar{\alpha}}}^{\chi } .
	\end{align}
	We remind that the parameter $\chi$ has been defined as $\chi=k\ma/2$ in the main inequality~\eqref{thm:main_theorem_area_law/main_ineq}.
\end{lemma}

{\bf Remark.} 
In the lemma, we adopt $0^0=1$ for $m=0$.
By taking the leading term with respect to $m$, we have 
\begin{align}
	\label{communtaro_bound_long_range_main_rewrite}
	\norm{ \ad_{\bar{H}}^m \br{O_{Z_0}} } \propto c^m m^{(\chi+1) m}
\end{align}
with $c$ an $\orderof{1}$ constant. 
Even in the case where the short-range (or exponentially decaying) interactions are considered, 
one can derive\footnote{In the short-range interacting cases, the summation in~\eqref{norm_ad_H_R_vec_s_summation__2} in the proof should be appropriately modified.}
\begin{align}
	\label{communtaro_bound_long_range_main_short}
	\norm{ \ad_{\bar{H}}^m \br{O_{Z_0}} } \propto \brr{ c \log^{\chi}(m) }^m m^m,
\end{align}
which still makes the imaginary time evolution $\norm{e^{\beta \ad_{\bar{H}}} \br{O_{Z_0}} }$ diverge to infinity. 

By applying this lemma to $H_\tc$, we immediately obtain the following corollary [The proof is immediately obtained by combining the inequality~\eqref{truncated_Hamiltonian_block_interaction} with the inequality~\eqref{communtaro_bound_long_range_main}.]:
\begin{corol} \label{corol:commutator_unbounded}
	Let us consider $h_{s,s+1}$ in Eq.~\eqref{def:truncated_Hamiltonian} for $s=[0,q]$, which is supported on $[-ql/2-l, ql/2+l]$ from the definition.
	Then, we obtain 
	\begin{align}
		\label{communtaro_bound_long_range_main_h_s_s+1}
		\norm{ \ad_{H_\tc}^m  (h_{s,s+1})} \le 
		c_0\bar{g}_{ql}  \br{2^{\bar{\alpha}} \br{\tilde{c}_1 m  \bar{g}_{m+ql}}^m +
			2\br{\frac{2+\bar{\alpha}}{\bar{\alpha}}}
			\br{ \tilde{c}_2g_1 m^{\chi +1}}^m } ,
	\end{align}
	where we use the definition of $\bar{g}_{ql}$ in~\eqref{truncated_Hamiltonian_block_interaction}. 
\end{corol}

In particular, when we consider a multi-commutator of $\ad_{H_\tc}^m \br{O_s}$ for an operator that commutes with $h_s$, we prove the following statement that meets our purpose (see~\ref{sec:Proof of Proposition_prop:commutator_unbounded} for the proof): 
\begin{prop}\label{prop:commutator_unbounded}
	For an arbitrary operator $O_s$ such that $[O_s,h_s]=0$, we prove the inequality of 
	\begin{align}
		\label{assump_kappa_form}
		&\norm{ \ad_{H_\tc}^m  (O_s)} \le  (T_m m)^m  \norm{O_s},\quad T_m = (2 c_0 \tilde{c}_3 +\tilde{c}_1)  \bar{g}_{m+ql } +\tilde{c}_2g_1 m^{\chi}  ,
	\end{align}
	where $\tilde{c}_3$ is defined as 
	\begin{align}
		\label{defi:tilde_c_3}
		\tilde{c}_3 = 2^{\bar{\alpha}}+ \frac{2(2+\bar{\alpha})}{\bar{\alpha}} . 
	\end{align}
\end{prop}

\subsubsection{Proof of Lemma~\ref{lemma:commutator_unbounded}} \label{sec:Proof of Lemma_lemma:commutator_unbounded}

We define the length scale $\ell_s$ ($s\ge -1$) as 
\begin{align}
	\ell_s = 2^s , \quad \ell_{-1}=0. 
\end{align}
Then, we consider a general region $[-R,R]$ and define the Hamiltonian $\bar{H}_R$ as follows: 
\begin{align}
	\label{Def_H_R_H_R,s}
	\bar{H}_R := \sum_{s=0}^\infty \bar{H}_{R,s} ,\quad  \bar{H}_{R,s}= \sum_{\substack{|Z|\le k, Z\subset  [-R,R] \\  \diam(Z)\in (\ell_{s-1}, \ell_s]}}\bar{h}_Z,
\end{align}
with
\begin{align}
	\sum_{\substack{Z\subset  [-R,R]: Z\ni i \\  \diam(Z)\in (\ell_{s-1}, \ell_s]}} \norm{\bar{h}_Z} 
	&\le \sum_{i': \dist_{i,i'} \ge \ell_{s-1}}\sum_{Z\subset  [-R,R]: Z\ni \{i,i'\} } \norm{\bar{h}_Z}  \notag \\
	&\le \bar{g}_R \sum_{r=\ell_{s-1}}^\infty  \sum_{i': \dist_{i,i'} =r}   \bar{J}(r) 
	\le 2 \bar{g}_R \eta_1(\ell_{s-1}+1) \bar{J}(\ell_{s-1}) ,
\end{align}
where we use the condition~\eqref{general_k-local_op_unbounded_cond} and the inequality~\eqref{def:eta_parameter} with $p=1$.

By using 
\begin{align}
	\bar{H} = \sum_{s=0}^\infty \bar{H}_{\infty, s} ,\quad  \bar{H}_{\infty, s}= \sum_{\substack{Z\subset \Lambda \\  \diam(Z)\in (\ell_{s-1}, \ell_s]}} \bar{h}_Z ,
\end{align}
we can write 
\begin{align}
	\label{norm_ad_H_R_vec_s_summation}
	\norm{\ad_{\bar{H}}^m(O_{Z_0})} 
	&\le\sum_{\bar{s}=0}^\infty  \sum_{\max(s_1,s_2 ,\ldots , s_m)=\bar{s}} \norm{\ad_{\bar{H}_{\infty, s_m}}\cdots\ad_{\bar{H}_{\infty, s_2}}\ad_{\bar{H}_{\infty, s_1}}  (O_{Z_0})}   \notag \\
	&=\sum_{\bar{s}=0}^\infty  \sum_{\max(s_1,s_2 ,\ldots , s_m)=\bar{s}}  \norm{\ad_{\bar{H}_{R_{\bar{s}}, s_m}}\cdots\ad_{\bar{H}_{R_{\bar{s}}, s_2}}\ad_{\bar{H}_{R_{\bar{s}}, s_1}}  (O_{Z_0})}
\end{align}
with $R_{\bar{s}}=m \ell_{\bar{s}}+ \ell=m2^{\bar{s}}+\ell$, where $\ad_{H_{\infty, s_m}}\cdots\ad_{H_{\infty, s_2}}\ad_{H_{\infty, s_1}}  (O_{Z_0})$ is supported on $[-R_{\bar{s}},R_{\bar{s}}]$ from the definition~\eqref{Def_H_R_H_R,s} and $Z_0\subseteq [-\ell,\ell]$.

Using Lemma~3 in Ref.~\cite{KUWAHARA201696}, we have 
\begin{align}
	\label{norm_ad_H_R_vec_s}
	\norm{\ad_{\bar{H}_{R_{\bar{s}}, s_m}}\cdots\ad_{\bar{H}_{R_{\bar{s}}, s_2}}\ad_{\bar{H}_{R_{\bar{s}}, s_1}}  (O_{Z_0})}
	&\le(2k)^m m! \prod_{j=1}^m  \brr{ 2 \bar{g}_{R_{\bar{s}}} \eta_1(\ell_{s_j-1}+1) \bar{J}(\ell_{s_j-1})}\notag \\
	&\le \br{2^{\chi +2} k\eta_1}^m  m! \br{\bar{g}_{m+\ell} + g_1 \bar{s}^{\chi } }^m\prod_{j=1}^m 2^{-\bar{\alpha} (s_j-1)} ,
\end{align}
where, from the conditions~\eqref{cond_r^2_1_decay} and \eqref{general_k-local_op_unbounded_cond}, we use
\begin{align}
	(\ell_{s-1}+1) \bar{J}(\ell_{s-1}) \le \frac{1}{\ell_{s-1}^{\bar{\alpha}}+1} \le 
	2^{-\bar{\alpha} (s-1)},
\end{align} 
and 
\begin{align}
	\bar{g}_{R_{\bar{s}}}
	= g_0+ g_1 \log^{\chi } \br{m2^{\bar{s}}+\ell+1} 
	& \le g_0 + g_1  \brr{ \log\br{m+\ell+1} +\bar{s}}^{\chi } \notag \\
	&\le  g_0 + g_1 \brr{2\log\br{m+\ell+1} }^{\chi } + g_1 \br{2\bar{s}}^{\chi }\notag \\
	&\le 2^{\chi }\br{\bar{g}_{m+\ell} + g_1 \bar{s}^{\chi } } .
\end{align} 
The above inequality is derived from $m2^{\bar{s}}+\ell+1\le (m+ \ell+1) e^{\bar{s}}$ and  $(y_1+y_2)^a \le (2y_1)^a + (2y_2)^a$ for $y_1,y_2,a>0$.

We then estimate the summation of  
\begin{align}
	\label{norm_ad_H_R_vec_s_sum_part}
	\sum_{\max(s_1,s_2 ,\ldots , s_m)=\bar{s}} \prod_{j=1}^m 2^{-\bar{\alpha} (s_j-1)}  
	&\le m  2^{-\bar{\alpha} (\bar{s}-1)} \br{\sum_{s=0}^{\infty} 2^{-\bar{\alpha} (s-1)}}^{m-1} \notag \\
	&= m  \br{\frac{2^{\bar{\alpha}}}{1- 2^{-\bar{\alpha}}}}^{m-1} 2^{-\bar{\alpha} (\bar{s}-1)}.
\end{align} 
By combining the inequalities~\eqref{norm_ad_H_R_vec_s} and \eqref{norm_ad_H_R_vec_s_sum_part} with the upper bound~\eqref{norm_ad_H_R_vec_s_summation}, we obtain 
\begin{align}
	\label{norm_ad_H_R_vec_s_summation__2}
	\norm{\ad_{\bar{H}}^m(O_{Z_0})} 
	&\le  \br{2^{\chi +2} k\eta_1}^m  m! \sum_{\bar{s}=0}^\infty \br{\bar{g}_{m+\ell} + g_1 \bar{s}^{\chi } }^m m\br{\frac{2^{\bar{\alpha}}}{1- 2^{-\bar{\alpha}}}}^{m-1} 2^{-\bar{\alpha} (\bar{s}-1)}  \notag \\
	&\le  \br{\frac{2^{\chi +\bar{\alpha}+3} 4k\eta_1 m}{1- 2^{-\bar{\alpha}}}}^m \sum_{\bar{s}=0}^\infty \brr{\bar{g}^m_{m+\ell} + \br{g_1 \bar{s}^{\chi }}^m  } 2^{-\bar{\alpha} (\bar{s}-1)}  ,
\end{align}
where we use $(y_1+y_2)^a \le (2y_1)^a + (2y_2)^a$ and $m\cdot m! \le m^m$ for $m\in \mathbb{N}$.

Finally, to upper-bound the summation, we use Supplemental Lemma~1 of Ref.~\cite{kuwahara2022optimal} to derive\footnote{The inequality 
	$\sum_{m=1}^\infty m^D e^{-c m} \le e^{-c} + e^{-c} 2^D D! [\max(1,1/c)]^{D+1}$ was derived. } 
\begin{align}
	\sum_{\bar{s}=1}^\infty 2^{-\bar{\alpha} (\bar{s}-1)}  \bar{s}^{\chi  m} 
	&\le 1+ 2^{\chi m} (\chi m)! \br{\frac{1}{\bar{\alpha} \log(2)} +1}^{\chi m+1} \notag \\
	&\le 2^{1+\chi m} (\chi m)^{\chi m} \br{\frac{2+\bar{\alpha}}{\bar{\alpha}}}^{\chi m+1}
	= 2 \br{\frac{2+\bar{\alpha}}{\bar{\alpha}}} \brr{ \frac{2\chi m (2+\bar{\alpha})}{\bar{\alpha}}}^{\chi m}.
\end{align}
Also, we have $\sum_{\bar{s}=0}^\infty 2^{-\bar{\alpha} (\bar{s}-1)} =2^{\bar{\alpha}} \br{1-2^{-\bar{\alpha} }}^{-1}$, and hence the main inequality is derived as follows:
\begin{align}
	\label{norm_ad_H_R_vec_s_summation__3}
	&\norm{\ad_{\bar{H}}^m(O_{Z_0})} 
	\le \br{\tilde{c}_1 m}^m
	\brrr{ 2^{\bar{\alpha}}  \bar{g}^m_{m+\ell}  + g_1^m \cdot 2 \br{\frac{2+\bar{\alpha}}{\bar{\alpha}}} \brr{ \frac{2\chi m (2+\bar{\alpha})}{\bar{\alpha}}}^{\chi m}} \notag \\
	&= 
	2^{\bar{\alpha}} \br{\tilde{c}_1 m  \bar{g}_{m+\ell}}^m +
	2\br{\frac{2+\bar{\alpha}}{\bar{\alpha}}}
	\br{ \tilde{c}_2g_1 m^{\chi +1}}^m ,
\end{align}
where we have defined $\tilde{c}_1$ and $\tilde{c}_2g_1$ as 
$\tilde{c}_1=\br{2^{\chi +\bar{\alpha}+3} 4k\eta_1}/\br{1- 2^{-\bar{\alpha}}}$ and 
$\tilde{c}_2= \tilde{c}_1 \brr{2\chi  (2+\bar{\alpha})/\bar{\alpha}}^{\chi }$, respectively. 
This completes the proof of Lemma~\ref{lemma:commutator_unbounded}.
$\square$

\subsubsection{Proof of Proposition~\ref{prop:commutator_unbounded}}\label{sec:Proof of Proposition_prop:commutator_unbounded}

Let us define $T_{0,m}$ as 
\begin{align}
	\label{def:T_0,m_tilde_c_3} 
	&T_{0,m} = \tilde{c}_1  \bar{g}_{m+ql } +\tilde{c}_2g_1 m^{\chi}   .   \quad 
\end{align}
Then, using $T_{0,m}$ and $\tilde{c}_3$ in Eq.~\eqref{defi:tilde_c_3}, 
we simplify the inequality~\eqref{communtaro_bound_long_range_main_h_s_s+1} in Corollary~\ref{corol:commutator_unbounded} as 
\begin{align}
	\label{communtaro_bound_h_s_s+1_simple}
	\norm{ \ad_{H_\tc}^m  (h_{s_0,s_0+1})} \le c_0\tilde{c}_3\bar{g}_{ql} \br{T_{0,m}m}^m \for \forall s_0 \in[0,q].
\end{align}

In the following, we aim to prove the inequality \eqref{assump_kappa_form}, i.e., 
\begin{align}
	&\norm{ \ad_{H_\tc}^m  (O_s)} \le  (T_m m)^m  \norm{O_s} =  \bigl[ (2 c_0 \tilde{c}_3 +\tilde{c}_1) m \bar{g}_{m+ql } +\tilde{c}_2g_1 m^{\chi+1} \bigr]^m \norm{O_s}   ,
\end{align}
by using the induction method.

For $m=1$, we prove the inequality~\eqref{assump_kappa_form} from the inequality~\eqref{truncated_Hamiltonian_block_interaction} as follows:
\begin{align}
	\label{ad_H_tc_O_s_up}
	\|\ad_{H_\tc} (O_s)\| = \|[h_{s-1,s} +h_{s,s+1}, O_s]\|  \le 2 \|h_{s-1,s} +h_{s,s+1}\| \cdot  \norm{O_s} \le 4c_0 \bar{g}_{ql}  \norm{O_s}
	\le T_1  \norm{O_s},
\end{align}
where we use $[h_s,O_s]=0$ and $\tilde{c}_3\ge 2$ from the definition in~\eqref{defi:tilde_c_3}.

We then assume the target inequality up to $m= m_0$ and consider $m=m_0+1$:
\begin{align}
	\|\ad_{H_\tc}^{m_0+1} (O_s)\| &= \|\ad_{H_\tc}^{m_0} \left( [h_{s-1,s} +h_{s,s+1}, O_s] \right)\|   \notag \\
	&\le 2 \sum_{m_1+m_2=m_0}  \binom{m_0}{m_1}\norm{\ad_{H_\tc}^{m_1}(O_s)} \cdot \norm{\ad_{H_\tc}^{m_2} \left( h_{s-1,s} +h_{s,s+1} \right)}  \notag \\
	&\le 2 \norm{O_s}  \sum_{m_1+m_2=m_0} \binom{m_0}{m_1} (T_{m_1} m_1)^{m_1} \norm{\ad_{H_\tc}^{m_2} \left( h_{s-1,s} +h_{s,s+1} \right)}.
	\label{induction_commutator_os}
\end{align}
By using the inequality~\eqref{communtaro_bound_h_s_s+1_simple} for $\ad_{H_\tc}^{m_2} ( h_{s-1,s} +h_{s,s+1})$, we have 
\begin{align}
	\norm{\ad_{H_\tc}^{m_2} ( h_{s-1,s} +h_{s,s+1})} \le 2 c_0\tilde{c}_3\bar{g}_{ql}  \br{T_{0,m_2}m_2}^{m_2} .
	\label{commutator_h_tc_h_s_s}
\end{align}
By applying the inequality~\eqref{commutator_h_tc_h_s_s} to \eqref{induction_commutator_os}, we obtain
\begin{align}
	\|\ad_{H_\tc}^{m_0+1} (O_s)\| &\le  
	2 \|O_s\| \cdot  2 c_0\tilde{c}_3\bar{g}_{ql} \sum_{m_1+m_2=m_0} \binom{m_0}{m_1}(T_{m_1} m_1)^{m_1}  \br{T_{0,m_2}m_2}^{m_2}   \notag \\
	&\le  \|O_s\| \cdot  4 c_0\tilde{c}_3\bar{g}_{ql} (T_{m_0} m_0)^{m_0} \sum_{m_1+m_2=m_0} \binom{m_0}{m_1} \frac{m_1^{m_1}  m_2^{m_2}}{(m_1+m_2)^{m_1+m_2}} \notag \\
	&\le   \|O_s\| \cdot  2T_{m_0} (T_{m_0} m_0)^{m_0} (m_0+1) \le   \brr{T_{m_0+1} (m_0+1)}^{m_0+1} \|O_s\|  ,
\end{align}
where, in the second inequality, we use $T_{0,m} \le T_m$ and the monotonic increase of $T_m$ for $m$,  
in the third inequality, we use Lemma~\ref{lem:binom_upper_bound_main} below and $2 c_0\tilde{c}_3\bar{g}_{ql}\le T_{m_0} $ from the definition in~\eqref{assump_kappa_form}, and in the last inequality, we use 
$m_0^{m_0} = (m_0+1)^{m_0} (1+1/m_0)^{-m_0} \le (m_0+1)^{m_0}/2$.   
This completes the proof of the main inequality~\eqref{assump_kappa_form} in Proposition~\ref{prop:commutator_unbounded}. $\square$

\subsubsection{Lemma on the binomial coefficient}

\begin{lemma}\label{lem:binom_upper_bound_main}
	Let us adopt $0^0=1$.  Then, for an arbitrary pair of integers $m_1$ and $m_2$, we have 
	\begin{align}
		\label{binom_upper_bound_main}
		\binom{m_1+m_2}{m_1} \le \frac{(m_1+m_2)^{m_1+m_2}}{m_1^{m_1} m_2^{m_2}}.
	\end{align}
\end{lemma}

\textit{Proof of Lemma~\ref{lem:binom_upper_bound_main}.}
Let us set $m_1\le m_2$ without loss of generality. 
We can trivially obtain the main inequality for the case of $m_1=0$, and hence, we only need to consider the case of $m_1\ge 1$.
For the proof, we utilize the following inequality in Ref.~\cite{785bbcf5-e12d-3b4f-9b4c-913b654991d7}:
\begin{align}
	\sqrt{2\pi n} \br{\frac{n}{e}}^n e^{1/(12n+1)}< n! < \sqrt{2\pi n} \br{\frac{n}{e}}^n e^{1/(12n)}.
\end{align}
Using it, we have 
\begin{align}
	m_1^{m_1}< \frac{e^{-1/(12m_1+1)}}{\sqrt{2\pi m_1}} e^{m_1} m_1!,\quad 
	m_2^{m_2}< \frac{e^{-1/(12m_2+1)}}{\sqrt{2\pi m_2}} e^{m_2} m_2!,
\end{align}
and 
\begin{align}
	(m_1+m_2)^{m_1+m_2} > \frac{e^{-1/(12m_1+12m_2)}}{\sqrt{2\pi (m_1+m_2)}} e^{m_1+m_2} (m_1+m_2)!.
\end{align}
By using them, we obtain 
\begin{align}
	\label{binom_proof_1}
	\frac{(m_1+m_2)^{m_1+m_2}}{m_1^{m_1} m_2^{m_2}} 
	> e^{1/(12m_1+1)+1/(12m_2+1)-1/(12m_1+12m_2)} \frac{2\pi \sqrt{ m_1 m_2}}{\sqrt{2\pi (m_1+m_2)}} \binom{m_1+m_2}{m_1}.
\end{align}
For $1\le m_1\le m_2$, we obtain 
\begin{align}
	\label{binom_proof_2}
	\frac{1}{12m_1+1}+\frac{1}{12m_2+1}- \frac{1}{12m_1+12m_2} \ge  \frac{2}{12m_1+1} - \frac{1}{12m_1+12} \ge 0 ,
\end{align}
and 
\begin{align}
	\label{binom_proof_3}
	\frac{m_1+m_2}{2\pi} = \frac{m_1}{2\pi} \br{1+ \frac{m_2}{m_1}} \le  \frac{m_1}{2\pi} (1+m_2)  \le \frac{2}{2\pi} m_1 m_2 \le m_1m_2.
\end{align}
By applying the inequalities~\eqref{binom_proof_2} and \eqref{binom_proof_3} to~\eqref{binom_proof_1}, we prove the main inequality~\eqref{binom_upper_bound_main}. 
This completes the proof.  $\square$

\subsection{Subtheorem on the energy distribution}  \label{sec:Key Proposition and difficult point}

Based on Proposition~\ref{prop:commutator_unbounded}, we aim to upper bound 
\begin{align}
	\label{O_s_Pi_I_z+theta_target}
	\norm{ \Pi_{\tc,I}  O_s  \Pi_{\tc,\le E}} ,
\end{align}
where $I \subset \mathbb{R}$ is arbitrary taken.
For our purpose, we consider  
\begin{align}
	\label{region_I_definition}
	I=[E+\theta, E+\theta+T_0),
\end{align}
where $T_0$ is defined from $T_m$ in~\eqref{assump_kappa_form}, i.e., $T_0=T_{m=0}=(2 c_0 \tilde{c}_3 +\tilde{c}_1)  \bar{g}_{ql }$. 

We prove the following subtheorem:
\begin{subtheorem} \label{subthm_energy_distribution}
	For an arbitrary $E$ and the region $I$ as in Eq.~\eqref{region_I_definition}, we prove the following upper bound: 
	\begin{align}
		\label{O_s_Pi_I_z+theta_fin}
		\norm{ \Pi_{\tc,I}  O_s  \Pi_{\tc,\le E}}  
		\le e \norm{O_s} \exp\brrr{- \min\brr{\frac{\theta}{4e \tilde{T}_{\theta/T_0}} , \br{\frac{\theta}{4e\tilde{c}_2g_1}}^{1/(1+\chi)}}   }   ,
	\end{align}
	where $O_s$ is arbitrarily chosen such that $[O_s,h_s]=0$, and the quantity $\tilde{T}_z$ ($z>0$) is defined as  
	\begin{align}
		\label{definition_tilde_T_x}
		\tilde{T}_z:= (2 c_0 \tilde{c}_3 +\tilde{c}_1)  \bar{g}_{z+ql } \longrightarrow T_m= \tilde{T}_m +\tilde{c}_2g_1 m^{\chi}. 
	\end{align} 
\end{subtheorem}

\subsubsection{Proof of Subtheorem~\ref{subthm_energy_distribution}}

As has been shown, the multi-commutator $\ad_{H_\tc}^m (O_s)$ has a norm that scales as $m^{m(1+\chi)}$, and hence we cannot rely on the original argument~\cite{Arad_2016}, which utilizes the imaginary time evolution as in~\eqref{upper_bound_norm_Pi_E'_O_s_Pi_E}.
To obtain a meaningful bound, we adopt analyses similar to those in Supplementary Lemma 39 of Ref.~\cite{Anshu_2021}.

We first derive 
\begin{align}
	\label{O_s_Pi_I_z+theta}
	\norm{ \Pi_{\tc,I}  O_s  \Pi_{\tc,\le E}} \le \frac{\norm{\Pi_{\tc,I}   O_s (E+\theta - H_\tc)^m }}{\theta^m},
\end{align}
where the inequality is derived from $ \Pi_{\tc,\le E} (E+\theta - H_\tc)^m  \succeq \theta^m$ as follows:
\begin{align}
	\norm{\Pi_{\tc,I}  O_s  (E+\theta - H_\tc)^m}
	& \ge \norm{\Pi_{\tc,I} O_s  (E+\theta - H_\tc)^m  \Pi_{\tc,\le E}}  \notag \\
	&\ge \norm{\Pi_{\tc,I} O_s \Pi_{\tc,\le E} \cdot \Pi_{\tc,\le E} (E+\theta - H_\tc)^m  \Pi_{\tc,\le E}}  \notag \\
	&\ge \theta^m \norm{\Pi_{\tc,I} O_s \Pi_{\tc,\le E}}.
\end{align}
Note that for arbitrary positive semi-definite operators $O$ and $O'$ that satisfy $O\succeq O' \succeq 0$ and $[O,O']=0$, we have $\norm{O_0O} \ge \norm{O_0O'}$, where the operator $O_0$ can be arbitrarily chosen.

We next calculate
\begin{align}
	(E+\theta - H_\tc)^m O_s  \Pi_{\tc,I}= \sum_{s=0}^m \binom{m}{s} \ad_{E+\theta - H_\tc}^s(O_s) \br{E+\theta - H_\tc}^{m-s} \Pi_{\tc,I} ,
\end{align}
which yields
\begin{align}
	\label{norm_H_tc-z_O_s_1}
	\norm{  (E+\theta - H_\tc)^m O_s  \Pi_{\tc,I} } \le  \sum_{s=0}^m \binom{m}{s} \norm{\ad_{H_\tc}^s(O_s)} T_0^{m-s} ,
\end{align}
where we use  $\norm{\br{E+\theta - H_\tc}^{m-s} \Pi_{\tc,I}} \le T_0^{m-s}$ and $\ad_{E+\theta - H_\tc}=-\ad_{H_\tc}$.

By applying the upper bound~\eqref{assump_kappa_form} to~\eqref{norm_H_tc-z_O_s_1}, we prove 
\begin{align}
	\label{norm_H_tc-z_O_s_2nd}
	\norm{  (E+\theta - H_\tc)^m O_s  \Pi_{\tc,I} } 
	&\le  \norm{O_s} \sum_{s=0}^m \binom{m}{s} (T_s s)^s T_0^{m-s} \notag \\
	&\le  \norm{O_s} (T_m m)^m  \sum_{s=0}^m \binom{m}{s} =  \norm{O_s} (2 T_m m)^m ,
\end{align}
where we use $T_s \le T_m$ for $0\le s\le m$. 
By combining the inequality~\eqref{norm_H_tc-z_O_s_2nd} with the upper bound \eqref{O_s_Pi_I_z+theta}, we have 
\begin{align}
	\label{O_s_Pi_I_z+theta_2}
	\norm{ \Pi_{\tc,I}  O_s  \Pi_{\tc,\le E}}  \le  \norm{O_s} \br{\frac{2 T_m m}{\theta} }^m  
	&\le \norm{O_s} \brr{\frac{2m (2 c_0 \tilde{c}_3 +\tilde{c}_1)  \bar{g}_{\theta/T_0+ql } +2\tilde{c}_2g_1 m^{1+\chi}}{\theta} }^m   \notag \\
	&= \norm{O_s} \br{\frac{2m \tilde{T}_{\theta/T_0} +2\tilde{c}_2g_1 m^{1+\chi}}{\theta} }^m ,
\end{align}
where the explicit form of $T_m$ was given in Eq.~\eqref{assump_kappa_form}, and we choose $m$ such that $m\le \theta/T_0$, which gives 
$\bar{g}_{m+ql } \le \bar{g}_{\theta/T_0+ql }$ since $\bar{g}_z$ monotonically increases with $z$, and in the last inequality, we use the notation of $\tilde{T}_z$ in Eq.~\eqref{definition_tilde_T_x}. 

We here set $m$ so that it satisfies both of the conditions 
\begin{align}
	&\frac{2m \tilde{T}_{\theta/T_0}}{\theta} \le \frac{1}{2e} ,\quad 
	\frac{2\tilde{c}_2g_1 m^{1+\chi}}{\theta} \le \frac{1}{2e} , \notag \\
	&\longrightarrow m=  \min\brr{\floor{\frac{\theta}{4e \tilde{T}_{\theta/T_0}}} , \floor{\br{\frac{\theta}{4e\tilde{c}_2g_1}}^{1/(1+\chi)}} } ,
\end{align}
which reduces the upper bound~\eqref{O_s_Pi_I_z+theta_2} to 
\begin{align}
	\norm{ \Pi_{\tc,I}  O_s  \Pi_{\tc,\le E}} \le  \norm{O_s} \br{\frac{2 T_m m}{\theta} }^m  
	\le e \norm{O_s} \exp\brrr{- \min\brr{\frac{\theta}{4e \tilde{T}_{\theta/T_0}} , \br{\frac{\theta}{4e\tilde{c}_2g_1}}^{1/(1+\chi)}}   }   .
\end{align}
We thus prove the main inequality~\eqref{O_s_Pi_I_z+theta_fin}. 
This completes the proof of Subtheorem~\ref{subthm_energy_distribution}. $\square$

\subsection{Proof of the Main Theorem} \label{sec:Proof of the Main Theorem_eff_ham}

In this section, we will prove Theorem~\ref{Effective Hamiltonian_multi_truncation}. 
For this purpose, we give a general theorem on the energy distribution of a subsystem $B_s \subset \Lambda$, given that the total energy lies within the interval $(-\infty, E]$. We will show the sub-exponential decay of the energy distribution of $h_s$ (see~\ref{Proof:thm:The energy distribution of the subsystem L_c} for the proof).

\begin{theorem} \label{thm:The energy distribution of the subsystem L_c}
	Let us set $E_{s,0}$ and $E_{\tc ,0}$ being the ground-state energies of $h_s$ and $H_\tc$. 
	Then, for $\tau_s=\tau + E_{s,0}$, 
	the overlap between the projections $\Pi^{(s)}_{> \tau_s}$ and $\Pi_{\tc,\le E}$ is bounded from above as:
	\begin{align}
		\left \| \Pi^{(s)}_{> \tau_s} \Pi_{\tc,\le E} \right\| \le  \mE_{\tau + E_{\tc,0}  - 4 c_0\bar{g}_{ql} - E -T_0} ,
		\label{Energy_dist_subsystem_Lc}
	\end{align}
	where $\mE_{y}$ for $\forall y>0$ is defined as 
	\begin{align}
		\label{def:varepsilon_y}
		\mE_{y}=\mu_1 \exp\br{- \frac{y}{8e \tilde{T}_{y/T_0}}}+  \mu_2  \exp\brr{-\frac{1}{2}\br{\frac{y}{4e\tilde{c}_2g_1}}^{1/(1+\chi)}},
	\end{align}
	and $\mu_1$ and $\mu_2$ are $\orderof{1}$ constants as follows:
	\begin{align}
		&\mu_1:=1+e \int_{0}^\infty (z+2) \exp\brr{- \frac{1}{8e} \cdot \frac{z}{1 + \log^\chi(z+3)}}dz ,  \label{defin:mu_1} \\
		&\mu_2:= 1+e  \int_{0}^\infty (z+2) \exp\brr{- \frac{1}{2}\br{\frac{ 2 c_0 \tilde{c}_3 +\tilde{c}_1}{4e\tilde{c}_2}  z}^{1/(1+\chi)}}  dz.  \label{defin:mu_2} 
	\end{align}
\end{theorem}

The proof of the main Theorem~\ref{Effective Hamiltonian_multi_truncation} 
comes from the combination of the above theorem and Lemma~\ref{lemma:effective_global}. 
For the convenience of readers, we show the theorem again:

{~}

\noindent 
{\bf Theorem~\ref{Effective Hamiltonian_multi_truncation}.} 
\textit{
	Let us define $\varepsilon_1$ and $\varepsilon_2$ as 
	\begin{align}
		\label{def_epsilon_1_eff_H_re}
		\varepsilon_1=  2q \mE_{\tau  - 4 c_0\bar{g}_{ql}  -T_0}  ,\quad 
		\varepsilon_2= \frac{\varepsilon_1}{1-\varepsilon_1}  2q \br{ \tau  +2 c_0\bar{g}_{ql} } ,
	\end{align}
	where $\mE_y$ ($y\ge 0$) is a sub-exponentially decaying function defined in Eq.~\eqref{def:varepsilon_y}, and $T_0$ is defined by $T_{m=0}$ using $T_m$ in Eq.~\eqref{assump_kappa_form}.
	Then, we obtain
	\begin{align}
		\label{thm:effective_global/main1_eff_re}
		\norm{ \ket{\Omega_\tc } - \ket{\tilde{\Omega}_\tc}} \le \sqrt{\frac{2\varepsilon_2}{\Delta_\tc}},
	\end{align} 
	and the spectral gap $\tilde{\Delta}$ is lower-bounded by
	\begin{align}
		\label{thm:effective_global_gap_eff_re}
		\tilde{\Delta}_\tc \ge \Delta_\tc - 3 \varepsilon_2    .
	\end{align} 
}

\textit{Proof of Theorem~\ref{Effective Hamiltonian_multi_truncation}.}
We here use Lemma~\ref{lemma:effective_global} with 
\begin{align}
	H\to H_\tc,\quad \ket{\Omega} \to \ket{\Omega_\tc},\quad  \Pi \to \bigotimes_{s=0}^{q+1} \Pi^{(s)}_{\le \tau_s} := \tilde{\Pi}. 
\end{align}
In the following, we aim to treat the parameter $\epsilon_{H_\tc}$ in Lemma~\ref{lemma:effective_global}: 
\begin{align}
	\epsilon_{H_\tc}=\frac{  \left \langle \Omega_\tc \abs{ \tilde{\Pi} ( H_\tc -E_{\tc,0})\tilde{\Pi} } \Omega_\tc \right \rangle}{\norm{\tilde{\Pi} \ket{\Omega_\tc} }^2 } .
\end{align}
We here introduce an additional parameter $\epsilon_{\Omega_\tc}$ as follows:
\begin{align}
	\epsilon_{\Omega_\tc}= \norm{\tilde{\Pi}^\co \ket{\Omega_\tc} } = \norm{\tilde{\Pi} \ket{\Omega_\tc} -  \ket{\Omega_\tc} }.
\end{align}

Our task is to derive the inequality of 
\begin{align}
	\label{thm:effective_global_cond_multi_truncate}
	\epsilon_{\Omega_\tc} \le 2q \mE_{\tau  - 4 c_0\bar{g}_{ql}  -T_0} =: \varepsilon_1 ,
\end{align}
and 
\begin{align}
	\label{thm2:effective_global_cond_multi_truncate}
	\epsilon_{H_\tc}
	\le \frac{\varepsilon_1}{1-\varepsilon_1}  2q \br{ \tau  +2 c_0\bar{g}_{ql} } =:\varepsilon_2 . 
\end{align}
After proving these inequalities, we prove the main inequalities of~\eqref{thm:effective_global/main1_eff_re}
and \eqref{thm:effective_global_gap_eff_re} from~\eqref{lemma:effective_global/main1} and \eqref{lemma:effective_global_gap}, respectively.

From the same inequality as~\eqref{norm_Pi_vec_q_ket_Omega}, we obtain 
\begin{align}
	\norm{\tilde{\Pi} \ket{\Omega_\tc} -\ket{\Omega_\tc}} 
	\le \sum_{s=0}^{q+1} \norm{ \Pi^{(s)}_{\le \tau_s} \ket{\Omega_\tc} -\ket{\Omega_\tc}} 
	=\sum_{s=0}^{q+1} \norm{ \Pi^{(s)}_{> \tau_s} \ket{\Omega_\tc}}  
	= \sum_{s=0}^{q+1} \norm{ \Pi^{(s)}_{> \tau_s}\Pi_{\tc, \le E_{\tc,0}}}  .
\end{align}
By applying the inequality~\eqref{Energy_dist_subsystem_Lc} with $E\to E_{\tc,0}$ to the RHS, we have 
\begin{align}
	\epsilon_{\Omega_\tc} =\norm{\tilde{\Pi} \ket{\Omega_\tc} -\ket{\Omega_\tc}} 
	\le  \sum_{s=0}^{q+1} \mE_{\tau  - 4 c_0\bar{g}_{ql}  -T_0}  
	= (q+2) \mE_{\tau  - 4 c_0\bar{g}_{ql}  -T_0} \le 2q \mE_{\tau  - 4 c_0\bar{g}_{ql}  -T_0},
\end{align}
where we use the condition $q\ge 4$ to get $q+2\le 2q$. 
We thus prove the first target inequality~\eqref{thm:effective_global_cond_multi_truncate} for $\epsilon_{\Omega_\tc}$. 

We next consider $\epsilon_{H_\tc}$ as 
\begin{align}
	\label{epsilon_H_tc_upp1}
	\epsilon_{H_\tc}
	&=\norm{\tilde{\Pi} \ket{\Omega_\tc} }^{-2} 
	\left \langle \Omega_\tc \abs{ \tilde{\Pi} ( H_\tc -E_{\tc,0}) \br{\tilde{\Pi}-1} } \Omega_\tc \right \rangle \notag \\
	&\le \norm{\tilde{\Pi} \ket{\Omega_\tc} }^{-1}  \norm{( H_\tc -E_{\tc,0}) \tilde{\Pi} }\cdot  \norm{\br{1-\tilde{\Pi}} \ket{ \Omega_\tc}} \notag \\
	&= \frac{\epsilon_{\Omega_\tc}}{1-\epsilon_{\Omega_\tc}} \norm{( H_\tc -E_{\tc,0}) \tilde{\Pi} }
	\le \frac{\varepsilon_1}{1-\varepsilon_1} \norm{( H_\tc -E_{\tc,0}) \tilde{\Pi} } ,
\end{align}
where we use the monotonic increase of $x/(1-x)$ for $0\le x <1$. 

We next estimate the norm of $\norm{( H_\tc -E_{\tc,0}) \tilde{\Pi} }$. Because of $[h_s, \tilde{\Pi} ]=0$, we have
\begin{align}
	( H_\tc -E_{\tc,0}) \tilde{\Pi}  = 
	\tilde{\Pi} \br{ \sum_{s=0}^{q+1}h_s  -E_{\tc,0} } \tilde{\Pi}  + \sum_{s=0}^{q} h_{s,s+1} \tilde{\Pi}  ,
\end{align}
which yields an upper bound of 
\begin{align}
	\label{epsilon_H_tc_upp_cal2} 
	\norm{( H_\tc -E_{\tc,0}) \tilde{\Pi} } 
	\le \norm{ \tilde{\Pi} \br{ \sum_{s=0}^{q+1}h_s  -E_{\tc,0} } \tilde{\Pi}}  + \sum_{s=0}^{q} \norm{h_{s,s+1} } .
\end{align}

We now aim to upper-bound
\begin{align}
	\norm{ \tilde{\Pi} \br{ \sum_{s=0}^{q+1}h_s  -E_{\tc,0} } \tilde{\Pi}} =\sup_{\psi} \brr{\abs{\bra{\psi}  \tilde{\Pi} \br{ \sum_{s=0}^{q+1}h_s  -E_{\tc,0} } \tilde{\Pi} \ket{\psi}} }. 
\end{align} 
In the following, we separately consider the expectations for $\sum_{s=0}^{q+1}h_s  -E_{\tc,0}$ and $E_{\tc,0}-\sum_{s=0}^{q+1}h_s$.
For an arbitrary quantum state $\ket{\psi}$, we have 
\begin{align}
	\bra{\psi} H_\tc \ket{\psi} 
	&= \sum_{s=0}^{q+1} \bra{\psi} h_s \ket{\psi} + \sum_{s=0}^{q} \bra{\psi} h_{s,s+1}  \ket{\psi} 
	\ge \sum_{s=0}^{q+1}  E_{s,0} - \sum_{s=0}^{q} \norm{h_{s,s+1}} , 
\end{align}
and hence
\begin{align}
	\label{E_tc_0_lower_bound}
	E_{\tc,0}= \inf_{\ket{\psi}} \br{ \bra{\psi} H_\tc \ket{\psi} } \ge  \sum_{s=0}^{q+1}  E_{s,0} - \sum_{s=0}^{q} \norm{h_{s,s+1}} , 
\end{align}
From the inequality~\eqref{E_tc_0_lower_bound} and $\norm{ \tilde{\Pi} h_s  \tilde{\Pi}} \le \tau_s$ for $\forall s$, 
we can derive 
\begin{align}
	\label{epsilon_H_tc_upp_cal2_plus} 
	\bra{\psi} \tilde{\Pi} \br{ \sum_{s=0}^{q+1}h_s  -E_{\tc,0} } \tilde{\Pi} \ket{\psi}  
	\le \sum_{s=0}^{q+1}\br{ \tau_s-  E_{s,0}} + \sum_{s=0}^{q} \norm{h_{s,s+1}} 
	= (q+2) \tau + \sum_{s=0}^{q} \norm{h_{s,s+1}}  , 
\end{align}
where we use $\tau_s= \tau+ E_{s,0}$ for $s\in [0,q+1]$. 

On the other hand, the ground energy $E_{\tc,0}$ is larger than the energy of the product state $\ket{\psi_0}=\bigotimes_{s=0}^{q+1} \ket{E_{s,0}}$:   
\begin{align}
	E_{\tc,0} \le \bra{\psi_0} H_\tc \ket{\psi_0} \le \sum_{s=0}^{q+1}  E_{s,0} + \sum_{s=0}^{q} \norm{h_{s,s+1}} , 
\end{align}
which yields 
\begin{align}
	\label{epsilon_H_tc_upp_cal2_minus} 
	\bra{\psi} \tilde{\Pi} \br{E_{\tc,0} - \sum_{s=0}^{q+1}h_s} \tilde{\Pi} \ket{\psi}  
	\le 
	\sum_{s=0}^{q+1}   \bra{\psi} \tilde{\Pi} \br{E_{s,0} - h_s} \tilde{\Pi} \ket{\psi} 
	+ \sum_{s=0}^{q} \norm{h_{s,s+1}} \le \sum_{s=0}^{q} \norm{h_{s,s+1}}  . 
\end{align}

From the inequalities~\eqref{epsilon_H_tc_upp_cal2_plus} and \eqref{epsilon_H_tc_upp_cal2_minus}, we prove 
\begin{align}
	\label{epsilon_H_tc_upp_cal2_fin} 
	\norm{ \tilde{\Pi} \br{ \sum_{s=0}^{q+1}h_s  -E_{\tc,0} } \tilde{\Pi}} \le  (q+2) \tau + \sum_{s=0}^{q} \norm{h_{s,s+1}}  . 
\end{align} 
This reduces the inequality~\eqref{epsilon_H_tc_upp_cal2} to 
\begin{align}
	\label{epsilon_H_tc_upp2}
	\norm{( H_\tc -E_{\tc,0}) \tilde{\Pi} } 
	\le (q+2) \tau  + 2 \sum_{s=0}^{q} \norm{h_{s,s+1}}  
	\le  2q \br{ \tau  +2 c_0\bar{g}_{ql} } ,
\end{align}
where, in the second inequality, we use the upper bound~\eqref{truncated_Hamiltonian_block_interaction}, i.e., $\| h_{s,s+1} \|\le c_0\bar{g}_{ql} $.
Therefore, by combining the above inequality with~\eqref{epsilon_H_tc_upp1}, we derive the second target  inequality~\eqref{thm2:effective_global_cond_multi_truncate}.
This completes the proof of Theorem~\ref{Effective Hamiltonian_multi_truncation}. $\square$

\subsection{Proof of Theorem~\ref{thm:The energy distribution of the subsystem L_c}} \label{Proof:thm:The energy distribution of the subsystem L_c}

We follow the approach from the proof of Proposition~8 of Ref.~\cite{Kuwahara2020arealaw}. First, consider an arbitrary normalized quantum state $\ket{\psi}$, and define the quantum state $\ket{\phi}$ as follows:
\begin{align}
	\ket{\phi} := \Pi_{> \tau_s}^{(s)} \Pi_{\tc,\le E} \ket{\psi}. \label{Def:of_phi}
\end{align}
Note that the state $\ket{\phi}$ may not be normalized. The norm of $\Pi_{> \tau_s}^{(s)} \Pi_{\tc,\le E}$ is then given by
\begin{align}
	\norm{ \Pi_{> \tau_s}^{(s)} \Pi_{\tc,\le E} } = \sup_{\ket{\psi}} \norm{\phi}, \label{Basic_theorem_for_effe_Hami}
\end{align}
where $\norm{\phi}$ denotes the norm of $\ket{\phi}$.

We aim to prove the following inequality (see~\ref{sec:Proof of the inequality_Norm_of_Psi_tilde} for the proof):
\begin{align}
	\norm{\phi} \leq \mE_{\langle H_\tc \rangle_{\phi} - E -T_0}           , \label{Norm_of_Psi_tilde}
\end{align}
where we use the definition~\eqref{def:varepsilon_y} for $\mE_y$ and 
\begin{align}
	\langle H_\tc \rangle_{\phi} = \frac{\bra{\phi}H_\tc \ket{\phi}}{\| \phi \|^2}. \label{Definition_H_phi_ave}
\end{align}

To obtain an explicit upper bound for $\norm{\phi}^2$ from Eq.~\eqref{Norm_of_Psi_tilde}, we must calculate a lower bound for $\langle H_\tc \rangle_{\phi}$:
\begin{align}
	\langle H_\tc \rangle_{\phi} = \langle h_s \rangle_{\phi} + \langle (h_{s,s+1} + h_{s-1,s}) \rangle_{\phi} + \langle \delta H_s \rangle_{\phi}, \label{explicit_H_tc_phi_ave}
\end{align}
where $\delta H_s := H_\tc - h_s - h_{s,s+1} - h_{s-1,s}$, acting on the sites $\Lambda_s := \Lambda \setminus B_s$. Denote the ground state and the ground-state energy of $\delta H_s$ by $\ket{E_{\Lambda_s,0}}$ and $E_{\Lambda_s,0}$, respectively.

From the definition of $\ket{\phi}$, we obtain:
\begin{align}
	&\langle h_s \rangle_{\phi} = \frac{1}{\norm{\phi}^2} \bra{\psi} \Pi_{\tc,\le E} \Pi_{> \tau_s}^{(s)} h_s \Pi_{> \tau_s}^{(s)} \Pi_{\tc,\le E} \ket{\psi} >\tau_s, \notag \\
	&\langle (h_{s,s+1} + h_{s-1,s}) \rangle_{\phi} \geq - (\| h_{s,s+1} \| + \norm{h_{s-1,s}}) \geq -2c_0\bar{g}_{ql} , \notag \\
	&\langle \delta H_s \rangle_{\phi} \geq E_{\Lambda_s,0} \geq E_{\tc,0} - E_{s,0} - 2c_0\bar{g}_{ql} , \label{delta_H_s_ave_phi}
\end{align}
where, in the second inequality, we use~\eqref{truncated_Hamiltonian_block_interaction}, and third inequality follows from:
\begin{align}
	E_{\tc,0} \leq \bra{E_{s,0}} \otimes \bra{E_{\Lambda_s,0}} H_\tc \ket{E_{s,0}} \otimes \ket{E_{\Lambda_s,0}} 
	&\leq E_{s,0} + E_{\Lambda_s,0} + \| h_{s,s+1} \| + \norm{h_{s-1,s}}  \notag \\
	&\leq E_{s,0} + E_{\Lambda_s,0} + 2c_0\bar{g}_{ql}  .
\end{align}
Thus, the inequalities in Eq.~\eqref{delta_H_s_ave_phi} yield the following lower bound for $\langle H_\tc \rangle_{\phi}$ from Eq.~\eqref{explicit_H_tc_phi_ave}:
\begin{align}
	\langle H_\tc \rangle_{\phi} \geq E_{\tc,0} + \tau_s - E_{s,0} - 4 c_0\bar{g}_{ql} = E_{\tc,0} + \tau - 4 c_0\bar{g}_{ql} 
	, \label{lower_bound_of_ave_of_H}
\end{align}
where we use $\tau_s= E_{s,0} +\tau$. 

Therefore, by combining the upper bound~\eqref{Norm_of_Psi_tilde} and the lower-bound \eqref{lower_bound_of_ave_of_H} with Eq.~\eqref{Basic_theorem_for_effe_Hami}, we prove 
\begin{align}
	\norm{ \Pi_{> \tau_s}^{(s)} \Pi_{\tc,\le E} } \le   \norm{\phi} \le  \mE_{\tau + E_{\tc,0}  - 4 c_0\bar{g}_{ql} - E -T_0} ,
\end{align}
where we use the monotonic decreasing of $\mE_y$ from Eq.~\eqref{def:varepsilon_y}.
This completes the proof of Theorem~\ref{thm:The energy distribution of the subsystem L_c}. $\square$

\subsubsection{Proof of the inequality~\eqref{Norm_of_Psi_tilde}} \label{sec:Proof of the inequality_Norm_of_Psi_tilde}

We now prove inequality~\eqref{Norm_of_Psi_tilde}.
In the following, we adopt the decomposition of 
\begin{align}
	[E+y,\infty) = \bigcup_{j=0}^\infty I_j ,\quad I_j = [E + y +T_0 j ,E +y + T_0 j+T_0) ,
\end{align}
where the parameter $y$ will be determined later. 
Note that $T_0$ is defined from $T_m$ in~\eqref{assump_kappa_form}, i.e., $T_0=T_{m=0}=(2 c_0 \tilde{c}_3 +\tilde{c}_1)  \bar{g}_{ql }$.
Then, we begin with the following equality:
\begin{align}
	\bra{\phi} H_\tc \ket{\phi} = \bra{\phi} \Pi_{\tc,< E+y} H_\tc \Pi_{\tc,< E+y} \ket{\phi} + \sum_{j=0}^{\infty} \bra{\phi} \Pi_{\tc,I_j} H_\tc  \Pi_{\tc,I_j} \ket{\phi}. 
\end{align}
We further calculate the upper bound of $\bra{\phi} H_\tc \ket{\phi}$ as 
\begin{align}
	\label{bra_phi_H_tc_ket_phi_upp}
	\bra{\phi} H_\tc \ket{\phi} &\leq (E+y) \| \Pi_{\tc,<E+y} \ket{\phi} \|^2 + \sum_{j=0}^{\infty} (E+y+jT_0+T_0) \norm{ \Pi_{\tc,I_j} \ket{\phi} }^2 \notag \\
	&= (E+y) \left( \norm{ \Pi_{\tc,<E} \ket{\phi} }^2 + \sum_{j=0}^{\infty} \norm{ \Pi_{\tc,I_j} \ket{\phi} }^2 \right) + T_0 \sum_{j=0}^{\infty} (j+1) \norm{\Pi_{\tc,I_j} \ket{\phi} }^2 \notag \\
	&= (E+y) \norm{\phi}^2 + T_0 \sum_{j=0}^{\infty} (j+1) \norm{\Pi_{\tc,I_j} \ket{\phi}}^2. 
\end{align}

By letting 
\begin{align}
	\label{def: bar_mE_y}
	\bar{\mE}_{y}:=  \sum_{j=0}^{\infty} (j+1) \norm{\Pi_{\tc,I_j} \ket{\phi}}^2 ,
\end{align}
we obtain 
\begin{align}
	\norm{\phi}^2 \cdot \langle H_\tc \rangle_{\phi}= \bra{\phi} H_\tc \ket{\phi} &\leq 
	(E+y) \norm{\phi}^2 
	+ T_0 \bar{\mE}_{y}  , 
\end{align}
where we use the definition in Eq.~\eqref{Definition_H_phi_ave} to get $\bra{\phi} H_\tc \ket{\phi} = \norm{\phi}^2 \cdot \langle H_\tc \rangle_{\phi}$.
Then, the above inequality reduces to:
\begin{align}
	\norm{\phi}^2 \leq \frac{T_0}{\langle H_\tc \rangle_{\phi} -E-y} \bar{\mE}_{y}  .
\end{align}
By choosing $y$ such that $\langle H_\tc \rangle_{\phi} -E-y =T_0$, we obtain:
\begin{align}
	\label{final_upp_without_<H_tc>}
	\norm{\phi} \leq \sqrt{ \bar{\mE}_{\langle H_\tc \rangle_{\phi} - E -T_0}} . 
\end{align}
Therefore, the remaining task is to estimate $ \bar{\mE}_{y}$ in Eq.~\eqref{def: bar_mE_y}. 

From the definition of $\ket{\phi}$ in Eq.~\eqref{Def:of_phi}, we first derive:
\begin{align}
	\norm{\Pi_{\tc,I_j} \ket{\phi}}^2 &= \norm{ \Pi_{\tc,I_j} \Pi_{> \tau_s}^{(s)} \Pi_{\tc,\le E} \ket{\psi}}^2 \leq \norm{ \Pi_{\tc,I_j} \Pi_{> \tau_s}^{(s)} \Pi_{\tc,\le E} }^2. \label{phi_2_Pi_jm1_j_norm}
\end{align}
To obtain an upper bound for $\norm{ \Pi_{\tc,I_j} \Pi_{> \tau_s}^{(s)} \Pi_{\tc,\le E}}^2$, we use the upper bound~\eqref{O_s_Pi_I_z+theta_fin} in Subtheorem~\ref{subthm_energy_distribution} with $\theta \to y +T_0 j$, which yields 
\begin{align}
	\norm{ \Pi_{\tc,I_j} \Pi_{> \tau_s}^{(s)} \Pi_{\tc,\le E} }\le 
	e  \exp\brrr{- \min\brr{\frac{y+T_0 j}{4e \tilde{T}_{y/T_0+j}} , \br{\frac{y+T_0 j}{4e\tilde{c}_2g_1}}^{1/(1+\chi)}}   }  .
\end{align}
Here, the RHS of the above inequality monotonically decreases with $y$.

Then, using the definition~\eqref{def:varepsilon_y}, we estimate the summation in Eq.~\eqref{def: bar_mE_y} to derive $\sqrt{\bar{\mE}_{y}} \le\mE_{y} $ as follows:  
\begin{align}
	\label{j+1_norm_I_j_tau_le_x_upp}
	\sqrt{\bar{\mE}_{y}} 
	&\le   \sum_{j=0}^{\infty} (j+1)^{1/2} \norm{\Pi_{\tc,I_j} \ket{\phi}}  \notag \\
	&\le  \sum_{j=0}^\infty (j+1)  \norm{ \Pi_{\tc,I_j} \Pi_{> \tau_s}^{(s)} \Pi_{\tc,\le E}}   \notag \\
	&\le           
	e  \exp\brrr{- \min\brr{\frac{y}{4e \tilde{T}_{y/T_0}} , \br{\frac{y}{4e\tilde{c}_2g_1}}^{1/(1+\chi)}}   }   \notag \\
	& \quad + 
	e \int_0^\infty (\theta+2) \brrr{\exp\brr{- \frac{y+T_0 \theta}{4e \tilde{T}_{y/T_0+\theta}}}  + 
		\exp\brr{-\br{\frac{y+T_0 \theta}{4e\tilde{c}_2g_1}}^{1/(1+\chi)}} }  d\theta \notag \\
	&\le \mu_1  \exp\br{- \frac{y}{8e \tilde{T}_{y/T_0}}}
	+  \mu_2  \exp\brr{-\frac{1}{2}\br{\frac{y}{4e\tilde{c}_2g_1}}^{1/(1+\chi)}}  =\mE_{y},                                   
\end{align}
where the constants $\mu_1$ and $\mu_2$ were defined in Eqs.~\eqref{defin:mu_1} and \eqref{defin:mu_2}, and in the last inequality, we use $T_0= (2 c_0 \tilde{c}_3 +\tilde{c}_1)  \bar{g}_{ql }$ from Eq.~\eqref{assump_kappa_form}, $\tilde{T}_z= (2 c_0 \tilde{c}_3 +\tilde{c}_1)  \bar{g}_{z+ql } $ from Eq.~\eqref{definition_tilde_T_x}, and $ \bar{g}_{ql }\ge \bar{g}_{2} \ge g_0+g_1\ge g_1$ [see Eq.~\eqref{def_bar_g_x}] to derive 
\begin{align}
	\label{Ineq:using_defin:mu_1}
	&e\int_0^\infty (\theta+2) \exp\br{- \frac{y+T_0 \theta}{4e \tilde{T}_{y/T_0+\theta}}} d\theta  \notag \\
	&\le e \exp\br{- \frac{y}{8e \tilde{T}_{y/T_0}}}  \int_0^\infty (\theta+2) \exp\brr{- \frac{(y/T_0 +\theta)T_0}{8e \tilde{T}_{y/T_0+\theta}}} d\theta  \notag \\
	&\le e\exp\br{- \frac{y}{8e \tilde{T}_{y/T_0}}}  \int_{y/T_0}^\infty (z+2-y/T_0)  \exp\brr{- \frac{(2 c_0 \tilde{c}_3 +\tilde{c}_1)  \bar{g}_{ql } z}{8e  \bar{g}_{z+ql } (2 c_0 \tilde{c}_3 +\tilde{c}_1)}} dz \notag \\
	&\le e^2\exp\br{- \frac{y}{8e \tilde{T}_{y/T_0}}}  \int_{0}^\infty (z+2) \exp\brr{- \frac{\bar{g}_{ql }}{8e  \bar{g}_{z+ql }}z } dz = (\mu_1-1) \exp\br{- \frac{y}{8e \tilde{T}_{y/T_0}}} ,
\end{align}
and 
\begin{align}
	& e \int_0^\infty (\theta+2) \exp\brr{- \br{\frac{y+T_0 \theta}{4e\tilde{c}_2g_1}}^{1/(1+\chi)}}   d\theta  \notag \\
	&\le e \exp\brr{-\frac{1}{2} \br{\frac{y}{4e\tilde{c}_2g_1}}^{1/(1+\chi)}}   \int_0^\infty  (\theta+2) \exp\brr{- \frac{1}{2} \br{\frac{(y/T_0+\theta)T_0}{4e\tilde{c}_2g_1}}^{1/(1+\chi)}}  d\theta \notag \\
	&\le e\exp\brr{-\frac{1}{2} \br{\frac{y}{4e\tilde{c}_2g_1}}^{1/(1+\chi)}}  
	\int_{y/T_0}^\infty \br{z +2- y/T_0}\exp\brr{-\frac{1}{2} \br{\frac{z }{4e\tilde{c}_2g_1} \cdot (2 c_0 \tilde{c}_3 +\tilde{c}_1)  \bar{g}_{ql }}^{1/(1+\chi)}} dz \notag \\
	&\le e\exp\brr{- \frac{1}{2}\br{\frac{y}{4e\tilde{c}_2g_1}}^{1/(1+\chi)}}   \int_{0}^\infty (z+2) \exp\brr{- \frac{1}{2}\br{\frac{ 2 c_0 \tilde{c}_3 +\tilde{c}_1}{4e\tilde{c}_2}  z}^{1/(1+\chi)}}  dz \notag \\
	&= (\mu_2-1) \exp\brr{-\frac{1}{2}\br{\frac{y}{4e\tilde{c}_2g_1}}^{1/(1+\chi)}}  .
\end{align}
Note that to reach the definition~\eqref{defin:mu_1} of $\mu_1$ in~\eqref{Ineq:using_defin:mu_1}, we use $ql\ge 2$ to obtain the upper bound of
\begin{align}
	\frac{\bar{g}_{ql }}{\bar{g}_{z+ql }}= \frac{g_0+g_1 \log^\chi(3)}{g_0+g_1 \log^\chi(z+3)} 
	\ge \frac{1}{1+\log^\chi(z+3)} .
\end{align}

Finally, by combining the upper bounds~\eqref{j+1_norm_I_j_tau_le_x_upp} with the inequality~\eqref{final_upp_without_<H_tc>}, we prove 
the target inequality~\eqref{Norm_of_Psi_tilde}.


\end{document}